\documentclass[a4paper,11pt,oneside,openright]{Classes/CUthesis}

\newenvironment{sistema}{%
    \left\lbrace\begin{array}{@{}l@{}}}
{\end{array}\right.}

\newcommand\figref[1]{Fig.~\ref{#1}}
\newcommand\figvref[1]{Fig.~\ref{#1}\vpageref{#1}}
\newcommand\figsref[1]{Figs.~\ref{#1}}
\newcommand\eqnref[1]{Eq.~\eqref{#1}}
\newcommand\eqnvref[1]{Eq.~\eqref{#1}\vpageref{#1}}
\newcommand\eqnsref[1]{Eqs.~\eqref{#1}}
\newcommand\secref[1]{\S\ref{#1}}
\newcommand\chapref[1]{Chapter~\ref{#1}}
\newcommand\chapsref[1]{Chapters~\ref{#1}}
\newcommand\appref[1]{Appendix~\ref{#1}}
\newcommand\refref[1]{Ref.~\cite{#1}}
\providecommand{\abs}[1]{\lvert#1\rvert}%
\providecommand{\Abs}[1]{\left\lvert#1\right\rvert}%
\providecommand{\smallERsum}[1]{{\sideset{}{^\varepsilon}{\textstyle\sum}\limits_#1}}%
\providecommand{\ERsum}[1]{{\sideset{}{^\varepsilon}{\sum}\limits_#1}}%

\providecommand{\bra}[1]{\langle#1\rvert}
\providecommand{\ket}[1]{\lvert#1\rangle}
\DeclareMathOperator{\Tr}{Tr}%
\DeclareMathOperator{\stdev}{stdev}%
\DeclareMathOperator{\mean}{mean}%

\def\omegac{\omega_\mathrm{C}}
\def\ellB{\ell_\mathrm{B}}
\def\Ef{E_\mathrm{F}}
\def\Ez{E_\mathrm{Z}}
\def\Ec{E_\mathrm{C}}
\def\Eex{E_\mathrm{ex}}
\def\Ead{E_\mathrm{AD}}
\def\Rad{R_\mathrm{AD}}
\def\nuAD{\nu_\mathrm{AD}}
\def\nuB{\nu_\mathrm{B}}
\def\dEsp{\Delta E_\mathrm{SP}}
\def\HSP{\hat{h}}
\def\bx{\mathbf{x}}
\def\esp{\varepsilon^\mathrm{SP}}
\def\ehf{\varepsilon^\mathrm{HF}}
\def\etaC{\eta_\mathrm{C}}
\def\Nup{N_\uparrow}
\def\Ndn{N_\downarrow}
\def\Nhup{N^\mathrm{h}_{\uparrow}}
\def\Nhdn{N^\mathrm{h}_{\downarrow}}
\def\Nh{N_\mathrm{h}}
\def\neupvec{\mathbf{n}^\mathrm{e}_{\uparrow}}
\def\nednvec{\mathbf{n}^\mathrm{e}_{\uparrow}}
\def\nhupvec{\mathbf{n}^\mathrm{h}_{\uparrow}}
\def\nhdnvec{\mathbf{n}^\mathrm{h}_{\downarrow}}
\def\nupvec{\mathbf{n}_\uparrow}
\def\ndnvec{\mathbf{n}_\downarrow}
\def\spinup{spin\nobreakdash-$\uparrow$}
\def\spindn{spin\nobreakdash-$\downarrow$}
\def\onetothree{1$\leftrightarrow$3}
\def\twotofour{2$\leftrightarrow$4}
\def\onetoone{1$\leftrightarrow$1}
\def\twototwo{2$\leftrightarrow$2}

\def\dsD{\overleftrightarrow{\mathbf{D}}}
\def\ainvec{\mathbf{a}^\mathrm{in}}
\def\aoutvec{\mathbf{a}^\mathrm{out}}
\def\finj{f_\mathrm{Inj}}
\def\fdet{f_\mathrm{Det}}
\def\fdettwo{f_\mathrm{Det2}}
\def\Gneq{G_\mathrm{Neq}}
\def\Gdrain{G_\mathrm{Drain}}
\def\Gdet{G_\mathrm{Det}}
\def\Gtot{G_\mathrm{Tot}}
\def\Gaup{G_{2\mathrm{T}\alpha}}
\def\Gap{G_{2\mathrm{T}\alpha^\prime}}
\def\G2t{G_\mathrm{2T}}
\def\Gup{G_\uparrow}
\def\Gdn{G_\downarrow}
\def\Vug{V_\mathrm{UG}}
\def\Vlg{V_\mathrm{LG}}
\def\Vad{V_\mathrm{AD}}
\def\Vex{V_\mathrm{EX}}
\def\Vsd{V_\mathrm{SD}}
\def\Vds{V_\mathrm{DS}}
\def\ils{\ell_\mathrm{SF}^{-1}}
\def\ilo{\ell_\mathrm{LL}^{-1}}
\def\lsf{\ell_\mathrm{SF}}
\def\lLL{\ell_\mathrm{LL}}
\def\aout{a^\mathrm{out}}
\def\nbd{\nobreakdash\textendash}
\def\GammaD{\Gamma^\mathrm{D}}
\def\GammaS{\Gamma^\mathrm{S}}
\def\gDup{\Gamma_{\mathrm{D}\uparrow}}
\def\gSup{\Gamma_{\mathrm{S}\uparrow}}
\def\gDdn{\Gamma_{\mathrm{D}\downarrow}}
\def\gSdn{\Gamma_{\mathrm{S}\downarrow}}
\def\muup{\mu_\uparrow}
\def\mudn{\mu_\downarrow}
\def\musf{\mu_\mathrm{SF}}
\def\eSOI{\varepsilon_\mathrm{SOI}}
\def\eHF{\varepsilon_\mathrm{HF}}
\def\Etherm{E_\mathrm{Therm}}
\def\mueV{\micro\electronvolt}
\def\dmuup{\Delta\mu_\uparrow}
\def\Npone{N\!+\!1}
\def\Nmone{N\!-\!1}
\def\Cs{C_\mathrm{S}}
\def\Cd{C_\mathrm{D}}
\def\Cg{C_\mathrm{G}}
\def\Cr{C_\mathrm{R}}
\def\Vs{V_\mathrm{S}}
\def\Vd{V_\mathrm{D}}
\def\Vg{V_\mathrm{G}}
\def\nG{n_\mathrm{G}}
\def\muD{\mu_\mathrm{D}}
\def\muS{\mu_\mathrm{S}}
\def\mudot{\mu_\mathrm{dot}}
\def\alphaG{\alpha_\mathrm{G}}
\def\alphaD{\alpha_\mathrm{D}}
\def\ksr{{k\sigma r}}
\def\kls{{k\ell\sigma}}
\def\ls{{\ell\sigma}}
\def\lps{{\ell^\prime\sigma}}
\def\sptos{{s^\prime\rightarrow s}}
\def\kptok{{\chi^\prime\rightarrow \chi}}
\def\llps{{\ell\ell^\prime\sigma}}
\def\sp{{s^\prime}}
\def\gSrssp{\gamma^\Sigma_{r,s\sp}}
\def\mussp{\mu_{s\sp}}
\def\dEs{\Delta E_S}
\def\dEt{\Delta E_{T_0}}
\def\Szz{S_{z0}}
\def\El{\varepsilon_L}
\def\Es{\varepsilon_S}
\def\Rs{R_\mathrm{S}}
\def\Rd{R_\mathrm{D}}
\def\nL{n_\mathrm{L}}
\def\nS{n_\mathrm{S}}
\def\Btot{B_\mathrm{Tot}}
\def\Bperp{B_\perp}

\ifpdf
    \pdfcatalog
        {/PageMode(/UseOutlines)
        /OpenAction(fitbh)}
\fi

\title{Probing Electron-Electron Interactions with a Quantum Antidot}

\ifpdf
    \author{\href{mailto:lcb36@cam.ac.uk}{Lee Christopher Bassett}}
    \collegeordept{King's College}
    \university{\href{http://www.cam.ac.uk}{University of Cambridge}}
    \crest{\includegraphics[width=30mm]{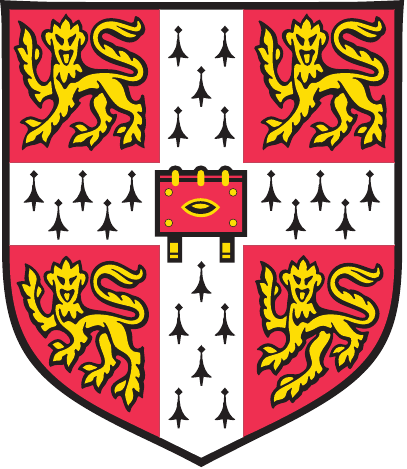}}
\else
    \author{Lee Christopher Bassett}
    \collegeordept{King's College}
    \university{University of Cambridge}
    \crest{\includegraphics[width=30mm]{ThesisFigures/UC_Shield}}
\fi

\renewcommand{\abstractname}{Summary} 

\degreename{Doctor of Philosophy}%
\degreedate{September 2009}%
\copyrightyear{2009}

\usepackage{geometry} 
\usepackage[T1]{fontenc} 
\usepackage{lmodern}
\usepackage[autolanguage]{numprint} 
\usepackage[cdot,mediumqspace,amssymb]{SIunits} 
\usepackage{varioref} 
\usepackage{setspace}
\usepackage{url} 
\usepackage{bm} 
\usepackage{StyleFiles/previousequation}
\EndPreamble

\begin{document}

\pagestyle{fancy}

\fncyfront

\frontmatter

\maketitle%

\tradcopyrightpage%

\setcounter{secnumdepth}{3}
\setcounter{tocdepth}{3}

%

\begin{dedication}
\emph{To my parents, who taught me to ask questions, \\
and to Dani, who is the only answer I need.}
\end{dedication}

%

\begin{summary}

In the integer quantum Hall (IQH) regime, an antidot provides a
finite, controllable `edge' of quantum Hall fluid which is an ideal
laboratory for investigating the collective dynamics of large
numbers of interacting electrons.  Transport measurements of single
antidots probe the excitation spectra of the antidot edge, and
gate-defined antidot devices offer the flexibility to vary both the
dimensions of the antidot and the couplings to the extended IQH edge
modes which serve as leads.  We can also use the spin-selectivity of
the IQH edge modes to perform spin-resolved transport measurements,
from which we can infer the antidot spin-structure.  This thesis
describes a combination of such transport experiments and related
computational models designed to investigate the effects of
electron-electron interactions in quantum antidots, with general
implications for the physics of spin and charge in IQH systems.

We focus on the regime of relatively low magnetic fields
($B\lesssim\unit{1}{\tesla}$) and antidot filling factor $\nuAD=2$,
in which the standard antidot transport experiments are
well-described by a single-particle (SP) model of antidot orbital
states in the lowest Landau level (LLL).  We find that the
\emph{orbital-excitation} energies observed in standard transport
experiments are well-described by SP physics but that the
\emph{spin-excitation} energy implied by spin-resolved measurements
are much smaller than that predicted by the SP model. By treating
the $\nuAD=2$ antidot as a `dot of holes' in the LLL and developing
a computational model for spin-resolved sequential transport, we
show that this observed spin-charge separation is consistent with
the edge-excitations predicted for a `maximum density droplet' (MDD)
of interacting holes in the LLL.

Our work provides a powerful example of the practical applications
of IQH edge modes for selective transport in mesoscopic quantum
electronics, which we have used to perform the first spin-resolved
measurements of $\nuAD=2$ transmission resonances. Our discovery of
spin-charge separation in the low-field antidot excitation spectrum
paints a picture of the antidot as a finite droplet of interacting
IQH fluid in the LLL, with all of the rich physics of exchange,
collective modes, spin textures, etc., which this entails. Our
results are therefore relevant not only for the physics of antidots,
but more broadly for the understanding of interacting electronic
systems of many particles in the IQH regime.

\end{summary}

%
%

\begin{acknowledgementslong}

As it is with just about everything I do, this thesis ended up
taking far longer to complete than I expected.  I think it was my
invaluable colleague Jon Griffiths who first taught me Hofstadter's
rule: that everything you try to do takes roughly 150\% of the time
you might expect, even when you account for Hofstadter's rule.
Still, this Ph.D., thesis writing included, has been an incredibly
enjoyable experience. Everyone should have the opportunity to be
paid to live in a fairy-tale town and play with expensive toys every
day. But unfortunately not everyone is so lucky, and so I am keenly
aware of the many people and organisations who have helped me to
live the life to which I have become accustomed, and which I shall
dearly miss in the coming years.

First, I acknowledge the unflinching support, valuable advice, and
almost unfathomable optimism of my supervisor Dr.\ Chris Ford.  He
always seemed to know that together we would make something of what
sometimes seemed to me to be an abyss of utter confusion, and
although it took longer than we hoped (thanks, Hofstadter), he was
right.  He is a fount of knowledge and experience on everything from
dilution refrigerators to rubber bungs, and I already realise that I
am leaving without having taken advantage of it nearly enough.

These types of experiments are necessarily the result of many
people's contributions, and several colleagues and collaborators
have been particularly helpful.  On the experimental side, Jon
Griffiths, Jon Prance, Abi Graham, Antonio Corcoles, Francois
Sfigakis, and Masaya Kataoka have all generously offered their help
and hard-earned advice in the lab, and Ken Cooper, David Anderson,
and Jon Griffiths have provided essential help with device
processing and electron-beam fabrication.  Masaya in particular was
a valuable resource for his deep knowledge of antidot experiments,
and taught by example with his skeptical and plodding style which
nearly always \emph{worked}. On the theoretical side, Crispin Barnes
provided many useful bits of insight, and provided the code for the
Green's function calculations I used for some of the modeling in
this thesis. Nigel Cooper of the Theory of Condensed Matter group
contributed many of the initial ideas regarding the importance of
exchange interactions for making sense of our spin-resolved
measurements, and gave a great deal of his time to several useful
discussions. Adam Thorn was nearly always `willing' to be distracted
to help with computer problems, and has also kindly offered to take
care of the printing/binding/submitting of this thesis in exchange
for beer.

I am grateful to the funding bodies which supported my study in
Cambridge, notably the Marshall Aid Commemoration Commission, the
National Science Foundation, and King's College, Cambridge.  I also
offer sincere thanks to my two examiners, Dr. Andrew Ferguson and
Prof. Steve Simon, for their patience as this ran behind schedule.

Many of my friends and family members offered their encouragement
and support over the years, and pretended not to mind that we lived
an ocean apart for five years.  My parents in particular have always
been supportive and understanding, and were even willing to
sacrifice their dining room table for three weeks as I raced to the
finish line which seemed to keep moving father away.

Finally, I owe a deepest debt of gratitude to my amazing wife
Danielle, mainly for just putting up with me.  She put up with the
late nights I spent at the lab, and the months of my experiments
where we lived like ships literally passing in the night, as I would
come to bed around the time she woke up to head to the river.  She
was even known to occasionally provide her company in the cryostat
lab in the wee hours, even during Christmas when there was no heat.
She put up with me being late for dinner every time my `five minute
job' ended up taking two hours.  And she put up with this thesis
taking a month longer to write than it was supposed to and then even
a bit longer (thanks again, Hofstadter), giving up her vacation for
the summer, and seems to be putting up with me keeping her awake by
typing these last sentences in the corner of our hotel room. Dani
has also acted as primary proofreader for my work, and has even
pretended to enjoy it.  Since I expect that very few people will set
eyes on this document who are not physicists (Mom, Dad, have you
made it this far?), and that most will be students, I'll offer a
piece of unsolicited advice: if you have a chance to marry another
physicist, take it, because nobody else will ever understand.

\end{acknowledgementslong}

\tableofcontents
\listoffigures
\chapter*{Preface\markboth{PREFACE}{}}
\addcontentsline{toc}{chapter}{Preface}

It is a great irony that we must work so hard to study interaction
effects in solid-state electronic devices.  In a typical
two-dimensional electron system (2DES) with carrier density around
$n_e=\numprint{2e11}$~\centi\metre\rpsquared, the Coulomb energy
between neighbouring electrons is
$e^2n_e^{1/2}/4\pi\epsilon\epsilon_0\approx\unit{5}{\milli\electronvolt}$,
which is almost always the largest relevant energy scale in a
mesoscopic device.  Especially given the small effective mass of
electrons in GaAs, $m^\ast = 0.067 m_e$, the forces associated with
such Coulomb interactions are immense; if an electron were given
this energy in free space, it would no longer be gravitationally
bound to the solar system!  In retrospect, it might therefore seem
quite surprising that early experiments with mesoscopic devices
failed to show evidence of interacting electrons, seeming instead to
reflect the `ballistic' dynamics of free particles
\citep[e.g.,][]{Ford1989a,Molenkamp1990,Spector1990}.  The
explanation for this behaviour rests in the theory of the Fermi
liquid which describes the two-dimensional electron gas, in which
the quasiparticle excitations are essentially equivalent to free
electrons.  Each electron becomes `dressed' by interactions with
other particles, and the collective excitations of the system have
well-defined charges, masses, momenta, etc., exactly as if the
system contained only a single electron \cite{Ashcroft1976}.  Of
course this does not mean that Coulomb effects are completely
absent, but rather that their experimental signatures are often more
subtle than might be expected.  It is also worth noting that in some
cases the description of the 2DES as a normal Fermi liquid breaks
down, and the quasiparticle excitations are entirely different; the
`composite fermions' which serve as quasiparticles in the fractional
quantum Hall regime are prime examples.

In this thesis we are interested in the effects of interactions in
the integer quantum Hall (IQH) regime, where the quasiparticles are
indeed `electron-like.'  We investigate these interactions by
studying electron transport through single quantum antidots, which
embody the physics of a finite edge of IQH fluid. Fundamentally, an
antidot is simply an `island' in the potential landscape of a 2DES
on which electrons are absent, the exact opposite of a quantum dot,
which is a small valley or `puddle' of electrons. Even so, magnetic
confinement leads to a quantised energy spectrum of zero-dimensional
antidot states which is in many ways analogous to the spectrum of
similarly-sized quantum dots.  Theoretically, it is in fact possible
to treat an antidot as a `dot of holes' in the 2DES, allowing for
comparisons between antidot experiments and the literature on
interactions in quantum Hall systems and in large quantum dots.
Since a full description of the dynamics of many interacting
particles remains a hard theoretical problem, experimental
investigations of these effects are of fundamental interest.

Many of the important consequences of the Coulomb interaction are
manifest through its interplay with electron spin. In particular,
the \emph{exchange interaction} results from the combination of
Coulomb repulsion and Pauli exclusion, and it can lead to very
complicated spin dynamics even for a system of just two electrons.
The development of few-electron quantum dots over the past few years
has opened promising avenues for research into coherent single-spin
dynamics and the controlled manipulation of electron spins for
quantum-information purposes (see \cite{Hanson2007} for a recent
review), most of which relies on the physics of exchange. For
systems of many particles, spin-charge interactions may lead to
exotic effects such as spin-density waves or Skyrmions, particularly
at the edge of the IQH fluid where nearby empty states make it
possible for particles to rearrange
\citep[e.g.,][]{Fertig1994,Lee1990,Sondhi1993}. Interest in such
`collective modes' has spawned a vast literature of theoretical
studies, but it has been difficult to find clear experimental
examples of such idealised constructions in the real world.  Some
progress has been made with large quantum dots
\cite{Ashoori1993,Hawrylak1999,Oosterkamp1999}, but it is difficult
to infer spin structure from transport experiments alone, and large
quantum dots are often complicated by the tendency of electrons to
rearrange between the `edge' and the `core' as experimental
parameters are varied.  This rearrangement is absent for antidots,
since the antidot `core' is always depleted, so we can be sure that
the features we observe are due to the behaviour of the edge alone.

In our experiments we also take advantage of the unique properties
of transport in the IQH regime, in which the `leads' are extended
edge modes whose topology we can control with appropriate gate
designs. Electrons propagate coherently through these edge modes
over mesoscopic distances, and the ability to control the topology
of edge mode networks using surface gates has led to the recent
realisation of electronic analogues to standard optical experiments,
such as Fabry-Perot \cite{Camino2007} and Mach-Zehnder \cite{Ji2003}
interferometers.  While IQH edge modes share many properties with
their optical counterparts, the addition of charge and spin
interactions makes such electronic experiments extremely
interesting, and has produced unexpected behaviour
\cite{Neder2006a,Zhang2009}. In this work we use the
spin-selectivity of these edge modes to perform spin-resolved
measurements of the antidot transport, from which we can infer
details about the spin-structure of the antidot edge.

Evidence for electron-electron interactions in antidots is certainly
not new. While standard transport experiments at low magnetic fields
($B\lesssim\unit{1}{\tesla}$) appear to be well-described by a
non-interacting model of single-particle (SP) antidot orbital states
\cite{Mace1995}, this model is known to fail at higher $B$
\cite{Ford1994,Kataoka2000,Kataoka2003,Sim2003}.  We expect that at
some point the structure of the antidot edge changes fundamentally,
as the characteristic orbital length scale $\ellB=\sqrt{\hbar/eB}$
shrinks and Coulomb interactions begin to dominate over the SP
energy scales, but the nature of the evolution between these two
regimes has not been well-understood up to this point.  The high-$B$
antidot edge probably takes the form of alternating compressible and
incompressible `stripes' for successive Landau levels, similar to
the self-consistent model of IQH edge modes predicted to minimise
the total energy along the extended edge \cite{Chklovskii1992}. The
recent review by Sim, Kataoka, and Ford \cite{Sim2008} provides a
useful discussion of previous antidot studies, treating particularly
those effects at higher fields which seem to require a
self-consistent model.

In this work we look more closely at the low-$B$ regime in which the
SP model seems to be valid.  We choose this regime for two main
reasons.  First, in order to develop a unified picture of antidot
physics which connects the SP model at low fields with the
self-consistent description at higher $B$, we seek evidence of
`inconsistencies' with SP physics at low $B$ which reveal the subtle
influence of interactions. In this sense our work follows directly
from the experiments of Chris Michael \cite{Michael2004}, who
studied antidot transport in the `crossover' regime
($B\approx1$\nbd\unit{2}{\tesla}) and noticed several intriguing
effects. We consider a few of his findings in detail, and
incorporate them into the general picture of low-$B$ antidot physics
developed in this work. Second, the simplicity of the SP antidot
model has suggested several potential applications for antidots,
such as a quantum Hall `pump' \cite{Simon2000}, or a `spin filter'
\cite{Zozoulenko2004}.  We consider the spin-filtering proposal in
detail with our spin-resolved measurements, and although we find
that interactions destroy the filtering ability of the antidots we
study, we do still believe that it is probably possible to design
devices in which the spin-selectivity is preserved for the purposes
of this application.  We hope that other members of the scientific
community, upon learning of our work, will recognise the flexibility
of antidot devices both for applications such as this and for the
experimental study of the fundamental physics of interactions in the
quantum Hall regime.

The experiments described in this thesis were carried out in the
Semiconductor Physics Group at the University of Cambridge. Except
where noted otherwise, all aspects of the experiments, including
device design, fabrication, testing, measurements, and analysis were
performed by myself, although I have gained invaluable help and
advice from many other members of the Semiconductor Physics group.
Most notably, all of this work was conceived and executed in close
collaboration with my supervisor, Dr.\ Chris Ford.  The theoretical
modeling presented herein is also predominantly my own, but I have
benefited greatly from the generous guidance of two experts in the
theory of mesoscopic electronic devices: Dr.\ Crispin Barnes of the
Semiconductor Physics group and Dr. Nigel Cooper of the Theory of
Condensed Matter group here at Cambridge.

\section*{Structure of this thesis}

Immediately following this preface begins the first of two
theoretical chapters designed to provide the background and
theoretical framework which is necessary for an understanding of the
results presented in the remainder of the work.  Much of the
information in these two chapters is widely understood and is
available elsewhere, but I have tried to unify the notation as much
as possible and to justify the major results used in the modeling
and analysis of later chapters.  Moreover, I have tried to address
in detail many of the issues which I found confusing at first (or
second or third) encounter, and to present some of the derivations
which I undertook because I could not find them printed elsewhere. I
hope that these chapters will be useful for future students
interested in mesoscopic electronic devices, and in quantum dots and
antidots in particular.

\chapref{chap:ADtheory} deals with the physics of single antidots. I
present the SP model in terms of the stationary solutions of
Schr\"{o}dinger's equation for a circularly-symmetric antidot in a
magnetic field, and discuss it in the context of the integer quantum
Hall effect for lowest Landau level (LLL) states in particular. I
try to develop an intuition for the interpretation of experimental
results in terms of these SP states by considering several
`observable' properties of these eigenstates and their classical
analogues in terms of `skipping orbit' trajectories around the
antidot. Finally, the Hartree-Fock mean-field theory for an antidot
is developed in terms of a system of interacting `holes' in the LLL,
which takes the form of a maximum density droplet (MDD) at low
fields.

\chapref{chap:TransportTheory} deals more generally with the theory
of transport in mesoscopic electronic systems.  I present the
Landauer-B\"{u}ttiker formalism for the analysis of linear-response
transport coefficients in the IQH regime, which I use throughout
this work to analyse edge-mode networks.  I then make the connection
between the Landauer-B\"{u}ttiker formula and a more general theory
of linear-response ballistic transport, formulated in terms of
time-independent Green's functions.  Such Green's function
calculations are used to study the effects of realistic device
geometries in \chapref{chap:Geometry}. Finally, I present the theory
of sequential transport through quantum dots, which is the basis for
the model developed in \chapref{chap:SpinTransportModel} to describe
transport through antidot states in the presence of interactions.

A discussion of the results of new measurements and modeling begins
in \chapref{chap:Geometry}. In this first study, I consider orbital
excitation spectra in the low-$B$ regime obtained through non-linear
transport measurements by Chris Michael \cite{Michael2004}, in which
the orbital excitation energy $\dEsp$ seems to decrease faster with
$B$ than the SP model predicts.  Using a simple model supported by
transport calculations based on the non-interacting Green's function
of a realistic device, I show that this behaviour may be attributed
to the inherent asymmetry of a real device rather than to
interactions.

\chapref{chap:SpinTransport} is the heart of this work, in which I
discuss the results of spin-resolved measurements of antidot
transport. Using a set of quantum point contacts as injectors and
detectors, the AD scattering coefficients for individual edge modes
are directly measured in the low-$B$ regime, with some surprising
implications for AD spin physics.  While the measured excitation
spectra fit the SP picture in agreement with the results of
\chapref{chap:Geometry}, the spin-dependent transport clearly does
not; whereas transport through individual Zeeman-split SP states
should be spin-polarised, experiments show that both spins are
transmitted through every resonance.  These measurements also
uncover spin-orbit mediated anticrossings between AD states and the
presence of `molecular' states resulting from an impurity in one of
the channels, further demonstrating the power of the selective
injection/detection technique to investigate details of quantum
transport.

I propose an explanation for these findings in
\chapref{chap:SpinTransportModel}.  The results are interpreted as
signatures of exchange interactions which lower the spin excitation
energy while preserving orbital SP energy scales.  By treating the
antidot theoretically as an MDD of holes in the LLL, I am able to
reproduce many aspects of the measurements within a computational
model of spin-resolved sequential transport.  I discuss the features
of this model and its limitations, and the implications of the
experiments for a general theoretical understanding of low-lying
excitations of the IQH edge.

Finally, I present in \chapref{chap:TiltedB} the results of several
additional measurements on the same device with the addition of an
\emph{in situ} rotating sample holder.  By changing the orientation
of the device relative to the magnetic field, I am able to vary the
Zeeman energy (which depends on the total field) independently of
the orbital energy scales (which depend only on the perpendicular
component of the field).  These data show clear evidence of crossing
orbital states, although a relatively high electron temperature
unfortunately means that the spin-selective technique used in
\chapref{chap:SpinTransport} would not add much useful information.
I also present evidence of coherently-coupled `antidot molecules'
which form as a result of unintentional impurities close to our
antidot.

The thesis concludes with a brief discussion of its central results
and their implications, and with suggestions for how these findings
could be extended or considered in more detail in the future.

\section*{Declaration of Originality}

This dissertation is the result of my own work and includes nothing
which is the outcome of work done in collaboration except where
specifically indicated in the text.  It contains less than 60,000
words.

\fncymain

\mainmatter
\chapter{Theoretical Background\label{chap:ADtheory}}

\ifpdf
    \graphicspath{{Chapter1/Figures/PNG/}{Chapter1/Figures/PDF/}{Chapter1/Figures/}}
\else
    \graphicspath{{Chapter1/Figures/EPS/}{Chapter1/Figures/}}
\fi

In many cases, the physics of quantum antidots is analogous to that
of comparably sized quantum dots at high magnetic fields, and so
much of the vast literature concerning transport experiments on
lithographically-defined quantum dots is directly relevant to the
study of antidots.  At first consideration this correspondence is
surprising, since fundamentally an antidot device is an open system
of an effectively infinite number of electrons, while a quantum dot
is a zero-dimensional object with a finite number of particles. But
as we will see, the effect of a magnetic field perpendicular to a
two-dimensional electron system (2DES) is to establish
zero-dimensional electronic states which encircle an antidot.
Analogous to the semi-classical picture of `skipping' cyclotron
orbits, these are the stationary solutions to the time-independent
Schr\"{o}dinger equation. Furthermore, because transport
measurements probe only those states near the surface of the
electronic Fermi sea, our experiments explore the properties of just
a few of these zero-dimensional states at any given situation and we
can generally ignore the continuum of electrons stretching away from
the antidot. In this chapter we consider the theoretical description
of such zero-dimensional antidot states, a framework which is
necessary for the interpretation of the experiments and models
presented in this thesis.

The chapter is divided into three sections.  In the first we solve
the single-particle (SP) Schr\"{o}dinger equation for the
non-interacting eigenstates of a parabolic antidot potential, i.e.\
the inverse of a parabolic quantum dot.  At the relatively low
magnetic fields ($B\lesssim$\unit{1}{\tesla}) studied in this work,
the SP wave functions are well-separated in both space and energy,
and so a non-interacting model is often sufficient to describe the
features observed in standard transport measurements.  So in the
second section, we further investigate the properties of these SP
states, hopefully to aid the reader's intuition about their physics.
Finally, in the third section we consider the treatment of
electron-electron interactions through Hartree-Fock theory, which we
will eventually use in \chapref{chap:SpinTransportModel} to explain
many of the features we observe which appear to be inconsistent with
the non-interacting SP model.

\section{Single-particle eigenstates\label{sec:SPstates}}
We begin with a consideration of the non-interacting SP eigenstates
for electrons in two dimensions subject to a perpendicular magnetic
field, since these form a useful starting point for discussion in
many aspects of this thesis.  The Hamiltonian for an electron
(charge $-e$) in the presence of a time-independent magnetic field
$\mathbf{B}$ and electric potential $\varphi(\mathbf{x})$ is
\begin{equation}\label{eq:HinBfield}
  \hat{H} = \frac{1}{2m^\ast}(\hat{\mathbf{p}}+e\mathbf{A})^2
  -e\varphi(\mathbf{x}),
\end{equation}
where $\mathbf{p}=-i\hbar\mathbf{\nabla}$ is the canonical momentum,
$\mathbf{A}$ is the magnetic vector potential ($\mathbf{B} =
\mathbf{\nabla}\wedge\mathbf{A}$), and $m^\ast=0.067m_e$ is the
effective electron mass in GaAs. In most cases we are interested in
problems with cylindrical symmetry, so it makes sense to choose the
symmetric gauge,
\begin{equation}\label{eq:SymGauge}
  \mathbf{A} = \frac{1}{2}\mathbf{B}\wedge\mathbf{x}.
\end{equation}

In a semiconductor heterostructure like the GaAs/AlGaAs structures
used in this work, the confinement along the growth ($z$) direction
is strong enough that, at the low temperatures used for
measurements, only a single energy level is populated.  Dynamics in
this dimension are then completely decoupled from those in the plane
of the 2DES, and the problem becomes purely two-dimensional.  We
therefore seek solutions to the Schr\"{o}dinger equation
\begin{equation}\label{eq:SWE}
  \hat{H}\psi = E\psi
\end{equation}
in the remaining $(x,y)$ coordinates.  To describe a
circularly-symmetric antidot potential $\varphi(r)$, we shift to
polar coordinates $(r,\phi)$ in which
$\mathbf{A}=\frac{1}{2}B_zr\hat{\mathbf{\phi}}$ to write
\eqnref{eq:SWE} in the form
\begin{equation}
  \label{eq:SWEpolar}
  \left[\frac{\hat{p}^2}{2m^\ast}+\left(\frac{\omegac}{2}\right)\hat{L}_z +
  \frac{1}{2}m^\ast\left(\frac{\omegac}{2}\right)^2r^2
  -e\varphi(r)\right]\psi = E\psi,
\end{equation}
where $\omegac = eB_z/m^\ast$ is the cyclotron frequency,
\begin{equation}
  \label{eq:psquared}
  \hat{p}^2 = -\hbar^2\left(\frac{1}{r}\frac{\partial}{\partial
  r}\left(r\frac{\partial}{\partial r}\right) +
  \frac{1}{r^2}\frac{\partial^2}{\partial\phi^2}\right),
\end{equation}
and
\begin{equation}
  \label{eq:Lz}
  \hat{L}_z = \hat{r}\wedge\hat{p} = -i\hbar\frac{\partial}{\partial\phi}
\end{equation}
is the canonical angular momentum operator.  Since the circular
symmetry implies that the eigenvalue of $\hat{L}_z$ is conserved, we
separate $\psi(r,\phi)$ by coordinates as
\begin{equation}\label{eq:SeparatePsi}
  \psi(r,\phi) = f(r)e^{im\phi},
\end{equation}
where $m = 0,\pm1,\pm2,\ldots$ is the azimuthal quantum number
($\hat{L}_z\psi = m\hbar\psi$).

\subsection{SP states in a parabolic antidot potential\label{sec:ParabolicAD}}

The particular case of a parabolic electric potential is frequently
encountered in the quantum dot literature, since it simply adds to
the quadratic effective potential from the magnetic field, and
\eqnref{eq:SWEpolar} has a well-known analytic solution
\cite{Landau1997}.  Since we are often interested only in AD states
near the Fermi level, where we may locally approximate the potential
by a suitable inverted parabola, these wave functions form a useful
basis for calculations, and so we explore some of their properties
below.

With a parabolic antidot potential with curvature determined by the
parameter $\omega_0$,
\begin{equation}
  \label{eq:ParabolicPot}
  -e\varphi(r) = -\frac{1}{2}m^\ast\omega_0^2r^2,
\end{equation}
and the definition of \eqnref{eq:SeparatePsi}, the radial part of
the Schr\"{o}dinger \eqnref{eq:SWEpolar} becomes
\begin{equation}
  \label{eq:SWEparabolic}
  \left[-\frac{\hbar^2}{2m^\ast}
      \left(\frac{1}{r}\frac{\partial}{\partial r}
          \left(r\frac{\partial}{\partial r}\right)
          -\frac{m^2}{r^2}\right) +
  \frac{m\hbar\omegac}{2} +
  \frac{1}{2}m^\ast\left(\frac{b\omegac}{2}\right)^2r^2
  \right]f(r) = Ef(r),
\end{equation}
where
\begin{equation}
  \label{eq:b}
  b = \sqrt{1-\left(\frac{2\omega_0}{\omegac}\right)^2}.
\end{equation}
Note that while we focus on the case of an antidot potential, the
analogous results for a quantum dot may be obtained by reversing the
sign under the square root in \eqnref{eq:b} (or through the
identification $\omega_0\rightarrow i\omega_0$), as well as those of
a free particle, for which $b=1$.

\begin{figure}[tb]
\begin{center}
\includegraphics[height=2.6truein,angle=270,trim=1truein 1truein 1truein 0truein]{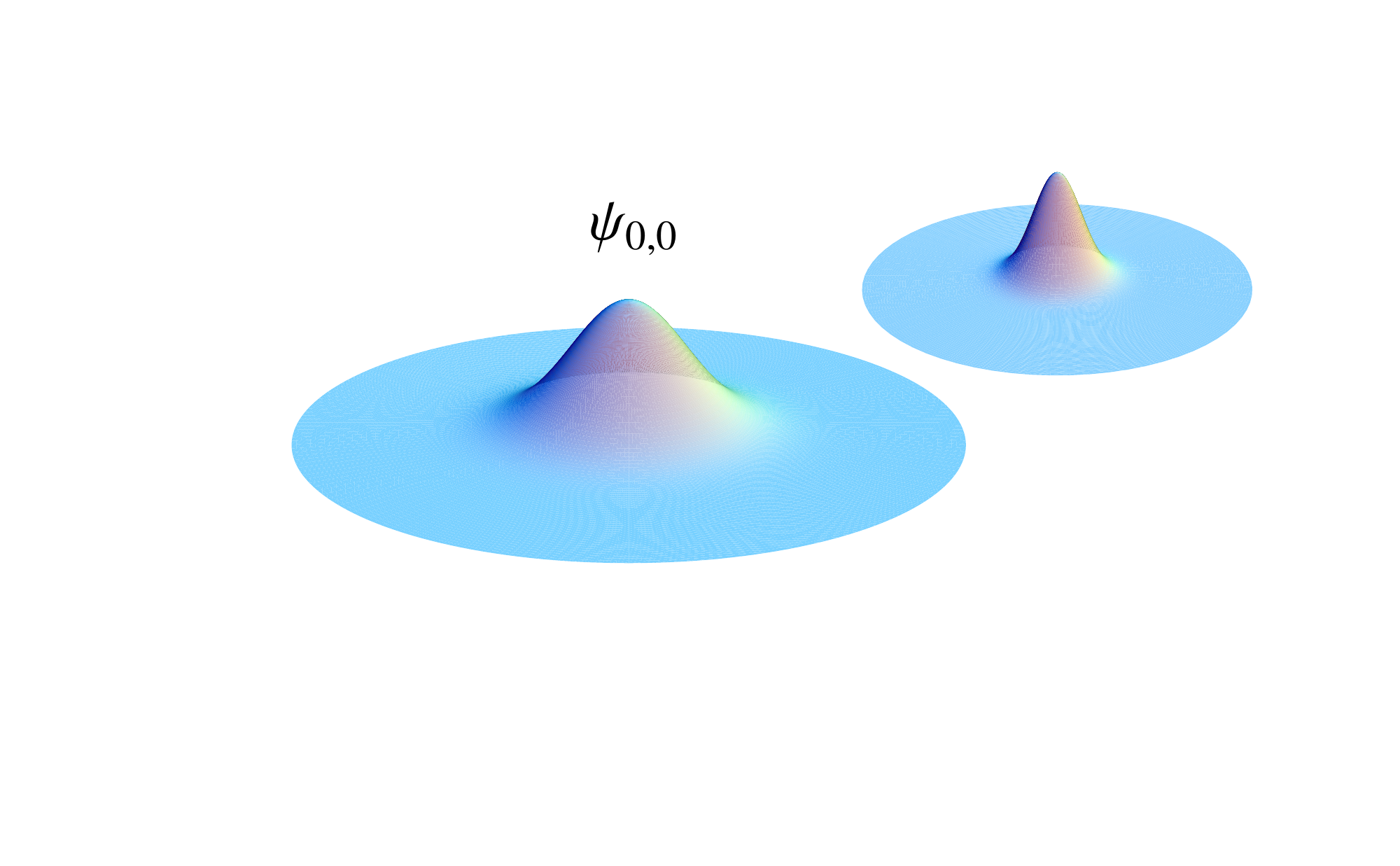}
\includegraphics[height=2.6truein,angle=270,trim=1truein 1truein 1truein 0truein]{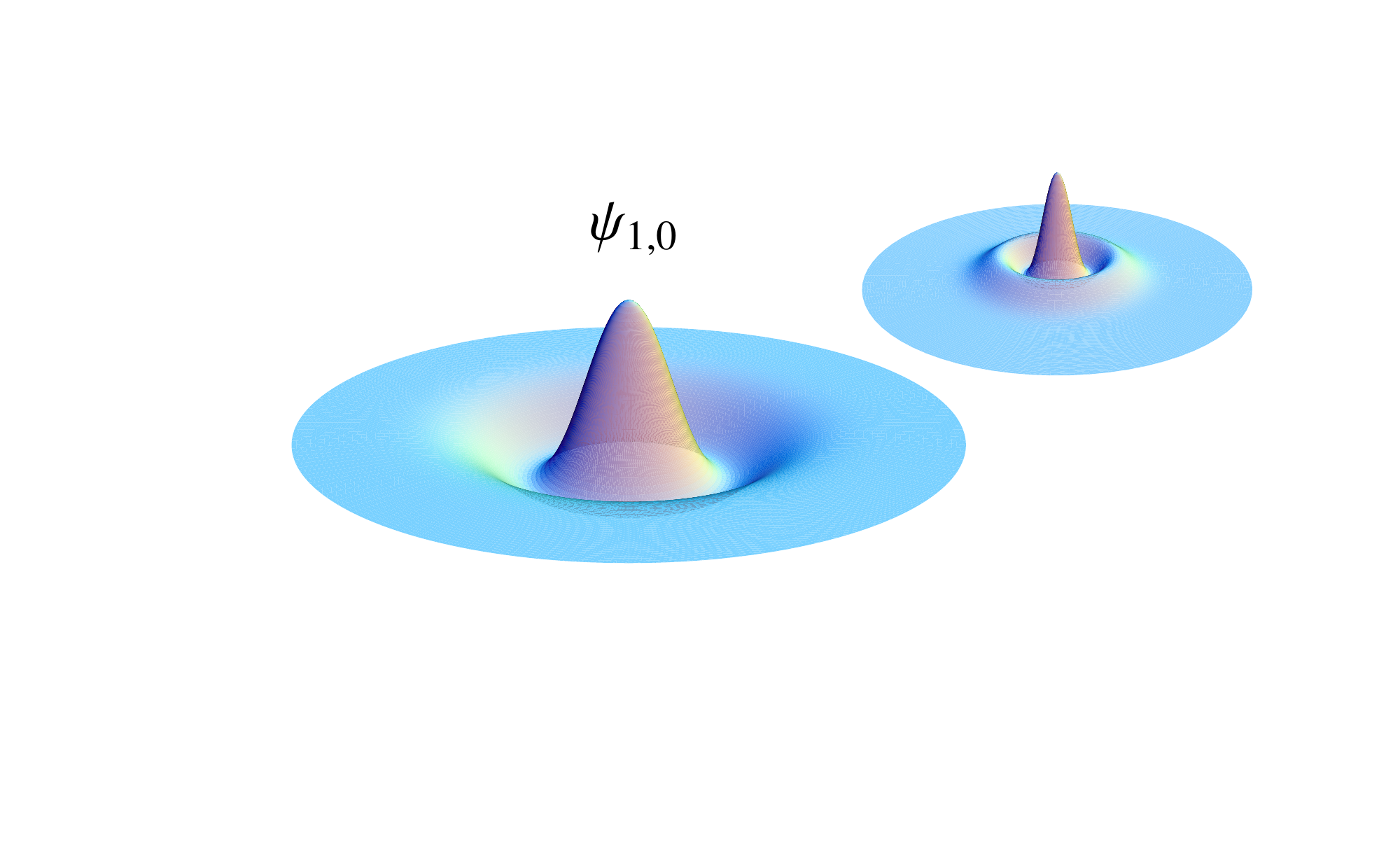}
\includegraphics[height=2.6truein,angle=270,trim=1truein 1truein 1truein 0truein]{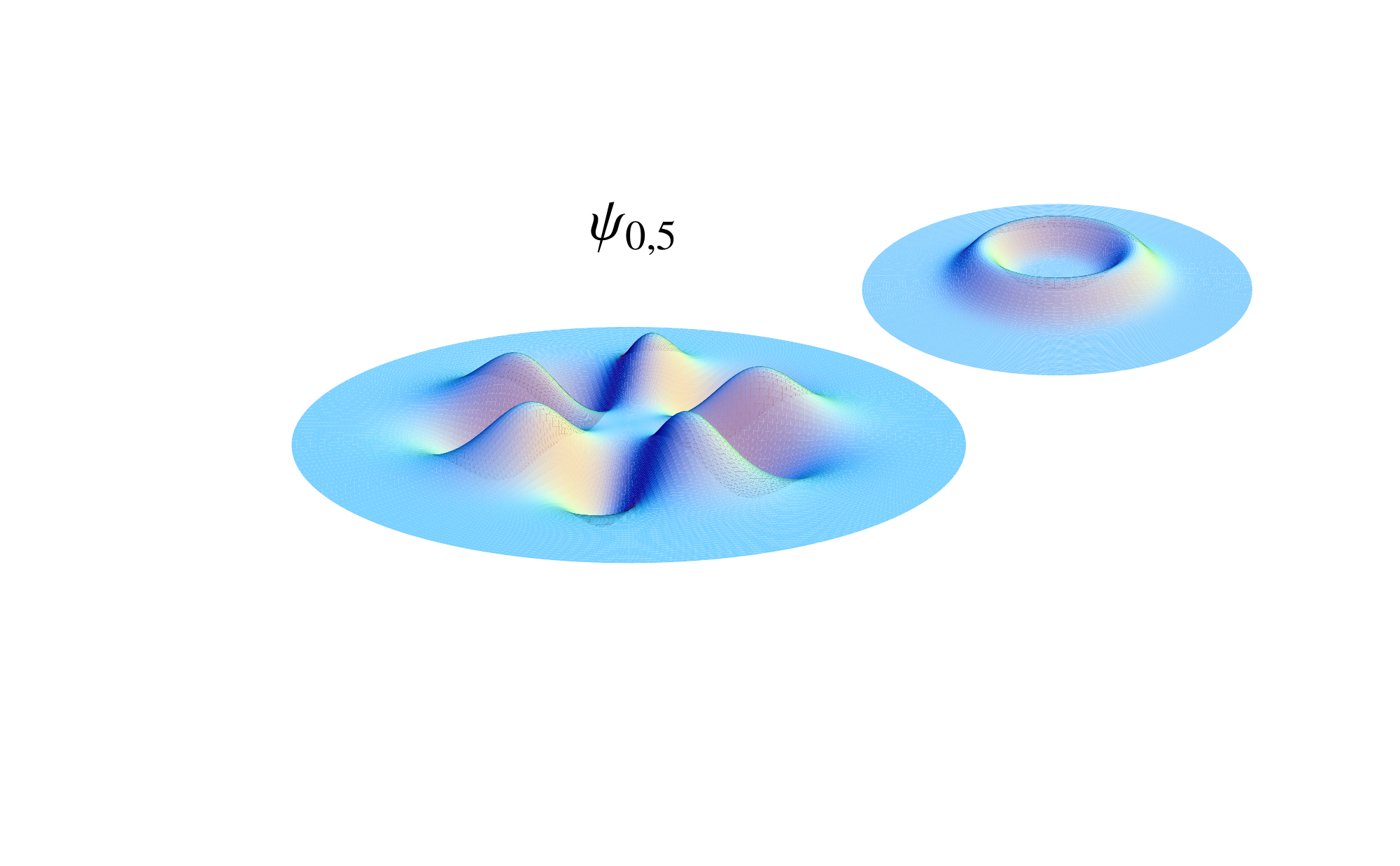}
\includegraphics[height=2.6truein,angle=270,trim=1truein 1truein 1truein 0truein]{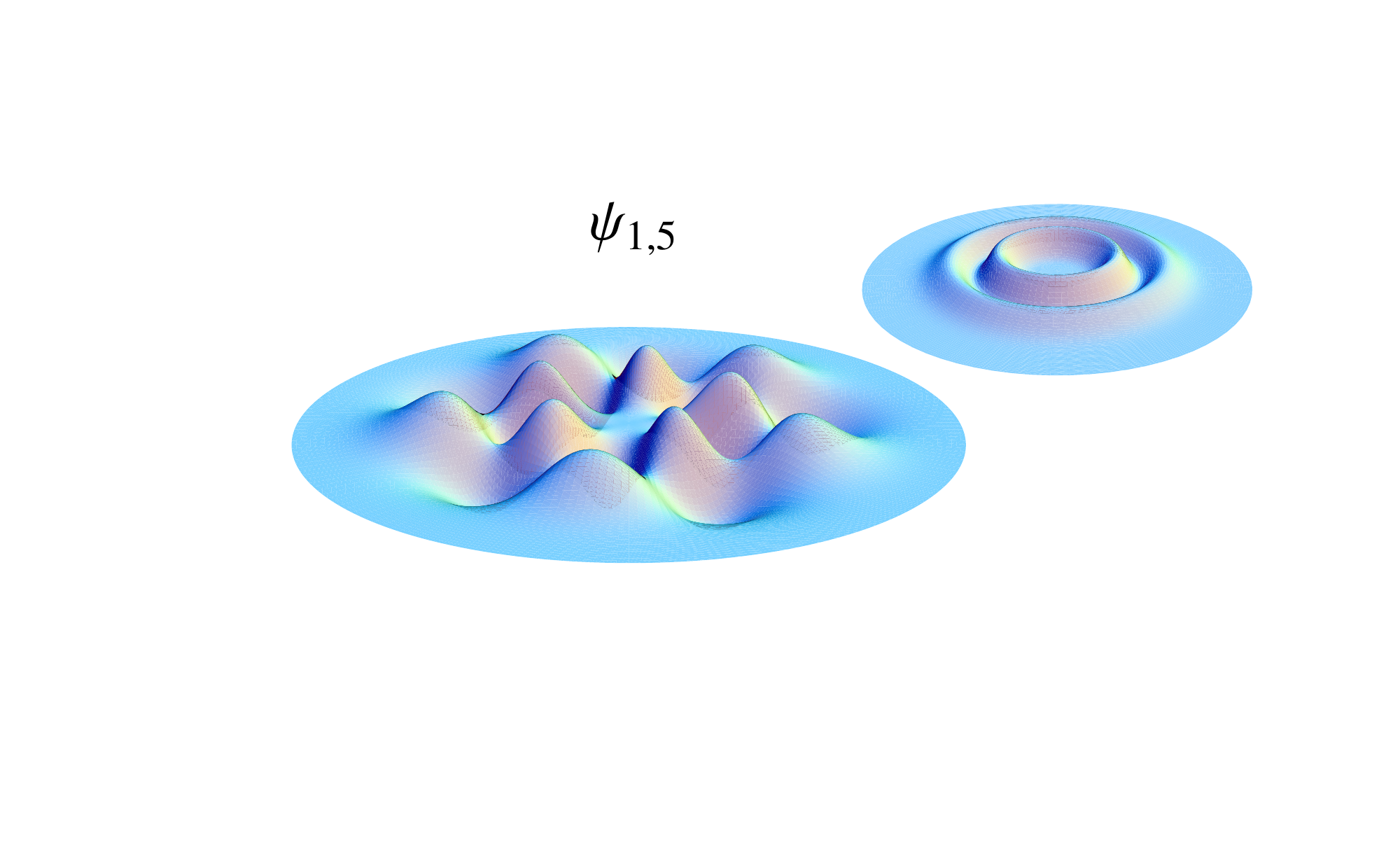}
\caption[Single-particle eigenstates]{Eigenstates
$\psi_{n,m}(r,\phi)$ for electrons subject to a parabolic potential
and a perpendicular magnetic field $B$. Each figure shows
$\Re\{\psi_{n,m}\}$ as well as $|\psi_{n,m}|^2$ as an inset.
\label{fig:SPstates}}
\end{center}
\end{figure}

We proceed to solve \eqnref{eq:SWEparabolic} in terms of the
dimensionless coordinate
\begin{equation}
  \label{eq:xi}
  \xi = \frac{bm^\ast\omegac r^2}{2\hbar} = \frac{1}{2}\left(\frac{r}{\ell}\right)^2,
\end{equation}
where $\ell = \sqrt{\hbar/(eB_zb)}$ is the natural length scale of
the system.  This leads to the ODE
\begin{equation}
  \label{eq:SWEODE}
  \xi f^{\prime\prime}+f^\prime +\left(-\frac{\xi}{4}+\beta -
  \frac{m^2}{4\xi}\right)f = 0,
\end{equation}
where
\begin{equation}
  \label{eq:beta}
  \beta = \frac{E}{\hbar b\omegac}-\frac{m}{2b}.
\end{equation}
This equation may be solved in a variety of ways, for example
through a power series expansion or by using raising and lowering
operators \cite{Ezawa2000}.  The method we present here rests on the
recognition that \eqnref{eq:SWEODE} is very similar to the Laguerre
ODE, given by
\begin{equation}
  \label{eq:LaguerreODE}
  xy^{\prime\prime}+(\nu+1-x)y^\prime + \lambda y = 0,
\end{equation}
for real constants $\nu$ and $\lambda$. By making the substitution
$f(\xi) = e^{a\xi}\xi^dg(\xi)$, we recast \eqnref{eq:SWEODE} as
\begin{equation}
  \xi g^{\prime\prime} +\bigl(2d+1+2a\xi\bigr)g^\prime +
  \left(\xi a^2 - \frac{\xi}{4} +  2ad + a  + \beta + \frac{d^2}{\xi} -
  \frac{m^2}{4\xi^2}\right)g = 0.
\end{equation}
Thus, by setting $a = -1/2$ and $d = \abs{m}/2$, we obtain the
Laguerre equation
\begin{equation}
  \xi g^{\prime\prime}+\bigl(\abs{m}+1-\xi\bigr)g^\prime +
  \left(\beta - \frac{\abs{m}}{2} - \frac{1}{2}\right)g = 0,
\end{equation}
in which we identify
\begin{equation}
\begin{sistema}
  \nu =  \abs{m} \\
  \lambda = \beta - \frac{\abs{m}}{2}-\frac{1}{2}.
\end{sistema}
\end{equation}
Solutions to \eqnref{eq:LaguerreODE} may be written in terms of the
associated Laguerre functions $L_\lambda^\nu(x)$, where $\lambda =
n$ must be a non-negative integer in order to satisfy the boundary
condition that
$\lim_{\xi\rightarrow\infty}f(\xi)=0$.\footnote{%
With this condition, the $L_n^{|m|}(x)$ are polynomials with a
finite number of terms.} %
This condition determines the energy eigenvalues
\begin{equation}
  \label{eq:Enm}
  E_{n,m} = b\hbar\omegac\left(n+\frac{\abs{m}}{2} +\frac{1}{2}\right)
  + \frac{m}{2}\hbar\omegac,
\end{equation}
with the corresponding radial wave functions
\begin{equation}
  \label{eq:fnm}
  f_{n,m}(\xi) = C_{n,m}
      e^{-\xi/2}\xi^{\abs{m}/2}L_n^{\abs{m}}(\xi),
\end{equation}
where the normalisation constants
\begin{equation}
  \label{eq:Cnm}
  C_{n,m} = \frac{1}{\ell}\sqrt{\frac{n!}{2\pi(n+\abs{m})!}}
\end{equation}
are easily computed with the help of the orthogonality relation for
the Laguerre polynomials:
\begin{equation}\label{eq:LaguerreOrth}
  \int_0^\infty e^{-x}x^k L_n^k(x)L_m^k(x)dx =
  \frac{(n+k)!}{n!}\delta_{nm}.
\end{equation}
A few examples of these wave functions are shown in
\figref{fig:SPstates}.

\subsection{Landau levels and the quantum Hall effect}
\begin{figure}[tb]
\centering
\includegraphics[]{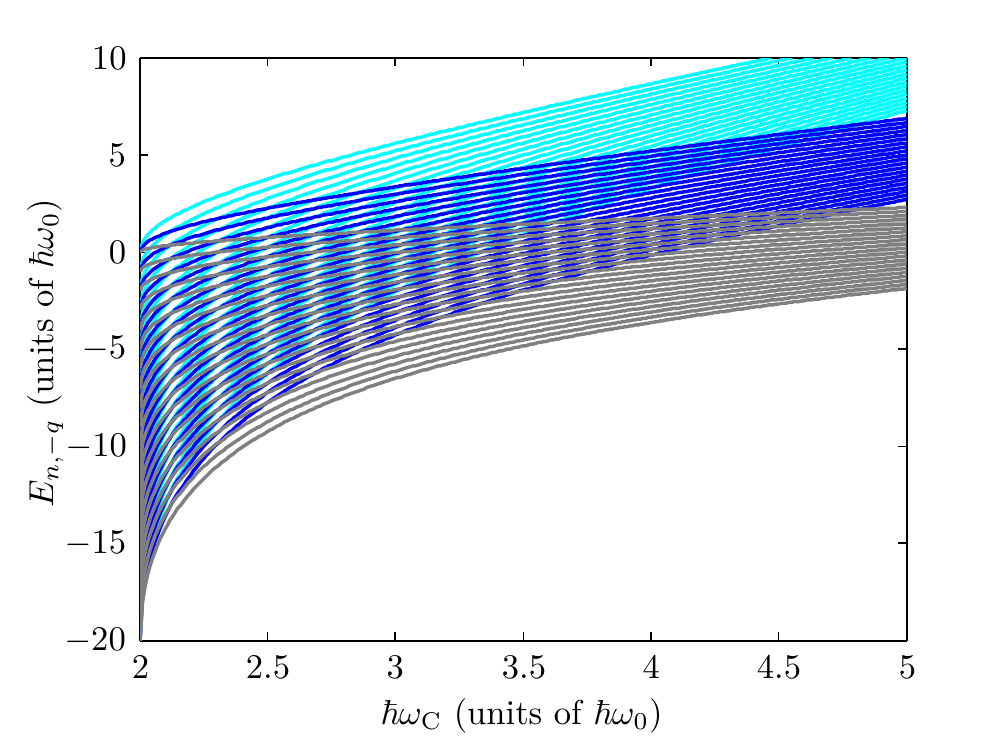}
    \caption[The Darwin-Fock spectrum]{%
Antidot eigenenergies $E_{n,-q}$ of states in the three lowest LLs,
with $n=0,1,2$ and $q=0$--20, as a function of magnetic energy
$\hbar\omegac$.  Note that no stationary states are defined for
$\omegac<2\omega_0$.  In a real device, saddle points in the
constrictions define a maximum $q$ for each LL, such that only
states from lower LLs exist below the Fermi energy.
\label{fig:DarwinFock}}
\end{figure}

In the context of quantum dots, it makes sense to consider the
evolution of states between the low-$B$ ($\omegac\ll\omega_0$) and
high-$B$ ($\omegac\gg\omega_0$) regimes, known as the Darwin-Fock
spectrum \cite{Darwin1930,Fock1928}.  For an antidot, however, the
low-$B$ case is not especially interesting, since for
$\omegac<2\omega_0$ the electrons are no longer confined, and so
experiments are always performed in the presence of a large magnetic
field.  Under these circumstances, the states begin to converge to a
set of Landau levels (LLs), which are the free electron eigenstates
obtained in the high-$B$ limit ($b\rightarrow 1$),
\begin{subequations}
  \label{eq:EnLL}
  \begin{align}
  E^\mathrm{free}_{n,m} & =
  \hbar\omegac\left(n+\frac{|m|+m}{2}+\frac{1}{2}\right) \\
    & = \hbar\omegac\left(n_\mathrm{LL}+\frac{1}{2}\right),
  \end{align}
\end{subequations}
where $n_\mathrm{LL} = 0,1,2,\ldots$ is the LL quantum number,
defined as
\begin{equation}
  \label{eq:nLL}
  n_\mathrm{LL} = \begin{cases}
    n & \text{if $m\leq0$},\\
    n+|m| & \text{if $m>0$}.
  \end{cases}
\end{equation}
The $B$-dependence of some of the eigenenergies in \eqnref{eq:Enm}
is shown in \figref{fig:DarwinFock}.

In a 2DES at low temperature, the occupied states are those below
the Fermi energy, $\Ef$, which is fixed by the two-dimensional
electron density of the sample, $n_e$.  At a fixed field, $n_e$ also
determines the \emph{filling factor},
\begin{equation}
    \nu=\frac{n_eh}{eB_\perp},
\end{equation}
as the number of electrons per magnetic flux quantum $h/e$, or
equivalently the number of filled (spin non-degenerate) LLs.\footnote{%
To see this relationship, consider the free electron wave function
given by \eqnref{eq:fnm}.  The envelope $e^{-\xi/2}\xi^{|m|/2}$ has
a maximum at $\xi=|m|$, and so the number of electrons within a
radius $\xi=|m|$ is $\approx|m|$ (this can be shown to be strictly
true for infinite systems).  Hence, each LL has density
$n^\mathrm{LL}_e = 1/(2\pi\ellB^2) = eB_\perp/h$. } %
Through the application of voltages to surface gates patterned above
the 2DES, $n_e$ (and hence $\nu$) may be varied throughout a device.
If this electric potential variation is small on the length scale of
free-electron wave functions (the magnetic length, $\ellB =
\sqrt{\hbar/(eB_\perp)}$), then it may be treated as a small
perturbation which adds a position-dependent offset to the LL energy
$(n_\mathrm{LL}+\frac{1}{2})\hbar\omegac$.

\begin{figure}[tb]
\centering
\includegraphics[]{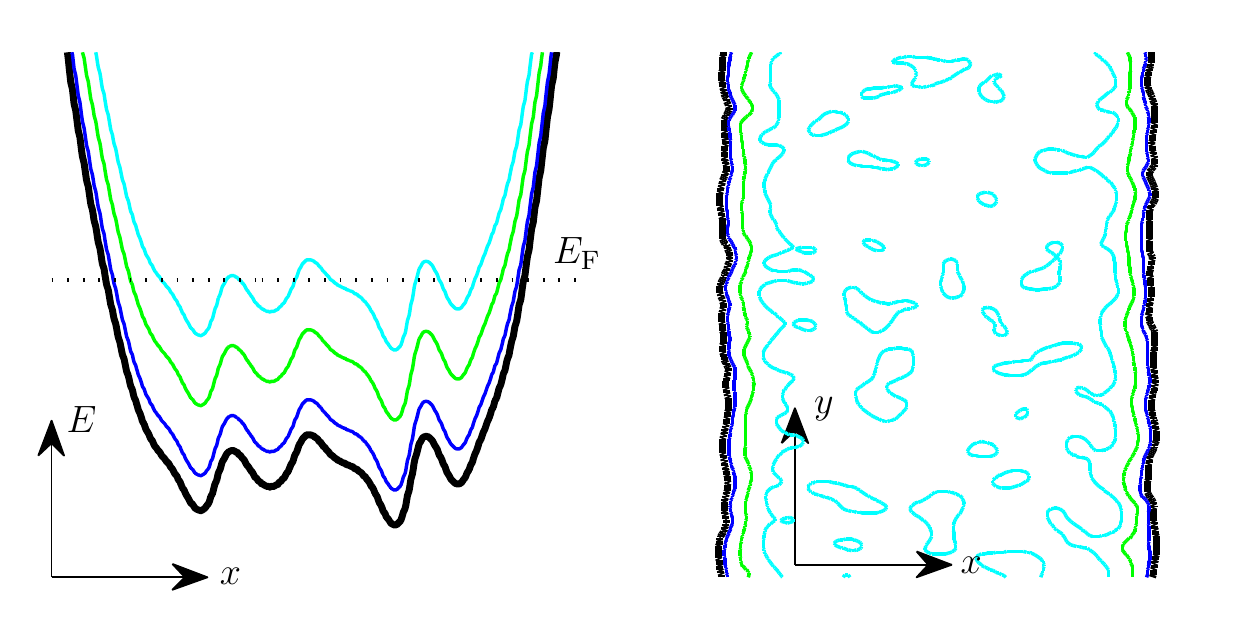}
    \caption[Landau levels in a real device]{%
Landau level energies as a function of position across a real device
containing random disorder.  The states in the bulk of the device
are localised around dots and antidots in the disorder potential,
such that only electrons at the edges of the sample propagate
macroscopic distances. \label{fig:QHE}}
\end{figure}

\begin{figure}[tb]
\begin{center}
\includegraphics[]{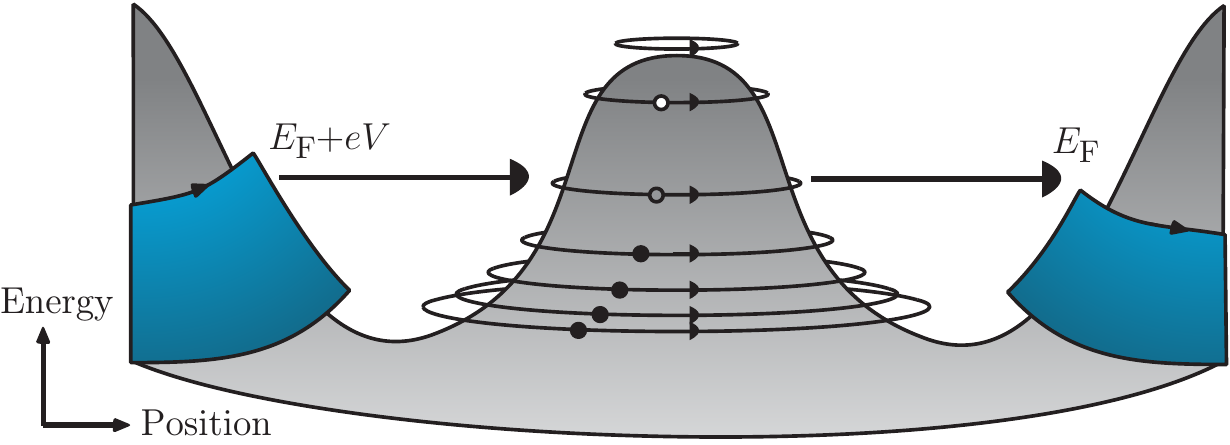}
\caption[Tunnelling through a single antidot]{Cross-sectional view
of an antidot at the centre of a split-gate, where tunnelling occurs
between the edge states acting as metallic leads and the quantised
antidot states. \label{fig:ADxsection}}
\end{center}
\end{figure}

The above picture of two-dimensional electron states leads to a
simple description of the quantum Hall effect
\cite{Klitzing1980,Buttiker1988} in terms of LLs, as depicted in
\figref{fig:QHE}.  Most of the electron states in the bulk 2D
regions are localised by the background disorder potential
(naturally-occurring dots and antidots) and so cannot contribute to
equilibrium transport. Near the edges of the sample, however, the
filled LLs follow the potential up through $\Ef$, creating a set of
extended \emph{edge states} capable of carrying current across the
device.  In contrast to the closed edge of a quantum dot or antidot
(with circumference smaller than the phase coherence length), these
edge states have a continuous spectrum, and so serve as metallic
leads with a density of states given by the Fermi distribution.  In
order to probe antidot states via transport measurements, these edge
states must be brought within tunnelling distance (of order $\ellB$)
of the antidot states, as shown in \figref{fig:ADxsection}. This is
accomplished by fabricating an antidot in the centre of a
split-gate, creating two constrictions in which the filling factor
$\nu_\mathrm{C}$ may be independently varied.  The minimum
$\nu_\mathrm{C}$ defines the highest LL with states fully encircling
the antidot, which we denote as $\nuAD$. This may be chosen anywhere
from zero to the bulk value $\nuB$, but in this work we are usually
interested in the case when only one spin-degenerate LL is occupied,
such that $\nuAD=2$.  We investigate the properties of these SP
states in further detail below.

\subsection{The single-particle picture of a $\nuAD=2$ antidot\label{sec:SPnu=2}}

When the constrictions on either side of an antidot are set to
$\nu_\mathrm{C}=2$, the only occupied antidot states are those in
the lowest Landau level (LLL).  These states are given by $n=0$,
$m\leq0$, and so with $q = -m$ as a non-negative integer (such that
the canonical angular momentum $L_z = -q\hbar$), they have energies
\begin{equation}
  \label{eq:ELLL}
  E^\mathrm{LLL}_q = \frac{1}{2}b\hbar\omegac -
  \frac{q}{2}(1-b)\hbar\omegac
  \qquad \text{($q=0,1,2,\ldots$),}
\end{equation}
and corresponding wave functions ($L_0^k(x) = 1$)
\begin{equation}
  \label{eq:psiLLL}
  \psi^\mathrm{LLL}_q =\frac{1}{\ell}\sqrt{\frac{1}{2\pi 2^{q}q!}}
  \left(\frac{r}{\ell}\right)^{q}e^{-r^2/4\ell^2}e^{-iq\phi}.
\end{equation}
From \eqnref{eq:ELLL} we see that the orbital energies are evenly
spaced, as expected for a system with harmonic confinement, and we
identify
\begin{equation}
  \label{eq:dEsp}
  \dEsp = (1-b)\frac{\hbar\omegac}{2}
\end{equation}
as the single-particle energy scale.  As we will see in
\chapref{chap:TransportTheory}, $\dEsp$ may be measured through
transport spectroscopy, and so by rearranging \eqnref{eq:dEsp} we
obtain
\begin{equation}
  \label{eq:effectiveb}
  b = 1-\frac{2\dEsp}{\hbar\omegac},
\end{equation}
as a useful relation for the effective harmonic parameter $b$ in a
given experiment.

Up to this point we have considered only the orbital part of the
electron wave function, but of course spin plays an important role
as well.  In the simplest case it enters the Hamiltonian only
through the Zeeman term:
\begin{equation}\label{eq:Zeeman}
  \hat{H}_\mathrm{Z} = g\mu_\mathrm{B}B\hat{s}_z,
\end{equation}
where $\mu_\mathrm{B}$ is the Bohr magneton, $g$ is the effective
electron $g$-factor,\footnote{%
Some confusion seems to exist in the literature about the sign of
the Zeeman term. The magnetic moment of the electron is
$\boldsymbol{\mu}=-g\mu_\mathrm{B}\boldsymbol{\sigma}$ (the sign
accounts for the negative electrostatic charge) and the Zeeman
energy is $-\boldsymbol{\mu}\cdot\mathbf{B}$, which leads to
\eqnref{eq:Zeeman}.  The effective $g$-factor in GaAs is negative
due to the strong effect of spin-orbit coupling ($g=-0.44$ in bulk
GaAs), and so the lower energy state has $s_z=+\frac{1}{2}$.
Normally the labeling of spin states is not important as long as one
sticks with a consistent definition, although the sign does matter
in the context of hyperfine and spin-orbit coupling between levels,
as discussed in \secref{sec:ZeroBiasExpts}.  We refer to the lower
(higher) energy state as spin-$\uparrow$ (spin-$\downarrow$)
throughout this thesis.} %
and $\hat{s}_z$ is the spin operator: $\hat{s}_z\psi_\sigma =
\sigma\psi_\sigma$, where $\sigma = \pm \frac{1}{2}$ is the electron
spin.  In the absence of any additional interactions which couple
the spin and orbital parts of the wave function (e.g., hyperfine
interactions with lattice nuclei or spin-orbit coupling), this
simply adds a spin-dependent constant to the energy eigenvalues:
\begin{equation}
  \hat{H}_\mathrm{Z}\psi_\sigma = -\sigma \Ez\psi_\sigma,
\end{equation}
where $\Ez = \abs{g}\mu_\mathrm{B}B$ is the Zeeman energy. Note that
it is the \emph{total} magnetic field $B$ which enters the
expression for $\Ez$, in contrast to the orbital part which responds
only to the perpendicular component $B_\perp$ through the vector
potential $\mathbf{A}$. This suggests a potentially useful
experimental handle to vary $\Ez$ independently from the orbital
wave functions, by changing both the total field $B$ and the angle
at which it is applied relative to the plane of the 2DES.  This
technique is explored in the experiments presented in
\chapref{chap:TiltedB} of this thesis.

\begin{figure}[tb]
\begin{center}
\includegraphics[]{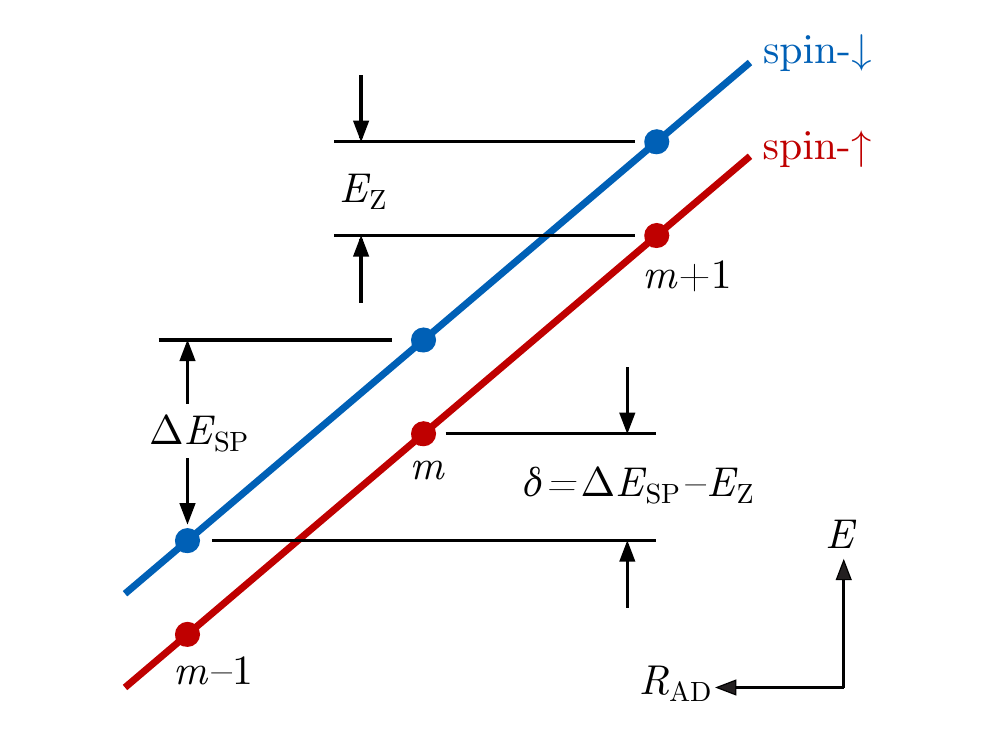}
\caption[Single-particle energies]{Single-particle energy spectrum
of a $\nuAD=2$ antidot.  When $\dEsp>\Ez$, as shown here, the
smallest excitations are $\Ez$ and $\delta = \dEsp-\Ez$ when the
highest-occupied state is spin-$\uparrow$ and spin-$\downarrow$,
respectively. \label{fig:SPladder}}
\end{center}
\end{figure}

Therefore, the primary effect of the electron spin is to split each
LL into two spin-polarised bands separated by $\Ez$, where
$\Ez\ll\hbar\omegac$ unless $\mathbf{B}$ is applied nearly in the
plane of the 2DES. The SP spectrum in the LLL thus consists of two
`ladders' of orbital states with spacing $\dEsp$, separated from
each other by $\Ez$, as shown in \figref{fig:SPladder}.  In this
model, particle and hole excitations are governed by these two
energy scales alone, with the possible values
\begin{equation}\label{eq:dEspSPmodel}
    E_\mathrm{ex} = \pm s\Ez + j\dEsp,
\end{equation}
where $s=0$ or 1 for spin-conserving or spin-flip transitions,
respectively, the upper (lower) sign is for an initial
spin-$\uparrow$ (spin-$\downarrow$) electron, and $j$ is any
integer.  In particular, when $\dEsp>\Ez$ as shown in
\figref{fig:SPladder}, the smallest particle excitation energies for
spin-$\uparrow$ and spin-$\downarrow$ electrons are $\Ez$ and
$\dEsp-\Ez$, respectively.  These are also the SP contribution to
the ground-state transition energies measured in equilibrium
transport.

\section{Properties of antidot eigenstates\label{sec:ADeigenstates}}

In this section we explore some of the observable physical
properties of the SP eigenstates defined above.  To begin, we
consider the physical meaning of the principal quantum numbers
$(n,m)$, since it is helpful to have a more intuitive understanding
of their role in relation to $n_\mathrm{LL}$ in light of the rather
strange identification in \eqnref{eq:nLL}.

The radial quantum number $n$ counts the number of nodes in the
radial wavefunction and so measures the degree of excitation. A
change of $n$ therefore corresponds to changing LLs. On the other
hand, the primary role of the azimuthal quantum number $m$ is to
determine the distance of the wave function peak from the origin, as
can be seen in fig.~\vref{fig:SPstates}.  Using standard
integrals\footnote{%
In particular, $\int_0^\infty e^{-x}x^{k+1}[L_n^k(x)]^2dx = \frac{(n+k)!}{n!}(2n+k+1)$.} %
it is straightforward to show that
\begin{equation}
  \label{eq:exprsq}
  \langle r^2\rangle = 2\ell^2(2n+\abs{m}+1),
\end{equation}
independent of the sign of $m$.  Still, the sign of $m$ is of
critical importance.  The transformation $m\rightarrow-m$
corresponds to taking the complex conjugate of $\psi$; this leaves
the radial part of the wavefunction unchanged, and yet these two
wave functions belong to different LLs, with vastly differing
energies when $\hbar\omegac$ is large.

To visualise this, consider that, from a classical perspective, an
`orbit' of radius $R$ can be achieved in two ways, in terms of a
guiding centre $X$ and cyclotron radius $a$, as depicted in
\figref{fig:orbits}:
\begin{figure}[tb]
\centering
\includegraphics[]{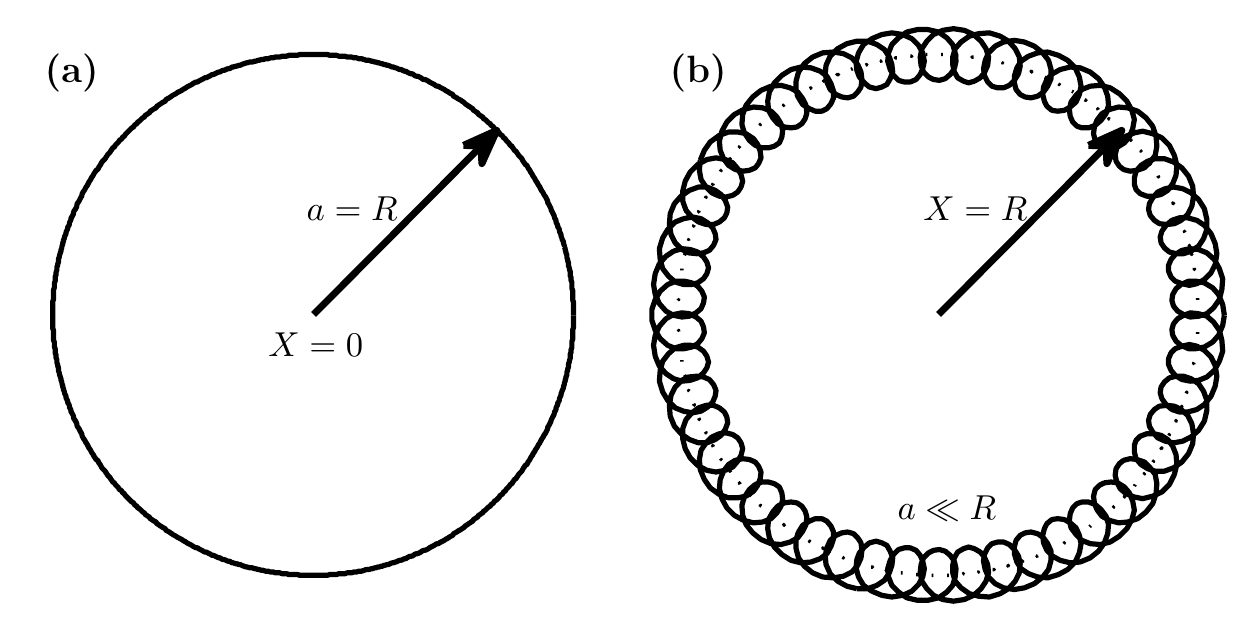}
    \caption[Classical electron orbits]{%
Examples of electron orbits at radius $R$, in terms of the guiding
centre $X$ and the cyclotron radius $a$. Case (a) corresponds to
states $\psi_{n,m}$ in higher LLs, with $m>0$, while case (b)
corresponds to lower LL states with $m<0$. \label{fig:orbits}}
\end{figure}
\begin{itemize}
  \item{$X\sim 0$, $a\sim R$ (\figref{fig:orbits}a):

    Classically, the kinetic energy of a particle with cyclotron
    radius $a$ is given by $K = \frac{1}{2}m^\ast\omegac^2a^2$, so these
    states will have energy $\propto R^2$, and therefore
    $n_\mathrm{LL}\propto \abs{m}$.  This description fits states with
    $m>0$, and although they are valid solutions to the Schr\"{o}dinger
    equation for states at radius $R$, they have energies $\gg\Ef$
    for typical experimental parameters, and so will never be
    populated.
  }
  \item{$X\sim R$, $a\ll R$ (\figref{fig:orbits}b):

    In this case, the energy of the state is small and determined by
    the background potential at radius $R$ rather than the cyclotron
    energy.  These states, which we identify with the $m<0$ case,
    are analogous to classical `skipping orbits' of electrons in lower
    LLs circling the antidot.  It is these states which we probe in
    transport experiments.
  }
\end{itemize}
The remainder of this section discusses the physical properties of
these $m<0$ states in greater detail, drawing parallels to the
classical picture of electron motion where applicable.

\subsection{Angular momentum}

As described above, the LLL states are labelled by their canonical
angular momentum, $L_z = -q\hbar$, where $q=0,1,2,\ldots$ measures
the distance from the origin.  The quantity $L_z$ is not the `real'
angular momentum, however, since it is not gauge invariant.  To see
this, we imagine piercing the centre of the antidot with a single
flux quanta of infinitesimal area.  This is equivalent to the gauge
transformation
\begin{subequations}
  \label{eq:GaugeT}
  \begin{eqnarray}
  \psi(r,\phi)  & \rightarrow & e^{-i\phi}\psi(r,\phi) \\
  A_\phi(r,\phi) &  \rightarrow & A_\phi(r,\phi) +
  \frac{\hbar}{er},
  \end{eqnarray}
\end{subequations}
which leaves the magnetic field (for $r\neq0$) and all other
observables unchanged. The canonical angular momentum, however,
clearly transforms as $L_z\rightarrow L_z-\hbar$ under this
transformation, and so cannot be an observable quantity.  The
observable, or \emph{kinetic} angular momentum is given by
\begin{equation}
  \label{eq:kineticAM}
  \mathbf{l} = \mathbf{r}\wedge m^\ast\mathbf{v},
\end{equation}
where $m^\ast\mathbf{v} = \mathbf{p}+e\mathbf{A}$ is the similarly
gauge-invariant kinetic momentum (as opposed to the canonical
momentum $\mathbf{p}$). With $\mathbf{L} =
\mathbf{r}\wedge\mathbf{p}$, we therefore have
\begin{equation}
    \label{eq:lvec}
  \mathbf{l}  = \mathbf{L} + e(\mathbf{r}\wedge\mathbf{A}),
\end{equation}
for which the $z$-component is
\begin{eqnarray}
  \label{eq:lz}
  l_z & = & L_z + erA_\phi \\
  & = & L_z + \frac{eB}{2}r^2,
\end{eqnarray}
and so we can write the quantum operator for kinetic angular
momentum in the form
\begin{equation}
  \hat{l}_z = \hat{L}_z + \frac{eB}{2}\langle r^2\rangle.
\end{equation}
From \eqnref{eq:lz} it is clear that the kinetic angular momentum is
invariant under the gauge transformation of \eqnsref{eq:GaugeT},
since the change in the magnetic vector potential cancels the change
in $L_z$. Furthermore, using \eqnref{eq:exprsq} we can evaluate
\begin{eqnarray}
  \hat{l}_z\psi_{n,-q} & = & \bigl[-q\hbar +
      \frac{\hbar}{b}(2n+\abs{q}+1)\bigr]\psi_{n,-q} \\
  & = &
  \frac{\hbar}{b}\bigl[1+2n_\mathrm{LL}+q(1-b)\bigr]\psi_{n,-q}.
  \label{eq:ADkineticAM}
\end{eqnarray}
So, within a LL, each successive state moving away from the AD
centre has increased angular momentum by an amount $\hbar(1-b)/b$,
as we would expect for classical orbits of increasing radius (and
constant angular frequency, as we show below).

\subsection{Current density}
Continuing our investigation of the classical analogues to antidot
SP states, we can find further insight into the `dynamics' of these
states\footnote{%
Of course these are stationary solutions to the time-independent
Schr\"{o}dinger equation and so are not dynamical in any quantum
sense.} %
by considering the circulation of charge around the ring, which is
described by the probability current density:
\begin{equation}
  \label{eq:currentdensity}
  \mathbf{J} = \frac{\hbar}{2m^\ast
  i}\left[\psi^\ast(\mathbf{D}\psi) -
  (\mathbf{D}\psi)^\ast\psi\right],
\end{equation}
where
\begin{equation}
  \label{eq:Dcovariant}
  \mathbf{D} = \mathbf{\nabla}+\frac{ie}{\hbar}\mathbf{A}
\end{equation}
is the gauge covariant derivative.\footnote{%
This is defined such that the quantity $(\mathbf{D}\psi)$ transforms
as $(\mathbf{D}\psi)\rightarrow e^{if(x)}(\mathbf{D}\psi)$ under the
gauge transformation
\begin{equation*}
  \begin{sistema}
    \psi \rightarrow e^{if(x)}\psi \\
    \mathbf{A}\rightarrow\mathbf{A}-\frac{\hbar}{e}\mathbf{\nabla}f.
  \end{sistema}
\end{equation*}
} %
In polar coordinates we have $\mathbf{\nabla} =
(\frac{\partial}{\partial
r}\hat{\mathbf{r}},\frac{\partial}{r\partial\phi}\hat{\mathbf{\phi}})$,
and so $J_r=0$ for SP states (since complex conjugation does not
affect the radial wave function), such that
\begin{equation}
  \label{eq:J}
  \mathbf{J}_{n,m}(r) = \frac{\omegac}{2}\left(r +
  \frac{2m\ellB^2}{r}\right)[f_{n,m}(r)]^2\hat{\mathbf{\phi}},
\end{equation}
where $\ellB = \sqrt{\hbar/eB} = b\ell$ is the magnetic length.

\begin{figure}[tb]
\centering
\includegraphics[]{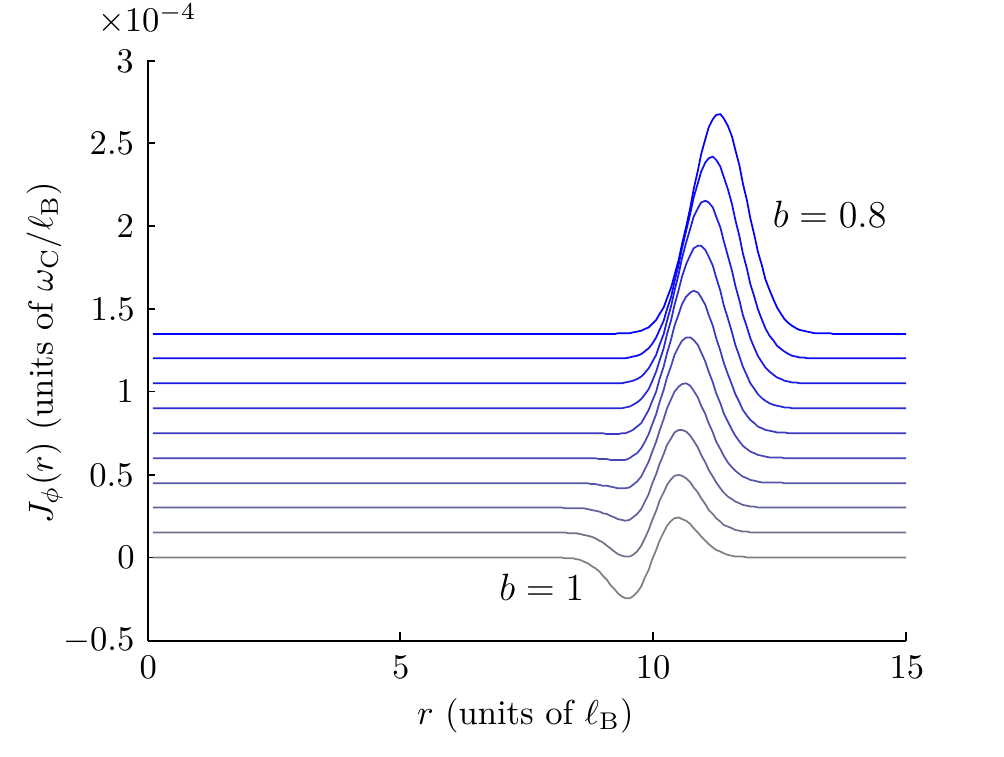}
    \caption[Current density]{%
Current density, \eqnref{eq:J}, of the LLL state $\psi_{0,-50}$ with
$b$ chosen in uniform steps from $b=1$ (the free electron state) to
0.8. The curves are vertically offset for clarity.
\label{fig:Jdensity}}
\end{figure}

This quantity is manifestly gauge invariant due to its definition in
terms of the covariant derivative, and so represents the physical
charge circulation of an electron in the state $\psi_{n,m}$. For our
antidot states with $q=-m>0$, note that $J_\phi$ changes sign from
negative to positive as $r$ increases through the critical radius
$r_\mathrm{c} = \sqrt{2q}\ellB$. As the electric field from the
antidot potential increases, either through a larger curvature
parameter $\omega_0$ or by considering states with larger $q$ (and
hence larger radius), the difference between the centre of the wave
function at $\sqrt{2q}\ell$ and $r_\mathrm{c}$ becomes larger, and
so a larger fraction of the state has $J_\phi>0$, as shown in
\figref{fig:Jdensity}.

Using the normalisation of Laguerre polynomials,
\eqnref{eq:LaguerreOrth}, and the additional integral relation
\begin{equation}
  \int_0^\infty e^{-x}x^{k-1}[L_n^k(x)]^2dx = \frac{(n+k)!}{kn!} \qquad
  (k\geq1),
\end{equation}
we evaluate the effective `orbital frequency' for SP states (with
$m\neq0$) to be
\begin{eqnarray}
  \label{eq:forbit}
  f_\mathrm{orbit} & = & \int_0^\infty J_\phi(r)dr \\
  & = & \frac{\omegac}{2}\int_0^\infty
  \left(r+\frac{2m\ellB^2}{r}\right)[f_{n,m}(r)]^2dr \nonumber\\
  & = & \frac{\omegac}{4\pi}\bigl(1+b\,\mathrm{sign}(m)\bigr), \nonumber
\end{eqnarray}
which is in fact independent of the values of $(n,m)$.  In
particular, we note that for free electrons ($b\rightarrow 1$), the
integrated current density equals zero, so
there is no net orbital velocity for these states.\footnote{%
Note, however, that the angular momentum, given by
\eqnref{eq:ADkineticAM}, is non-zero:
$l_z = (1+2n_\mathrm{LL})\hbar$.} %
Thus, one may interpret the free-electron state $\psi_{n,-q}$ as a
superposition of $\psi_{n,0}$ cyclotron states distributed about a
ring of radius $r_\mathrm{c}=\sqrt{2\abs{q}}\ellB$. With this
interpretation it is clear that the free-electron wave functions
within a LL are indeed all the same up to a spatial translation, a
point which is obvious for the wave functions obtained in other
gauges but less so in the symmetric gauge.

In the presence of the antidot potential, the same interpretation of
the LLL states in terms of translated cyclotron ground states
clearly holds, but now the electric field generated by the potential
causes the electrons to gain a net orbital velocity.  Classically,
this is described by $\mathbf{E}\wedge\mathbf{B}$ drift, in which
the orbital velocity is
\begin{equation}
  \mathbf{v}_\mathrm{drift} = \frac{\mathbf{E}\wedge\mathbf{B}}{\abs{B}^2} =
  \frac{\abs{E}}{\abs{B}}\hat{\mathbf{\phi}}.
\end{equation}
With the electric field given by
\begin{equation}
  \mathbf{E}(r) = -\mathbf{\nabla}\varphi(r) =
  -\frac{m\omega_0^2r}{e}\hat{\mathbf{r}},
\end{equation}
this corresponds to an orbital frequency
\begin{equation}
  f_{E\wedge B} = \frac{\langle v_\mathrm{drift} \rangle}{2\pi\langle r\rangle}
      = \frac{\omega_0^2}{2\pi\omegac}.
\end{equation}
In the high-$B$ limit ($\omegac\gg\omega_0$) of \eqnref{eq:forbit},
we indeed have
\begin{equation}
  f_\mathrm{orbit} =
  \frac{\omegac}{4\pi}\left(1-\sqrt{1-\frac{4\omega_0^2}{\omegac^2}}\right)
      \simeq \frac{\omega_0^2}{2\pi\omegac},
\end{equation}
in agreement with this estimate.

\subsection{The Aharonov-Bohm effect and its relation to antidots\label{sec:ABeffect}}

The Aharonov-Bohm (AB) effect \cite{Ehrenberg1949,Aharonov1959} is
probably the most well-known example of a \emph{geometrical phase}
(or Berry phase, after Michael Berry who generalised the concept
\cite{Berry1984}), and concerns the observable effects of
electromagnetic potentials.  In classical electrodynamics, Maxwell's
equations are formulated in terms of the fields
$(\mathbf{E},\mathbf{B})$ and the potentials $(\varphi,\mathbf{A})$
are not physically observable, since they may be changed by any
gauge transformation, leaving the fields invariant.  But it is the
potentials which directly enter the quantum-mechanical Hamiltonian,
as in \eqnref{eq:HinBfield}, and this can lead to physical
consequences for a particle subject to nonzero potentials, even in a
region where $\mathbf{E}$ and $\mathbf{B}$ are zero.

In general, observable quantities may depend on the path dependent
quantity $e^{i\gamma}$, where
\begin{equation}
  \gamma = \frac{q}{\hbar}\int_P^Q(\varphi dt - \mathbf{A}\cdot
  d\mathbf{x})
\end{equation}
is the geometrical phase acquired by a particle of charge $q$ moving
between spacetime points $P$ and $Q$ \cite{Wu1975}.  In particular,
if a particle may travel from $P$ to $Q$ along two different paths,
the phase difference $\Delta\gamma$ between the resulting wave
functions is given by
\begin{equation}\label{eq:ABphase}
  \Delta\gamma = \frac{q}{\hbar}\oint(\varphi dt - \mathbf{A}\cdot
  d\mathbf{x}),
\end{equation}
with the integral evaluated around the spacetime loop created by the
two paths, since topologically path $\mathcal{P}_1$ equals path
$\mathcal{P}_2$ plus the loop $(\mathcal{P}_1 - \mathcal{P}_2)$.
This relative phase is observable when the two paths form part of an
interference experiment, and such studies have confirmed
\eqnref{eq:ABphase} for both electric \cite{Oudenaarden1998} and
magnetic \cite{Osakabe1986} potentials.

The `magnetic' part in particular is recognised as an important
property of electronic devices in two-dimensions, since it is
straightforward to construct electron interferometers such as
Aharonov-Bohm rings \cite{Timp1987}, in which electrons may take
either of two paths in traversing a loop in the presence of a
magnetic field. Since the electric potential in this case is
constant (or at least equal for the two paths) there is no
contribution from the first term of \eqnref{eq:ABphase}, and the
remaining loop integral in two spatial dimensions may be rewritten
using Stokes' theorem as
\begin{equation} \label{eq:ABphaseB}
  \Delta\gamma =
  \frac{e}{\hbar}\iint_\mathcal{S}\mathbf{B}\cdot d\mathbf{S} =
  2\pi\frac{\Phi}{\phi_0},
\end{equation}
where $\Phi$ is the magnetic flux through the surface $\mathcal{S}$
enclosed by the loop and $\phi_0 = h/e$ is the quantum unit of
magnetic flux.

In the literature on quantum antidots, this result is often used to
justify the statement that orbital radii (and hence energies via the
background antidot potential) are quantised by the condition
\begin{equation}\label{eq:fluxquant}
\pi \langle r^2\rangle B_\perp= m\phi_0
\end{equation}
for integer $m$. Indeed, from \eqnref{eq:exprsq} we see that the
free-electron eigenstates $\psi_{n,m}^\mathrm{free}$ satisfy
\begin{equation}\label{eq:expAfree}
  \pi\langle r^2\rangle = \frac{h}{eB_\perp}(2n+\abs{m}+1),
\end{equation}
in agreement with this argument.  But while convenient, this
explanation is misleading for several reasons.  First, from
\eqnref{eq:exprsq} we see that the quantity $\pi\langle r^2\rangle$
for parabolic antidot states differs from the free-electron case by
a non-negligible factor $b^{-1}$.  A similar result holds for any
antidot potential: the SP states are pushed outwards relative to the
free-electron states by the repulsive electric field.  Furthermore,
it is straightforward to show that,
even for the free electron states,\footnote{%
From \eqnref{eq:expAfree} we have $\Delta r \simeq \ellB^2/r$
for $|m|\gg 1$.} %
the spacing $\Delta r$ of the wave function maxima for successive
states $(|m|,|m|+1)$ satisfies $\Delta r\ll \ellB$ for large $|m|$.
Since the states have width $\approx\ellB$, it is clear that the
electrons are not in any sense `prohibited' from existing at a
radius which does not satisfy the flux-quantisation condition.

The confusion on this point seems to lie with the interpretation of
\eqnref{eq:ABphaseB}.  The AB phase represents the phase difference
between two wave packets which follow \emph{different} paths through
spacetime.  While this has a natural interpretation for a particle
traversing an AB ring (at low $B$ such that both paths are allowed)
or other interferometer, it is difficult to identify two paths which
may interfere for a particle in a chiral antidot state, or in any
stationary state for that matter.  An electron cannot, for example,
interfere with itself simply by circling the antidot --- it encloses
no area in spacetime and so there can be no interference term.  It
is therefore preferable to think of the SP eigenstates simply as the
natural solutions to Schr\"{o}dinger's equation in the presence of
magnetic confinement rather than a consequence of the AB effect.

Understandably, an argument based on flux quantisation via the AB
effect is usually used to describe the periodicity of so-called
Aharonov-Bohm oscillations which are observed as a function of
magnetic field.  AB oscillations constituted the first experimental
observations of single quantum antidots \cite{Smith1989}, and the
details of their periodicity and line-shapes continue to yield
important insight into the underlying physics in these systems.  It
is well-established experimentally that an antidot with filling
factor $\nuAD=f$ shows periodic conductance resonances in $B$, with
$f$
resonances occurring per base period $\Delta B$,\footnote{%
Note that this does \emph{not} mean that the resonances necessarily
have
period $\Delta B/f$.} %
for $f$ up to at least six \cite{Goldman2008}.  The AB argument is
straightforward: if the flux-quantisation condition of
\eqnref{eq:fluxquant} holds, then it is easy to show that a state
with area $A$ enclosing $m$ flux quanta will be replaced by the
adjacent ($m+1$) state when the field has changed by
\begin{equation} \label{eq:ABperiod}
  \Delta B_\perp = \frac{h}{eA}.
\end{equation}
Since the area is fixed by the Fermi energy and the antidot
potential, we therefore expect periodic resonances for each of the
$f$ spin-polarised LLs as the SP states pass through $\Ef$ with
period $\Delta B$, and hence $f$ resonances per cycle.

In many experiments, \eqnref{eq:ABperiod} is a valid approximation,
although the reasoning used to obtain it should concern the SP
states rather than the AB effect.  In particular, if the antidot
potential varies on a scale much larger than $\ellB$, then it is a
good approximation to use the free particle wave functions
$\psi_{n,m}^\mathrm{free}$ and treat the electric potential as a
small perturbation which simply modifies the SP energies:
\begin{equation}\label{eq:Espapprox}
  E_{nm\sigma}\simeq
  \hbar\omegac\left(n_\mathrm{LL}
      +\frac{1}{2}\right)
      -\sigma\Ez+\langle\varphi(r)\rangle.
\end{equation}
For the parabolic potential we have been considering in this
chapter, this corresponds to the limit $\omega_0\ll\omegac$
($b\approx 1$), and in many experiments this is indeed the
case.\footnote{%
For example, with a typical value of
$\dEsp\approx$~\unit{100}{\micro\electronvolt} at
$B=$~\unit{1}{\tesla}, we have $b\approx 0.9$
from \eqnref{eq:effectiveb}. } %
If so, then we know from \eqnref{eq:expAfree} that the
flux-quantisation does approximately hold, and so the AB period will
in fact be given by \eqnref{eq:ABperiod}.  For stronger antidot
potentials and/or weaker magnetic fields, however, this perturbative
approximation is not valid, and the AB period will be modified.  For
the parabolic antidot, for example, it is in general given by
\begin{equation}
  \Delta B_\perp = \frac{h}{ebA},
\end{equation}
which may be significantly larger than that predicted by
\eqnref{eq:ABperiod}.

Thus far we have been considering a purely non-interacting model,
but as we will see in the next section, Coulomb interactions affect
antidot physics in readily observable ways.   Coulomb blockade, in
particular, plays a fundamental role in the description of
sequential transport of electrons through an antidot.  In the
context of AB oscillations, it is therefore important to consider
the effects of changing magnetic fields on the charge of the system.
If the electrons around the antidot do not rearrange as the field
changes (for instance, if the ground state can be written as a
`maximum density droplet' of holes occupying a fixed set of SP
orbitals --- see \secref{sec:MDDs}), then the excess
charge\footnote{%
Note that although the physical `antidot charge' (i.e.\ number of
holes) remains fixed, the `excess charge' of \eqnref{eq:qexcess} is
unbalanced by the system since the chemical potential, rather than
the number of electrons, is fixed in the 2DES.  We can therefore
think of $\delta q$ as effectively a `gate charge' induced
by the magnetic field.} %
accumulated at the perimeter of the antidot may be approximated as
\begin{equation}\label{eq:qexcess}
  \delta q = (2\pi r\delta r)e n_e ,
\end{equation}
where $n_e$ is the electron density, given by
\begin{equation}
  n_e = \frac{eB}{h}\nuAD .
\end{equation}
For the SP orbitals we have from \eqnref{eq:exprsq} that
\begin{equation}
  \delta r = -\frac{r}{2B}\delta B,
\end{equation}
and so
\begin{equation}
  \delta q = -e\nuAD \frac{r^2}{2\ellB^2}\frac{\delta B}{B}.
\end{equation}
The charging condition $\Delta q = \pm e$ therefore implies a
resonance period
\begin{equation}
  \Delta B = \frac{1}{\nuAD}\frac{h}{e\pi r^2},
\end{equation}
in agreement with our estimate based on the non-interacting energy
levels.  Such periodic charging as a function of $B$ has been
directly measured experimentally, using the conductance of a quantum
point contact as a capacitively-coupled charge sensor
\cite{Kataoka1999}, and in many cases the energy associated with
charging dominates over the SP energy scale.

\section{Theoretical treatment of electron interactions \label{sec:ADIntTheory}}

The preceding sections of this chapter concern a non-interacting
description of antidot electronic states.  In some cases the physics
of the SP model, with the important addition of a charging energy to
reflect Coulomb blockade, is sufficient to describe the major
features of antidot transport measurements.  Especially at
relatively low fields ($B\lesssim$ \unit{1}{\tesla}), the SP model
provides a good description of both equilibrium conductance
experiments (such as the AB oscillations discussed in the previous
section) and the excitation spectra obtained from non-equilibrium
transport measurements.  By including the variability of the tunnel
couplings between the SP states and the current-carrying edge
states, even very complicated conductance traces may be reproduced
accurately within a non-interacting picture \cite{Mace1995}.  At
larger fields, additional structure begins to appear, such as
`double-frequency' AB oscillations
\cite{Ford1994,Sachrajda1994,Kataoka2000,Kataoka2003} and Kondo
resonances \cite{Kataoka2002}, which cannot be described within the
SP framework.  A recent review covering these effects and relevant
theoretical descriptions is given in \refref{Sim2008}.

The experiments discussed in this thesis are almost all in the
low-$B$ regime, and indeed \chapsref{chap:Geometry} and
\ref{chap:TiltedB} present examples of effects which can mostly be
described within a non-interacting model.  Still, when one looks
more deeply, for example through the spin-resolved transport
measurements described in \figref{chap:SpinTransport},
inconsistencies with the SP model emerge which may only be explained
through the introduction of additional physics.  In this section we
consider a microscopic picture of electron-electron interactions
based on Hartree-Fock theory, in which we transform the electronic
system into a `maximum density droplet' of holes in the LLL.  This
model forms an appropriate description of the physics at low to
intermediate $B$, and similar methods have been used recently to
describe the ground-state transitions responsible for the Kondo
effect \cite{Sim2004,Hwang2004}.  An alternate phenomenological
description in terms of capacitive charging interactions has
successfully explained observations at high-$B$ \cite{Sim2003}, as
outlined in the review \cite{Sim2008} cited above.

\subsection{Hartree-Fock theory\label{sec:HFtheory}}

The full Hamiltonian for a system of $N$ interacting electrons,
within the standard Born-Oppenheimer approximation in which the
electronic degrees of freedom are decoupled from those of the
lattice \cite{Ashcroft1976}, can be written in the form
\begin{equation}
  \label{eq:Hinteracting}
  \hat{H} = \sum_i^N\HSP_i +
  \frac{e^2}{4\pi\epsilon\epsilon_0}\sum_{i>j}^N\frac{1}{\abs{\mathbf{x}_i-\mathbf{x}_j}},
\end{equation}
where $\HSP_i$ is the SP Hamiltonian acting on the $i^\mathrm{th}$
electron, which in our case is given by
\begin{equation}\label{eq:Hi}
  \hat{h}_i = \frac{1}{2m^\ast}(-i\hbar\mathbf{\nabla}_i+e\mathbf{A})^2
  -e\varphi(\mathbf{x}_i)-g\mu_\mathrm{B}B\hat{s}_{zi}.
\end{equation}
No analytic solutions are known for a general Hamiltonian of this
form with more than one electron, and much of solid-state physics
concerns various methods for approximating the effect of the
interaction term in \eqnref{eq:Hinteracting}.

The Hartree-Fock (HF) method is one of several `mean field'
approaches to this problem, in which each electron in a system is
influenced by an effective potential due to the charge density of
all the other electrons. In particular, we assume that each electron
in the system is described by its own SP wave function, such that
the multielectron wave function may be written as a Slater
determinant of orthonormal SP spin orbitals $\psi_i$:
\begin{equation}
  \label{eq:SlaterDet}
  \Psi =\frac{1}{\sqrt{N!}}
      \begin{vmatrix}
            \psi_1(\xi_1) & \psi_1(\xi_2) & \cdots & \psi_1(\xi_N) \\
            \psi_2(\xi_1) & \psi_2(\xi_2) & \cdots & \psi_2(\xi_N) \\
            \vdots & \vdots & \ddots & \vdots \\
            \psi_N(\xi_1) & \psi_N(\xi_2) & \cdots & \psi_N(\xi_N)
      \end{vmatrix},
\end{equation}
where $\xi_i$ represents both the position
and spin projection of the $i^\mathrm{th}$ particle.\footnote{%
It is assumed that the orbital and spin parts of the wave function
are separable, i.e.\ that $\psi_i(\xi) =
\psi_{n_i,m_i}(\mathbf{x})\chi_i(s)$, where $\chi=\bigl(
\begin{smallmatrix}
  1 \\ 0
\end{smallmatrix}\bigr)$
or $\bigl(
\begin{smallmatrix}
  0 \\ 1
\end{smallmatrix}\bigr)$ in terms of the argument $s=1,2$ in spin
space. Matrix elements between spin orbitals therefore imply
integration of spatial coordinates and summation
over spin projections.} %
By construction, this wave function satisfies the antisymmetry
requirement for fermions, that is
\begin{equation}\label{eq:Pauli}
  \Psi(\xi_1,\xi_2,\ldots,\xi_i,\ldots,\xi_j,\ldots,\xi_N) =
      -\Psi(\xi_1,\xi_2,\ldots,\xi_j,\ldots,\xi_i,\ldots,\xi_N),
\end{equation}
since the interchange of particles $i$ and $j$ corresponds to the
interchange of two columns of the determinant, and hence a change of
sign. Thus the Pauli exclusion principle is satisfied: the wave
function $\Psi$ vanishes when $\xi_i=\xi_j$ for any $i\neq
j$.\footnote{%
For similar reasons, clearly $\Psi=0$ identically if any
$\psi_i=\psi_j$ for $i\neq j$, enforcing the condition that the SP
spin orbitals chosen must be distinct.} %

General expressions for matrix elements between determinantal wave
functions like \eqnref{eq:SlaterDet} are well known, given for
example in the book by Bethe and Jackiw \cite{Bethe1986}.  In the
expectation value for the energy of $\Psi$, many of the terms vanish
due to the orthogonality of the $\psi_i$, leaving
\begin{equation}
  \langle \Psi \lvert \hat{H}\rvert\Psi\rangle =
      \sum_i\langle i\lvert\hat{h}\rvert i\rangle +
      \sum_{i<j}\Bigl(\langle ij\lvert V_\mathrm{C}\rvert ij\rangle -
          \langle ij\lvert V_\mathrm{C}\rvert ji\rangle\Bigr),
\end{equation}
in terms of the SP energies
\begin{equation}
  \langle i\rvert\hat{h}\lvert i\rangle = \int d\bx\,
      \psi_i(\bx)\hat{h}_i\psi_i(\bx) = \esp_i,
\end{equation}
and two-particle matrix elements of the Coulomb operator
\begin{equation}
  \langle ij\lvert V_\mathrm{C}\rvert kl\rangle  =
      \frac{e^2}{4\pi\epsilon\epsilon_0}
      \delta_{\sigma_i\sigma_k}\delta_{\sigma_j\sigma_l}
      \iint d\bx d\bx^\prime\,
      \psi_i^\ast(\bx)\psi_j^\ast(\bx^\prime)
      \frac{1}{\abs{\bx-\bx^\prime}}
      \psi_k(\bx)\psi_l(\bx^\prime),
\end{equation}
where we have completed the spin summations using the identity
\begin{equation}
  \sum_s\chi_i^\dagger(s)\chi_j(s) = \delta_{\sigma_i\sigma_j}.
\end{equation}
The total energy of the state $\Psi$ may therefore be written in the
form\footnote{%
We have used the facts that $J$ and $K$ are symmetric in $i,j$ and
that $J_{ii}=K_{ii}$ in rewriting the sum in \eqnref{eq:HFEtot}.} %
\begin{equation}\label{eq:HFEtot}
  E = \sum_i\esp_i +
  \frac{1}{2}\sum_{ij}(J_{ij}-\delta_{\sigma_i\sigma_j}K_{ij}),
\end{equation}
where
\begin{eqnarray}
  J_{ij} & = & \frac{e^2}{4\pi\epsilon\epsilon_0}
      \iint d\bx d\bx^\prime\,
      \abs{\psi_i(\bx)}^2
      \frac{1}{\abs{\bx-\bx^\prime}}
      \abs{\psi_j(\bx^\prime)}^2, \label{eq:Jij}\\
  K_{ij} & = & \frac{e^2}{4\pi\epsilon\epsilon_0}
      \iint d\bx d\bx^\prime\,
      \psi_i^\ast(\bx)\psi_j^\ast(\bx^\prime)
      \frac{1}{\abs{\bx-\bx^\prime}}
      \psi_j(\bx)\psi_i(\bx^\prime) \label{eq:Kij}
\end{eqnarray}
are the so-called `direct' and `exchange' Coulomb matrix elements,
respectively.

The part of \eqnref{eq:HFEtot} due to the direct term $J$ is exactly
the Coulomb `overlap' energy we expect between particles occupying a
charge distribution given by $-e\sum_i|\psi_i|^2$, and if we had
used a simple product wavefunction of these orbitals instead of the
Slater determinant, it would be the only contribution from the
Coulomb interaction.  The nonlocal exchange term $K$ therefore
reflects the effects of the wave function antisymmetry introduced
through the Slater determinant. Due to the Pauli exclusion
principle, \eqnref{eq:Pauli}, electrons of the same spin `avoid each
other' more in the antisymmetric $\Psi$ than they would in a simple
product wave function.  This means that the direct Coulomb
interaction actually overestimates the configuration energy, and the
exchange term may be thought of as a correction accounting for the
indistinguishability and antisymmetry of fermions. Slightly more
rigorously, it can be shown \cite{Bethe1986} that the effective
Coulomb interaction of an electron with $N$ other electrons of the
same spin (therefore including both the direct and exchange terms)
is equivalent to the potential from a charge distribution containing
total charge $-e(N-1)$, i.e.\ from one less than the total number of
other electrons. Thus it is as if each electron carries with it a
hole (often called a Fermi hole) which affects its interaction with
other electrons of the same spin.

By the variational principle, the energy $E$ thus obtained for our
choice of wave function $\Psi$ represents an upper bound on the
ground-state energy of the system.  The best approximation of a
single Slater determinant like \eqnref{eq:SlaterDet} to the true
ground state of the system can therefore be found from the
variational condition $\delta\langle \hat{H}\rangle=0$ under
arbitrary variations $\delta\psi_i$.  Using the method of Lagrange
multipliers to enforce the condition that the $\psi_i$ are
normalised,\footnote{%
It can be shown that the $\psi_i$ remain orthogonal as a result of
this calculation \cite{Bethe1986}.} %
this may be written
\begin{equation}
  \frac{\delta}{\delta\psi_i}\left[
      \langle\Psi\rvert\hat{H}\lvert\Psi\rangle
      + \sum_i\varepsilon_i\left(\int\abs{\psi_i(\bx)}^2d\bx - 1\right)
  \right] = 0.
\end{equation}
This procedure leads to the Hartree-Fock equations
\begin{equation}
  \label{eq:HFeqn}
  \begin{split}
  \varepsilon_i\psi_i(\bx)  = \hat{h}\psi_i(\bx)
      & +\sum_j\int d\bx^\prime
          \frac{\abs{\psi_j(\bx^\prime)}^2}{\abs{\bx-\bx^\prime}}\psi_i(\bx)\\
      & \qquad
      -\sum_i\delta_{\sigma_i\sigma_j}\int d\bx^\prime
          \frac{\psi_j^\ast(\bx^\prime)\psi_i(\bx^\prime)}{\abs{\bx-\bx^\prime}}
          \psi_j(\bx),
  \end{split}
\end{equation}
which resemble a set of SP Schr\"{o}dinger equations. Indeed, by
taking the inner product of \eqnref{eq:HFeqn} with
$\psi_i^\ast(\bx)$, we see that the eigenvalue $\varepsilon_i$
represents the part of the total energy $E$ involving the
$i^\mathrm{th}$ electron:
\begin{equation}\label{eq:HFei}
  \varepsilon_i =
  \esp_i+\sum_j(J_{ij}-\delta_{\sigma_i\sigma_j}K_{ij}).
\end{equation}
Notice, however, that the `operators' in the last two terms of
\eqnref{eq:HFeqn} involve the solutions $\psi_i$. The HF equations
must therefore be solved self-consistently, which is normally
accomplished by writing the orbitals $\psi_i$ as an expansion over a
set of basis orbitals $\phi_k$,
\begin{equation}
    \psi_i = \sum_k c_{ik}\phi_k,
\end{equation}
and then solving the resulting linear algebra problem (the
Roothaan-Hall equations \cite{Roothaan1951,Hall1951}) for the
coefficients $c_{ik}$ through an iterative procedure.

\subsection{The particle-hole transformation}

In \chapref{chap:SpinTransport} we will use the Hartree-Fock method
to treat electron-electron interactions in an antidot at filling
factor $\nuAD=2$.  Neglecting mixing with higher
LLs,\footnote{\label{fn:LLmixing}%
Such mixing is not necessarily negligible. With the addition of the
Coulomb term to the Hamiltonian, the LL index $n$ is no longer a
good quantum number, so the true eigenstates will be mixtures of
orbitals with definite $z$-projections of angular momentum, $L_z =
M\hbar$, i.e.\ $\Psi_{M\sigma} =
\sum_{n,m}c_{nm\sigma}\psi_{nm\sigma}$.  The strength of this mixing
depends on the dimensionless ratio $\kappa=E_\mathrm{C}/\hbar\omegac
\sim 1/\sqrt{B}$ between the Coulomb energy scale
$E_\mathrm{C}=e^2/4\pi\epsilon\epsilon_0\ell$ and the LL spacing
$\hbar\omegac$.  For experimentally accessible magnetic fields in
GaAs, $\kappa$ is $O(1)$ (for example, $\kappa \approx 2.5$ for
$B=\unit{1}{\tesla}$), suggesting that LL mixing should not be weak
in these systems.  However, several studies concerned with the
effects of LL mixing with regards to fractional quantum Hall states
\citep[e.g.,][]{Melik-Alaverdian1999,Murthy2002} have concluded that
LL mixing has only a small ($\lesssim 5$\%) effect on measurable
properties such as transport gaps.  More generally, Bishara and
Nayak \cite{Bishara2009} have recently shown that the
\emph{renormalised} effective interactions due to LL mixing in the
first and second LLs are indeed smaller than na\"{i}vely expected,
of order $\lesssim 0.1\kappa$.  We therefore believe that neglecting
this mixing forms a reasonable, if not perfect, approximation.
Although it is beyond the scope of this work, it would certainly be
of interest to theoretically investigate the implications of this
assumption in the future.} %
the natural basis states to use are the LLL wave functions
$\psi_{m\sigma}$ given by \eqnref{eq:psiLLL}\footnote{%
Note we have made the identification $m=q>0$ from
\eqnref{eq:psiLLL}.} %
\begin{equation}
  \psi_{m\sigma} =\frac{1}{\ell}\sqrt{\frac{1}{2\pi 2^{m}m!}}
  \left(\frac{r}{\ell}\right)^{m}e^{-r^2/4\ell^2}e^{-im\phi}\chi_\sigma,
\end{equation}
As we have seen, these wave functions are exact solutions of the SP
Schr\"{o}dinger equation for a parabolic antidot potential, in which
case $\ell = \ellB/\sqrt{b}$ and the SP energies are
\begin{equation}
  \esp_{m\sigma} = \frac{1}{2}b\hbar\omegac -
  \frac{m}{2}(1-b)\hbar\omegac - \sigma \Ez.
\end{equation}
In the case of a non-parabolic but slowly-varying (on the scale of
$\ellB$) potential, the $\psi_{m\sigma}$ are suitable approximations
to the true wave functions, with $\ell\approx\ellB$ and SP energies
given to first order in the antidot perturbation by
\eqnref{eq:Espapprox}.  The Coulomb matrix elements in
\eqnsref{eq:Jij} and \eqref{eq:Kij} between these LLL states may be
evaluated numerically \cite{Stone1992} or analytically \cite{Tsiper2002}.

We therefore have an intuitive picture of an isolated $\nuAD=2$
antidot in terms of a set of electronic occupation vectors
$(\neupvec,\nednvec)$, with components $n^\mathrm{e}_{m\sigma}=0$ or
1 for each orbital $m$.  Since the orbitals extend infinitely into
the bulk, in practice we must choose a cutoff orbital which is far
enough away from the antidot states we are considering not to
influence the calculations.  Alternatively, we may describe the same
configuration in terms of an infinite, spin-split LL containing
finite integer numbers $(\Nhup,\Nhdn)$ of `holes' in the
spin-$\uparrow$ and spin-$\downarrow$ parts, respectively. The
equivalent hole-occupation vectors are therefore
$\mathbf{n}^\mathrm{h}_\sigma = \mathbf{1} -
\mathbf{n}^\mathrm{e}_{\sigma}$, and as we show below, a description
in terms of interactions between holes is essentially equivalent to
the corresponding electron picture, but it removes the need to worry
about the convergence of sums over an infinite number of particles
in calculations.  At this point it is also important to note that,
in general, the states of this `fermionic basis' (characterised by
the occupation numbers of fermion orbitals) are not eigenstates of
the interacting Hamiltonian.  They form a natural basis for
calculations, however, and can be related to the true eigenstates by
a standard procedure, which we discuss further in the next section.

First we consider the electron description at $\nuAD=2$.  Consider
the case of an infinite, fully-filled LL ($n^\mathrm{e}_{m\sigma}=1
\,\forall\; m,\sigma$). According to \eqnref{eq:HFei}, the energy of
the $i^\mathrm{th}$ electron in a Slater-determinant wave function
composed of these SP states is
\begin{equation}\label{eq:Emi}
  E_{m_i\sigma_i} = \esp_{m_i\sigma_i}+U_\mathrm{Sheet}(m_i),
\end{equation}
where
\begin{equation}
    \label{eq:Usheet}
  U_\mathrm{Sheet}(m) = \sum_{n}\left(2J_{m,n}-
      K_{m,n}\right)
\end{equation}
is the Coulomb energy required to add an electron to orbital $m$ (of
either spin) in an otherwise filled LL. The sum over $J_{m,n}$ in
\eqnref{eq:Usheet} does not converge to a finite value, reflecting
the infinite self-energy
of a 2-dimensional sheet of charge.\footnote{%
The matrix element $J_{m,m+q}\sim 1/\sqrt{q}$ for $q\gg1$. } %
We can make the calculations finite by adding a uniform positive
background charge to the system which supplies the potential
$V_\mathrm{BG}(m)$ for an electron in orbital $m$ such that
\begin{equation}
  \Delta U_\mathrm{S}(m) =U_\mathrm{Sheet}(m) - eV_\mathrm{BG}(m)
\end{equation}
is finite.  Specifically, since the Hartree product wave function of
a fully-filled LL has uniform probability distribution,\footnote{%
This is easier to see in another gauge (e.g.\ the Landau gauge in
which the wave functions are `stripes' in one Cartesian coordinate
with arbitrary translations in the other), but since the probability
density must be gauge invariant we know the result holds in the
symmetric gauge as well.} %
we can choose the background potential to precisely cancel the
contribution from the direct term, such that
\begin{equation}\label{eq:deltaUs}
  \Delta U_\mathrm{S}(m) =
  -\sum_nK_{m,n}.
\end{equation}
This is a convergent sum, but it is still over infinitely many
orbital states, and so in using the electron description we must
always be careful that we take the sum far enough to reach
convergence for a given $m$.  To avoid this ambiguity, it is
preferable to transform the infinite electron system into the one
containing a finite number of holes discussed above.

The transformation proceeds as follows.  If for the moment we leave
out the neutralising background charge, we see from \eqnref{eq:Emi}
that by removing (adding) the $1^\mathrm{st}$ electron (hole) from
(to) the state $(m_1,\sigma_1)$, the total configuration energy
changes by $\Delta E_1 = -E_{m_1\sigma_1}$.  If we then remove a
second electron, the resulting change in energy is
\begin{eqnarray}
  \Delta E_2 & = &  -\esp_{m_2\sigma_2} - \sum_{i>1}\left(J_{m_2,m_i}-
      \delta_{\sigma_2,\sigma_i}K_{m_2,m_i}\right) \\
  & = & -\esp_{m_2\sigma_2} - U_\mathrm{Sheet}(m_2) + J_{m_2,m_1}-
      \delta_{\sigma_2,\sigma_1}K_{m_2,m_1}.
\end{eqnarray}
Following this pattern, the state with $N_\mathrm{h}$ holes has
energy (relative to the fully-filled state)
\begin{equation}
\label{eq:Un}
  U(N_\mathrm{h})  = -\sum_i^{N_\mathrm{h}} \esp_{m_i\sigma_i} -
      \sum_i^{N_\mathrm{h}}U_\mathrm{Sheet}(m_i) +
      \sum_{i>j}^{N_\mathrm{h}}\left(J_{m_i,m_j}-
      \delta_{\sigma_i,\sigma_j}K_{m_i,m_j}\right).
\end{equation}
Now, choosing the background potential as in \eqnref{eq:deltaUs} to
cancel the infinite part of $U_\mathrm{Sheet}$, we can write the
configuration of the $N$-hole state as (rewriting the sum in
\eqnref{eq:Un} using the symmetry of $J$ and $K$)
\begin{equation}
  U(N_\mathrm{h}) = - \sum_i^{N_\mathrm{h}} \tilde{\varepsilon}_{m_i\sigma_i} +
      \frac{1}{2}\sum_{i,j}^{N_\mathrm{h}}(J_{m_i,m_j}-
      \delta_{\sigma_i,\sigma_j}K_{m_i,m_j}),
\end{equation}
where
\begin{equation}\label{eq:effesp}
  \tilde{\varepsilon}_{m_i\sigma_i}=\esp_{m_i\sigma_i}
      -\sum_n K_{m_i,n}.
\end{equation}
In practice it is usually a good approximation to treat the exchange
corrections to $\esp_{m\sigma}$ as constant for all $m$ (in which
case they can be absorbed into the definition of the background
charge), since the variation of \eqnref{eq:deltaUs} for different
$m$ is generally orders of magnitude smaller than the variation in
$\esp_{m\sigma}$. In terms of the hole orbital-occupation vectors
$\mathbf{n}^\mathrm{h}_\sigma$, the total HF configuration energy of
a state in the fermionic basis may be written in the form convenient
for calculations,
\begin{multline}\label{eq:ConfigEmatform}
  U(\nhupvec,\nhdnvec) = -\tilde{\mathbf{\varepsilon}}\cdot(\nhupvec+\nhdnvec) \\
      + \frac{1}{2}\left[(\nhupvec+\nhdnvec)^\mathrm{T}\mathrm{J}(\nhupvec+\nhdnvec) -
      (\nhupvec)^\mathrm{T}\mathrm{K}(\nhupvec) -
      (\nhdnvec)^\mathrm{T}\mathrm{K}(\nhdnvec)\right].
\end{multline}
Aside from the sign of the SP energies, reflecting the confining
property of the antidot potential for holes, and the aforementioned
exchange correction, this is entirely equivalent to the interaction
energy of a finite system of electrons.

\subsection{Maximum density droplets\label{sec:MDDs}}

As alluded to above, the fermionic basis states discussed thus far
are in general not eigenstates of the interacting Hamiltonian,
\eqnref{eq:Hinteracting}.  This is because the SP angular momentum
operators $\hat{L}_{zi}$, acting on the $i^\mathrm{th}$ electron
only, do not commute with the electron-electron interaction term.
The total system is still rotationally symmetric, however, and so
the total angular momentum projection  $M = \sum_{m\sigma}
mn^\mathrm{h}_{m\sigma}$ is a good quantum number of the
multiparticle state. Similarly, although the individual spin
operators $\hat{s}_{zi}$ do commute with the Hamiltonian, our choice
of a Slater-determinant wave function couples the individual spins
to the spatial symmetry (and hence the energy) of the state.  Hence
we must consider instead the total spin projection $S_z =
\frac{1}{2}\sum_m
(n^\mathrm{e}_{m\uparrow}-n^e_{m\downarrow})$.\footnote{%
Note that in our convention $S_z$ is defined in terms of electron
numbers, such that $S_z =
\frac{1}{2}(N^\mathrm{h}_\downarrow-N^\mathrm{h}_\uparrow)$.  Total
angular momentum, however, is defined in terms of hole occupation
$M=\sum_{m\sigma}mn^\mathrm{h}_{m\sigma}$ such that it will be
finite (the sign is not important since we are considering
excitations from the MDD state, which depend on $\abs{\Delta M}$).} %
The eigenenergies of the system may therefore be obtained by
diagonalising the matrix Hamiltonian constructed from the subspace
of fermionic basis states with a given $(M,S_z)$, using the rules
for addition of angular momentum.  This process leads to a `bosonic'
basis \cite{Stone1992}, in which the neutral excitations are
described by a spectrum of `edge waves' similar to the
one-dimensional Tomonaga-Luttinger liquid model
\cite{Luttinger1963,Tomonaga1950}.

\begin{figure}[tb]
    \centering
    \includegraphics[]{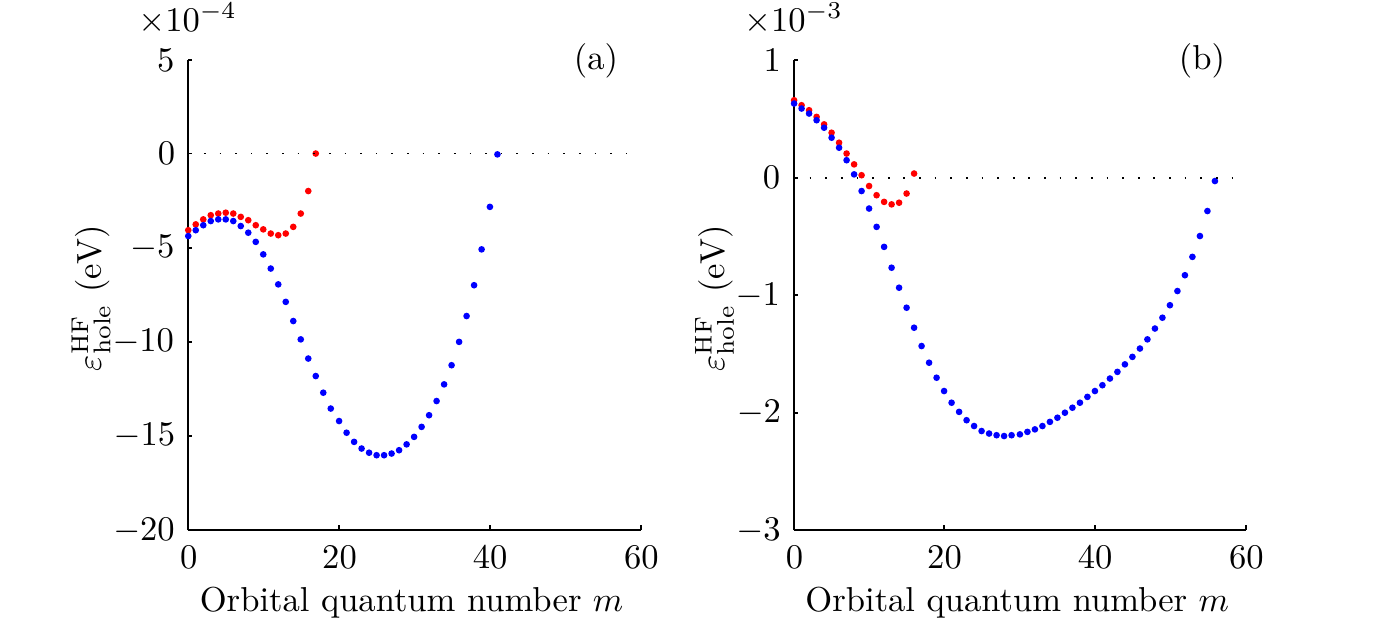}
    \caption[MDD stability]{%
Self-energy of holes in an MDD confined by a parabolic antidot
potential computed according to \eqnref{eq:HFei2}, showing both
stable (a) and unstable (b) configurations.  The MDD is stable if
all of the spin-$\uparrow$ (red) and spin-$\downarrow$ (blue) holes
have lower self-energy than the outermost states, which is shown by
a dotted line.  The spin $S_z$ used in each plot is the ground-state
spin for an MDD with $N_\mathrm{h}$ holes. As seen above, typically
the antidot becomes highly spin polarised before the MDD
configuration becomes unstable. Calculation parameters are chosen to
match the device used in \chapref{chap:SpinTransport}:
$\Rad=\unit{400}{\nano\metre}$, $B_\perp=\unit{0.6}{\tesla}$, $\Ez =
\unit{30}{\micro\electronvolt}$, with $\dEsp =
\unit{350}{\micro\electronvolt}$ in (a) and
\unit{300}{\micro\electronvolt} in (b).  The Coulomb interaction is
at its full value ($\etaC=1$). \label{fig:MDDstability}}
\end{figure}

In the particular case of a `maximum density droplet' (MDD),
however, the fermionic basis states we have been considering are
actually exact eigenstates of the interacting Hamiltonian (within
the approximation that mixing from higher LLs can be neglected ---
see footnote~\vref{fn:LLmixing}).  A hole MDD is defined by the
total number of holes, $N_\mathrm{h}$, and its spin, $S_z$, such
that (recall that $m=0,1,2,\ldots$)
\begin{equation}
  n^\mathrm{h}_{m\sigma} =
  \begin{cases}
    1 & \text{for $m\leq N^\mathrm{h}_\sigma-1$},\\
    0 & \text{otherwise},
  \end{cases}
\end{equation}
where
$N^\mathrm{h}_{\uparrow/\downarrow}=\frac{1}{2}(N^\mathrm{h}\mp
2S_z)$.  The total angular momentum of this state,
$M=\Nhup+\Nhdn-2$, is the minimum value allowed by the Pauli
exclusion principle, and this is the \emph{only} such configuration
with spin $S_z$ and total angular momentum $M$, so it is therefore
an eigenstate of both $L_z$ and the Hamiltonian. If this
configuration is stable (i.e.\ any states with higher $M$ have
greater energy), then it must be the ground state of the system for
given $S_z$.   The stability of the MDD is controlled by the
interplay between the repulsive Coulomb interaction and the
attractive (for holes) antidot potential.  Since hole states at
higher $m$ occur at larger radii, larger antidot confinement favours
the MDD as the ground state, but of course holes farther away from
the antidot experience a reduced Coulomb potential from the
remaining holes, favouring a different configuration. For a given
set of parameters, we can test the stability of the MDD by
considering the energy associated with each particle in the system,
\begin{equation}\label{eq:HFei2}
  \ehf_{m_i\sigma_i} =- \tilde{\varepsilon}_{m_i\sigma_i} +
      \sum_{j}^{N_\mathrm{h}}(J_{m_i,m_j}-
      \delta_{\sigma_i,\sigma_j}K_{m_i,m_j}).
\end{equation}
If all of these satisfy the condition
\begin{equation}\label{eq:MDDstablecondition}
  \ehf_{m_i\sigma_i}\leq \ehf_{(N^\mathrm{h}_\sigma-1)\sigma},
\end{equation}
then the MDD will be the stable ground state for the spin $S_z$.
Examples of both stable and unstable ground states are shown in
\figref{fig:MDDstability}.

In calculations, we control the strength of the confining potential
through the SP energy spectrum, and add an additional parameter
$\etaC$ multiplying the Coulomb term in the Hamiltonian, which we
can then use as a `knob' to control the strength of
electron-electron interactions. In comparing directly with
experiments, we find it is often necessary to set $\etaC\approx
0.1$.  This is certainly unsatisfying, and reflects the notorious
difficulties involved with quantitative comparisons between
theoretical predictions of energy gaps and real experimental data
\citep[e.g.,][]{Shukla2000}.  Several factors likely contribute to
the `softening' of the Coulomb interactions in real devices,
including LL mixing, screening by nearby gates and compressible
regions of the 2DES, and the finite thickness of the electron wave
functions in the growth direction, all of which are ignored in our
calculations.  We believe that it is not unreasonable to believe
that a combination of these factors might reduce the Coulomb
interaction between electrons by an order of magnitude from its bare
value.

\chapter{Transport Theory\label{chap:TransportTheory}}

\ifpdf
    \graphicspath{{Chapter2/Figures/PNG/}{Chapter2/Figures/PDF/}{Chapter2/Figures/}}
\else
    \graphicspath{{Chapter2/Figures/EPS/}{Chapter2/Figures/}}
\fi


From an experimental point of view, we have frustratingly little
access to the physics of microscopic quantum systems at the bottom
of a dilution refrigerator.  Without the addition of complicated
additional equipment such as optical or scanning probes, we are
limited to transport experiments, through which we can measure two
things: electrical currents and voltages.  It is therefore the
physicist's task to make connections between the device physics we
would like to understand and a measurable current or voltage. To
accomplish this task, we have a wide variety of experimental `knobs'
to vary, in the form of device parameters, electric and magnetic
fields, and temperature.  In this chapter we briefly review the
theory of transport in two-dimensional quantum electronic systems,
which we shall need to interpret our experimental results in terms
of the antidot physics presented in \chapref{chap:ADtheory}.

\section{The Landauer-B\"{u}ttiker formalism \label{sec:LBformalism}}

It is often convenient to treat transport in mesoscopic electronic
devices through a scattering framework, in which currents injected
to and from the active region are resolved into a set of known
eigenfunctions, or `modes', in the leads which connect the system to
the outside world.  If these lead modes form a complete basis, then
the transport properties of the device may be described in terms of
an $S$-matrix, composed of scattering amplitudes which give the
`connections' between each mode in every lead of the system.  In
many cases the form of these lead modes is obvious, such as when a
coherent device is probed through a set of quantum point contacts
(QPCs) which are naturally modelled as one-dimensional quantum
wires. We have already seen in \chapref{chap:ADtheory} how a
perpendicular magnetic field splits the density of states of a 2DES
into a set of LLs, which form localised states throughout the bulk
of a device and chiral edge states along the sample edges.  These
edge modes provide a set of one-dimensional eigenfunctions for the
scattering states, and, as pointed out by B\"{u}ttiker
\cite{Buttiker1986,Buttiker1988}, their chirality often leads to
enormous simplifications of the $S$-matrix, such that the transport
properties of even very complicated devices may be evaluated
straightforwardly with only a little algebra.

\begin{figure}[tb]
\begin{center}
\includegraphics[width=5truein]{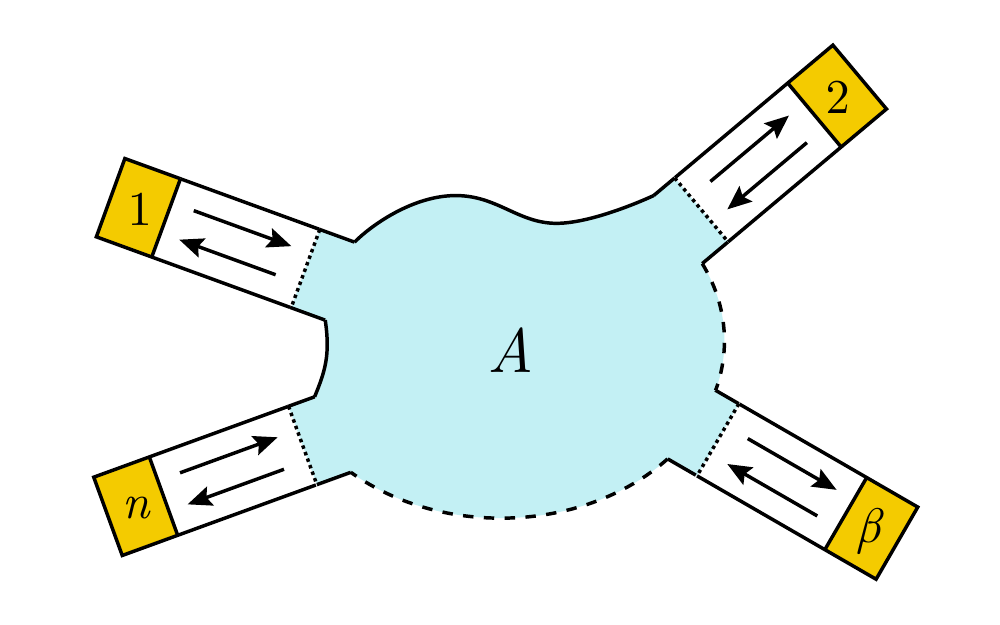}
\caption[Multi-terminal device]{A general multi-terminal device, in
which a set of $n$ ideal leads connect to an arbitrary elastic
scattering centre, or `active region,' $A$, which is described by a
unitary $S$-matrix. We consider the case where all of the leads and
the bulk have the same filling factor $N$, although this
simplification is not strictly necessary.
\label{fig:MultiprobeDevice}}
\end{center}
\end{figure}

The work of B\"{u}ttiker resulted from efforts to explain the
experimental observations of the quantum Hall effect
\cite{Klitzing1980}, in which four-terminal electrical measurements
can provide exactly quantised resistance/conductance values even in
macroscopic samples which are clearly not phase-coherent throughout.
He extended the earlier work by Landauer
\cite{Landauer1957,Landauer1970} (reformulated around the same time
as B\"{u}ttiker's own work in Ref.~\cite{Stone1988}) to consider a
multi-terminal geometry, in which a set of perfect leads connect to
an arbitrary elastic scattering centre,\footnote{%
Additional features of real devices may be incorporated as well. For
example, non-ideal ohmic contacts which produce non-equilibrium
populations of edge modes may be modelled as ideal contacts
separated from the lead by appropriate elastic scattering centres,
and inelastic scattering between modes along the same edge can be
included by adding additional `ideal voltage probes' to the system,
which act as inelastic scatters and fully equilibrate the
populations within an edge.} %
as shown in \figref{fig:MultiprobeDevice}.
Here we consider the case where the active region and all of the
leads are characterised by the same filling factor $N$, although it
is straightforward to treat situations (such as the
selective-injection device studied in \chapsref{chap:SpinTransport},
\ref{chap:SpinTransportModel} and \ref{chap:TiltedB}) in which the
number of modes in each lead is different.

The eigenfunctions of the leads are the solutions to the
Schr\"{o}dinger equation for two-dimensional electrons in the
presence of a perpendicular magnetic field $B$, with the SP
Hamiltonian given by \eqnref{eq:HinBfield}.  In contrast to our
approach to zero-dimensional antidot states in
\chapref{chap:ADtheory}, the one-dimensional states are best
described in the Landau gauge $\mathbf{A} = (-By,0,0)$, where we
choose Cartesian coordinates with $x$ along the direction of the
lead,\footnote{%
It is in general possible to choose a consistent gauge for the
system as a whole such that $\mathbf{A}$ is given in the Landau
gauge asymptotically for each
lead, as described in Appendix~E of Ref.~\cite{Baranger1989}.} %
such that
\begin{equation}
  \hat{H} = \frac{1}{2m^\ast}\left[ (\hat{p}_x-eBy)^2+\hat{p}_y^2\right] -
  e\varphi(y),
\end{equation}
where $\varphi(y)$ is the confining potential of the lead.  The
solutions in this gauge are propagating waves $\psi_n =
e^{ikx}\phi_n(y)$, where $\phi_n(y)$ solves the one-dimensional
eigenvalue equation
\begin{equation}\label{eq:SWELandau}
  \left[-\frac{\hbar}{2m^\ast}\frac{\partial^2}{\partial
  y^2}+\frac{1}{2}m^\ast\omegac^2(y-y_0)^2 -
  e\varphi(y)\right]\phi_{n,k}(y) = E_{n,k}\phi_{n,k}(y),
\end{equation}
with the $k$-dependent parameter $y_0$ given by
\begin{equation}
  y_0 = \frac{\hbar k}{eB} = \ellB^2k.
\end{equation}
In a region with a uniform potential, $\varphi\equiv0$,
\eqnref{eq:SWELandau} has harmonic oscillator solutions with
\begin{equation}
  E_{n,k} = \hbar\omegac\left(n+\frac{1}{2}\right),
\end{equation}
where $n$ is the LL index as expected, and $\phi_{n,k}$ is localised
near $y=y_0$ with a length scale given by $\ellB$.  Exactly as we
found in \chapref{chap:ADtheory}, if the potential $\varphi(y)$
varies slowly on the scale of $\ellB$, then the true solutions will
be similar to the free-electron wave functions, with eigenenergies
$E_{n,k}$ given as a function of $y_0$:
\begin{equation}
  E_{n,k} = E(n,B,y_0(k)).
\end{equation}
The total number $N$ of occupied modes is then determined by the
Fermi energy, $\Ef$, through the maximum value of the LL index $n$
for which $E_{n,k}<\Ef$ at the centre of the lead.  These states are
analogous to the semiclassical `skipping orbits' we considered for
the antidot SP states in \secref{sec:ADeigenstates}, except that the
spectrum within each LL $n$ is continuous, labelled by the
wavevector $k$, rather than discrete.  As for the antidot states,
they have a group velocity along the edge as a result of the
confining potential,
given by\footnote{%
Semiclassically, for a perturbative confining potential such that
$E_{n,k}\simeq-e\varphi(y_0)+\hbar\omegac(n+\frac{1}{2})$,
\eqnref{eq:vnk} gives the expected result for
$\mathbf{E}\wedge\mathbf{B}$ drift:
$v_{n,k}\simeq \abs{\mathbf{\nabla}\varphi}/\abs{B}$.} %
\begin{equation}\label{eq:vnk}
  v_{n,k} = \frac{1}{\hbar}\frac{dE_{n,k}}{dk} =
      \frac{1}{\hbar}\left(\frac{dE_{n,k}}{dy_0}\right)\left(\frac{dy_0}{dk}\right) =
      \frac{1}{eB}\frac{dE_{n,k}}{dy_0}.
\end{equation}
The current carried by mode $n$ within an energy window $\Delta\mu$
(at zero temperature) is then given in terms of the group velocity
and the density of states $dn/dE$,
\begin{equation}
  I_n = ev_{n,k}\left(\frac{dn}{dE}\right)\Delta\mu,
\end{equation}
and since $dn/dk = 1/2\pi$ in one dimension,\footnote{%
For a system of length $L$, each state occupies a length $2\pi/L$ in
$k$-space, such that the total number of states is $N = k(L/2\pi)$,
and hence $dn/dk = 1/2\pi$.} %
we have
\begin{equation}
  \frac{dn}{dE} =
  \left(\frac{dn}{dk}\right)\left(\frac{dk}{dE}\right)
      = \frac{1}{2\pi\hbar v},
\end{equation}
such that
\begin{equation}
  I_n = \frac{e}{h}\Delta\mu.
\end{equation}
Each mode in the lead therefore carries a current proportional to
its chemical potential.

This result allows us to directly connect the experimentally
measured currents and voltages (related to the chemical potential
through $\mu = -eV$) to a simple scattering problem.  B\"{u}ttiker's
central result is an expression for the current in lead $\alpha$ as
a function of the chemical potentials in each lead:
\begin{equation}
  \label{eq:LBcurrent}
  I_\alpha = \frac{e}{h}\biggl[(N-R_\alpha)\mu_\alpha -
      \sum_{\beta\neq\alpha}T_{\alpha\beta}\,\mu_\beta\biggr],
\end{equation}
where $T_{\alpha\beta}$ are the transmission coefficients for
current to pass from lead $\beta$ to lead $\alpha$, and $R_\alpha =
T_{\alpha\alpha}$ is the reflection coefficient for current to
return to lead $\alpha$ after entering the device.  As we will see
in more detail in the next section, these coefficients represent the
summed scattering probabilities of all the modes in the system,
given by
\begin{equation}\label{eq:TabDefn}
  T_{\alpha\beta} = \sum_{n,m}^N\Abs{t_{\alpha\beta,nm}}^2,
\end{equation}
where $t_{\alpha\beta,nm}$ is the scattering probability amplitude
(an element of the $S$-matrix) for a transition from mode $m$ in
lead $\beta$ to mode $n$ in lead $\alpha$, evaluated at the Fermi
energy.

In an experiment, the leads are part of a larger electronic circuit,
and are typically used in one of two ways.  Either we fix the
voltage (and hence $\mu$) in order to source or measure a current,
or we fix the current by attaching the lead to a known current
source. Voltage probes are a special case of the latter method, for
which $I=0$, and hence \eqnref{eq:LBcurrent} gives
\begin{equation}\label{eq:Vvprobe}
  V_\alpha =
  \frac{\sum_{\beta\neq\alpha}T_{\alpha\beta}V_\beta}{N-R_\alpha}.
\end{equation}
Alternatively, current-measuring probes are typically set to ground
potential ($\mu=0$), and therefore measure a current given by
\begin{equation}\label{eq:Iiprobe}
  I_\alpha =
  \frac{e^2}{h}\sum_{\beta\neq\alpha}T_{\alpha\beta}V_\beta,
\end{equation}
which is independent of the probe's reflection coefficient.

As mentioned previously, the key simplification of the quantum Hall
regime results from the \emph{chirality} of the edge modes.  Unless
the edge modes originating from different contacts are brought
within direct tunnelling distance in the active region of the
device, there is a very low probability that an electron can leave a
given edge, even though it experiences many scattering events
between the various modes on that edge.  If all the edges in
\figref{fig:MultiprobeDevice} remain separated, all of the current
in the $N$ modes leaving a contact will enter the next one, such
that, for clockwise-circulating edge modes, $T_{\alpha+1,\alpha}=N$,
and all other $T_{\alpha\beta}=0$ (including the $R_\alpha$).
Combined with \eqnref{eq:LBcurrent}, this scenario leads directly to
the observed four-terminal resistances which define the quantum Hall
effect \cite{Buttiker1988}.  For an arbitrary device in the QH
regime, it is usually straightforward to identify the transmission
coefficients in a similar manner, and then to work out the desired
currents and voltages algebraically from \eqnref{eq:LBcurrent} using
the known currents and voltages of the experimental circuit.

\begin{figure}[tb]
\begin{center}
\includegraphics[]{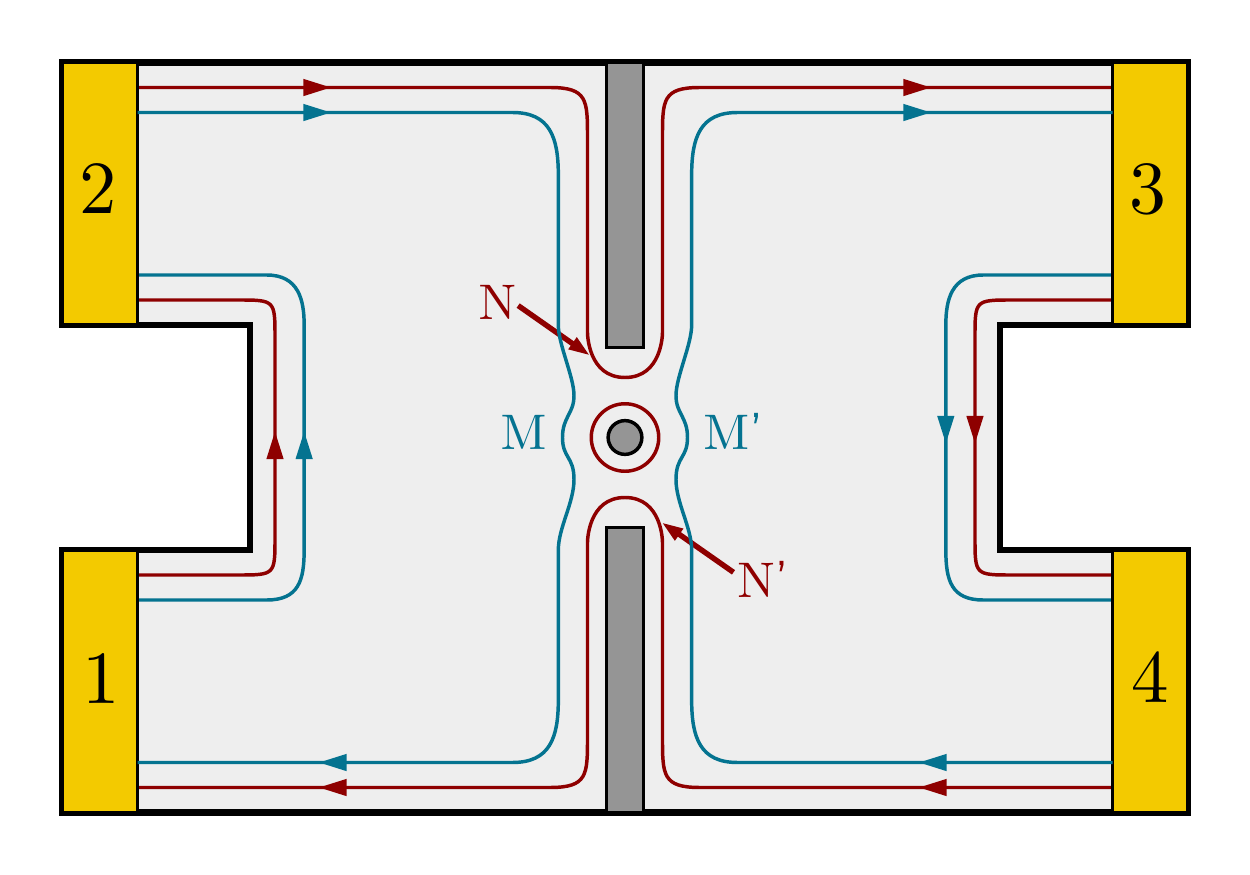}
\caption[Antidot edge-mode network]{Schematic diagram showing the
edge-mode network for a four-terminal measurement of a single
antidot. \label{fig:4TAD}}
\end{center}
\end{figure}

Consider as an example a typical antidot device as sketched in
\figref{fig:4TAD}.
We split the $\nuB$ total modes of the upper (lower) edge into the
$N$ ($N^\prime$) which flow through the upper (lower) constriction
and the $M$ ($M^\prime$) which are reflected. We consider the case
in which the constrictions are tuned symmetrically, i.e.\ $N =
N^\prime = \nuAD$ (the primed and unprimed notation allows us to
distinguish between the modes at the antidot which originated from
contacts 4 and 2, respectively). In a standard four-terminal
measurement, we apply a voltage $V_0$ to lead 1 to drive a current
through the device, measure the current flowing out of lead 3
(setting $V_3=0$), and measure the voltage difference $V_{24} =
V_2-V_4$ between the contacts in leads 2 and 4. We first consider
the voltage probes 2 and 4, described by \eqnref{eq:Vvprobe}.
Considering the topology of the edge modes, we immediately see that
\begin{equation}
  T_{2\alpha} =
      \begin{cases}
        \nuB  & \text{for $\alpha=1$}\\
        0     & \text{otherwise},
      \end{cases}
\end{equation}
and similarly $T_{43}=\nuB$ with all other $T_{4\alpha}=0$.  Thus
\eqnref{eq:Vvprobe} implies that $V_2=V_0$ and $V_4 = 0$, such that
the measured voltage $V_{24}=V_0$.  Then, to compute the
four-terminal conductance of the device,
\begin{equation}
  G_{13,24} = \frac{I_{3}}{V_{24}} \qquad \text{(with current source at 1),}
\end{equation}
we need only work out the current in probe 3, which is given by
\eqnref{eq:Iiprobe} with only one nonvanishing term,
\begin{equation}
  I_3 =\frac{e^2}{h}T_{32}V_0.
\end{equation}
The conductance is therefore given by\footnote{%
N.B., although the scattering probabilities for any
\emph{particular} scattering event depend on the phase information
included in the scattering amplitudes $t_{\alpha\beta,nm}$, the
measured quantities $I$ and $V$ are time-averaged, and so depend
only on the probabilities $\abs{t_{\alpha\beta,nm}}^2$ because the
relative phases of the input states are uncorrelated.} %
\begin{equation}
  G_{13,24} = \frac{e^2}{h}\sum_{n,m}^{\nuB}\Abs{t_{32,nm}}^2,
\end{equation}
in terms of the individual scattering probabilities between the
modes originating from contact 2 and those which eventually enter
contact 3.

We can express this result in a slightly more transparent form by
defining a set of antidot scattering probabilities $p_{m^{\prime}n}
= p_{nm^\prime}$ for transitions between modes $n\in(N,M)$ and
$m^\prime\in(N^\prime,M^\prime)$.  Recognising that, in the absence
of scattering through the antidot, $T_{32}=N=\nuAD$, and using the
fact that $T_{32}+T_{12}=T_{14}+T_{34}=\nuB$, we obtain
\begin{equation}\label{eq:GAD4T}
  G_{13,24} = \frac{e^2}{h}\left(\nuAD+
      \sum_{n\in M, m^\prime\in M^\prime}p_{nm^\prime} -
      \sum_{n\in N,m^\prime\in N^\prime}p_{nm^\prime}\right).
\end{equation}
With this formulation, we have a natural interpretation of the first
term as the base conductance of the constrictions, and the second
and third as `forward-scattering' between states $(M,M^\prime)$ and
`back-scattering' between states $(N,N^\prime)$ which result in
increased and decreased conductance, respectively.\footnote{%
Note that several potential scattering events, such as transitions
between modes in states $(N,M)$ or $(N^\prime,M^\prime)$, do not
appear in \eqnref{eq:GAD4T}, since they do not change the current
which flows through the device.  They do still occur, of course, and
are measurable with the selective injection/detection technique
presented in \chapref{chap:SpinTransport}.} %
In the specific case of $\nuAD=2$, if we assume spin-conservation
and that only the modes of the second LL ($\nu=3,4$) contribute to
forward-scattering, then the antidot conductance is given by
\begin{equation}
  G_{13,24} =
  \frac{e^2}{h}\left(2+p_{33^\prime}+p_{44^\prime}-
      p_{11^\prime}-p_{22^\prime}\right).
\end{equation}
Thus, the Landauer-B\"{u}ttiker formalism reduces the problem of
modeling electrical transport through an antidot to the evaluation
of the individual antidot scattering probabilities, which reflect
the antidot physics in which we are interested.  Several approaches
exist for this purpose, a few of which we will discuss in the
remainder of this chapter. Natural extensions of the procedure
described in this example allow the deconstruction of more
complicated devices and measurement setups, such as those we
consider in \chapref{chap:SpinTransport} of this work.

\section[Green's functions]{Green's functions and linear-response
theory\label{sec:GFs}}

We have seen in the previous section how the linear-response
conductance of the device may be written in terms of the scattering
probabilities between the eigenfunctions of the leads through the
Landauer-B\"{u}ttiker formula, \eqnref{eq:LBcurrent}.  Around the
same time as B\"{u}ttiker's own work, Baranger and Stone
\cite{Baranger1989} developed a more general theory of electrical
linear-response for mesoscopic devices, through which the individual
scattering amplitudes $t_{\alpha\beta,nm}$ are expressed in terms of
the time-independent Green's function for the device, evaluated at
the Fermi energy. As we shall discuss below, efficient algorithms
exist for the calculation of such Green's functions for arbitrary
potentials, and so this provides a flexible method for modelling
ballistic transport in coherent electronic devices, which we will
use in \chapref{chap:Geometry} to investigate the effects on the
antidot SP spectrum of the asymmetry introduced by the QPC in which
the antidot is embedded.

\subsection{Definitions}

Green's functions, or propagators, provide a powerful and versatile
approach to scattering problems.  They supply a framework within
which to include arbitrary potentials and many types of more
complicated interactions.  Here we consider only a time-independent,
non-interacting theory in the linear-response regime, since this
will be useful to us in \chapref{chap:Geometry}, but applications of
the general technique
are in principle much broader.\footnote{%
See, for example, the recent review by Sim et al.\ on the topic of
interactions in antidots \cite{Sim2008} for a somewhat more general
formulation of the antidot scattering problem.} %
For example, Green's functions including interactions through local
spin-density functional theory have been used recently
\cite{Ihnatsenka2006a} to investigate the formation of compressible
regions around antidots. We do not attempt a systematic review of
the properties of Green's functions here, and simply quote many of
the results that we shall need, generally following the treatment of
Baranger and Stone \cite{Baranger1989}.  Many introductory quantum
mechanics textbooks  \citep[e.g.,][]{Griffiths1995} treat
time-independent scattering theory through a Green's function
approach, and for more detail on Green's functions in particular,
the book by Economou \cite{Economou2006a} is quite a good reference.

Fundamentally, a time-independent Green's function is defined as the
solution to the differential equation
\begin{equation}\label{eq:GFdefn}
  \bigl[z-\mathcal{L}(\bx)\bigr]G(\bx,\bx^\prime;z) =
  \delta(\bx-\bx^\prime),
\end{equation}
where $z = \lambda\pm i\eta$ is a complex variable, and
$\mathcal{L}$ is a linear Hermitian operator with eigenvalues
$\lambda_n$ and eigenfunctions $\phi_n(\bx)$,
\begin{equation}
  \mathcal{L}(\bx)\phi_n(\bx) = \lambda_n\phi_n(\bx),
\end{equation}
which are orthonormal and complete, i.e.,
\begin{equation}\label{eq:OrthAndComplete}
  \int \phi_n^\ast(\bx)\phi_m(\bx)d\bx = \delta_{n,m}
      \qquad\text{and}\qquad
      \sum_n \phi_n(\bx)\phi_n(\bx^\prime) =
      \delta(\bx-\bx^\prime).
\end{equation}
From \eqnref{eq:GFdefn} it is clear that, if $u(\bx)$ solves the
inhomogeneous equation
\begin{equation}\label{eq:inhomoeqn}
  \bigl[z - \mathcal{L}(\bx)\bigr]u(\bx) = f(\bx),
\end{equation}
then $u(\bx)$ is given by\footnote{%
Note that a slightly different definition must be used if
$z=\lambda_n$ where $\lambda_n$ is one of the eigenvalues of
$\mathcal{L}$.  See, for example, \eqref{eq:LippmannSchwinger} later
in this section.} %
\begin{equation}\label{eq:GFinhomosoln}
  u(\bx) = \int G(\bx,\bx^\prime;z) f(\bx^\prime)\,d\bx^\prime,
\end{equation}
since acting on the above equation with
$\bigl[z-\mathcal{L}(\bx)\bigr]$ returns \eqnref{eq:inhomoeqn}
through the definition of $G$ in \eqnref{eq:GFdefn}.  So we can
think of $G(\bx,\bx^\prime;z)$ as the inverse of the differential
operator $\bigl[z-\mathcal{L}(\bx)\bigr]$, in the sense given by
\eqnref{eq:GFinhomosoln}.

We consider the same general scattering problem as in the previous
section, depicted in \figref{fig:MultiprobeDevice}, in which a set
of ideal metallic leads connect to the active region $A$, where the
potential is given by $U(\bx) = -e\varphi(\bx)$.  The Green's
function we need is that defined by the full Hamiltonian for the
active region,
\begin{equation}
  \mathcal{L} = \hat{H}(\bx) =  \hat{H}_0 + U(\bx),
\end{equation}
composed of both the potential $U(\bx)$ and the free-particle part,
\begin{equation}
  \hat{H}_0 = \frac{1}{2m^\ast}(\hat{\mathbf{p}}+e\mathbf{A})^2.
\end{equation}
The eigenfunctions $\psi_a$ of this Hamiltonian are the wave
functions of the complete problem, i.e., the solutions to
Schr\"{o}dinger's equation
\begin{equation}\label{eq:SWE2}
  \hat{H}\psi_a(\bx) = \varepsilon_a\psi_a(\bx),
\end{equation}
where $a$ represents a complete set of quantum numbers.\footnote{%
For simplicity we ignore spin as in the previous section.  In the
absence of any spin-mixing, spin is easily included by adding a
spin-dependent constant to the potential $U(\bx)$ corresponding to
the Zeeman energy.} %
If we rewrite the Schr\"{o}dinger equation using
\eqnref{eq:GFinhomosoln}, we obtain
\begin{equation}
  \psi_a(\bx) = \int
  G(\bx,\bx^\prime;z)(z-\varepsilon_a)\psi_a(\bx^\prime)d\bx^\prime,
\end{equation}
and then using the orthonormality of the eigenfunctions,
\eqnref{eq:OrthAndComplete}, we find that we can express the Green's
function in terms of the wave functions,
\begin{equation}\label{eq:GFfrompsi}
  G(\bx,\bx^\prime;z) =
  \sum_a\frac{\psi_a(\bx)\psi^\ast_a(\bx^\prime)}{z-\varepsilon_a}.
\end{equation}
From this definition it is obvious that $G(\bx,\bx^\prime;z)$ is not
analytic if $z$ is equal to any of the eigenvalues $\varepsilon_a$
(which may have a continuous spectrum), and so we define instead the
limits
\begin{equation}\label{eq:GFpmdefn}
  G^\pm(\bx,\bx^\prime;E) =
  \lim_{\eta\rightarrow0^+}G(\bx,\bx^\prime;E\pm i\eta),
\end{equation}
where $E$ and $\eta$ are real, which are known as the retarded $(+)$
and advanced $(-)$ Green's functions, respectively.

Besides the link between the Green's function of a device and the
transport properties we discuss below, $G^\pm(\bx,\bx^\prime;E)$ is
also directly related to the density of states and in particular the
\emph{local density of states}, given by
\begin{equation}\label{eq:LDOSdefn}
  \rho(\bx;E) = \sum_{a}\delta(E-E_a)\psi_a^\ast(\bx)\psi_a(\bx).
\end{equation}
By expressing $G^\pm$ through \eqnref{eq:GFfrompsi} and applying the
identity\footnote{%
P denotes the principal value, understood to mean that an integral
$\int\mathrm{P}(\frac{1}{x})dx$ will exclude the singularity at
$x=0$.} %
\begin{equation}
  \lim_{y\rightarrow 0^+}\frac{1}{x\pm iy} =
      \mathrm{P}\frac{1}{x} \mp i\pi\delta(x),
\end{equation}
we obtain
\begin{equation}
  G^\pm(\bx,\bx;E) = \mathrm{P}\sum_a
      \frac{\psi_a(\bx)\psi_a^\ast(\bx)}{E-E_a}
      \mp i\pi\sum_a\delta(E-E_a)\psi_a(\bx)\psi_a^\ast(\bx),
\end{equation}
from which we identify the local density of states,
\begin{equation}
  \label{eq:LDOSfromGF}
  \rho(\bx;E) = \mp \frac{1}{\pi}\Im\bigl[G^\pm(\bx,\bx;E)\bigr].
\end{equation}
We can also obtain the full density of states $N(E) =
\sum_a\delta(E-E_a)$ by integrating over space,
\begin{equation}
  N(E) = \int \rho(\bx;E)\,d\bx = \mp\frac{1}{\pi}\Im\bigl[\Tr
  G^\pm(E)\bigr].
\end{equation}
When modelling transport in mesoscopic structures, the local density
of states is very useful as a visualisation tool, allowing one to
`see' the scattering states at a given energy, which helps to
identify the source of features in the conductance.

\subsection{Connection to scattering theory\label{sec:GFscattering}}

To establish the scattering framework more concretely, we want to
expand the wave functions $\psi_a$ in terms of the eigenfunctions in
the leads.  Following Baranger and Stone \cite{Baranger1989}, we
choose the Landau gauge to describe the lead eigenstates, and a set
of local coordinates $(x_\beta,y_\beta)$ for lead $\beta$, such that
\begin{equation}\label{eq:LeadStates}
  \xi_a^\pm(\bx_\beta) =
      \frac{1}{\sqrt{\theta_a}}e^{\pm ik_ax}\phi^\pm_{n_a,k_a}(y_\beta),
\end{equation}
where $\phi_{n,k}(y_\beta)$ are the transverse wave functions
determined by the lead confining potential, as in
\eqnref{eq:SWELandau}.  We have explicitly separated the outgoing
$(+)$ and incoming $(-)$ states, and the normalisation factor
$\theta_a$ is the outgoing `particle flux' through the lead, related
to the current by $I_a = -e\theta_a$. This somewhat unusual
normalisation ensures that the $S$-matrix linking these states will
be unitary. The current is computed in terms of the probability
current-density operator given by \eqnvref{eq:currentdensity}.  To
simplify notation, we write the matrix elements of the probability
current-density operator as
\begin{equation}
  [\hat{\mathbf{J}}(\bx)]_{ab} =
  \frac{\hbar}{2m^\ast i}\bigl[\psi_a^\ast(\bx)\dsD\psi_b(\bx)\bigr],
\end{equation}
in terms of the double-sided derivative operator defined by
\begin{equation}
  f\dsD g = f(\bx)\mathbf{D}g(\bx) - g(\bx)\mathbf{D}^\ast f(\bx),
\end{equation}
where $\mathbf{D}$ is the gauge covariant derivative defined in
\eqnvref{eq:Dcovariant}. Then the normalisation $\theta_a$ is
computed by integrating the current passing through the
cross-section $\mathcal{C}_\beta$ of lead~$\beta$, i.e.,
\begin{equation}
  \theta_a = \frac{\hbar}{2m^\ast i}\int_{\mathcal{C}_\beta} dy_\beta
      \bigl[e^{+ik_ax}\phi^+_{n_a,k_a}\bigr]^\ast
      (\dsD\cdot \hat{\bx}_\beta)
      e^{+ik_ax}\phi^+_{n_a,k_a},
\end{equation}
where $\hat{\bx}_\beta$ is the unit normal vector perpendicular to
$\mathcal{C}_\beta$.

Current conservation provides a set of identities for lead
eigenstates at the same energy which are similar to standard
orthogonality
relations:\footnote{%
The states $\phi_{n,k}^\pm(y)$ are not orthogonal at fixed energy
$\varepsilon_{n,k}=E$ due to the dependence of \eqnref{eq:SWELandau}
on $k$ (which defines the minimum of the effective magnetic
potential through $y_0$), so it is not possible to resolve the
scattering state into lead eigenstates by simply projecting onto the
set $\phi_{n,k}^\pm(y)$, as one could do at $B=0$.  Instead, we compute
matrix elements with the current-density operator and use the `orthogonality'
relations of \eqnsref{eq:CDidentities}.} %
\begin{subequations}\label{eq:CDidentities}
\begin{align}
    \int_{\mathcal{C}_\beta} dy_\beta\,
          \xi_a^{\pm\ast}(\bx_\beta)
          (\dsD\cdot \hat{\bx}_\beta)
          \xi_b^{\pm}(\bx_\beta)
          & =  \pm\frac{2m^\ast i}{\hbar}\delta_{ab},\quad \varepsilon_a=\varepsilon_b, \\
    \int_{\mathcal{C}_\beta} dy_\beta\,
          \xi_a^{+\ast}(\bx_\beta)
          (\dsD\cdot \hat{\bx}_\beta)
          \xi_b^{-}(\bx_\beta)
          & =  0,\\
    \int_{\mathcal{C}_\beta} dy_\beta\,
          \xi_a^{-\ast}(\bx_\beta)
          (\dsD\cdot \hat{\bx}_\beta)
          \xi_b^{+}(\bx_\beta)
          & =  0.
          \end{align}
\end{subequations}
Now, we can expand a scattering wave $\psi_{\beta,a}(\bx)$, which
originates in state $a$ of lead $\beta$, in terms of these lead
eigenstates, such that asymptotically in each lead,
\begin{equation}\label{eq:PsiAsymptotic}
    \psi_{\beta,a}(\bx)\rightarrow
        \begin{cases}
           \xi_a^-(\bx_\beta) + \smallERsum{c} t_{\beta\beta,ca}\xi^+_c(\bx_\beta),
               \quad \text{$\bx$ in lead $\beta$,} \\
           \smallERsum{c} t_{\gamma\beta,ca}\xi_c^+(\bx_\gamma),\quad \text{$\bx$ in lead $\gamma$,}
        \end{cases}
\end{equation}
where the energy-restricted sum $\smallERsum{c}$ is defined as
\begin{equation}
  \ERsum{c} \equiv \int
  dc\,\delta(\varepsilon-\varepsilon_c).
\end{equation}
The energy-dependent $t_{\alpha\beta,ab}$ are the elements of the
$S$-matrix, which gives the asymptotic components of an output state
\begin{equation}
  \psi^\mathrm{out}(\bx) =
      \sum_{\beta}\ERsum{a}c^\mathrm{out}_{\beta,a}\xi^+_a(\bx_\beta)
\end{equation}
resulting from an arbitrary input state
\begin{equation}
  \psi^\mathrm{in}(\bx) =
      \sum_{\beta}\ERsum{a}c^\mathrm{in}_{\beta,a}\xi^-_a(\bx_\beta)
\end{equation}
through the matrix equation
\begin{equation}
  c^\mathrm{out}_{\beta,a} =
  \sum_{\gamma}\ERsum{c}t_{\beta\gamma,ac}c^\mathrm{in}_{\gamma,c}.
\end{equation}
Current conservation means that the total current carried by the
scattering wave $\psi_{\beta,a}(\bx)$ in \eqnref{eq:PsiAsymptotic},
integrated across all leads, should be zero. Using the
current-operator identities of \eqnsref{eq:CDidentities}, we easily
obtain the identity
\begin{equation}
  \sum_{\gamma}\ERsum{c}t_{\alpha\gamma,ac}^\ast t_{\gamma\beta,cb} =
  \delta_{\alpha\beta}\delta_{ab},
\end{equation}
which is an expression of the unitarity of the $S$-matrix.  From
this immediately follows the relationship stated in
\eqnvref{eq:TabDefn}, recast for energy-dependent scattering
amplitudes as
\begin{equation}
  T_{\alpha\beta}(\varepsilon) =
      \ERsum{{a,b}}\abs{t_{\alpha\beta,ab}}^2,
\end{equation}
making explicit the connection to the Landauer-B\"{u}ttiker
formalism.

Finally, to determine the conductance properties of a given device,
we compute the scattering amplitudes using the lead eigenfunctions
and the Green's function for the active region, which naturally
provides the `connections' between states at different points in
space at a given energy.  Using the properties of
$G^{\pm}(\bx,\bx^\prime;\varepsilon)$, Baranger and Stone
\cite{Baranger1989} arrive at the expression
\begin{equation}
  \psi_{\beta,a}(\bx) = -\frac{\hbar^2}{2m^\ast}
      \int_{\mathcal{C}_\beta} dy_\beta^\prime\,
      G^+(\bx,\bx_\beta^\prime;\varepsilon)
      (\dsD^\prime\cdot\hat{\bx}_\beta)
      \xi_a^-(\bx^\prime_\beta),\qquad \text{$\bx$ in $A$.}
\end{equation}
By taking the projection with $\xi^{+\ast}_{b}(\bx_\alpha)$ through
the current-density operator, we extract the $S$-matrix elements to
reach the desired result
\begin{equation}\label{eq:tfromGF}
  t_{\alpha\beta,ab} = -\frac{i\hbar^3}{4m^{\ast 2}}
      \int_{\mathcal{C}_\alpha} \!\!dy_\alpha
      \int_{\mathcal{C}_\beta} \!\!dy_\beta^\prime\,
      G^+(\bx_\alpha,\bx_\beta^\prime;\varepsilon)
      (\dsD^\ast\cdot\hat{\bx}_\alpha)
      (\dsD^\prime\cdot\hat{\bx}_\beta)
      \xi_b^{+\ast}(\bx_\alpha)
      \xi_a^{-}(\bx_\beta^\prime).
\end{equation}
This formalism also provides a natural way to incorporate the
effects of finite temperature.  The nonzero contributions to the
current still come from the scattering amplitudes near the Fermi
level, but we must average over the range of energies which are
`partially occupied' according to the Fermi distribution
\begin{equation}
    f(\varepsilon) = \frac{1}{1+\exp(\frac{\varepsilon-\Ef}{kT})}.
\end{equation}
This leads to an expression for the current flowing out of lead
$\alpha$,
\begin{equation}\label{eq:ILBform}
  I_\alpha = -\frac{e^2}{h}\biggl[N_\alpha V_\alpha -
      \sum_{\beta} V_{\beta}
      \int d\varepsilon\,
      \bigl[-f^\prime(\varepsilon)\bigr]
      T_{\alpha\beta}(\varepsilon)\biggr],
\end{equation}
where the sum runs over all leads (including $\alpha$), and
$N_\alpha$ is the number of modes in lead $\alpha$.  This expression
is equivalent to the Landauer-B\"{u}ttiker formula,
\eqnref{eq:LBcurrent}, in the limit of zero temperature, or in the
case of energy-independent transmission coefficients.

\subsection{Calculating time-independent Green's functions}

Given the connection presented in the previous section between
Green's functions and the currents and voltages measured in
transport experiments, we have a powerful framework with which to
model ballistic transport in mesoscopic devices. All that remains is
the calculation of the Green's function
$G^+(\bx,\bx^\prime;\varepsilon)$ in \eqnref{eq:tfromGF}.  In some
sense, the calculation of the full Green's function for the
Hamiltonian $\hat{H}_0+U(\bx)$ (with appropriate boundary conditions
in the leads) is equivalent to solving the time-independent
Schr\"{o}dinger equation,\footnote{%
Essentially this is because the poles of $G(\bx,\bx^\prime;z)$
coincide with the eigenvalues of $\hat{H}$, as seen in
\eqnref{eq:GFfrompsi}, and so knowledge of the poles of $G$ gives us
the energy spectrum of $\hat{H}$.} %
but methods exist for the Green's function calculation which are far
more efficient than diagonalising the full Hamiltonian.  These
mainly rely a slightly indirect method, starting with the known
Green's function for a simplified Hamiltonian (in this case, the
free-particle Hamiltonian $\hat{H}_0$), and then building up the
full Green's function through an iterative process.

This procedure takes advantage of the easily quantifiable change to
$G(\bx,\bx^\prime;z)$ which results from the addition of a small
perturbation to the Hamiltonian.  Consider the free-particle Green's
function defined by
\begin{equation}
  \bigl[z-\hat{H}_0(\bx)\bigr]G_0(\bx,\bx^\prime;z) =
  \delta(\bx-\bx^\prime),
\end{equation}
which is given by \eqnref{eq:GFfrompsi} in terms of the known
free-particle wave functions $\phi_n(\bx)$ (LL eigenstates).  Now
suppose we add a perturbing potential $\tilde{U}(\bx)$, and write
Schr\"{o}dinger's equation for the new eigenstates $\psi_n(\bx)$ in
the form
\begin{equation}\label{eq:SWEforpsin}
  \bigl[\varepsilon_n - \hat{H}_0(\bx)\bigr]\psi_n(\bx) =
  \tilde{U}(\bx)\psi_n(\bx).
\end{equation}
Comparing this to \eqnsref{eq:inhomoeqn} and
(\ref{eq:GFinhomosoln}), we obtain the Lippmann-Schwinger equation,
\begin{equation}
  \label{eq:LippmannSchwinger}
  \psi_n(\bx) = \phi_n(\bx) + \int\!
  G_0^\pm(\bx,\bx^\prime;\varepsilon_n)\tilde{U}(\bx^\prime)\psi_n(\bx^\prime)\,d\bx^\prime,
\end{equation}
in which it is necessary to use the retarded/advanced Green's
functions which are defined at $z = \varepsilon_n$, and to include
the term $\phi_n(\bx)$ to satisfy \eqnref{eq:SWEforpsin} as
$\tilde{U}(\bx)\rightarrow 0$.  The Green's function for the
perturbed Hamiltonian is then defined by
\begin{equation}
  \bigl[z-\hat{H}_0(\bx) - \tilde{U}(\bx)\bigr]G(\bx,\bx^\prime;z) =
  \delta(\bx-\bx^\prime),
\end{equation}
from which we obtain the related integral expression, similar to
\eqnref{eq:LippmannSchwinger},
\begin{equation}\label{eq:LS2}
  \psi_n(\bx) = \phi_n(\bx) + \int\!
      G^\pm(\bx,\bx^\prime;\varepsilon_n)\tilde{U}(\bx^\prime)\phi_n(\bx^\prime)\,d\bx^\prime.
\end{equation}
By inserting \eqnref{eq:LS2} into the Lippmann-Schwinger equation,
and comparing the result again with \eqnref{eq:LS2}, we obtain the
Dyson equation relating $G^\pm$ to $G_0^\pm$,
\begin{equation}
  G^\pm(\bx,\bx^\prime;\varepsilon) =
      G_0^\pm(\bx,\bx^\prime;\varepsilon) +
      \int d\bx^{\prime\prime}\,
          G_0^\pm(\bx,\bx^{\prime\prime};\varepsilon)
          \tilde{U}(\bx^{\prime\prime})
          G^\pm(\bx^{\prime\prime},\bx^{\prime};\varepsilon),
\end{equation}
or, in abstract matrix form,
\begin{equation}
  \label{eq:DysonEqn}
  G = G_0 + G_0 \tilde{U} G.
\end{equation}
The solution for $G$ from the Dyson equation is obviously still
implicit, but in practice we can choose the `perturbation'
$\tilde{U}$ such that the calculation remains computationally
efficient at every step, and thereby iteratively build up the full
potential $U(\bx)$.

In calculations, we divide the active region of the device into a
lattice, for which $\hat{H}$ is the well-known tight-binding
Hamiltonian, in matrix form.\footnote{%
Baranger and Stone provide the lattice form of the continuum results
quoted in \secref{sec:GFscattering} in Appendix B of their paper
\cite{Baranger1989}.} %
We must choose the lattice spacing to be less than the smallest
length scale of the problem, which for mesoscopic devices is usually
the magnetic length,
$\ellB=$~\unit{26}{\nano\metre\usk\reciprocal\tesla}.  Therefore, in
modelling micron\nobreakdash-scale devices at fields
$\approx$~\unit{1}{\tesla}, we can easily have lattice dimensions of
order $\approx 1000$.  If the computational region $A$ is
rectangular, with lattice dimension $N\times M$, then the
Hamiltonian matrix will be of size $NM\times NM$, and thus solving
the eigenvalue problem for the Hamiltonian directly is often
computationally prohibitive.  Instead, we use the Dyson equation to
build up the full Green's function iteratively, dealing with only a
small portion of the system at each step, such that the matrices
involved remain manageable.  The particular computation we use in
\chapref{chap:Geometry} is based on the work of MacKinnon
\cite{MacKinnon1985} and implemented in a program written by
C.~H.~W.~Barnes.  Essentially, the $N\times M$ rectangular
computational domain is divided in to $N$ `slices' each containing
$M$ lattice points.  Tight-binding versions of the Dyson equation
lead to a relationship between the Green's function of a system with
$n+1$ slices and that of the system with $n$ slices.  Starting with
the free-particle $G_0$ determined by the boundary conditions at the
first slice, we can add successive slices to the system in this way,
dealing only with the Green's functions of individual slices, which
are size $M\times M$.  Clearly, this method is particularly
well-suited to anisotropic systems with $M\ll N$, such as quantum
wires, but the method is suitable for any system with dimensions
such that the inversion of $M\times M$ matrices is computationally
feasible.

We have used Green's functions calculated through this method to
investigate several important effects of non-interacting particles
which are ignored in the idealised model presented in
\chapref{chap:ADtheory}.  For example, the effects of
tunnel-coupling between the quasi-zero-dimensional antidot states
and the extended edge-modes in the two-dimensional regions are
naturally incorporated in the Green's function of the full region,
and we can easily investigate the dependence of these couplings on
the potential gradients, antidot dimensions, and applied magnetic
field. In \chapref{chap:Geometry} we use the Green's function
technique to explore the effects of the split gates; these gates
bring the edge modes within tunneling distance of the antidot but
also break the circular symmetry of the system, which we show has
measurable effects on the single-particle energy levels.

For the investigation of non-interacting effects like these, the
approach presented here is relatively easy-to-use and extremely
flexible. In order to incorporate additional physics due to
interactions, however, a great deal of additional theoretical
machinery is necessary.  Methods do exist to compute Green's
functions self-consistently which include spin and charge
interactions, but these are beyond the scope of this work.  For
example, in a recent study antidot Green's functions were computed
using spin density functional theory \cite{Ihnatsenka2006a}, in
order to investigate the formation of `compressible regions' at the
antidot edge, which have been predicted to exist at high magnetic
fields \cite{Kataoka2000,Sim2003}.  In many experimental regimes,
however, antidot transport is well-described by the model of
sequential tunneling commonly used to represent transmission through
a quantum dot.  The quantum-Hall energy gaps provide natural tunnel
barriers between the zero-dimensional antidot states and the
`leads,' which in this case are nearby edge-modes.  This description
separates the relatively well-understood quantum Hall physics of the
leads from the properties of the antidot states, and so we can
include the effects of spin and charge interactions on the antidot
energy spectrum and tunneling selection-rules in a straightforward
manner. By treating the antidot as an isolated zero-dimensional
system with capacitive couplings to the gates and leads, we can also
move beyond the linear-response regime to explore the effects of
finite drain-source bias.  We discuss this `dot model' of antidot
transport in the following section.

\section{Sequential transport\label{sec:SequentialTransport}}

In the theory of sequential transport, we consider a
zero-dimensional island which is weakly connected to a set of
metallic leads by tunneling barriers. In this section we refer to
the island as a dot, although the description of an antidot is
entirely equivalent as long as the couplings are weak.  Transitions
between different occupation states of the dot are modeled as a
Markov chain, in which a `master equation' describes the stochastic
evolution of the system.  The solution of the master equation
provides the equilibrium occupation probability for each state, from
which the transport current may be computed.  For a given set of dot
states, we must therefore compute the transition rates which make up
the master equation.  These are described by a tunneling Hamiltonian
including energy conservation, dot selection rules, and the
(possibly energy- and/or spin-dependent) lead-dot tunnel couplings.
In many cases, it is sufficient to consider only the lowest-order
terms in the perturbation theory for tunneling, in which the rates
are given by Fermi's golden rule for the transition rate between
individual eigenstates and the continuum in the leads.  This theory
was developed from earlier descriptions of small metallic islands
\citep[e.g.,][]{Averin1986} to include the quantised levels in a
quantum dot \cite{Averin1991,Beenakker1991,Meir1991} and has now
become an essential tool in the study of quantum dots.  Here we
generally follow the review of Kouwenhoven, Sch\"{o}n, and Sohn
\cite{Kouwenhoven1997a}, with a few minor alterations to clarify
intermediate steps and to express results in the form we use to
model spin-resolved antidot transport in
\chapref{chap:SpinTransportModel}.

\subsection{Coulomb blockade}

\begin{figure}[tb]
\begin{center}
\includegraphics[]{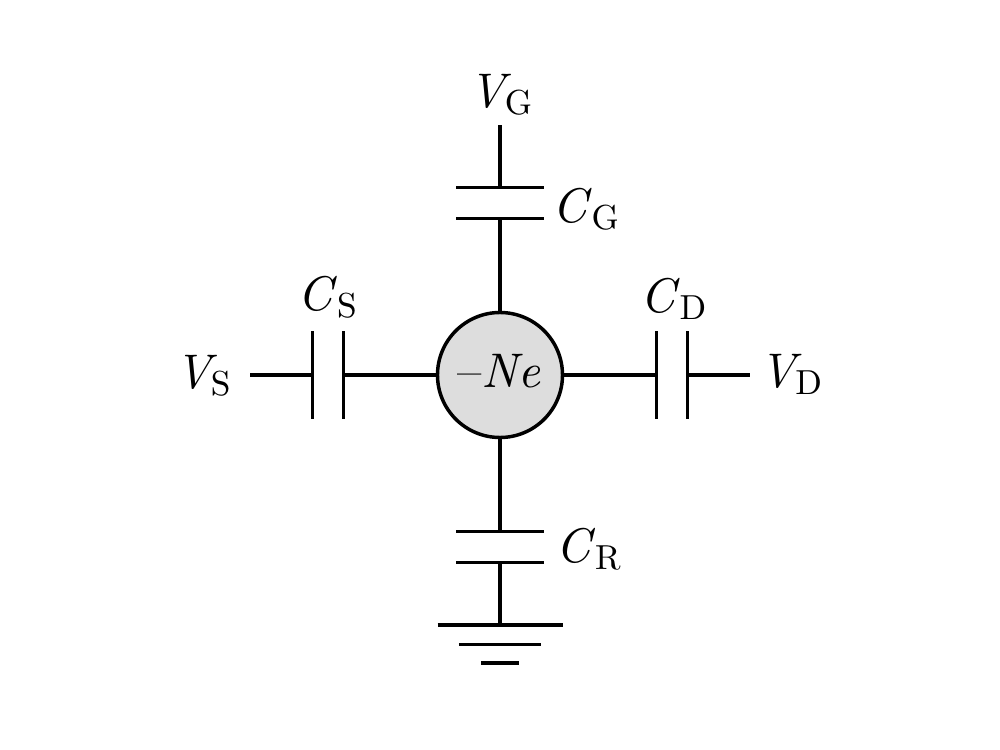}
\caption[Equivalent capacitor network for quantum dot
transport]{%
Equivalent capacitor network for transport through a quantum dot.
The charge on the dot is quantised by the electron charge, with $Q =
-Ne$, and the potential $\phi$ is determined by the capacitive
couplings $(\Cs,\Cd,\Cg)$ to the source, drain and gate voltages
$(\Vs,\Vd,\Vg)$, respectively.  We model any remaining coupling to
other elements of the device which have fixed voltages as an
additional capacitance $\Cr$ to ground potential.
\label{fig:CapNetwork}}
\end{center}
\end{figure}

We begin with a brief review of Coulomb blockade and the
interpretation of non-linear conductance measurements of quantum
dots in the constant-interaction model.  The electrostatics of the
system may be represented by an
equivalent capacitor network,\footnote{%
Tunnel couplings between the dot and the leads may more
realistically be modeled as a capacitor and resistor connected in
parallel, to account for the finite current which flows through
these connections.} %
as shown for example in \figref{fig:CapNetwork}.
From the capacitor network we can relate the charge $Q$ on the dot
with its potential $\phi$ through the equation
\begin{equation}
  Q = \Cs(\phi - \Vs) + \Cd(\phi-\Vd)+\Cg(\phi-\Vg) + \Cr\phi,
\end{equation}
such that
\begin{equation}
  \phi = \frac{Q}{C}+\phi_\mathrm{ext},
\end{equation}
where $C = \Cs+\Cd+\Cg+\Cr$ is the total capacitance and
\begin{equation}
  \phi_\mathrm{ext} =\frac{\Cs\Vs+\Cd\Vd+\Cg\Vg}{C}
\end{equation}
is the potential due to the external voltages. The dot charge $Q$ is
quantised by the electron charge, and when the dot contains $N$
electrons, its energy is
\begin{equation}\label{eq:UofN1}
  U(N) = \int_0^{-Ne}\!\phi \;dQ = \frac{(Ne)^2}{2C} -
  Ne\phi_\mathrm{ext}.
\end{equation}
On the other hand, $\phi_\mathrm{ext}$ may be varied continuously
through the external voltages, with a corresponding effective charge
\begin{equation}
  Q_\mathrm{ext} = C\phi_\mathrm{ext} = +e\nG.
\end{equation}
Therefore, \eqnref{eq:UofN1} may be rewritten as
\begin{equation}\label{eq:UofN2}
  U(N) = \Ec(N-\nG)^2 + \nG^2\Ec,
\end{equation}
where $\Ec = \frac{e^2}{2C}$ is the charging energy.

In the constant-interaction model, we assume that the effect of the
electron charge is entirely described through the electrostatic
energy of \eqnref{eq:UofN2}, and we can simply add to this the
single-particle eigenenergies $\varepsilon_i$ of the dot. A general
$N$\nobreakdash-electron state is characterised by a set of
occupation numbers $\lbrace n_i \rbrace$, where $n_i=0$ or 1 due to
Fermi exclusion and $\sum n_i = N$, such that
\begin{equation} \label{eq:UofN3}
  U(\lbrace n_i\rbrace) = \Ec(N-\nG)^2 + \nG^2\Ec + \sum_i
  n_i\varepsilon_i.
\end{equation}
Often we are concerned only with transitions between ground states,
in which only states $i=1,2,\dotsc,N$ are filled, which have energy
\begin{equation}
  U_0(N) = \Ec(N-\nG)^2 + \nG^2\Ec + \sum_{i=1}^N \varepsilon_i.
\end{equation}
Tunneling between the dot and the leads conserves energy, so we need
to compare the chemical potentials of the leads, e.g., $\muD =
-e\Vd$, with that of the dot.  For transitions between ground
states, the dot chemical potential is given by
\begin{equation}
  \mudot(N) = U(N+1)-U(N) = 2\Ec(N-\nG) + \Ec + \varepsilon_{N+1}.
\end{equation}
Transport requires an occupied state in the source and an unoccupied
state in the drain (or vice versa) with $\muS = \muD = \mudot$, and
so at zero bias transport only occurs for $\mudot\approx0$ within a
few $kT$. Outside this regime, transport is forbidden due to the
`Coulomb blockade' resulting from the quantisation of electronic
charge.  The dependence of $\nG$ on the gate voltage $\Vg$ therefore
leads to a set of conductance peaks as a function of $\Vg$, as the
chemical potentials of subsequent transitions align with those of
the leads.

\begin{figure}[tp]
\begin{center}
\includegraphics[]{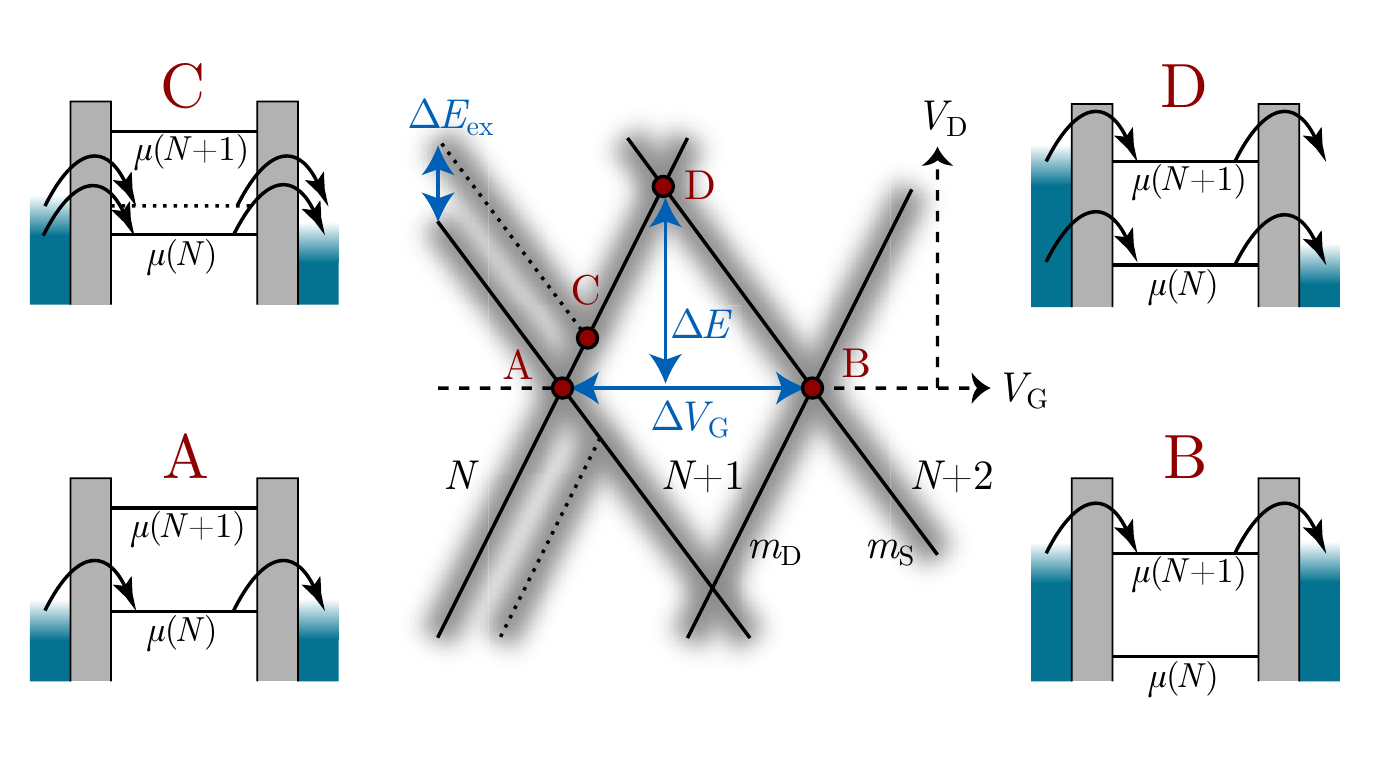}
\caption[Schematic of non-linear transport and Coulomb blockade]{%
Schematic diagram of non-linear conductance through a quantum dot,
as a function of $\Vg$ and $\Vds$ (with bias applied to the drain
contact).  Transport occurs when the chemical potential for a dot
transition sits in the `window' between $\muS = 0$ and $\muD =
-e\Vd$.  Peaks in conductance are observed when one of the dot
chemical potentials passes either $\muS$ or $\muD$, with a
broadening proportional to the electron temperature in the leads.
Inside the diamond, transport is forbidden due to Coulomb blockade,
and the number of electrons remains fixed at $N+1$ as shown.  The
capacitances of the system and energy scales of the dot may be
extracted from such plots as explained in the text, by measuring the
slopes $(m_\mathrm{S},m_\mathrm{D})$ of the source and drain lines,
and the spacings of peaks in either gate voltage or bias, which is
proportional to energy through $\Delta E = -e\Delta\Vd$.  The
$N\leftrightarrow N+1$ excited state drawn as a dotted line and
shown in the transport diagram labeled `C' corresponds to electrons
tunneling through the same orbital state as the ground-state
$N+1\leftrightarrow N+2$ transition, but without the charging energy
$\Ec$ associated with adding the $N+2^\mathrm{nd}$ electron to the
dot.  Identification of such lines allows us to separate the
charging and single-particle contributions to the addition spectrum.
\label{fig:CBdiamonds}}
\end{center}
\end{figure}

If we vary the source and/or drain potentials, current can flow
through the dot when $\mudot(N)$ lies in the transport window
defined by $\muS$ and $\muD$.  In the plane of $\Vg$ and $\Vds = \Vd
- \Vs$, this produces a pattern of `Coulomb diamonds' within which
transport is blockaded, as shown in \figref{fig:CBdiamonds}.
By analysing the slopes and spacings of the various lines in such
data, we can extract the values of the capacitances and, more
importantly, the `lever-arm' scaling factor
\begin{equation}
  \alphaG = \frac{\partial \phi_\mathrm{ext}}{\partial\Vg} =
  \frac{\Cd}{C},
\end{equation}
which allows us to convert adjustments of $\Vd$ into changes in
$\mudot$.  This factor is determined by the dimensions of the
Coulomb diamonds, as shown in \figref{fig:CBdiamonds}, as the ratio
of the spacing in $\Vg$ between subsequent peaks and the value of
$\Vds$ at which they intersect,
\begin{equation}
  \alphaG = \frac{\Delta\Vg}{\Delta\Vds}.
\end{equation}
Assuming that bias is applied only to the drain
contact,\footnote{%
We choose this convention since, in our experiments, we typically
bias the input offset of the current preamplifier which acts as the
drain for our device.} %
such that $\Vs = 0$ always, the slopes of the `source lines'
($\mudot = \muS = 0$) and `drain lines' ($\mudot = \muD = -e\Vd$)
are given by
\begin{subequations}\label{eq:CBslopes}
  \begin{align}
    m_\mathrm{S} & = -\frac{\alphaG}{\alphaD} = -\frac{\Cg}{\Cd} \\
    m_\mathrm{D} & = \frac{\alphaG}{1-\alphaD} = \frac{\Cg}{C-\Cd},
  \end{align}
\end{subequations}
where $\alphaD = \partial\phi_\mathrm{ext}/\partial\Vd$ is the lever
arm factor for changes in $\Vd$.

The addition energy $\Delta E$ shown in \figref{fig:CBdiamonds}
contains both electrostatic and quantum-mechanical contributions,
\begin{equation}\label{eq:deltaEdot}
\begin{split}
  \Delta E & = \mudot(N+1) - \mudot(N) \\
   & = 2\Ec + \varepsilon_{N+1}-\varepsilon_N.
\end{split}
\end{equation}
The quantum contribution $\Delta\Eex =
\varepsilon_{N+1}-\varepsilon_N$ is often observable as the first of
several extra lines which appear outside the Coulomb blockade
region, as shown in \figref{fig:CBdiamonds}.  By measuring such
excitations we can isolate the charging energy $\Ec$ from
\eqnref{eq:deltaEdot} and then solve \eqnsref{eq:CBslopes} for the
individual capacitances $C$, $\Cg$, and $\Cd$.  If we swap the
source and drain contacts for additional measurements, we can
extract $\Cs$ and $\Cr$ through a similar analysis.

\subsection{Master equation approach\label{sec:MasterEqn}}

In this section we consider a quantum-mechanical description of the
dot in terms of a set of `fermionic' states $\lvert s\rangle =
\lvert \lbrace n_{\ell,\sigma}\rbrace\rangle$, labeled by occupation
numbers $n_{\ell,\sigma}=0,1$ for the state with orbital and spin
quantum numbers $\ell$ and $\sigma$, respectively.  These are
suitable states when electron-electron interactions are ignored, and
provide the most convenient basis for calculations, but they do not
correctly reproduce the energies or degeneracies of an interacting
system.  We discuss this issue further and consider other choices
for basis states in \chapref{chap:SpinTransportModel}.

We split the system into three parts, such that the total
Hamiltonian is $H = H_\mathrm{dot} + H_\mathrm{res} +
H_\mathrm{tun}$, representing the physics of the dot, the
reservoirs, and the tunneling between them. The Hamiltonian for the
dot is
\begin{equation}
  H_\mathrm{dot} = \sum_s E_s \lvert s\rangle\langle s\rvert,
\end{equation}
where, within the constant interaction model,
\begin{equation}
  E_s = \sum_{\ell\sigma}\varepsilon_{\ell\sigma}n_{\ell\sigma} +
  \Ec(N-\nG)^2,
\end{equation}
as in \eqnref{eq:UofN3}.\footnote{%
We have dropped the $N$-independent term $\nG^2\Ec$ from
\eqnref{eq:UofN3} since it affects all states equally and hence does
not appear in the chemical potential.} %
Similarly, the Hamiltonians describing the reservoirs and tunneling
to and from the dot are given in second-quantised form by
\begin{subequations}
\begin{align}
  H_\mathrm{res} & = \sum_{r = \mathrm{S,D}}\left[
      \sum_{k\sigma}\varepsilon_\ksr a_\ksr^\dag a_\ksr + \mu_r
      \hat{n}_r \right], \\
  H_\mathrm{tun} & = \sum_{r=\mathrm{S,D}}\left[\sum_\kls
      T^r_\kls a^\dag_\ksr a_\ls + \text{h.c.}\right],
\end{align}
\end{subequations}
where the leads (assumed to be non-interacting) are labeled by
reservoir $r$, wave vector $k$, and spin $\sigma$.  The operators
$a_\ksr$ and $a_\ls$ annihilate particles in the lead states $\lvert
k\sigma\rangle$ of reservoir $r$ and dot states $\lvert\ls\rangle$,
respectively, and $\hat{n}_r$ is the particle-number operator for
lead $r$, with chemical potential $\mu_r= -eV_r$.

Assuming the couplings to the leads $T^r_\kls$ are small relative to
$kT$, such that thermal fluctuations dominate over
quantum-mechanical fluctuations, we can use Fermi's golden rule to
write the tunneling rates for the transition between dot states
$\sptos$ and reservoir states $\kptok$ to first order as
\begin{equation}\label{eq:FGRrate}
  W^p_{s^\prime\chi^\prime\rightarrow s\chi}\simeq
      \frac{2\pi}{\hbar}\Bigl\lvert\langle \chi s\rvert H_\mathrm{tun}
          \lvert \chi^\prime s^\prime\rangle\Bigr\rvert^2
      \delta(E_s - E_{s^\prime} + E_\chi - E_{\chi^\prime}+p\mu_r),
\end{equation}
where $p = \pm1$ denotes the change of electron number on the dot,
and $E_\chi$ is the energy of the reservoir state $\chi$. We are
interested in the rates between individual dot states which are
obtained by summing out the contributions from all lead states,
\begin{equation}\label{eq:sumoutleads}
  \gamma^p_\sptos = \mspace{-18mu} \sum_{\substack{\chi\chi^\prime \\ N(\chi^\prime) =
  N(\chi)+p}}
      \mspace{-18mu} W^p_{s^\prime\chi^\prime\rightarrow s\chi}
      \rho^\mathrm{eq}_\mathrm{res}(\chi^\prime),
\end{equation}
where $\rho^\mathrm{eq}_\mathrm{res}$ is the equilibrium density of
states in the reservoirs.  This calculation is outlined in
\appref{app:TunRates}, from which we obtain the result
\begin{subequations}\label{eq:gammapm}
  \begin{align}
    \gamma^+_{r,\sptos} & = \sum_\llps \Gamma^r_\llps(E_s-E_\sp)
        \bra{s}a^\dag_\ls\ket{\sp} \bra{\sp} a_\lps \ket{s}
        f_r(E_s - E_\sp), \label{eq:gammap}\\
    \gamma^-_{r,\sptos} & = \sum_\llps \Gamma^r_\llps(E_\sp - E_s)
        \bra{s}a_\ls\ket{\sp} \bra{\sp}a^\dag_\lps\ket{s}
        \Bigl[1-f_r(E_\sp - E_s)\Bigr],\label{eq:gammam}
  \end{align}
\end{subequations}
where the spectral function is defined as
\begin{equation}\label{eq:spectralfn}
  \Gamma^r_\llps(E) = \frac{2\pi}{\hbar}\sum_k T^r_\kls
  T^{r\ast}_{k\ell^\prime\sigma} \delta(E-\varepsilon_\ksr),
\end{equation}
and
\begin{equation}
  f_r(E) = \frac{1}{1+e^{(E-\mu_r)/kT}}
\end{equation}
are the Fermi functions which describe the occupation of states in
the reservoirs.

With the total transition rate given by\footnote{%
Note that at least one of the terms in \eqnref{eq:gammaT} will be
zero due to selection rules (the matrix elements in \eqnsref{eq:gammapm}).} %
\begin{equation}
  \label{eq:gammaT}
  \gamma_{\sp s} = \sum_{r = \mathrm{S,D}}\bigl(\gamma^+_{r,\sptos} +
  \gamma^-_{r,\sptos}\bigr),
\end{equation}
we proceed to construct the master equation to solve for the
equilibrium occupation probabilities $P(s)$.  In equilibrium, the
total evolution `out' of state $s$ must equal the total evolution
`in,' i.e.,
\begin{equation}\label{eq:ratebalance}
  0 = \sum_s\bigl[ \gamma_{\sp s} P(\sp)  -
  \gamma_{s\sp}P(s)\bigr].
\end{equation}
By combining \eqnref{eq:ratebalance} with the normalisation
condition $\sum_i^n P(s_i) = 1$, for $n$ available states, we obtain
the master equation in matrix form:
\begin{equation}
  \label{eq:MasterEqn}
  \begin{pmatrix}
  \sum_i\gamma_{s_i s_1} & -\gamma_{s_1s_2} & \cdots & -\gamma_{s_1 s_n} \\
  -\gamma_{s_2s_1} & \sum_i\gamma_{s_is_2} & \cdots & -\gamma_{s_2s_n} \\
  \vdots & \vdots & \ddots & \vdots \\
  -\gamma_{s_ns_1} & -\gamma_{s_Ns_2} & \cdots & \sum_i\gamma{s_is_n} \\
  1 & 1 & \cdots & 1
  \end{pmatrix}
  \begin{pmatrix}
    P(s_1) \\
    P(s_2) \\
    \vdots \\
    P(s_n)
  \end{pmatrix}
   =
  \begin{pmatrix}
    0 \\
    0 \\
    \vdots \\
    0 \\
    1
  \end{pmatrix}.
\end{equation}
Once the master equation has been solved for the probabilities
$P(s)$, we can compute the current flowing out of each lead from the
expression
\begin{equation} \label{eq:Ir1}
\begin{split}
  I_r & = e\sum_{s\sp} \bigl[ \gamma^+_{r,\sptos}P(\sp)
      - \gamma^-_{r,s\rightarrow\sp}P(s)\bigr] \\
      & = e\sum_{s\sp}\bigl[\gamma^+_{r,\sptos} -
      \gamma^-_{r,\sptos}\bigr] P(\sp),
\end{split}
\end{equation}
where we have used \eqnref{eq:ratebalance} to simplify the second
term.  Using the relation
\begin{equation}
  \sum_r\bigl[\gamma^+_{r,\sptos} -
      \gamma^-_{r,\sptos}\bigr] = \Bigl(N(s)-N(\sp)\Bigr)\gamma_{\sp
      s},
\end{equation}
it is straightforward to show that $\sum_r I_r = 0$, i.e.\ that the
total current is conserved.

We can further simplify the expression for the current by noting
that the rates satisfy the `detailed balance' condition\footnote{%
From this point we adopt the convention $N(s) = N(\sp)+1$.} %
\begin{equation}\label{eq:detailedbal}
  \frac{\gamma^-_{r,s\rightarrow\sp}}{\gamma^+_{r,\sptos}} =
  \frac{1-f_r(\mussp)}{f_r(\mussp)}
   = e^{(\mudot - \mu_r)/kT},
\end{equation}
where $\mussp = E_s - E_\sp$ is the chemical potential for the dot
transition. In terms of the quantity
\begin{equation}
  \gSrssp =\gamma^+_{r,\sptos} + \gamma^-_{r,s\rightarrow\sp},
\end{equation}
\eqnref{eq:detailedbal} gives
\begin{subequations}
  \begin{align}
    \gamma^+_{r,\sptos} & = \gSrssp f_r(\mussp), \\
    \gamma^-_{r,s\rightarrow\sp} & = \gSrssp \bigl[1 -
    f_r(\mussp)\bigr],
  \end{align}
\end{subequations}
and by comparison with \eqnsref{eq:gammapm} we see that $\gSrssp$ is
independent of the reservoir chemical potentials $\mu_r$.  If the
states $\ket{s}$ do in fact represent non-interacting configurations
of single-particle orbital states, then the matrix elements in
\eqnsref{eq:gammapm} simplify through
\begin{equation}
  \bra{s}a^\dag_\ls\ket{\sp}\bra{\sp}a_\lps\ket{s} =
  \delta_{\ell\ell^\prime}\bigl\lvert
  \bra{s}a^\dag_\ls\ket{\sp}\bigr\rvert^2,
\end{equation}
and we obtain
\begin{equation}
  \gSrssp = \sum_\ls\Gamma^r_{\ell\ell\sigma}(\mussp)
  \bigl\lvert\bra{s}a^\dag_\ls\ket{\sp}\bigr\rvert^2.
\end{equation}
In general the tunnel couplings $\Gamma^r_\ls(\mussp)$ may depend on
the dot states $\ket{\ls}$ and on the transport chemical potential
$\mussp$.  In \chapref{chap:SpinTransportModel} we consider in
detail the case of spin-dependent tunnel barriers, but for now we
drop the dependence on $\ket{\ls}$ to write \eqnref{eq:Ir1} in the
form
\begin{equation}\label{eq:Ir2}
  I_r = e\sum_{s\sp}\Gamma_r(\mussp) M_{s\sp}\Bigl[f_r(\mussp)P(\sp)
  - \bigl(1-f_r(\mussp)\bigr)P(s)\Bigr],
\end{equation}
where
\begin{equation}
  M_{s\sp} = \sum_\ls
  \bigl\lvert\bra{s}a^\dag_\ls\ket{\sp}\bigr\rvert^2
\end{equation}
represents the selection rules for transitions between states
$s^\prime\leftrightarrow s$.  Using current conservation we can
derive the relation
\begin{equation}
  \sum_{s\sp}M_{s\sp}P(s) = \sum_{s\sp
  r}\frac{\Gamma_r}{\Gamma}M_{s\sp}\bigl[ P(\sp)+P(s)\bigr]
  f_r(\mussp),
\end{equation}
where $\Gamma = \sum_r \Gamma_r$ and we have suppressed the
dependence of the $\Gamma$'s on $\mussp$. We can use this relation
to eliminate the term independent of $f_r$ in \eqnref{eq:Ir2} to
obtain our final expression for the current out of lead $r$,
\begin{equation}
  I_r = e\sum_{s\sp r^\prime}
  \frac{\Gamma_r\Gamma_{r^\prime}}{\Gamma}
      M_{s\sp}\bigl[P(\sp)+P(s)\bigr]
      \times\bigl[f_r(\mussp) - f_{r^\prime}(\mussp)\bigr].
\end{equation}
We use this expression to calculate the current transmitted through
our antidot in the model described in
\chapref{chap:SpinTransportModel}, and compute the conductance at
finite bias by
\begin{equation}
  G(\Vd) = \frac{I(\Vd + \delta\Vd) - I(\Vd -
  \delta\Vd)}{2\delta\Vd},
\end{equation}
which is typically a good approximation if $e\delta\Vd\lesssim kT$.

To make a connection to the linear-response theory discussed in the
first two sections of this chapter, we can set $\mu_r = \mu -
e\delta V_r$ and solve for the conductance coefficients defined by
\begin{equation}
  I_r = \sum_{r^\prime}G_{rr^\prime}(\delta V_r - \delta
  V_{r^\prime}).
\end{equation}
This was first considered by Beenakker \cite{Beenakker1991} with the
well-known result
\begin{equation}
  G_{rr^\prime} = -e^2\sum_{s\sp}\frac{\Gamma_r\Gamma_{r^\prime}}{\Gamma}
      M_{s\sp}\bigl[ P(s)+P(\sp)\bigr]
      f^\prime_\mu(\mussp).
\end{equation}
In the special case of sequential tunneling through a single level,
we have only two states to consider (occupied and unoccupied), which
satisfy $P(1)+P(2)=1$, and so the conductance through the dot is
given by
\begin{equation}\label{eq:Gsinglelevel}
  G_\mathrm{SD} = -2e^2\left(
  \frac{\Gamma_\mathrm{S}\Gamma_\mathrm{D}}{\Gamma_\mathrm{S}+\Gamma_\mathrm{D}}\right)
  f^\prime_\mu(\mudot).
\end{equation}
If the tunnel couplings are energy-independent, the line-shape of
this resonance as a function of $\mudot$ will be determined by the
derivative of the Fermi function,
\begin{equation}\label{eq:FermiDerivative}
  -f^\prime_\mu(\mudot) =
  \frac{1}{4kT}\cosh^{-2}\left(\frac{\mudot-\mu}{2kT}\right).
\end{equation}
By comparing \eqnref{eq:Gsinglelevel} with \eqnvref{eq:ILBform}, we
can also identify the transmission coefficient of this single level
within the Landauer-B\"{u}ttiker formalism, given by
\begin{equation}
  T(\varepsilon) = h\Gamma_\mathrm{eff}\delta(\varepsilon - \mudot),
\end{equation}
where
\begin{equation}
  \Gamma_\mathrm{eff} =
\frac{2\Gamma_\mathrm{S}\Gamma_\mathrm{D}}{\Gamma_\mathrm{S}+\Gamma_\mathrm{D}}
\end{equation}
is the effective tunnel coupling.


\chapter{Geometrical Effects on Single-Particle Excitation Energies\label{chap:Geometry}}
\chaptermark{Geometrical Effects}

\ifpdf
    \graphicspath{{Chapter3/Figures/PNG/}{Chapter3/Figures/PDF/}{Chapter3/Figures/}}
\else
    \graphicspath{{Chapter3/Figures/EPS/}{Chapter3/Figures/}}
\fi


The precise role of electron-electron interactions in antidot
transport experiments has remained elusive.  Although the Coulomb
energy scale $(e^2/4\pi\epsilon\epsilon_0)n_e^{1/2}$ dominates all
other relevant energies in a typical antidot device with electron
density $n_e\approx1$\nbd\numprint{3e11}~\centi\metre\rpsquared, we
find that many of the experimental observations, particularly at low
magnetic fields ($B\lesssim\unit{2}{\tesla}$), may be fully
understood through the non-interacting single-particle (SP) picture
described in \secref{sec:SPstates}.  This is presumably due to the
nature of the Fermi liquid which constitutes the two-dimensional
electron system (2DES), in which the quasiparticle excitations are
essentially equivalent to free electrons.  At higher fields this
picture no longer seems to suffice, since observations of
`double-frequency' Aharonov-Bohm oscillations
\cite{Ford1994,Kataoka2002a} and Kondo-like effects
\cite{Kataoka2004a} have no straightforward interpretation within
the SP model.

Michael~et~al.~\cite{Michael2004,Michael2006} performed antidot
transport experiments at intermediate fields
($B\approx1$\nbd\unit{2}{\tesla}) in an attempt to clarify the
nature of the transition away from non-interacting physics.  They
noticed several interesting effects, but chief among them was an
observed `softening' of the single-particle orbital energy spacing
$\dEsp$ at higher magnetic fields.  They suggested that this
observation might indicate the breakdown of the SP model in favour
of an interacting picture of alternating compressible and
incompressible strips, in which the orbital excitation energy would
be suppressed. This compressible-region (CR) model has been proposed
to explain experimental observations at higher fields
\cite{Kataoka2000,Sim2003}, but the details of its emergence from
the SP-like physics at lower fields has not been well-understood.
There have also been disagreements within the community concerning
the implications of experiments in this regime
\cite{Kataoka2000,Karakurt2001,Kataoka2004,Goldman2004}. Despite the
observed reduction in $\dEsp$, several other features of the
experiments by Michael~et~al.\ seem to be inconsistent with the CR
picture.  For example, it is not clear that well-defined excitation
energies should be observable at all in a CR picture, in which a
`band' of partially-occupied states is available for transport near
the Fermi energy.  In this chapter we reconsider the observations of
Michael~et~al.\ within a non-interacting model, and we find an
explanation for the suppression of $\dEsp$ by taking into account a
realistic geometry for the full antidot device, including the
split-gate within which it is embedded.

This conclusion does not rule out the formation of CRs at high
fields, or the importance of interactions with regards to some of
the more subtle effects observed in the intermediate-$B$ regime.  In
\chapsref{chap:SpinTransport} and \ref{chap:SpinTransportModel} of
this thesis we discuss several of these observations in more detail,
and we propose a model for $\nuAD=2$ antidots in the low- to
intermediate-$B$ regime in terms of a `maximum density droplet' of
holes in the lowest Landau level (LLL), in which interactions
influence many aspects of the antidot spin-structure through
exchange effects, while preserving the SP-like excitations we
observe in non-linear transport measurements. The results of this
chapter therefore serve as an example of the flexibility of
non-interacting physics and the importance of a full consideration
of device geometry for precise comparisons with experimental
measurements.  Clearly, when one is looking for evidence of new
physics based on the disparity between measurements and the
predictions of an accepted model, it is important to carefully
consider whether certain features are completely incompatible with
the model at hand, or if they could be explained through minor
alterations.

\section{Background and motivation}

\begin{figure}[tb]
    \centering
    \includegraphics[]{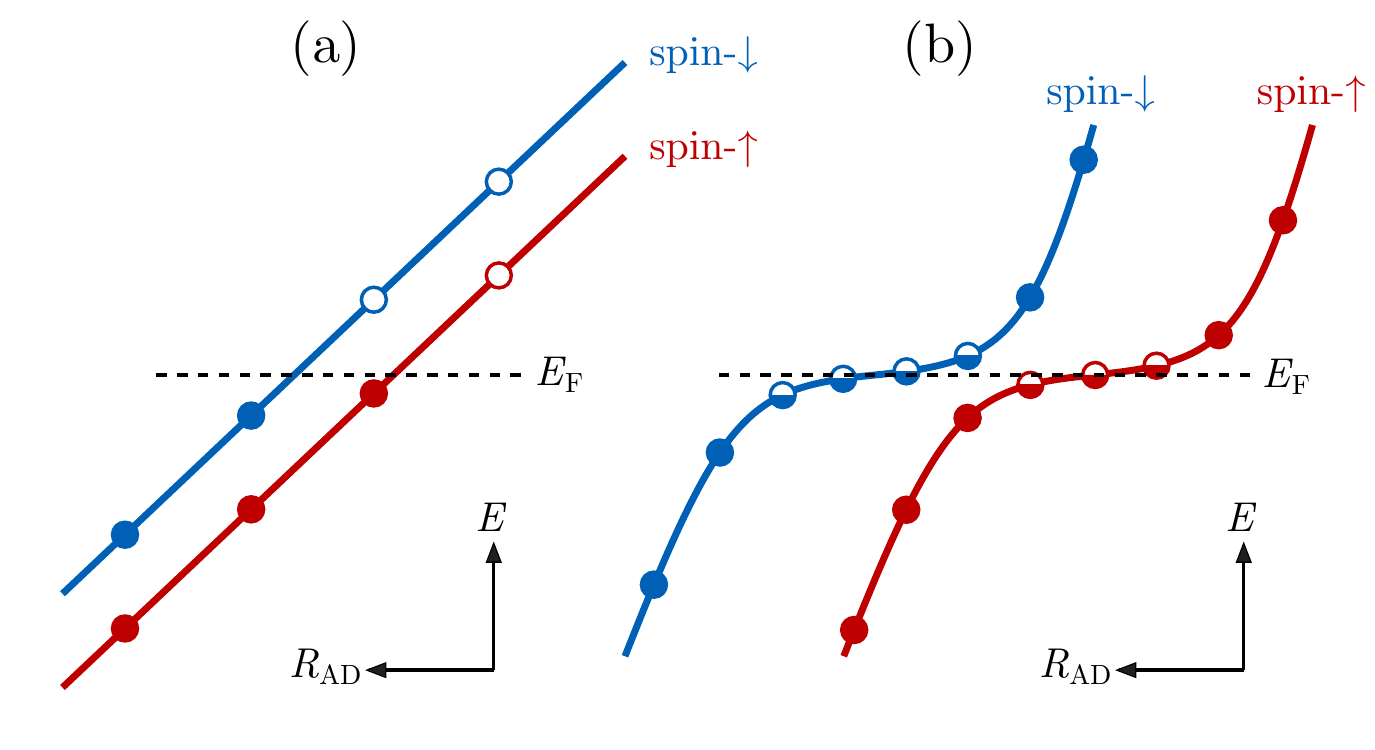}
    \caption[Compressible regions]{%
Schematic of antidot orbital energy-levels within the LLL for the SP
model (a) and in a self-consistent model including compressible
regions (b).  The CRs consist of partially-occupied orbitals which
screen the background antidot potential, separated by incompressible
regions in which the self-consistent potential varies rapidly
between LLs. \label{fig:antidotCRs}}
\end{figure}

The CR model of antidot transport is based on the self-consistent
model of quantum Hall edge-modes due to Chklovskii, Shklovskii, and
Glazman \cite{Chklovskii1992}, as depicted in
\figref{fig:antidotCRs}.
In order to avoid the large Coulomb-energy cost associated with the
abrupt change in electron density at the antidot edge which is
predicted by the SP model, the spin-split LLL could rearrange as
shown in the right-hand side of \figref{fig:antidotCRs}.  The
orbital states in each CR are partially occupied, such that they
screen the antidot potential and allow $n_e$ to vary smoothly
between LLs ($n_e$ changes by $eB/h$ in each CR).  This ability of
CRs to screen electric charge results in experimentally observable
effects.  In a similar manner to a set of concentric cylinders
acting as a capacitor, the outer CR, composed of the $\nu=2$
(\spindn) electrons, is able to screen the charge which accumulates
in the inner $\nu=1$ (\spinup) states.  Transport through the
antidot then occurs entirely through the \spindn\ CR with the
resonance condition $\mudn=0$, determined the by gate- and
field-dependent chemical potential
\begin{equation}
  \mudn(\Nup,\Ndn) = E(\Nup,\Ndn+1)-E(\Nup,\Ndn),
\end{equation}
where $E(\Nup,\Ndn)$ is the configuration energy of an antidot
containing $\Nup$ and $\Ndn$ particles in the \spinup\ and \spindn\
states, respectively.  Such a system may be naturally modeled as a
classical capacitor network, and it is straightforward to show that
strong coupling between the two rings results in periodic transport
resonances for the outer ring at twice the Aharonov-Bohm frequency,
in agreement with experimental observations at high fields
\cite{Sim2003}.  Kondo resonances in antidots may also be explained
through the capacitive model, as a second-order transport process
which occurs at $\musf^\pm = 0$, where
\begin{equation}
  \musf^\pm = E(\Nup\pm1,\Ndn\mp1) - E(\Nup,\Ndn),
\end{equation}
but this requires nearly equal antidot-lead couplings for both
\spinup\ and \spindn\ tunneling, which seems unlikely given the
spatial separation of CRs in the self-consistent model.  The Kondo
effect also emerges naturally from the microscopic model of antidot
maximum density droplets \cite{Sim2004,Hwang2004}, and it is
probably that regime in which it is most relevant. Experimentally,
we find that Kondo resonances disappear as $B$ increases, leaving
behind pure double-frequency oscillations; this may reflect the
breakdown of the maximum density droplet and the separation of spins
into separate CRs.

As mentioned above, the proposal of a self-consistent antidot model
including CRs has provoked some disagreement within the transport
community. Karakurt~et~al.\ \cite{Karakurt2001} made careful
measurements of the temperature dependence of individual antidot
resonances, and concluded that they were consistent with a model of
thermally-broadened sequential transport through an individual
quantum state.  Noting that the temperature dependence of
charging-dominated transport through a uniform continuum of states
would be quite different (temperature independent, in fact), they
concluded that CRs were absent in their device. Since then it has
been pointed out \cite{Kataoka2004} that the density of states for a
CR is highly non-uniform, being strongly peaked near the Fermi
energy, and thus that the temperature dependence of transport
resonances would be similar to that of a single state. At present,
exactly what measurement could constitute an irrefutable proof or
disproof for the presence of CRs in antidots is still an open
question.  On the theoretical side, Ihnatsenka and Zozoulenko have
performed density functional theory calculations, including spin, to
investigate the self-consistent quantum structure of antidots
\cite{Ihnatsenka2006a}.  They found that, for an antidot of radius
\unit{200}{\nano\metre}, a CR forms only for the outer (\spindn)
state at fields above $\approx\unit{4}{\tesla}$.  It is worth
pointing out that an outer CR is all that is necessary within the
capacitive model for double-frequency oscillations, as long as it
may efficiently screen the \spinup\ charge which accumulates within
it.

In their measurements of an antidot at intermediate fields,
Michael~et~al.\ \cite{Michael2006} were the first to obtain clear
examples of antidot \emph{excitation spectra} from non-linear
transport measurements.  Such excitations are routinely observed in
quantum dot measurements as a function of source-drain bias,
allowing for the identification of quantum numbers and comparisons
with theoretical calculations \citep[e.g.,][]{Kouwenhoven1997}, but
have been elusive in antidot measurements.  On the whole, the
observed antidot excitation spectra offer further support for the SP
model, reflecting a pair of excitation energies which are readily
identified with the Zeeman energy $\Ez$ and the SP spacing between
adjacent orbital levels, $\dEsp$.  Charging effects are incorporated
within the constant-interaction model \cite{McEuen1991}, and the
charging energy $\Ec$ may also be extracted from the non-linear
transport measurements.

In the perturbative limit discussed in \secref{sec:ABeffect}, where
the antidot potential varies on a length scale much larger than the
magnetic length $\ellB = \sqrt{\hbar/eB}$, we can use the
free-particle LLL eigenstates as approximate antidot wave functions.
The radial positions of these states are governed by the
flux-quantisation condition $\pi\langle r^2\rangle B = mh/e$, where
$m$ is an integer, and so we can approximate the energy-separation
between adjacent states as
\begin{equation} \label{eq:dEspapprox1}
  \dEsp \simeq
  \Delta r\left.\frac{dU}{dr}\right\rvert_{R_\mathrm{AD}}
  \simeq   \left(\frac{\hbar}{eBR_\mathrm{AD}}\right)
      \left.\frac{dU}{dr}\right\rvert_{R_\mathrm{AD}},
\end{equation}
in terms of the antidot potential $U(r)$ and the change $\Delta r$
required to add one flux quantum $h/e$ to the loop.  The antidot
radius $R_\mathrm{AD}$ is determined by the antidot gate potential
and the electron density, so we expect it to be roughly independent
of $B$.  Since the potential slope at $R_\mathrm{AD}$ is also
independent of $B$, we therefore expect to observe $\dEsp\propto
1/B$.  Michael~et~al.\ observed a significant suppression of $\dEsp$
below the expected $1/B$ dependence at higher fields, and they
suggested that this could imply a reorganisation of states into a
CR.  Here we propose an alternate explanation, namely that the
presence of the potential due to the split-gate in which the antidot
is embedded breaks the circular symmetry of the problem, and can
lead to a significant reduction of $\dEsp$ for states near the
saddle point of each constriction.  We were guided to this theory by
a realisation that the observed suppression of $\dEsp$ seemed to
coincide more with the gate-dependent position of the high-$B$ end
of the $\nuAD=2$ plateau, at which states begin to be reflected
across the antidot constrictions, than with any fixed value of $B$.

\begin{figure}[tb]
    \centering
    \includegraphics[]{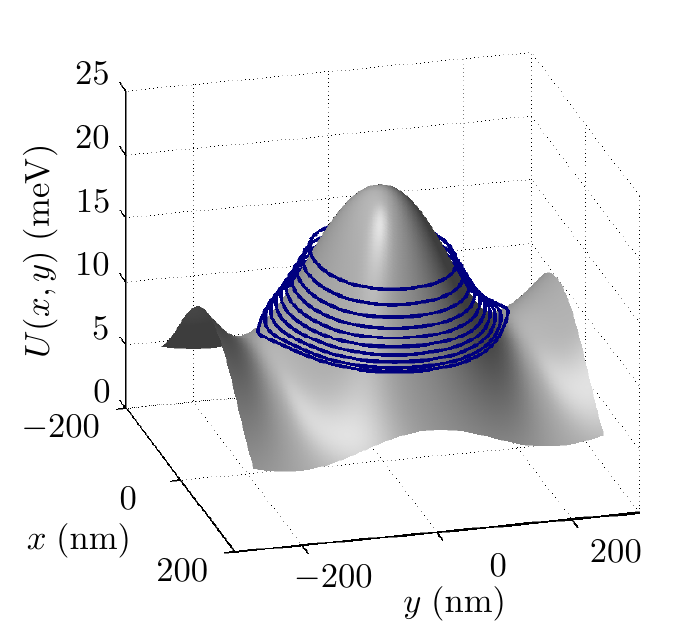}
    \includegraphics[]{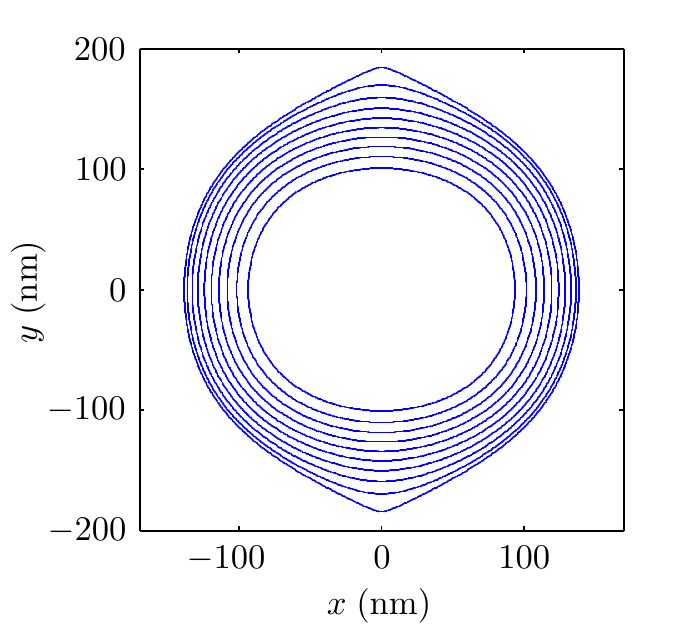}
    \caption[`Bulging' of SP orbitals]{%
Left panel --- Bare potential created by an antidot at the centre of
a split gate (computed as in \cite{Davies1995}), with contours
computed according to \eqnref{eq:dEspfromcontour} at
$B=\unit{0.8}{\tesla}$.  Right panel
--- top view of the contours, showing the `bulging' effect for
states close to the saddle point, which results in a reduced energy
spacing. \label{fig:SPorbitals}}
\end{figure}

The idea may be easily understood from a consideration of the
antidot states shown in \figref{fig:SPorbitals}.
Assuming that the states are well-approximated by `loops' of width
$\ellB$, with contours determined by the flux-quantisation
condition, we observe that the `bulging' of states near saddle point
in each constriction accounts for a relatively large amount of
additional enclosed area. This results in a reduced spatial
separation around the remainder of the antidot, and assuming a
roughly linear potential gradient in these regions, requires $\dEsp$
to be much less than \eqnref{eq:dEspapprox1} predicts.  We can
generalise \eqnref{eq:dEspapprox1} to account for this asymmetry, by
computing $\dEsp$ from the contours obtained for a given antidot
potential at the Fermi energy $\Ef$:
\begin{equation}
\dEsp = -\frac{h}{eB}\left[ \int_0^{2\pi}\left(\frac{dU}{dr}
\right)_{(\mathcal{C},\theta)}^{-1}\mathcal{C}(\theta)d\theta\right]^{-1},
\label{eq:dEspfromcontour}
\end{equation}
where $\mathcal{C}(\theta)$ is the contour defined by
$U_\mathrm{eff}(\mathcal{C},\theta) = \Ef$, using the effective
potential
\begin{equation}
  U_\mathrm{eff} = U(r,\theta) + E_\mathrm{cyc} + \Ez,
\end{equation}
where $E_\mathrm{cyc}=\hbar\omegac/2$ and $\Ez
=g\mu_\mathrm{B}B\sigma$ are the cyclotron and Zeeman energies for
the LLL states in the constrictions (with spin
$\sigma=\pm\frac{1}{2}$), as described in \secref{sec:SPstates} of
this thesis.  In the following discussion, we find that this model
is highly successful at describing the observed $B$-dependence of
$\dEsp$, and we will justify the flux-quantisation assumption on
which it is based by comparing its predictions with a calculation of
the full non-interacting Green's function for a realistic `antidot
$+$ split gate' geometry.

\section{Results}

\begin{figure}[tp]
    \centering
    \includegraphics[]{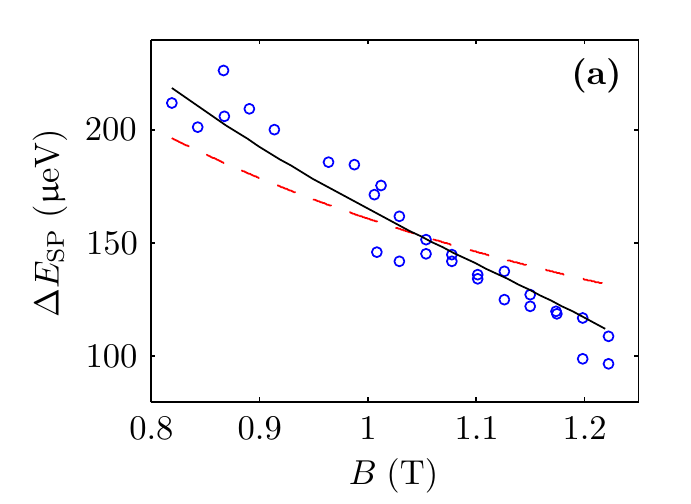}
    \includegraphics[]{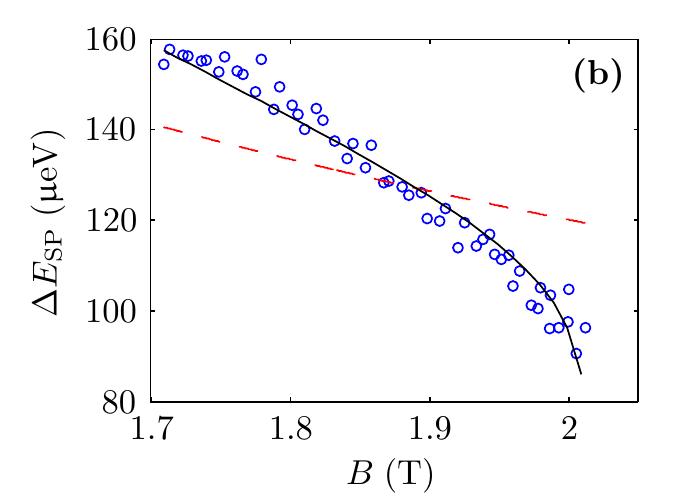}\\
    \includegraphics[]{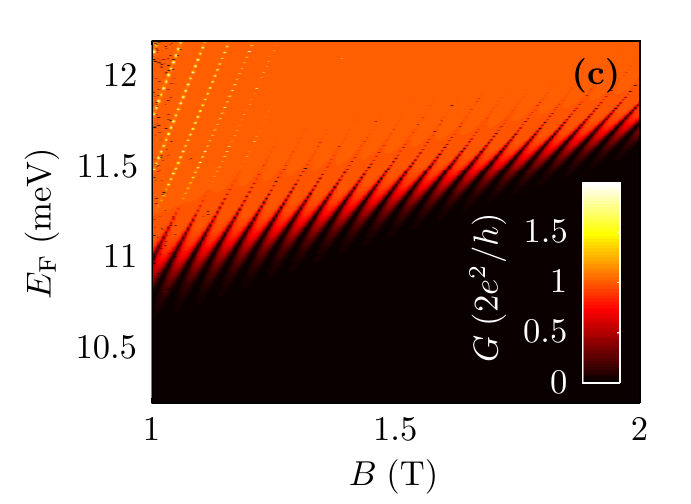}
    \includegraphics[]{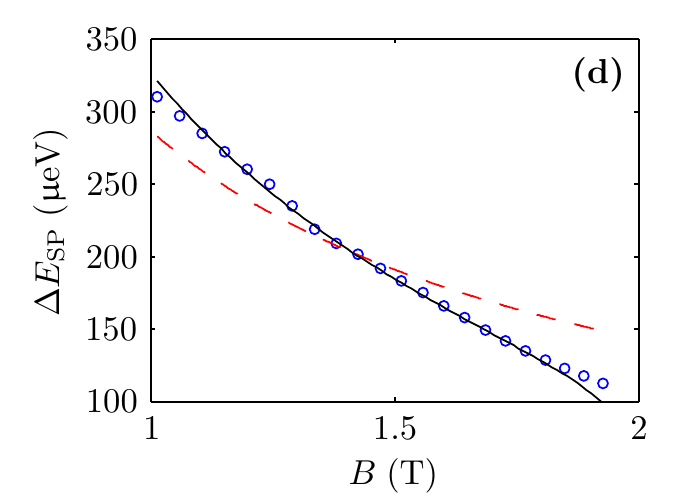}
    \caption[Suppression of $\dEsp$]{%
Top panels ---  Single-particle energy $\dEsp$ (circles) extracted
from DC-bias measurements in different ranges of magnetic field
(with different gate voltages). A fit $\propto 1/B$ (dashed red
curve) fails to match the data while our model (solid black curve)
predicts the reduction of $\dEsp$ at higher fields. Bottom panels
--- Conductance (c) in units of $2e^2/h$ as a function of $B$ and
$\Ef$, calculated from the noninteracting Green's function for the
antidot potential shown in \figref{fig:SPorbitals}, and the
corresponding energy spacing (d), calculated from (c) at
$\Ef=\unit{11.7}{\milli\electronvolt}$. As for the experimental data
in (a) and (b), our model based on \eqnref{eq:dEspfromcontour}
(solid curve) accounts for the discrepancy of the calculated values
of $\dEsp$ from a $1/B$ dependence (dashed curve).\label{fig:fits}}
\end{figure}

The measurements of $\dEsp$ versus $B$ shown in panels (a) and (b)
of \figref{fig:fits} were obtained by Chris Michael, having been
extracted from a series of non-linear transport measurements as
described in \refref{Michael2006}. The device consists of an AD gate
\unit{200}{\nano\metre} in diameter centred in a split gate of width
\unit{1}{\micro\metre}. Complete device details may be found in
\refref{Michael2006}.
These two sets of measurements were taken at different settings of
the antidot and split-gate voltages, to tune the $\nuAD=2$ plateau
to different values of $B$, and are representative of measurements
taken throughout the range $B\approx0.5$\nbd\unit{2.5}{\tesla}.  It
is clearly observed that $\dEsp$ changes faster than expected with
$B$ (the dashed red lines represents best-fit functions
$\propto1/B$).  Our model based on \eqnref{eq:dEspfromcontour}
(black curves in \figref{fig:fits}) performs much better.  It is
based on the bare potential produced by the lithographic arrangement
of gates, computed as the solution to the Laplace equation at the
position of the 2DES, as described in \refref{Davies1995}.  We allow
for a constant screening factor to account for the ionised donor
layer, but this is completely determined by the measured
Aharonov-Bohm period and $\dEsp$ at the low-$B$ end of the plateau.
We can also estimate $\Ef$ from the field at which the $\nu=2$ state
is depopulated in the channel (the midpoint of the transition
between the $\nuAD=1$ and $\nuAD=2$ plateaux).  The remaining
parameters $E_\mathrm{cyc}$ and $\Ez$ are determined solely by $B$,
so the functions shown in \figref{fig:fits} actually have no free
parameters for fitting.  This calculation includes no effects of
tunneling in the constrictions, which results in an artificial drop
to zero as the saddle point reaches $\Ef$ and closed orbits no
longer exist.

For additional comparison with this essentially classical model, and
to justify the flux-quantisation condition which leads to
\eqnref{eq:dEspfromcontour}, we have calculated the full
non-interacting Green's function for an AD $+$ split-gate geometry
using an iterative procedure \cite{MacKinnon1985}.  The connection
between the time-independent Green's function of an open geometry
and the quantities measured in transport experiments is discussed in
\secref{sec:GFs}.  The calculation does not include spin, but the
orbital spacing $\dEsp$ is an orbital effect, a consideration of
spin is not really necessary.  We note that, although the values of
$\dEsp$ may actually be slightly different for \spinup\ and \spindn\
states at the Fermi energy due to this geometric effect, the
lowest-energy excitations of an antidot with total spin $S_z =
\frac{1}{2}$ and $S_z=0$ are $\Ez$ and $\dEsp-\Ez$ respectively,
where $\dEsp$ is the value for \spindn\ only, so it is a suitable
approximation to consider spinless electrons in this calculation.
Panel (c) of \figref{fig:fits} shows the calculated antidot
conductance as a function of $\Ef$ and $B$, from which we extract
$\dEsp$ as the vertical spacing between resonances along a
horizontal line at constant $\Ef$, as shown in \figref{fig:fits}d.
The agreement between this calculation and our model is highly
satisfactory, and since the calculation includes no effects of
electron interactions, we can conclude that a self-consistent model
(e.g., including CRs) is not necessary to explain these
observations.

\begin{figure}[tb]
    \centering
    \includegraphics[]{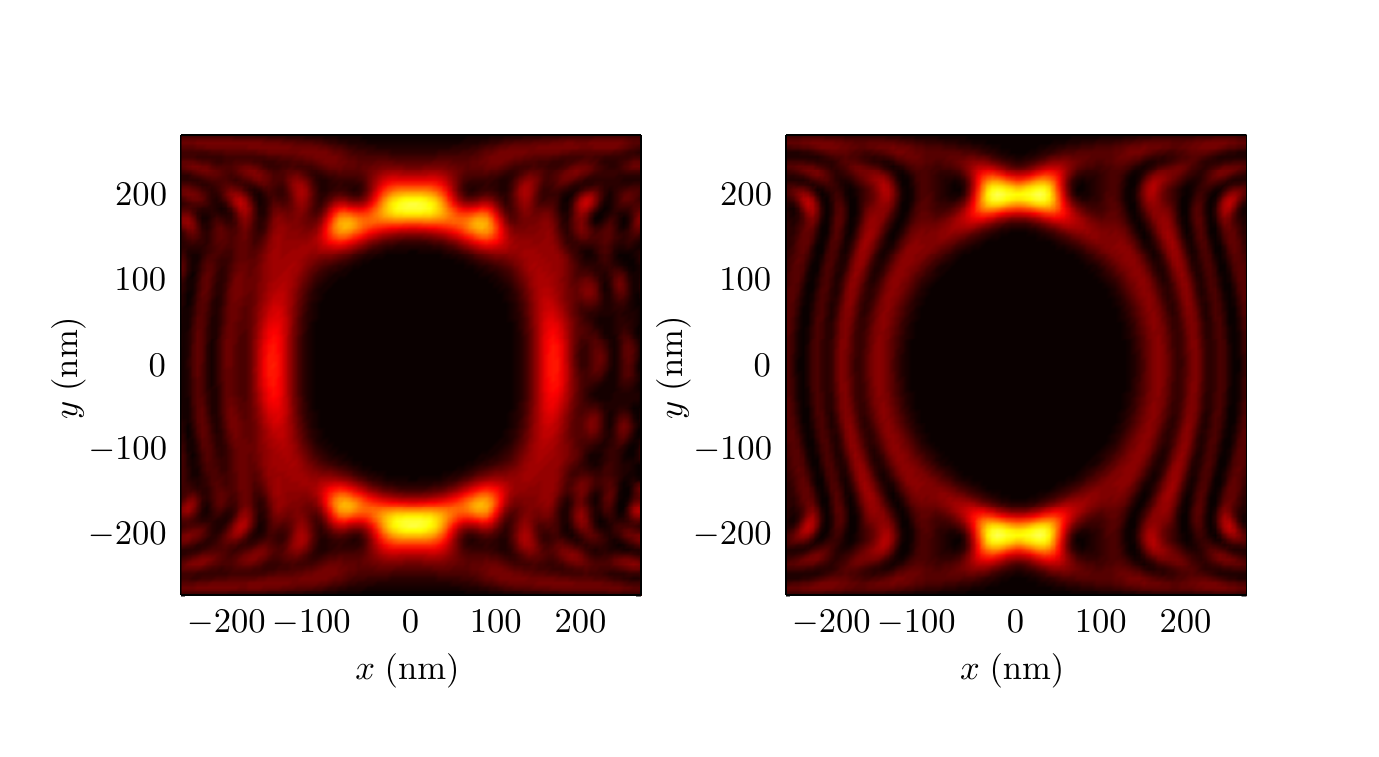}
    \caption[Local density of states]{%
Local density of states calculated from the noninteracting Green's
function for the AD shown in \figref{fig:SPorbitals}.  Note that the
transmission resonance (left panel) at low field
($\approx\unit{1}{\tesla}$) is more circularly symmetric than the
reflection resonance (right panel) at higher field
($\approx\unit{1.5}{\tesla}$), which appears vertically elongated
due to the `bulging' of the SP states into the constrictions.
\label{fig:GF}}
\end{figure}

From the Green's function we can also compute the local density of
states, derived in \secref{sec:GFs} as
\begin{equation}
  \rho(\bx;E) = \mp \frac{1}{\pi}\Im\bigl[G^\pm(\bx,\bx;E)\bigr].
\end{equation}
This is useful as a visualisation tool, showing the spatial
structure of states near $\Ef$.  In \figref{fig:GF}, we plot the
calculated local density of states at the locations of both a
transmission and reflection resonance, in the left and right panels
respectively.  The departure from circular symmetry can be easily
observed in these plots, particularly for the reflection resonance
which is due to an antidot state very near to the saddle point.

\section{Conclusions}

We therefore conclude that the observed suppression of $\dEsp$ is a
simple result of the potential profile in our experimental geometry,
rather than a signature of a reorganisation of states into a CR.
Although we know that interaction effects become essential for an
understanding of AD resonances at high fields, this study
demonstrates the ability of the SP model to explain relatively
complicated features of the excitation spectrum of ADs in the
low-field regime. An understanding of these effects is critical for
the design of antidot-based applications which seek to utilise SP
processes in a specific regime, or for the study of other aspects of
transport which we do believe are signatures of electron-electron
interactions, as discussed in \chapsref{chap:SpinTransport} and
\ref{chap:SpinTransportModel}.

\chapter[Spin-Resolved Transport: Signatures of Interactions]{Spin-Resolved Transport: \\
Signatures of Interactions\label{chap:SpinTransport}}
\chaptermark{Spin-Resolved Transport: Experiments}

\ifpdf
    \graphicspath{{Chapter4/Figures/PNG/}{Chapter4/Figures/PDF/}{Chapter4/Figures/}}
\else
    \graphicspath{{Chapter4/Figures/EPS/}{Chapter4/Figures/}}
\fi


The spin structure of a quantum antidot is intimately connected to
the details of the Coulomb interactions between electrons through
the exchange effect which results from Pauli exclusion, as we have
seen in \chapref{chap:ADtheory}.  Furthermore, electron spin is
generally conserved for quantum Hall (QH) transport experiments in
GaAs; of the two relevant spin-mixing processes, spin-orbit coupling
is relatively weak, and the hyperfine interaction requires a close
degeneracy (in both energy and space) between electrons of opposite
spins, which is usually absent in QH systems. The antidot SP
spectrum described in \chapref{chap:ADtheory}, being composed of
pure spin states, clearly imposes strict spin selection rules for
tunnelling events between the antidot and the leads.  Even after
including interactions, the total spin projection $S_z$ of the
antidot remains a good quantum number, as long as the spin-mixing
effects mentioned above are negligible, and so similar selection
rules remain.  We can therefore gain valuable information about the
underlying structure of the antidot states by measuring the spin of
the electrons involved in transport.  Experimentally, this is made
possible by the unique properties of QH systems in which the `leads'
of different spins actually correspond to spatially distinct QH edge
modes, which we can physically separate in a device by means of
quantum point contacts (QPCs).  By isolating the current in each
mode before and after they reach the antidot, we identify the spin
of the relevant electrons.  In this chapter, we describe how such
measurements lead to a detailed picture of the energy spectrum at
the $\nuAD=2$ antidot edge, including spin.  Our surprising result
is that, while standard transport measurements seem broadly
consistent with the non-interacting model, the observed
spin-resolved transport is not. We interpret these results as the
signatures of electron-electron interactions which lead to a
separation of scales for orbital and spin excitations of the
many-body antidot ground state.  The results are consistent with the
ground state being a maximum-density-droplet (MDD) of `holes' in the
LLL, and so these measurements are of general interest as a method
of experimentally probing the many-body physics of such systems.

\section[Motivation]{Motivation and Previous Studies}

To date, most studies of electron-electron interactions in single
antidot structures have concerned the region of intermediate
($B\approx$~1--\unit{3}{\tesla}) to high
($B\gtrsim$~\unit{3}{\tesla}) magnetic field, because it is in this
regime that the SP picture clearly breaks down, failing to describe
even equilibrium conductance measurements.  The AB reflection
resonances between $\nuAD=$1--2 have received particular attention,
since this is the `simplest' regime in which to study
interacting electrons of both spins.\footnote{%
The transmission resonances observed above the $\nuAD=2$ plateau
disappear at relatively low fields, since the inter-LL spacing
cannot be controlled with gate voltages as can the intra-LL distance
in the constrictions, and the inter-LL tunnel couplings drop off
rapidly with the decreasing magnetic length.} %
At intermediate fields, extra resonances appear (in addition to the
two resonances per period expected from the SP model) which have
many of the features of Kondo resonances in quantum dots
\cite{Kataoka2002,Kataoka2004a}.  Then, at even higher fields, these
additional resonances disappear and the frequency of the
oscillations doubles \emph{exactly}, in the sense that no modulation
remains at the base AB period observed at lower fields
\cite{Ford1994}.  This frequency-doubling is naturally explained by
a model in which Coulomb (plus exchange) interactions drive the
spin-$\uparrow$ and spin-$\downarrow$ edges apart spatially,
eventually leading to the formation of a set of concentric
compressible regions similar to those believed to exist for bulk
edge states \cite{Chklovskii1992}.  The classical electrostatic
interaction between the edges of opposite spin results in charging
resonances at twice the AB frequency, as the outer state screens the
charge accumulated in the inner ring \cite{Kataoka2000}. Although
there has been some disagreement about the observable properties of
CRs and the fields at which they should form
\cite{Karakurt2001,Kataoka2004,Goldman2004}, it is generally
accepted that a self-consistent description including
electron-electron interactions is necessary to describe antidot
physics at these high fields.  The generalised charging model
presented by \refref{Sim2003} captures most of these features well,
and is also sufficient to explain the pattern of Kondo resonances
observed at intermediate fields.  Aharonov-Bohm `subperiods' have
also been observed in large quantum dots, and have been explained
through similar arguments based on charging-dominated transport
\cite{Rosenow2007}.  A recent review of both experimental
observations and relevant theoretical descriptions of antidots in
this regime is given by \refref{Sim2008}.

In magnetic fields below \unit{1}{\tesla}, it has been generally
assumed that Coulomb interactions play only a minor role,
contributing a charging energy to the resonance condition as in the
constant interaction model described in
\secref{sec:SequentialTransport} but otherwise leaving the SP
spectrum unaltered.  Equilibrium conductance measurements as a
function of both gate voltages and $B$ are well-described by this
model, with even complicated lineshapes and amplitude modulation
explained through a consideration of the geometry and variation of
the antidot-lead tunnel couplings, as in the edge-state model of
Mace~et~al.\ \cite{Mace1995} (hereafter, the Mace-Barnes model).
Recently, Michael et al.\ performed a series of non-linear
conductance measurements in this regime, obtaining the first
unambiguous examples of antidot excitation spectra
\cite{Michael2004,Michael2006}.  Especially on the $\nuAD=2$
plateau, the excitation spectrum thus observed qualitatively fits
the SP model, reflecting a pair of energy scales which are
straightforwardly identified with the bare orbital and Zeeman
energies $\dEsp$ and $\Ez$, respectively.

The major open question, then, is the nature of the evolution of the
SP antidot states at low $B$ to a more complicated self-consistent
arrangement dominated by electron-electron interactions at higher
fields.  It is of course possible that interactions are important
throughout the entire field range, but a model including them needs
to preserve the observed structure of the SP model at low fields and
explain the changes which occur as $B$ increases. Suspiciously,
beyond the basic picture of SP excitations in the observations of
Michael \cite{Michael2004} are several intriguing details which do
not seem consistent with non-interacting physics. We considered one
of these in \chapref{chap:Geometry} --- the observed suppression of
$\dEsp$ with increasing $B$ --- and showed that a realistic
consideration of the antidot geometry is sufficient to explain it
within the SP model. Still, several other puzzles remain. The
`competition' between transmission and reflection resonances at the
high-$B$ edge of the $\nuAD=2$ plateau is particularly intriguing,
since associated peaks and dips occasionally appear at slightly
different positions and with different widths, resulting in
asymmetric lineshapes in the transition region. This behaviour is
clearly inconsistent with the Mace-Barnes model, in which the
transmission and reflection resonances result respectively from
inter- and intra-LL
tunnelling through the same antidot state.\footnote{%
Similar features were reported at the analogous position on the
$\nuAD=1$ plateau, which is particularly surprising since the
antidot states (and transport electrons) should be fully
spin-polarised.} %
The behaviour of the Zeeman splitting poses another puzzle.  As a
function of increasing $B$, the spacing of peaks/dips within each
pair of resonances initially increases
linearly as expected within the SP picture,\footnote{%
The slope of this increase, however, reflects a somewhat enhanced
value for the Land\'{e} $g$-factor of $\abs{g}\approx 0.6$\nbd0.7
compared to the bulk GaAs value of $\abs{g}=0.44$
\cite{Michael2006}.} %
but then appears to saturate, fluctuating about a value close to
half of the full AB period before locking at exactly that value in
the double-frequency regime at high field, hinting at a crossover
regime between non-interacting and self-consistent behaviour.

Most of the open questions described above concern the rearrangement
of spins in the $\nuAD=2$ antidot as a function of $B$ and the
coupling of these states to the various bulk edge modes which convey
the transport electrons.  Information about the spin(s) involved in
each resonance in this regime is therefore highly desirable, which
we present in this Chapter through a set of spin-resolved transport
experiments. Spin-resolved experiments have been performed before on
antidots, as pioneered by Kataoka et al.\ \cite{Kataoka2003}, but
they were limited to the high-$B$ regime by the equilibration
between edge modes which prevents selective injection/detection at
lower fields. We circumvented this restriction through a careful
selection of device parameters as described in the next section.
Besides the desire for a more complete understanding of electron
interactions in antidot systems, we were motivated to perform
spin-selective measurements in the low-$B$ regime by the proposal
\cite{Zozoulenko2004} to use a single antidot as a spin filter, or a
system of two antidots in series as a `spin switch,' in which a
current of either spin polarisation can be turned on and off at will
through minute adjustments of the antidot gate voltages.  The
proposal relies essentially on the Mace-Barnes picture of reflection
resonances in SP antidots, which we have found is somewhat flawed,
but the concept is still feasible with transmission resonances
instead, requiring only an antidot operating in the SP regime with
well-separated pure-spin states which serve as polarisers for the
transmitted current.  Such a tunable spin injector/detector could be
useful as a component in larger devices which use QH edge channels
for coherent electron transport.

\section{Experimental methods}

The zero-field transport mean free path of two-dimensional electrons
at the Fermi energy in typical high-mobility GaAs-AlGaAs
heterostructures is on the order of 10--\unit{100}{\micro\metre},
which is already larger than the active region of the mesoscopic
devices we are considering. In the QH regime, the equilibration
length for electrons traveling in edge modes is larger still,
reaching $\approx$~\unit{1}{\milli\metre} in fields of several Telsa
\cite{Wees1989,Muller1992}.  The suppression of scattering events is
due to a small spatial overlap of the states --- for scattering
between LLs this is naturally explained by an increased spatial
separation (through the LL spacing $\hbar\omegac$) and decreased
width (the magnetic length $\ellB$) at higher $B$.  The situation of
spin-flip scattering within LLs is more complicated, since the
Zeeman energy due to the bare $g$-factor is too small to separate
opposite-spin states significantly,\footnote{%
In GaAs, $\Ez/\hbar\omegac = gm_e^\ast/2m_e \approx 0.01$.} %
but both spin-orbit and exchange effects lead to an enhanced
effective $g$-factor for electrons at the sample edge
\cite{Muller1992}, which reduces wave function overlap sufficiently
to suppress spin-flip scattering through the available mechanisms
(spin-orbit and/or hyperfine coupling) mentioned in the introduction
to this chapter.

\begin{figure}[p]
\begin{center}
\includegraphics[]{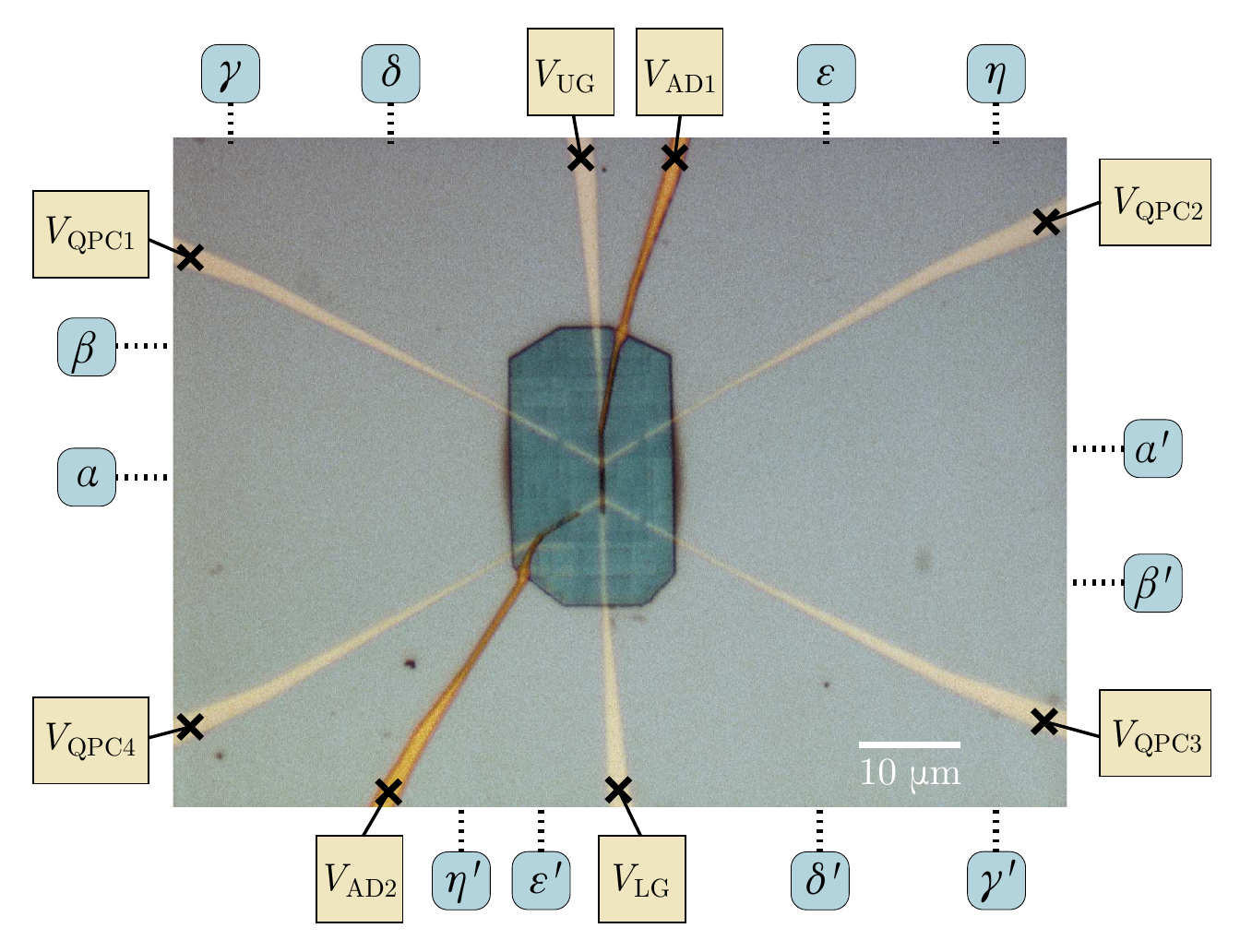}
\caption[Spin injection/detection device]{Optical microscope
photograph of the device used for the spin-selective measurements in
this chapter.  Ohmic contacts to the 2DES are labelled by Greek
letters.  The first layer of metallic Schottky gates consists of the
upper (UG), lower (LG), and four QPC gates, as well as isolated
circular antidot gates (AD1 at the centre and AD2 in QPC4).  The
active region is covered by a layer of crosslinked polymer, through
which metal bridging gates contact the antidots individually
\cite{Ford1989}.  A larger view of the first layer alone is shown in
the scanning electron micrograph of \figref{fig:SEMandCircuit}
\label{fig:Device}}
\end{center}
\end{figure}

Our measurements take advantage of this extraordinarily long
equilibration length to selectively introduce and subsequently
measure small non-equilibrium populations of electrons in the bulk
edge states, allowing us to extract the individual elements of the
antidot scattering probability matrix described in
\chapref{chap:TransportTheory}, rather than just the trace of this
matrix provided by the equilibrium conductance. Our device, shown in
\figref{fig:Device}, is largely based on the design of the original
spin-selective experiments of Kataoka et al.\ \cite{Kataoka2003},
with a few minor alterations as described below to facilitate its
low-field operation. QPCs are added to each of the incoming and
outgoing edges, behind which are ohmic contacts which allow us to
determine the current or voltage of the modes allowed through each
QPC.  The device can be operated in a variety of modes, depending on
whether the ohmic contacts are used to source or probe currents or
voltages, but all of these configurations are easily analysed within
the Landauer-B\"{u}ttiker formalism presented in
\secref{sec:LBformalism}.  Throughout this chapter, we refer to the
spin-resolved LLs by their `index' as counted from the edge of the
sample.  Thus modes (1,2) make up the LLL, (3,4) the next LL, etc.,
such that all odd (even) numbered modes consist of \spinup\
(\spindn) electrons.  In treating the SP model of antidot transport,
we adopt the same labelling of LLL states in the $\nuAD=2$ antidot,
referring to the spin-preserving transmission and reflection
resonances as, \onetothree/\twotofour\ and \onetoone/\twototwo\,
respectively.

\subsection{Design considerations}

The spin-selective experiments of Kataoka et al.\ \cite{Kataoka2003}
proved that the double-frequency $\nuAD=1$--2 oscillations observed
at high $B$ are fully spin-polarised, resulting from only
``\twototwo'' tunnelling across the constrictions.  This is in
agreement with the self-consistent model in which the Coulomb
blockade is lifted for \spindn\ tunnelling twice per AB cycle due to
the capacitive interaction between spatially-separated rings of
opposite spin \cite{Sim2003}.  For fields below
$B\approx$~\unit{3}{\tesla}, however, both inter-LL and spin-flip
equilibration became much more pronounced, obscuring the
interpretation of the selective injection/detection measurements. In
order to extend the method to lower fields, we made two major
alterations to the device.  First, we used a wafer\footnote{%
T792 --- full specifications are in \appref{app:Wafers}.
Measurements on T792 were performed in the dark, since the device
became unstable after illumination.  This is in contrast to most
previous antidot experiments, in which strong AB oscillations were
only observed after heavy illumination.  Indeed, we tested several
other wafers with similar basic parameters to T792 and found
unsatisfactory results in the dark.  Unfortunately, it is still
unclear what constitutes a `good' wafer for antidot experiments, and
it is often the case that a given wafer will only operate
satisfactorily either in the dark or after illumination since the
changes to the band structure produced by the rearrangement of donor
charges is so severe.} %
with a significantly reduced carrier density of $n_e =
\numprint{1.1e11}$~\centi\metre\rpsquared, compared to $n_e =
\numprint{3e11}$~\centi\metre\rpsquared\ in the experiments of
Ref.~\cite{Kataoka2003}. With a reduced density, the filling factor
in the bulk is lower for a given value of $B$, and the potential
slope at the edges is shallower, causing the spacing between edge
modes to be larger. The exchange effect is also stronger at lower
densities \cite{Smith1992}, helping to reduce spin-flip scattering
in edge transport as discussed above.  Second, we redesigned the
device used in Ref.~\cite{Kataoka2003} to eliminate the corners in
the contours of the main edge between the QPCs and the central
antidot, which are thought to contribute to inter-edge scattering
\cite{Olendski2005,Palacios1992}. The edge contours at the Fermi
energy were calculated using the
\texttt{GatesCalc} program,\footnote{%
Written and maintained by Adam Thorn, available within the
Semiconductor Physics Group at
\url{http://spz.sp.phy.cam.ac.uk/wiki/index.php/GatesCalc}.} %
which computes the bare electrostatic potential due to the gates as
the solution to Laplace's equation at the level of the 2DES.
Examples of contours calculated through this method may be seen in
\figvref{fig:SEMandCircuit}.

In addition to the modifications discussed above, we included a
second antidot (AD2) into one of the QPCs, in the lower-left of the
device as shown in \figref{fig:Device}.  This serves a dual purpose:
the two antidots in series reproduce the topology of the spin-filter
proposed in Ref.~\cite{Zozoulenko2004}, allowing us to explore the
behaviour of such a device, and, by making AD2 lithographically
smaller than the central antidot (AD1), we provide a method of
exploring antidot size-effects in a single device. When necessary,
we can still use AD2 as a simple QPC by applying a large negative
bias to the gate QPC4, fully depleting one of the channels around
the antidot.  We can then use the AD2 gate to control the remaining
QPC for injection/detection measurements of AD1.  The antidot size
plays an important role in determining the SP spacing between
orbital states, which in the perturbative limit discussed in
\chapref{chap:ADtheory} is given by
\begin{equation}\label{eq:dEspapprox}
  \dEsp \simeq
      \left(\frac{\hbar}{eBR_\mathrm{AD}}\right)
      \left.\frac{dU}{dr}\right\rvert_{R_\mathrm{AD}}.
\end{equation}
Assuming that the slope of the potential at the Fermi level is
relatively constant, this means that $\dEsp$ is roughly inversely
proportional to antidot size. We have also found that the antidot
size has a significant effect on the inter-LL
(\onetothree/\twotofour) coupling, with the transmission resonances
disappearing at much lower $B$ for smaller antidots.  This is
presumably related to either the reduced circumference over which
this tunnelling may occur in a smaller antidot, or to the smoother
deformation of the $\nu=3$,4 edge contours, or both.  In either
case, it presents a conflicting goal in designing a spin-filter
device based on these transmission resonances, since a smaller
antidot may provide a larger $\dEsp$ and hence larger level
selectivity, but may not operate at a large enough field to sustain
non-equilibrium populations of edge modes.  In fact, the
\onetothree/\twotofour\ transmission resonances we wish to enhance
at the antidot are essentially the same inter-LL scattering events
we want to avoid along the edges.  With careful design and tuning,
we have shown that this is possible in practice, but it is
nonetheless a delicate balance to strike.

\subsection{Calibration of edge equilibration\label{sec:EdgeScattering}}

\begin{figure}[tb]
\begin{center}
\includegraphics[]{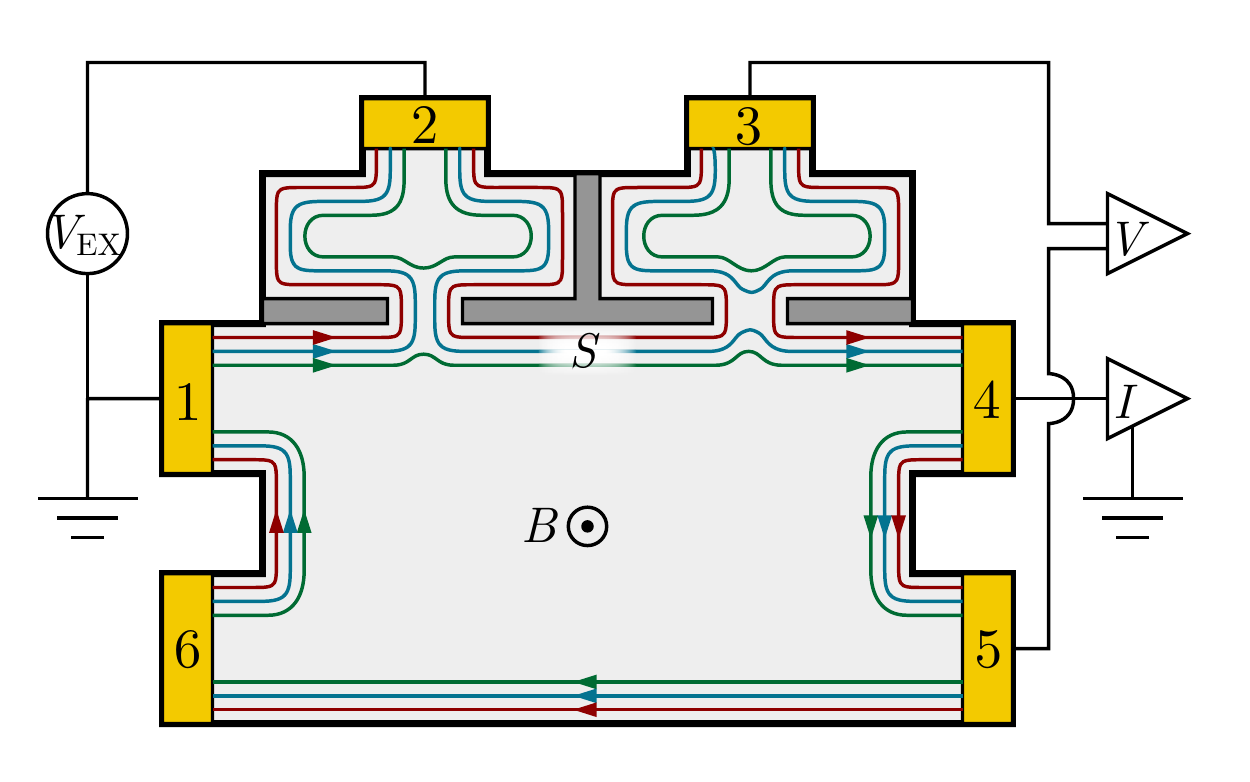}
\caption[Calibration of edge equilibration]{Schematic of the
edge-mode network and experimental circuit for the calibration of
scattering along the injector and detector edges.  Contact 2 serves
an the injector, and is subject to a small applied voltage
$V_\mathrm{ex}$. The filling factor of the injector QPC determines
the number of modes which are initially `populated' with chemical
potential $-eV_\mathrm{ex}$ along the main edge.  After scattering
amongst each other, a subset $\nu=1$ to $\fdet$ of modes are allowed
into the detector. In a four-terminal measurement, we measure the
current at contact 4 (the drain), and the potential difference
between contacts 3 and 5. \label{fig:EdgeCalibration}}
\end{center}
\end{figure}

The non-equilibrium measurements we make in this chapter are easily
analysed using the Landauer-B\"{u}ttiker formalism presented in
\secref{sec:LBformalism}.  In general, however, the transmission
coefficients $T_{\alpha\beta}$ we measure result from a combination
of scattering events at the antidot and inter-edge-mode scattering
along the edges separating the injector and detector QPC from the
antidot.  In order to extract the contribution from antidot
scattering alone, we must account for, or preferably eliminate, the
edge scattering.  To accomplish this, we first perform selective
injection/detection measurements on the edges alone, by setting gate
voltages appropriately to obtain the edge-mode topology shown
schematically in \figref{fig:EdgeCalibration}.   For example, to
calibrate the scattering due to the full length of the top edge, we
ground the voltages of the lower half of the device, including AD1,
and use gates QPC1 and QPC2 to perform selective
injection/detection. Or, by setting $V_\mathrm{BG}$ to a large
negative value, it is possible to deplete the lower antidot
constriction completely and use $V_\mathrm{AD1}$ to control the
remaining QPC, in order to further isolate the scattering
contributions from the injector and detector sides alone.  Analogous
settings allow for the calibration of the bottom edge.

Consider the experiment described in \figref{fig:EdgeCalibration}.
After calibrating the injector and detector QPCs such that the gate
voltages needed to produce integer filling factors $\finj$ and
$\fdet$ are known for given settings of $V_\mathrm{CG}$ and
$B$,\footnote{%
The QPCs are calibrated with standard two- or four-terminal
conductance measurements.  Our device was sufficiently stable, and
the integer plateaux were wide enough, that linear relationships
between the QPC gate voltages needed for each filling factor and
$V_\mathrm{CG}$, $B$ were sufficient for us to select the desired
QPC filling factors accurately on demand throughout the experiment.} %
we set up the four-terminal measurement shown.  The small AC
excitation voltage $V_0$ applied to ohmic contact 2 in the injector
leads to an initial population of the edge modes $\nu\leq\finj$ to
the chemical potential $\mu_\mathrm{ex}=-eV_0$, while the remaining
$(\nuB-\finj)$ modes have $\mu=0$, since they originated from ohmic
1, at ground potential. We denote the initial population by a vector
$\ainvec$, where
\begin{equation}
  a_i^\mathrm{in} =
      \begin{cases}
        1\quad\text{for $i\leq\finj$,}\\
        0\quad\text{otherwise.}
      \end{cases}
\end{equation}
After propagating along the edge, the resulting population at the
entrance to the detector is the vector $\aoutvec$, where
\begin{equation}
  a_i^\mathrm{out} = \sum_{i=1}^{\nuB}p_{ij}a_j^\mathrm{in}
      = \sum_{i=1}^{\finj}p_{ij},
\end{equation}
in terms of the elements of the scattering probability matrix,
$p_{ij}$.  These are the probabilities $\abs{t_{\alpha\beta,ij}}^2$
from the $S$-matrix of the full device, where $\alpha,\beta$ are the
contacts in which the modes $i,j$ terminate and originate,
respectively. Note that since the initial phases of the modes are
uncorrelated, the time-averaged quantities we measure do not reflect
any of the phase information included in the $S$-matrix amplitudes
$t_{\alpha\beta,ij}$.

The modes are then split at the detector QPC, with $\nu\leq\fdet$
passing through to contact 3, and the remaining modes reflected to
contact 4.  Thus the transmission coefficient $T_{32}$ for current
passing from the injector to the detector is given by
\begin{equation}
  T_{32} = \sum_{i=1}^{\fdet}a_i^\mathrm{out} =
  \sum_{i=1}^{\fdet}\sum_{j=1}^{\finj} p_{ij}.
\end{equation}
At this point, one could simply measure the current flowing out of
contact 3 to extract $T_{32}$ through the Landauer-B\"{u}ttiker
formula,
\begin{equation}
  I_3 = \frac{e^2}{h}V_0T_{32},\qquad\text{for $V_3=0$},
\end{equation}
but this would be a two-terminal measurement, and so we would need
to account for series resistance in the circuit to determine $V_0$.
Instead, we make the four-terminal measurement shown in
\figref{fig:EdgeCalibration}. For this configuration, the
Landauer-B\"{u}ttiker formalism gives the relations
\begin{subequations}
  \begin{align}
    I_4 & = -I_2 = \finj\frac{e^2}{h}V_0, \\
    V_3 & = \frac{T_{32}}{\fdet}V_0, \\
    V_5 & = V_6 = 0,
  \end{align}
\end{subequations}
and hence the non-equilibrium conductance is given by
\begin{equation}\label{eq:Gneqdefn}
\begin{split}
  \Gneq & = \frac{I_4}{V_3-V_5} =
      \frac{e^2}{h}\left(\frac{\finj\fdet}{T_{32}}\right) \\
        & = \frac{e^2}{h}\finj\fdet\left[
            \sum_{i=1}^{\fdet}\sum_{j=1}^{\finj} p_{ij}\right]^{-1},
\end{split}
\end{equation}
independent of the excitation voltage $V_0$.

If the propagation along the edge is perfectly adiabatic, i.e.\
$p_{ij} = \delta_{ij}$, then \eqnref{eq:Gneqdefn} reduces to
\begin{equation}
  \Gneq = \frac{e^2}{h}\max(\finj,\fdet),
\end{equation}
so we infer the presence of non-zero off-diagonal scattering
probabilities when the measured value deviates from this result.
Moreover, by making systematic measurements of $\Gneq(\finj,\fdet)$
for $\finj,\fdet = 1,2,\dotsc,N$, we can extract the individual
scattering probabilities $p_{ij}$ for $i =1,2,\dotsc,N$ by solving
\eqnref{eq:Gneqdefn} to obtain\footnote{%
In order to use \eqnref{eq:pijfromGneq} for $i,j=1$, we define
$\Gneq(m,n)=1$ for $m,n=0$.} %
\begin{equation}\label{eq:pijfromGneq}
  p_{ij} = \frac{e^2}{h}\left[
      \frac{ij}{\Gneq(i,j)} + \frac{(i-1)(j-1)}{\Gneq(i-1,j-1)}
      -\frac{(i-1)j}{\Gneq(i-1,j)}-\frac{i(j-1)}{\Gneq(i,j-1)}
      \right].
\end{equation}
Scattering probabilities for propagation along the entire upper edge
as a function of $B$ and $\Vug$ extracted through this method are
shown in \figref{fig:EdgePij}.  Several important features are
immediately apparent:
\begin{figure}[p]
    \centering
    \includegraphics[]{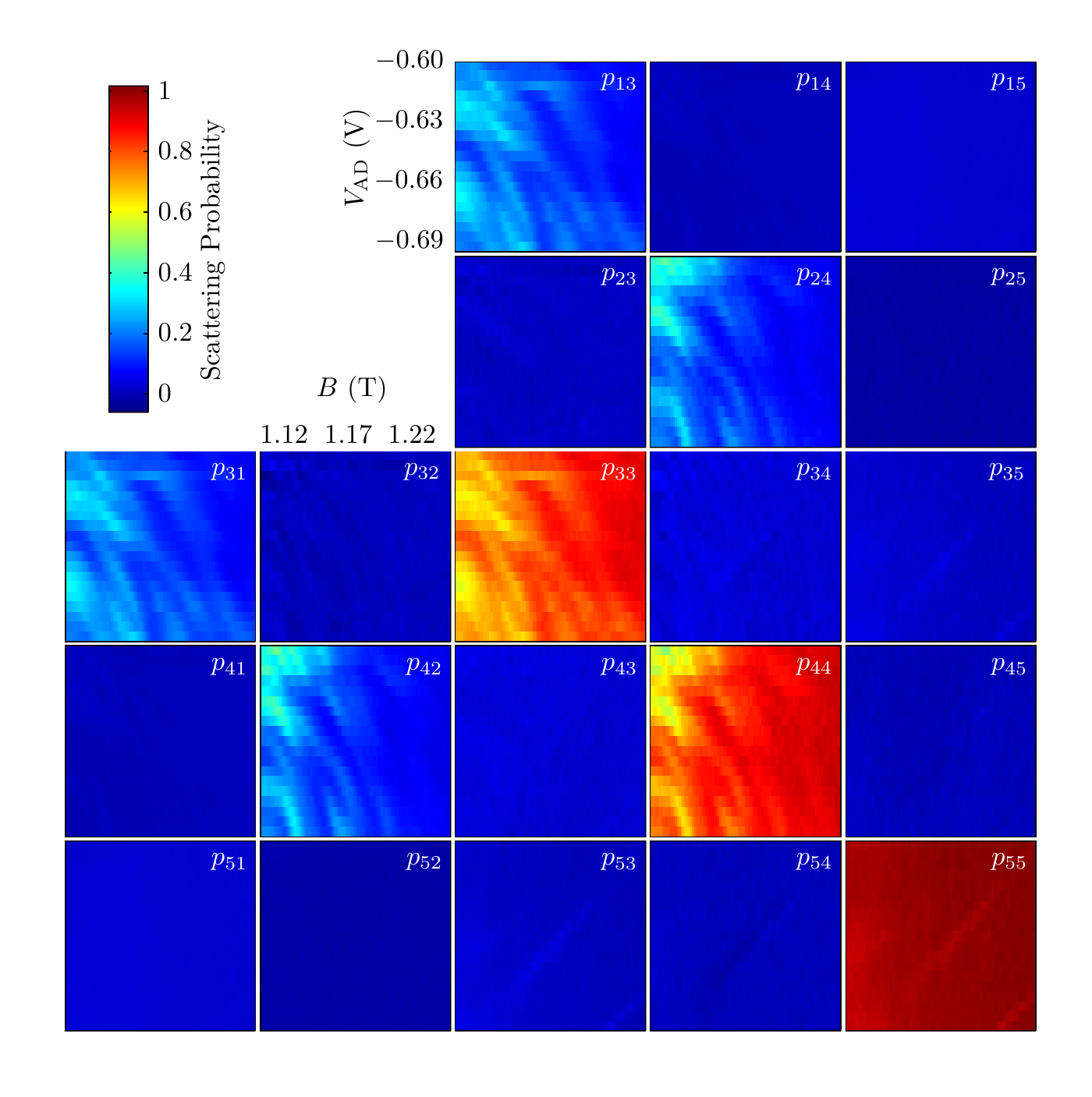}
    \caption[Edge-mode scattering probabilities]{Scattering
    probabilities $p_{ij}$ for propagation along the entire upper edge,
    from QPC1 to QPC2, as a function of $B$ and $V_\mathrm{UG}$.  The
    elements $i,j=5$ correspond to all bulk states
    $\nu>4$.  In the range of $B$ considered, the bulk filling factor
    varies through $\nuB=6$\nobreakdash--7, but the 5$^\mathrm{th}$
    column and row above demonstrate that these states have
    negligible coupling to states $\nu=1$\nbd4 throughout.  Matrix
    elements for $i,j\leq2$ were not computed in this case because
    QPC1 could not reach $\finj=1$ at these values of $V_\mathrm{UG}$.
    The probability matrix is symmetric, so some data above is
    redundant, but it is included to assist visual perception of the
    patterns in scattering between modes. \label{fig:EdgePij}}
\end{figure}
\begin{itemize}
  \item{Scattering between modes $\nu=1$\nobreakdash--4 and $\nu>4$
  is strongly suppressed in the range of $B$ shown in
  \figref{fig:EdgePij}.  The elements for which $i,j=5$ actually
  correspond to all states $\nu>4$, measured by fully opening the
  injector and/or detector QPC.  The bulk filling factor in this
  field range varies from $\nuB=7$ to 6, but we conclude that the
  $\nu>4$ modes remain largely decoupled from the two lowest LLs.
  This is expected for $\nuB\leq6$, as the $n_\mathrm{LL}=2$ LL
  moves into the bulk and begins to depopulate, but the degree of
  separation when $\nuB=7$ is somewhat more surprising.}
  \item{Spin-flip scattering (e.g., $p_{23}$, $p_{34}$) is
  suppressed throughout the range of $B$ considered.  As discussed
  above, this is a prerequisite for spin-selective measurements.}
  \item{Inter-LL scattering between the lowest-two LLs is
  non-negligible ($p_{13}$ and $p_{24}$ nearly reach their maximum
  value of 0.5 for some settings of $B$ and $\Vug$), but drops
  rapidly with increasing $B$ in the region shown.  As mentioned
  previously, in order to study antidot transmission resonances
  through non-equilibrium measurements we must find a regime in
  which the \onetothree/\twotofour\ scattering is allowed at the
  antidot, but suppressed along the edge.  For this device, this
  transition regime occurs for the range of $B$ shown in
  \figref{fig:EdgePij}.\footnote{%
As described in \secref{sec:TiltingB}, these measurements were taken
with the sample tilted at $\approx\unit{60}{\degree}$ relative to
the direction of $B$ in order to enhance $\Ez$.  This means that the
perpendicular component $B_\perp$ is roughly one-half of the
total field $B$ used throughout this chapter.} %
  Below $B\approx\unit{1.1}{\tesla}$, strong inter-LL scattering
  renders non-equilibrium measurements impossible, while for
  $B\gtrsim\unit{1.25}{\tesla}$, the antidot
  transmission resonances disappear.
  The patterns observed in $p_{13}$ and $p_{24}$ as a function of
  $B$ and $\Vug$ in \figref{fig:EdgePij} are similar but not identical,
  probably reflecting impurities which lie along the edge contour
  which have slightly different effects on each mode.
  By mapping the scattering through this method, we can identify
  settings for which both types of inter-LL scattering are strongly
  suppressed, even at the low-$B$ end of the transition regime.}
\end{itemize}
For our purposes, it is enough to show experimentally that we can
suppress both spin-flip and inter-LL scattering along the edge at
low-$B$, but it is clear that much further investigation into the
details of the edge-scattering is possible through this method.
Similar studies have been performed before,
\citep[e.g.,][]{Alphenaar1990,Komiyama1992,Muller1992,Hirai1995,Wurtz2002},
but to our knowledge none has considered the detailed structure of
the scattering probabilities as shown in \figref{fig:EdgePij}.

\begin{figure}[tb]
    \centering
    \includegraphics[]{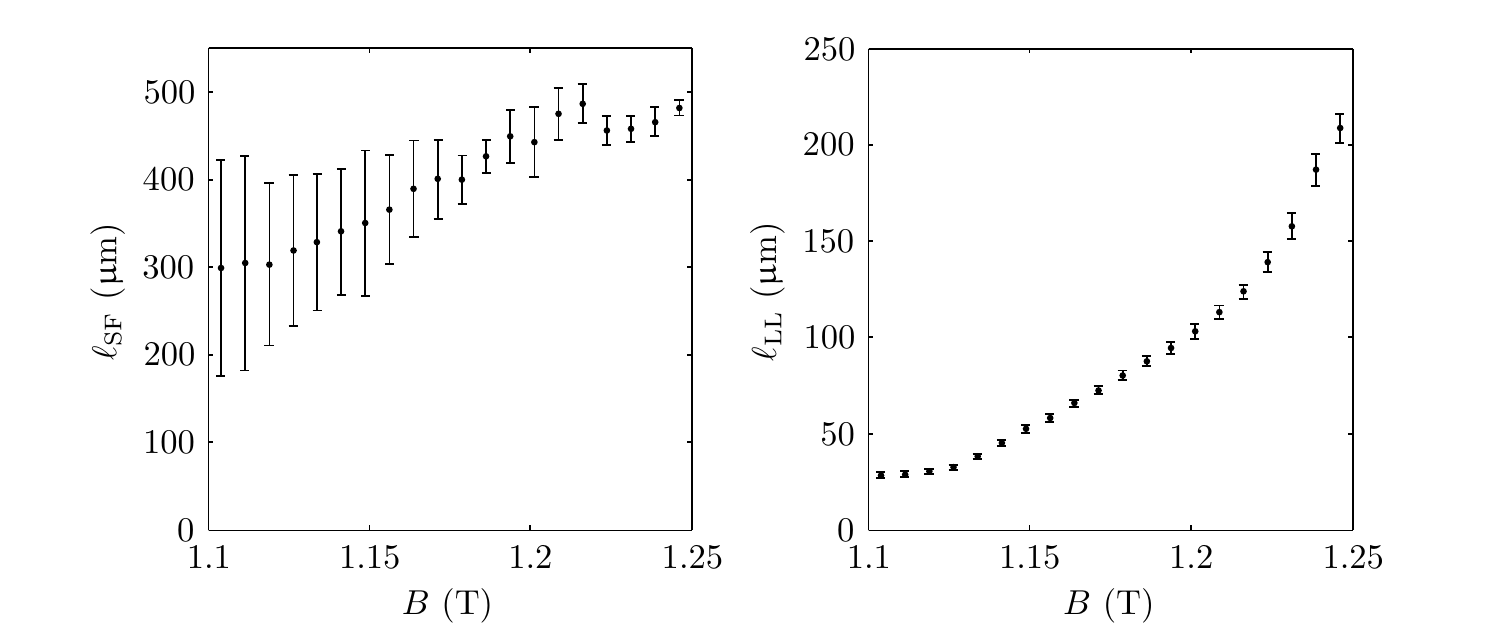}
    \caption[Equilibration length]{%
Equilibration lengths $\ell_\mathrm{SF}$ and $\ell_\mathrm{LL}$, for
spin-flip and inter-LL scattering, respectively, calculated from the
scattering probabilities shown in \figref{fig:EdgePij} assuming no
relaxation to $\nu>4$. At lower $B$, this neglected scattering is of
the same order as $\ell_\mathrm{SF}$, leading to the increased
uncertainty in that measurement.  In the case of inter-LL
scattering, the uncertainty increases with $B$ since
$\ell_\mathrm{LL}$ becomes larger than the path length
$L=\unit{10}{\micro\metre}$. \label{fig:Lequilibration}}
\end{figure}

As an additional quantitative measure of edge-mode scattering, and
for comparison with the previous studies cited above, we estimate
the `equilibration length' for both spin and inter-LL scattering, as
shown in \figref{fig:Lequilibration}.
These are computed through an extension of the procedure employed by
Ref.\ \cite{Muller1992}. We model the edge mode populations
$\mathbf{a}(x)$ as a function of propagation distance through a set
of coupled rate equations
\begin{equation}\label{eq:ScatteringRateEqn}
  \frac{d\mathbf{a}}{dx} = \mathrm{R}\mathbf{a},
\end{equation}
where the rate matrix $\mathrm{R}$ gives the scattering rates per
unit length for the various transitions.  Assuming no scattering to
$\nu>4$, a simple approximation\footnote{%
We have considered additional complications to this model, such as
an account of the different scattering distances between pairs of
opposite-spin modes.  The algebra is more tedious, but leads broadly
to the same result (within the error bars of
\figref{fig:Lequilibration}).} %
to these rates is given by
\begin{equation}
  \mathrm{R} = \left(%
\begin{array}{cccc}
  -(\ilo+2\ils) & \ils & \ilo & \ils \\
  \ils & -(\ilo+2\ils) & \ils & \ilo \\
  \ilo & \ils & -(\ilo+2\ils) & \ils \\
  \ils & \ilo & \ils & -(\ilo+2\ils) \\
\end{array}%
\right),
\end{equation}
where $\ilo$ and $\ils$ are the scattering rates between any pair of
modes of the same or opposite spin, respectively.  By solving
\eqnref{eq:ScatteringRateEqn} using this rate matrix, we can find
expressions for $\lsf$ and $\lLL$ in terms of the occupations
$\mathbf{a}^\mathrm{out}$ measured at $x=L$, where
$L=\unit{10}{\micro\metre}$ is the path length between the QPCs.
Several such relations may be obtained by considering different
initial populations (through $\finj$) and subsets of the final mode
occupations.  For example,
\begin{subequations}
    \begin{align}
      \ils & =
      -\frac{1}{4L}\log(\aout_1+\aout_3-\aout_2-\aout_4),
      \qquad\text{for $\finj=3$,}\\
      \ilo & =
      -\ils-\frac{1}{2L}\log(\aout_1+\aout_2-\aout_3-\aout_4),
      \qquad\text{for $\finj=2$.}
    \end{align}
\end{subequations}
The equilibration lengths plotted in \figref{fig:Lequilibration} are
computed according to these relations, using the mean occupation
vector $\bar{\mathbf{a}}^\mathrm{out}$, averaged over the range of
$\Vug$ considered.  Changing $\Vug$ corresponds to physically moving
the edge contour relative to the background impurity potential, and
so averaging in this way allows us to obtain the mean equilibration
lengths for a given wafer as a function of $B$, using only a single
pair of QPCs.  Clearly, these results reflect our previous
conclusions concerning the relative weakness of spin-flip scattering
and the rapid change in inter-LL scattering as $B$ increases. In
particular, for $B\lesssim\unit{1.15}{\tesla}$, we see that the
average equilibration length $\lLL$ is similar in magnitude to $L$,
but we have shown that by carefully tuning gate voltages we can
avoid scattering centres along the edge contour and so enhance the
equilibration length significantly above the mean value.

\subsection[Tuning the Zeeman energy]{Tuning the
Zeeman energy: tilted field measurements\label{sec:TiltingB}}

In this chapter we are interested in the antidot transmission
resonances at filling factor $\nuAD=2$.  These have several
practical advantages over the $\nuAD=1$\nobreakdash--2 reflection
resonances which have been the focus of several previous studies
\cite{Ford1994,Kataoka2002,Kataoka2003}, for example:
\begin{itemize}
  \item{For the purposes of studying spin- and charge-driven
interactions in the $\nu=2$ quantum Hall fluid, it is only on the
$\nuAD=2$ plateau that we can be reasonably sure that the antidot is
fully surrounded by a filled LL.  Once reflection resonances start
to occur, a theoretical description may need to account for the
complicated energetics of a depopulating LL in the constrictions. }
  \item{From a more practical perspective, the tunnel couplings
which govern reflection resonances depend on the width of the side
constrictions, and so are strongly affected by changes in gate
voltages.  Sweeping the magnetic field has a much weaker effect on
these tunnel couplings, and so has been the preferred technique in
previous experiments, but as we showed in the previous section,
changes in $B$ strongly affect edge equilibration, which can obscure
the interpretation of non-equilibrium measurements.  We prefer to
sweep instead the antidot gate voltage at a fixed magnetic field
chosen to minimise edge scattering.  The tunnel couplings between
co-propagating edge modes, on the other hand, which control
transmission resonances, depend only weakly (through the potential
slope and background impurity potential) on the gate voltage.}
  \item{In addition to their strong dependence on gate voltages, the
tunnel couplings for reflection resonances of different spins are
strongly asymmetric.  In fact, we show in the next section that the
$\nuAD=1$\nbd 2 resonances occur \emph{only} through \spindn\
(\twototwo) tunnelling, even in the low\nobreakdash-$B$ regime.
Transmission resonances, on the other hand, are characterised by
roughly equal tunnel couplings for \spinup\ and \spindn\ electrons,
and so provide more obvious information about the spin selection
rules for the antidot states themselves.}
\end{itemize}
While these advantages motivate our decision to study transmission
resonances, we have already mentioned in the previous section that
such experiments require substantial amounts of fine tuning, in
order to balance the enhanced edge equilibration at low $B$ with the
disappearance of the transmission resonances at higher fields. These
two constraints leave a relatively narrow range of $B_\perp$ within
which non-equilibrium measurements are possible for a given
device,\footnote{%
For small antidots, such an overlap region may not even exist.} %
which may not be ideal for particular applications.  In the
spin-filter device, for example, we require both of the SP
excitation energies $\Ez$ and $\dEsp-\Ez$ (see \secref{sec:SPnu=2})
to be larger than the thermal broadening of the leads ($\approx2kT$)
in order to resolve individual states. At fixed $B$, we have some
degree of control over $\dEsp$ through the gate voltage, which
changes the antidot size and hence $\dEsp$ through
\eqnref{eq:dEspapprox}.  To gain independent control over $\Ez$, we
can tilt the \emph{direction} in which the magnetic field is
applied, since both the tunnel couplings and $\dEsp$ depend only on
the perpendicular component of the field, while $\Ez$ depends on the
full magnitude of $B$.

It would be ideal to have \emph{in situ} control of the field
direction, for example through the rotating sample holder used in
\chapref{chap:TiltedB}, but unfortunately, technical difficulties
with the rotating sample holder prevented its use for these
experiments. We were, however, able to mount the device at a fixed
orientation relative to the magnetic axis for a given cooldown. From
initial measurements of our device in the standard perpendicular
orientation, we determined that the desirable perpendicular field
was around $B_\perp\approx\unit{0.6}{\tesla}$, and that the electron
temperature\footnote{%
The electron temperature is determined through an analysis of
thermally-broadened peak lineshapes as a function of mixing-chamber
temperature.} %
in the device was $T=60$\nbd\unit{70}{\milli\kelvin}
($kT=5$\nbd\unit{6}{\micro\electronvolt}). The Zeeman energy at this
field, in terms of the bare $g$\nobreakdash-factor in GaAs, is only
$\Ez\approx\unit{15}{\micro\electronvolt}$, which is at the lower
limit of the energy scales we can expect to resolve. Excitation
spectra suggested that the SP energy was around $\dEsp\approx
50$\nbd\unit{60}{\micro\electronvolt}, and so we chose to raise the
sample and remount it, tilted at an angle of
$\approx\unit{60}{\degree}$ to the magnetic axis, to approximately
double the Zeeman energy.

Our measurements suggest\footnote{%
This is especially clear in the measurements of
\chapref{chap:TiltedB}, taken with a rotating sample holder.} %
that the $g$\nobreakdash-factor is actually significantly enhanced
(by a factor of $\approx 1.5$) from the bare value of $g=-0.44$ in
our device, so this seems to have been a slight over-rotation,
resulting in a situation for the central antidot in which
$\Ez\approx\dEsp$. In a sense this was fortuitous, since it meant
that we could observe state crossings as a function of $B$ and
$\Vad$ within the parameter space available to us at this fixed
angle.  As we show later in this chapter, comparisons of the
excitation spectra near these crossings with our theoretical models
help to isolate the effects due to electron interactions and the
underlying antidot energy spectrum from those due to other aspects
of the experiment, such as applied bias, asymmetric tunnel barriers,
and temperature.

\section{Experimental results\label{sec:NeqExpResults}}

\begin{figure}[tp]
\centering
\includegraphics[]{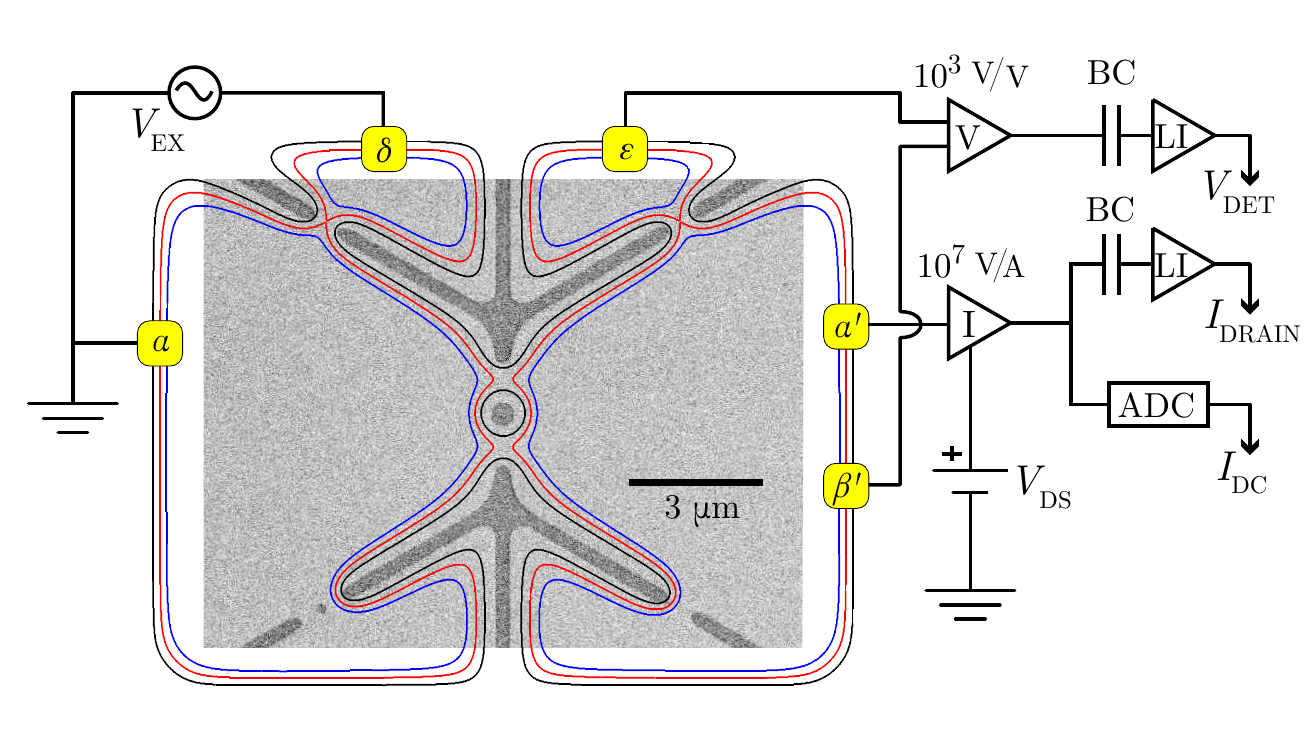}
\caption[Spin-selective measurement circuit]{Scanning electron
micrograph of the first metallic layer of a device similar to that
used for the experiments in this chapter, with edge-modes and the
non-equilibrium measurement circuit overlaid. The four-terminal
non-equilibrium conductance $\Gneq$ is measured using lock-in
techniques, with an AC excitation voltage $\Vex$ which has amplitude
$e\Vex\approx\unit{3}{\micro\electronvolt}\lesssim kT$ and frequency
$\approx\,\unit{70}{\hertz}$.  The output of the current
preamplifier is fed to both a lock-in amplifier (LI) and an
analogue-to-digital converter (ADC), to measure the AC and DC
components of the current, respectively.  Blocking capacitors (BC)
with a capacitance $C=\unit{40}{\micro\farad}$ prevent DC components
from overloading the LI amplifiers. A DC source-drain bias $\Vsd$,
applied to the drain through the current preamplifier, serves to
cancel the input offset of the preamplifier for linear-response
measurements, or to supply the source-drain bias for nonlinear
transport experiments.  Edge-mode contours were computed from the
bare electrostatic potential of the gates using the
\texttt{GatesCalc} program.  The black contour represents both of
the LLL edge modes ($\nu=1,2$), and the $\nu=3$ and 4 modes are
represented by red and blue contours, respectively.
\label{fig:SEMandCircuit}}
\end{figure}

Non-equilibrium measurements of the central antidot proceed
analogously to the edge-scattering experiments described in
\secref{sec:EdgeScattering}.  We use one QPC as an injector, to
selectively populate a subset of source edge modes, and then use a
second QPC as a detector, to individually measure the population of
each mode in the drain.  A typical measurement setup is shown in
\figref{fig:SEMandCircuit}, although there are many variations.  For
example, by attaching the excitation voltage to a contact inside the
injector as shown, we populate modes $\nu=1$ to $\finj$ of the
source, but we can also populate the outer modes $\nu=\finj+1$ to
$\nuB$ by reversing the connections to earth and $\Vex$ on the
left-hand side of the device.  The second antidot in the
constriction at the lower-left can also be operated as a detector
QPC, allowing us to measure the populations of the reflected edge
modes.

\subsection{Spin-selective measurements in linear response\label{sec:ZeroBiasExpts}}

\begin{figure}[p]
    \centering
    \includegraphics[]{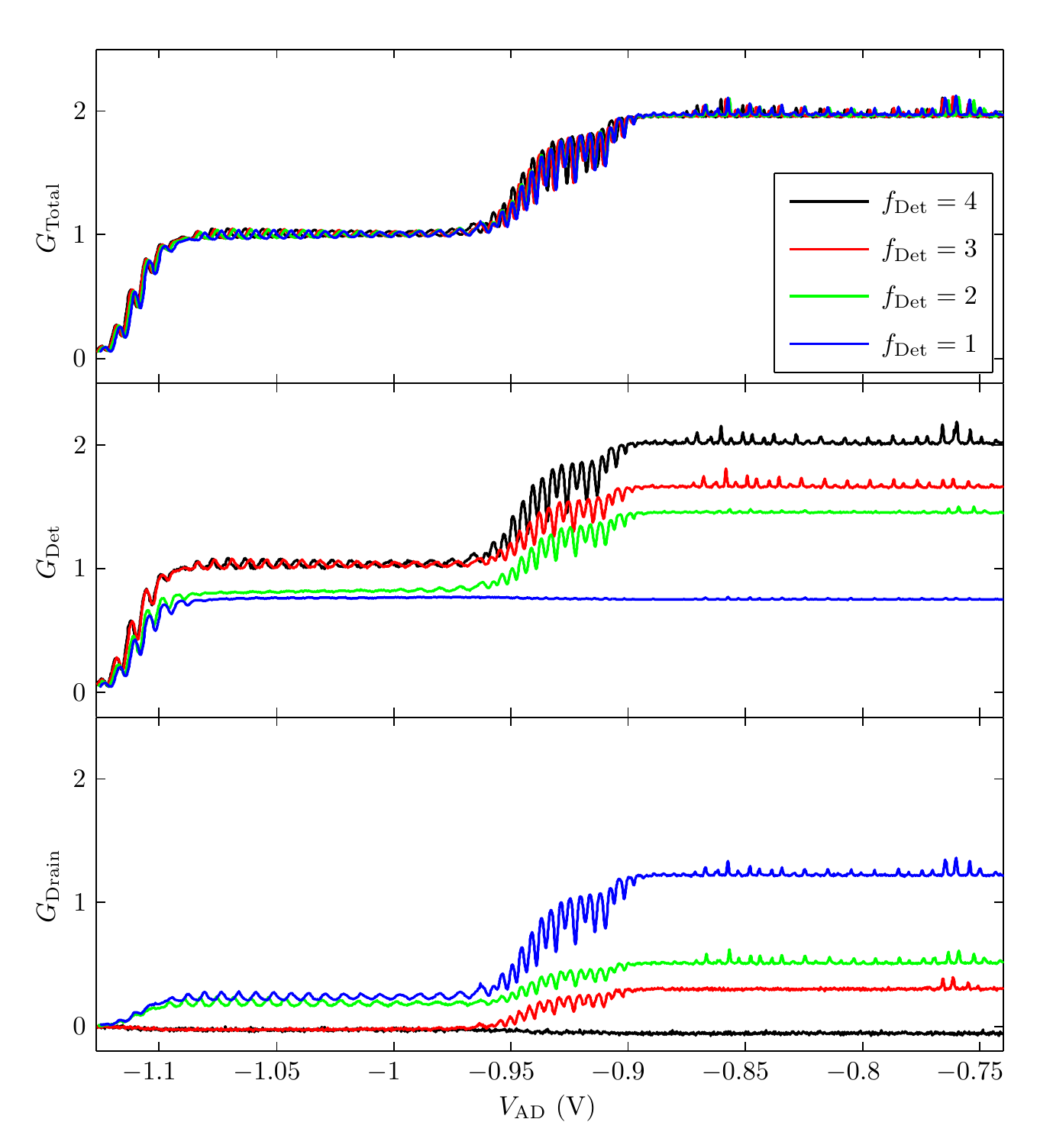}
    \caption[Selective detection measurements]{%
Selective detection measurements of the central antidot at
$B_\perp=\unit{0.6}{\tesla}$, obtained by separately measuring the
current flowing out of the detector (ohmic~$\varepsilon$, middle
panel) and drain (ohmic~$\alpha^\prime$, bottom panel) contacts,
with $\fdet=1$\nbd 4 as labelled above, and the injector QPC fully
open. The total conductance $G_\mathrm{Total}$ shown in the upper
panel is the sum of the detector and drain measurements.
Two-terminal conductances are given in units of $e^2/h$, with no
corrections for series resistances in the circuit.  These data were
taken in a separate cooldown of the device than the rest of the
measurements presented in this chapter. \label{fig:DetMeasurements}}
\end{figure}

Selective detection measurements of the central antidot for the
entire range of filling factor $\nuAD=0$\nbd 2 are shown in
\figref{fig:DetMeasurements}.
For such long sweeps, slow device drift on the scale of the fine
resonance structure makes it impractical to simply subtract separate
traces as described in \secref{sec:EdgeScattering} to extract the
antidot scattering probabilities.  Where necessary, we can still
perform this subtraction for shorter, higher-resolution sweeps, or
we can quantitatively extract scattering probabilities for
resonances individually through peak-fitting procedures, but most of
the important conclusions are readily observable by eye without any
complicated analysis.

\begin{figure}[tb]
    \centering
    \includegraphics[]{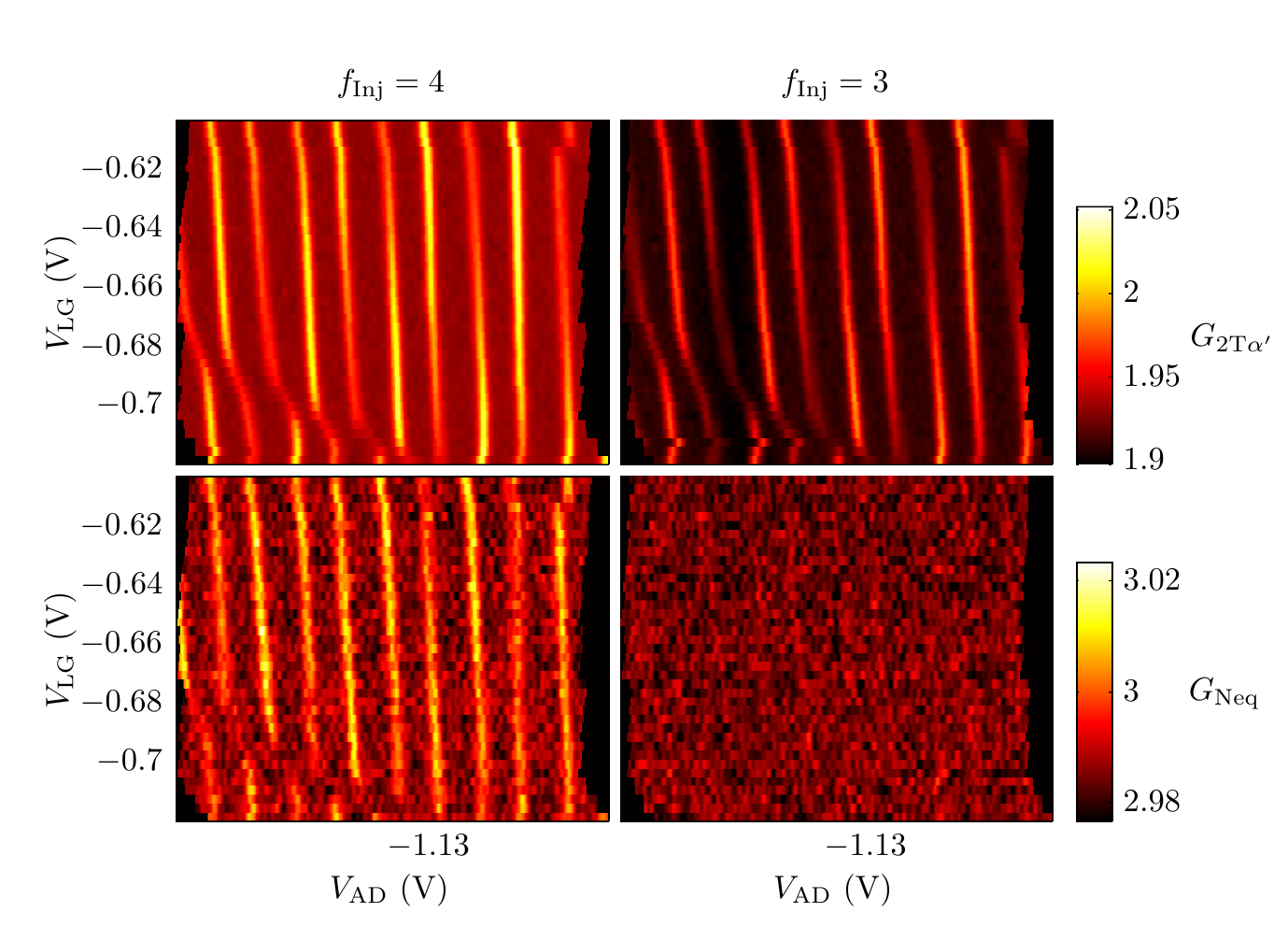}
    \caption[Spin conservation]{%
Non-equilibrium measurements of transmission resonances on the
$\nuAD=2$ plateau obtained by sweeping $\Vad$ and stepping $\Vlg$.
The top and bottom rows respectively show the two-terminal antidot
conductance $\Gap = I_{\alpha^\prime}/\Vex$ and the non-equilibrium
conductance $\Gneq = I_{\alpha^\prime}/V_{\varepsilon\beta^\prime}$
in units of $e^2/h$, with $\fdet=3$ throughout and $\finj=4$ and 3
in the left and right columns, respectively. The disruption to the
resonances in the bottom-left corner is the result of a nearby
impurity, as described in \secref{sec:ADmolecule}. Sweeps are
individually shifted horizontally to correct for device drift and to
clarify the resonance pattern. \label{fig:SpinConservation}}
\end{figure}

Since it is easier to interpret visually, the data presented in
\figref{fig:DetMeasurements} are the result of two-terminal
measurements, rather than the non-equilibrium four-terminal
measurement shown in \figref{fig:SEMandCircuit}.  The injector QPC
is fully open, so all source modes are populated, and we separately
measure the currents\footnote{%
We plot two-terminal conductance $G=I/\Vex$ instead of current since
this is roughly independent of $\Vex$ (in our case the series
resistance of $\approx\unit{300}{\ohm}$ is almost negligible
compared to the $\approx\unit{10}{\kilo\ohm}$ resistance of the
device).} %
flowing out of contacts $\varepsilon$ in the detector and
$\alpha^\prime$ in the drain, which reflect the populations of the
edge modes at the entrance to the detector QPC.  When the detector
is set at $\fdet=4$ (black curves in \figref{fig:DetMeasurements}),
all of the current appears in the detector as expected, since we
have already shown that scattering to $\nu>4$ modes is negligible.
For $\fdet<4$, the signal is divided between $\Gdet$ and $\Gdrain$,
and we can determine which features of the trace are carried by
individual modes by subtracting the curves `by eye.' Two important
features are immediately apparent:
\begin{itemize}
    \item{%
Perhaps the most striking detail in \figref{fig:DetMeasurements} is
the nearly complete lack of structure in $\Gdet$ when $\fdet=1$
(middle panel, blue curve). This suggests that the $\nu=1$ mode does
not couple to the antidot at all,\footnote{%
To prove this statement, we must rule out any `cancellation' due to
scattering between other modes by checking the situation when
$\finj<4$.  In such experiments, the $\fdet=1$ trace is indeed
featureless for all settings of $\finj$.} %
except where the $\nuAD<1$ oscillations begin around
$\Vad=\unit{-1.1}{\volt}$.  This means that the $\nuAD=2$ reflection
resonances are due solely to \spindn\ (\twototwo) tunnelling, with
no contribution from the \spinup\ ($\nu=1$) state, in contradiction
to the non-interacting Mace-Barnes model. This is essentially the
same behaviour observed by Kataoka~et~al.~\cite{Kataoka2003} in the
`double-frequency' regime at much higher fields, which was
interpreted as evidence for compressible regions. Although the
resonance spacings appear rather uniform in our experiments, we
observe a clear odd-even modulation in amplitude apart from a few
phase slips (e.g., around $\Vad=\unit{-0.91}{\volt}$), and so we are
clearly not in the double-frequency regime here.}
    \item{%
While the $\nuAD=2$ reflection resonances appear only in \spindn\
channels (a signal appears in the $\nu=4$ mode due to equilibration
along the detector edge), the transmission resonances show nearly
opposite behaviour.  Consider for example the measurement of
$\Gdrain$ at $\fdet=3$ (bottom panel, red curve), representing the
population of the $\nu=4$ state at the detector. Except for a few
resonances near $\Vad=\unit{-0.76}{\volt}$, the $\nuAD=2$ plateau is
nearly featureless.  When $\finj$ is reduced to 2 (green curve),
however, the transmission resonances all appear in $\Gdrain$ with
nearly their full amplitude.  This implies that these conductance
peaks are mostly the result of \spinup\ (\onetothree) tunnelling,
with \twotofour\ only becoming significant within a
small `envelope' encompassing 4\nbd 5 resonances.\footnote{%
As seen in \figref{fig:SpinConservation}, detailed measurements show
a small contribution from \spindn\ tunnelling to each peak, but the
\spinup\ transmission is generally around an order of
magnitude greater.} %
Again this is in direct contrast to the simple non-interacting model
which predicts peaks of alternating polarisation, even though the
resonances show clear odd-even modulation suggestive of the SP
model.  It is also inconsistent with a model based on compressible
regions, which would predict tunnelling only through the outer
\spindn\ state for transmission as well as reflection resonances.}
\end{itemize}
Thus we find that the antidot resonances around $\nuAD=2$ are
spin-polarised, but not in the way predicted by the non-interacting
model. Reflection resonances are fully \spindn\ (\twototwo\
tunnelling), and transmission resonances are dominated by \spinup\
(\onetothree).

This unexpected result suggests that the spin of electrons involved
in transport is determined by tunnel couplings to the leads, rather
than by intrinsic selection rules of the antidot states.  For
reflection resonances this is not particularly surprising.  The
distance across the constriction for \twototwo\ tunnelling is likely
to be much greater than that for \onetoone\ transport, especially
since the long spin-equilibration length measured in
\secref{sec:EdgeScattering} suggests that the $\nu=1$ and 2 LLs are
spatially separated. Our measurements suggest that the $\nu=1$ state
does not couple to the antidot at all in this regime, except for the
reflection resonances at $\nuAD<1$ where the $\nu=2$ mode is fully
excluded from the constrictions.  We might hope to observe a
stronger amplitude modulation of alternate reflection resonances if
the antidot states were spin-selective, but when sweeping $\Vad$ the
\twototwo\ coupling quickly becomes large enough to produce
significant lifetime broadening, within the first few dips from the
$\nuAD=2$ plateau, from which point we would not expect to preserve
the energy-selectivity required to probe individual antidot states.

The observed details of the transmission resonances, however, are
harder to explain.  Although the $\nu=3$ mode is physically `closer'
to the antidot, we expect the \onetothree\ and \twotofour\
tunnel-couplings to be nearly identical based on the edge-mode
topology. It is possible that details of the potential slope and
edge-mode structure could lead to different tunnelling distances for
these two processes, but we suspect instead that the most likely
explanation for the dominance of \onetothree\ tunnelling is that an
impurity on the drain side of the potential disrupts the $\nu=4$
mode, decreasing the \twotofour\ coupling $\gDdn$ on that side. We
have extensive experimental evidence for such nearby impurities, as
described in \secref{sec:ADmolecule} and believe that
$\gDdn\ll\gSdn$ for several reasons explained later in this chapter.
The odd-even amplitude modulation observed for the most part in
\figref{fig:DetMeasurements} therefore suggests that the antidot
states do have a spin `character,' in the sense that alternate
resonances are more or less transparent to the predominant \spinup\
tunnelling, but the selectivity is clearly incomplete. These
resonances do not suffer from significant lifetime broadening; in
the region of magnetic field we study, their lineshape is
well-matched by the Fermi-derivative function expected for
thermally-broadened resonances, so we expect the transport to be
energy-selective within the range of a few $kT\approx
10$\nbd\unit{15}{\micro\electronvolt}.  We concentrate on explaining
the details of these transmission resonances for the remainder of
this chapter.

The incomplete spin-selectivity we observe may be explained in two
ways. Either the antidot spectrum is composed of `pure' spin-states
with an energy-spacing less than the thermal resolution of the
leads, or the antidot states themselves are spin-hybridised as the
result of the spin-orbit or hyperfine spin-mixing interactions.  In
either case we might expect that \spinup\ would couple differently
to alternate states, producing the amplitude modulation observed in
\figref{fig:DetMeasurements}, but there is a crucial observable
difference between the two models.  Spin is \emph{conserved} during
transport through nearly-degenerate pure-spin states, but this is
not the case for hybridised states for which $S_z$ is not a good
quantum number; the spin of each electron passing through such a
state precesses to an angle which depends on the length of time it
remains there, and so the time-averaged output spin will be
independent of the input spin. We can differentiate between these
two mechanisms by using the injector QPC to populate only one of the
$\nu=3$ or 4 modes, as shown in \figref{fig:SpinConservation}.

The data in \figref{fig:SpinConservation} were obtained through the
non-equilibrium experimental setup shown in
\figref{fig:SEMandCircuit}, with the detector set to $\fdet=3$ to
separate the $\nu=3$ and 4 edge modes.  We plot both the two
terminal conductance $\Gap = I_{\alpha^\prime}/\Vex$ (corresponding
to $\Gtot$ in \figref{fig:DetMeasurements}) and the non-equilibrium
conductance $\Gneq = I_{\alpha^\prime}/V_{\varepsilon\beta^\prime}$.
Through the Landauer-B\"{u}ttiker formalism, we may write $\Gneq$ in
the form
\begin{equation}
  \Gneq = \fdet \frac{e^2}{h}\left[
      1 + \frac{\sum_{i=\fdet+1}^{\nuB}a^\mathrm{Det}_i}
          {\sum_{i=1}^{\fdet}a^\mathrm{Det}_i}\right],
\end{equation}
where the vector $\mathbf{a}^\mathrm{Det}$ gives the populations of
the edge modes at the entrance to the detector.  For $\fdet=3$, the
second term in brackets above is dominated by variations of the
numerator, so we can approximate
\begin{equation}
  \Gneq \simeq \frac{3e^2}{h}\left[
  1+\frac{1}{A}a^\mathrm{Det}_4
  \right],
\end{equation}
where $A\simeq a^\mathrm{Det}_1+a^\mathrm{Det}_2\approx2$ is a
constant depending only on the edge-equilibration.\footnote{%
Equilibration along the top edge is independent of both $\Vad$ and
$\Vlg$.} %
Features observed in $\Gneq$ therefore reflect changes in the
population of the $\nu=4$ mode at the detector.

When $\finj=4$, such that all source modes are populated (left
column of \figref{fig:SpinConservation}), we see a strong signal in
$\Gap$, and a much weaker corresponding signal in $\Gneq$, implying
that electrons passing through the antidot may enter either the
\spinup\ ($\nu=3$) or the \spindn\ ($\nu=4$) mode of the drain,
although they are much more likely to take the \spinup\ channel due
to the asymmetric tunnel couplings. We also observe an odd-even
pattern of amplitudes for these resonances, and, interestingly, the
pattern systematically reverses as a function of $\Vlg$.  With a
closer look, it is also clear that the `brighter' resonances in
$\Gap$, which is dominated by \spinup, are correspondingly `dimmer'
in the \spindn\ signal of $\Gneq$, again suggesting that the antidot
states do have a `preferred' spin, even though the selection is
incomplete.

\begin{figure}[tb]
    \centering
    \includegraphics[]{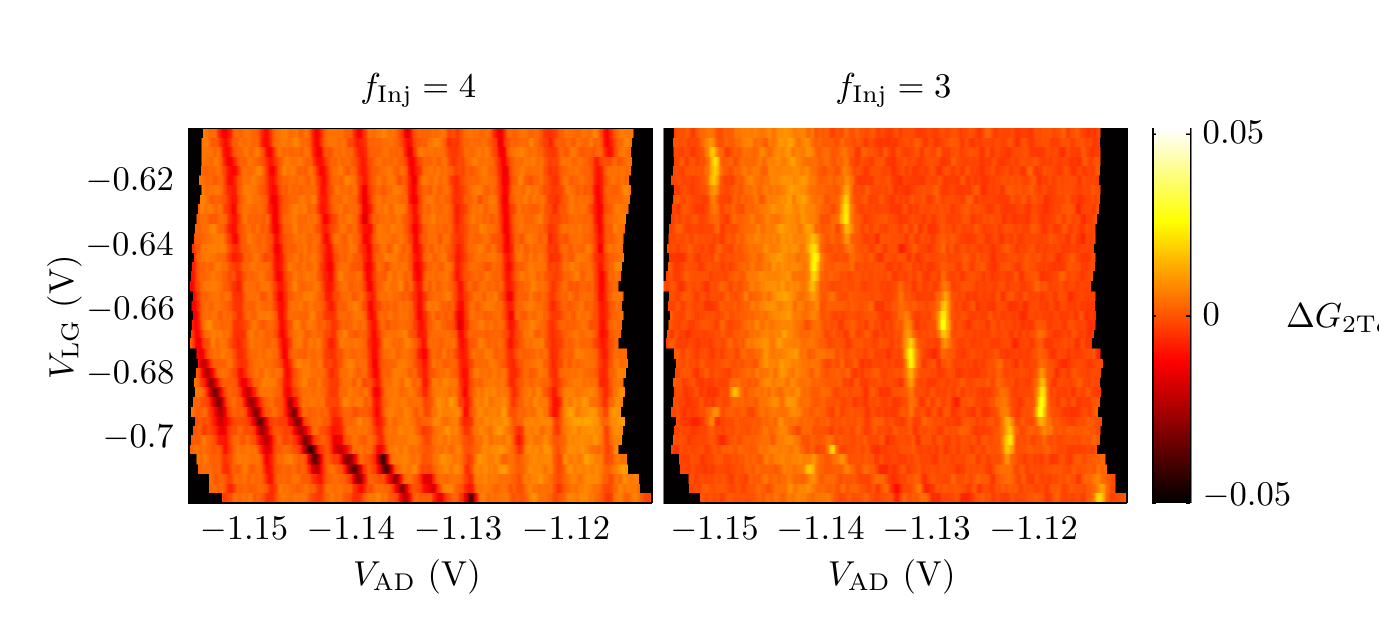}
    \caption[Reflected mode populations]{%
Non-equilibrium measurements of the \emph{reflected} population of
the $\nu=4$ state, taken together with the data of
\figref{fig:SpinConservation}, for $\finj=4$ (left panel) and 3
(right panel).  Since the background level is strongly affected by
changing edge-equilibration as a function of $\Vlg$, we plot the
variation $\Delta\Gaup$ obtained by subtracting the mean of each
horizontal sweep, in order to clarify the antidot resonance
structure. \label{fig:Greflected}}
\end{figure}

By setting $\finj=3$ (right column of
\figref{fig:SpinConservation}), we no longer populate the $\nu=4$
source channel, and a different picture emerges.  We still observe a
strong signal in $\Gap$, but now $\Gneq$ is essentially
featureless.\footnote{%
The resonances in $\Gap$ appear to darken in
\figref{fig:SpinConservation} but this mainly reflects the decreased
background resulting from a small amount of equilibration along the
injector edge.  This equilibration also explains the extremely faint
resonance signal remaining in $\Gneq$.} %
Taken together, these measurements imply that electron spin is
\emph{conserved} in passing from the source to the drain, since by
removing the source of \spindn\ electrons, we measure only the
\spinup\ signal in the drain.  This suggests an interpretation of
the results based on pure, nearly-degenerate spin-states, rather
than a hybridised-spin model.  To be absolutely sure of this
interpretation, we must check also the \emph{reflected} current when
$\finj=3$, since highly asymmetric couplings $\GammaS_4\gg\GammaD_4$
will result in the majority of any `spin-flip' current appearing in
the $\nu=4$ mode on the source side of the device.  This is
accomplished by using one constriction of the smaller antidot in
QPC4 as a second detector. We measure the current flowing out of
contact $\varepsilon^\prime$, with the filling factor of QPC4 also
set to three, but we want the contribution from the $\nu=4$ mode
which is reflected by this constriction and arrives in the grounded
contact $\alpha^\prime$. By current conservation, this is given by
\begin{equation}
  \Gaup = \finj \frac{e^2}{h} - \Gap - G_{2\mathrm{T}\varepsilon^\prime},
\end{equation}
up to a constant offset due to series resistance.  These
measurements are shown in \figref{fig:Greflected}.
When $\finj=4$, we observe weak dips in $\Gaup$ corresponding to the
transmitted \spindn\ current detected in $\Gneq$.  When $\finj=3$
the resonance pattern mostly disappears, and we are left with a
clear pattern of bright `spots,' representing spin-flip processes in
which a \spinup\ electron enters the antidot from the $\nu=3$ source
state and then returns to the source $\nu=4$ state as \spindn.  With
closer inspection, we see that the locations of these spin-flip
events correspond to the places in which the bright-dark pattern of
resonance pairs reverses in \figref{fig:SpinConservation}, and
further analysis shows that they also correspond to the `closest
approach' in $\Vad$ of each pair of resonances.  This behaviour is
highly suggestive of crossings between the energy levels of
opposite-spin states.  There is faint evidence for spin-flip
features matching those of \figref{fig:Greflected} in the
transmitted $\Gneq$ data shown in the lower-right panel of
\figref{fig:SpinConservation}, but they are clearly much weaker than
those seen in the reflected current.  This observation implies that
$\gDdn\ll \gSdn$, which we believe to be the result of a nearby
impurity which disrupts the $\nu=4$ edge mode near the antidot.

\subsection{Discussion of linear response data}

From the linear-response measurements presented in the previous
section, we obtain a picture of antidot transport with the following
important characteristics:
\begin{itemize}
  \item{
The antidot resonances we observe are not spin-selective. Amplitude
modulations suggest that states have a `preferred' spin, but since
the modulation is rarely larger than a factor of $\approx 2$, it is
clear that the electrons of the `un-preferred' spin may pass through
the antidot easily.\footnote{%
If the tunnel couplings were equal for both spins, a factor of two
modulation in spin-selectivity would result in observed
polarisations of only $\approx 30 \%$.} %
The spin polarisation we observe in the transmitted current is
driven instead by the antidot-lead couplings, which are much
stronger for \spinup\ tunnelling than for \spindn.}
  \item{
Generally, spin is conserved during transport.\footnote{%
The data shown in \figsref{fig:SpinConservation} and
\ref{fig:Greflected} were chosen to highlight the observed pattern
of spin-flip events.  Throughout the entire range of parameters we
studied, such events are even more rare than they appear
in \figref{fig:Greflected}.} %
Note that this does not preclude Kondo-like \emph{cotunnelling}
events in which an electron on the antidot of one spin is replaced
by an electron of the opposite spin.  These processes would not be
detected as spin-flips in the time-averaged quantities we measure,
since such an event must eventually be followed by the reverse
tunnelling process in order for the antidot to maintain an
equilibrium spin state.}
  \item{
With careful tuning of external fields, we can find specific regimes
in which spin is not conserved for individual electrons on the
antidot.  These locations appear to correspond to `crossings' of
opposite-spin states of the antidot addition spectrum.}
\end{itemize}
These features may be broadly described in terms of three important
energy scales. The first is the thermal energy $\Etherm\approx 2kT$,
which defines the energy window of allowed tunnelling events about
$\Ef$. The others, which we discuss below, are the energy separation
of antidot states with differing (pure) spin, and the energy scale
for hybridisation of opposite-spin states, which we attribute to the
spin-orbit interaction based on an analysis of the anticrossings
observed in \figref{fig:Greflected}.

\subsubsection{Spin-conserved transport}

Considering first the `usual' case in which spin is conserved during
transport, such that the total antidot spin-projection $S_z$ is a
good quantum number, sequential transport is governed by the
chemical potentials $\mu_\sigma(N,S_z)$, given by
\begin{subequations}
  \begin{align}
    \muup(N,S_z) & = U(N+1,S_z+\tfrac{1}{2}) - U(N,S_z) \\
    \mudn(N,S_z) & = U(N+1,S_z-\tfrac{1}{2}) - U(N,S_z),
  \end{align}
\end{subequations}
in terms of the configuration energy $U(N,S_z)$ of an antidot with
$N$ particles\footnote{%
For an antidot, $N$ corresponds either to the number of electrons
within some `cutoff' orbital, beyond which all states are filled, or
the number of `holes' in the
otherwise-occupied LLs, as described in \secref{sec:HFtheory}.} %
and total spin $S_z$.  At a resonance condition between
configurations with $N$ and $N+1$ particles, transport of both
\spinup\ and \spindn\ electrons will be energetically allowed if one
or both of the following conditions hold:
\begin{subequations}
\begin{gather}
  \bigl\lvert\muup(N,S_{z0})-\mudn(N,S_{z0})\bigr\rvert \lesssim
      \Etherm\\
  \text{or}\quad\bigl\lvert\muup(N,S_{z0}^\prime-\tfrac{1}{2})-\mudn(N,S_{z0}^\prime+\tfrac{1}{2})\bigr\rvert\lesssim \Etherm,
\end{gather}
\end{subequations}
where $S_{z0}$ and $S_{z0}^\prime$ are the ground-state spins of the
antidot states with $N$ and $N+1$ particles, respectively.  These
are equivalent to the conditions
\begin{equation}\label{eq:musfCondition}
  \musf^\pm(N,S_{z0})\lesssim \Etherm\quad\text{or}\quad
      \musf^\pm(N+1,S_{z0}^\prime)\lesssim \Etherm,
\end{equation}
written in terms of the `spin-flip' chemical potentials defined by
\begin{equation}\label{eq:musfDefn}
  \musf^\pm(N,S_z) = U(N,S_z\pm1)-U(N,S_z).
\end{equation}
Thus we identify $\musf$, together with $\Etherm\approx 2kT$ as the
relevant energy scale for spin-selective tunnelling. The observed
lack of spin-selectivity in our experiments implies that
\eqnref{eq:musfCondition} must hold for each $N\leftrightarrow N+1$
transition.  Within the non-interacting theory described in
\secref{sec:SPnu=2}, this corresponds to a near-degeneracy between
pairs of states in the opposite-spin `ladders' of the SP spectrum.
In particular, within the SP model we find
\begin{equation}\label{eq:musfSPmodel}
  \musf^\pm(N,S_z) = (1\pm 2S_z)\dEsp - \Ez,
\end{equation}
and \eqnref{eq:musfCondition} will be satisfied whenever $\Ez -
n\dEsp\lesssim \Etherm$ for any nonnegative integer $n$.

\subsubsection{Spin-flip transport\label{sec:SpinFlipTransport}}

Next, we consider the regime in which spin is \emph{not} conserved,
as at the locations of the bright `spots' on the right-hand panel of
\figref{fig:Greflected}.  Spin relaxation may occur either through
hyperfine coupling between electrons and GaAs nuclei, or through the
phonon-mediated spin-orbit interaction.  A useful overview of these
interactions in the context of quantum dots is provided in the
review of Hanson~et~al.~\cite{Hanson2007}.  These two mechanisms are
characterised by very different time- and energy-scales, which we
can use to decouple their contributions to our experimental
observations.

We can describe the hyperfine interaction in terms of an effective
magnetic field $B_\mathrm{HF}$, known as the Overhauser field, which
accounts for the randomly-oriented ensemble of nuclear spins
interacting with a given electron \cite{Merkulov2002}. The component
of this field perpendicular to the externally applied $\mathbf{B}$
will lead to mixing between antidot states with spin projection
$S_z$ and $S_z+1$, but due to energy-conservation requirements, this
mixing is only strong when these states satisfy $\musf^\pm \lesssim
\varepsilon_\mathrm{HF}$, where $\varepsilon_\mathrm{HF}=
g\mu_\mathrm{B}B_\mathrm{HF}$ is the energy scale for the hyperfine
interaction. The magnitude of the Overhauser field resulting from
statistical variations of $N$ nuclei\footnote{%
The scale for nuclear magnetic moments is set by the nuclear
magneton, $\mu_\mathrm{N} = \hbar\abs{e}/2m_p =
\unit{3.2}{\nano\electronvolt\per\tesla}$, which is much smaller
than the Bohr magneton for electrons, $\mu_\mathrm{B} =
\hbar\abs{e}/2m_e=\unit{60}{\micro\electronvolt\per\tesla}$, and so
the nuclear Zeeman energy $E_\mathrm{Z,N} = g\mu_\mathrm{N}B$ always
satisfies $E_\mathrm{Z,N}\ll kT$ for experimentally accessible
temperatures and
fields.} %
is given by $B_\mathrm{N,max}/\sqrt{N}$, where
$B_\mathrm{N,max}=\unit{5.3}{\tesla}$ is the field that would be
produced by the full polarisation of all GaAs nuclei. The dominant
hyperfine contribution is a contact interaction proportional to the
overlap between the electron and nuclear spatial probability
densities, so we can estimate $N$ from the spatial size of the
electron wave function and the nuclear density of the GaAs crystal,
$n_\mathrm{GaAs} = \unit{44}{\nano\metre\rpcubed}$.  For typical
antidot wave functions with radius
$r_\mathrm{AD}\approx\unit{400}{\nano\metre}$ and width
$\ellB\approx\unit{30}{\nano\metre}$ at
$B_\perp=\unit{0.6}{\tesla}$, we have $(2\pi r_\mathrm{AD}\ellB)
n_\mathrm{GaAs}\approx\numprint{3e6}\,\reciprocal{\nano\metre}$, and
with the quantum well thickness in the $z$-direction of
$\approx\unit{10}{\nano\metre}$, we estimate that each electron in
the antidot couples to $N\approx\power{10}{7}$\nbd\power{10}{8}
nuclei, and therefore sees an Overhauser field with rms magnitude
$B_\mathrm{HF}\approx\unit{1}{\milli\tesla}$.  This corresponds to
an extremely small coupling energy of
$\varepsilon_\mathrm{HF}\approx\unit{25}{\nano\electronvolt}$, and
so spin-flips can only occur through the hyperfine mechanism when
antidot spin-configuration energies are very nearly degenerate. When
such coupling occurs, electron spins will precess about the
Overhauser field with a rate
$\varepsilon_\mathrm{HF}/h\approx\unit{10}{\mega\hertz}$, which is
comparable to the rate at which electrons pass through the antidot,
estimated from typical transport currents of
$\approx\unit{1}{\pico\ampere}=e(\unit{6}{\mega\hertz})$, so we
could expect hyperfine coupling to produce spin-flips in our
experiments. Besides the spin-precession frequency, hyperfine
effects are characterised by a very long time-scale, measured in
minutes to hours, associated with the relaxation of nuclear spins
which have been `pumped' through hyperfine interactions with
electrons, and with the complicated feedback loops which may result
from such pumping.  This long time-scale often serves as an
experimental signature of the hyperfine effect, but we did not
observe any effects of this nature in our experiments.

Spin-orbit coupling, on the other hand, results from electric fields
in the semiconductor material due to both the band offset at the
GaAs-AlGaAs heterojunction (the Rashba contribution) and the bulk
inversion asymmetry of the GaAs crystal (the Dresselhaus
contribution). The strength of the interaction, $\eSOI$, depends
both on the magnitude of these electric fields and on the electron
momentum vector\footnote{%
Note, however, that the Rashba and linear-in-$k$ Dresselhaus terms
(the Dresselhaus contribution also has a cubic term, although this
is usually suppressed relative to the linear terms by confinement in
the $z$-direction) produce spin-rotations over a \emph{length}
$\ell_\mathrm{SOI}$ which is $k$-independent, since the increased
spin-precession rate at higher $k$ is compensated for by the larger
linear velocity.  This description is useful for ballistic electron
motion in two dimensions, but less so for edge states or
zero-dimensional systems.} %
$\mathbf{k}$, so the spin-orbit interaction is sensitive to the
electron density $n_e=k_\mathrm{F}^2/2\pi$.  The
spin-orbit strength is therefore highly sample-dependent\footnote{%
The Rashba contribution, in particular, depends on details of the
heterostructure which influence the electric field perpendicular to
the 2DES.} %
and precise measurements of its value are not trivial.  For
two-dimensional ballistic motion, Miller~et~al.~\cite{Miller2003}
used a model of weak (anti)localisation measurements to separately
extract the different spin-orbit contributions in a GaAs
heterostructure as a function of electron density.  At the lowest
density they considered,
$n_e=\numprint{1.4e11}$~\centi\metre\rpsquared, they found roughly
similar magnitudes for the Rashba and linear Dresselhaus coupling
strengths, of $\eSOI\approx \unit{20}{\micro\electronvolt}$.  We
would expect this value to be slightly lower in our device, which
has $n_e=\numprint{1.1e11}$~\centi\metre\rpsquared, but even so it
is clear that the spin-orbit interaction is a much stronger effect
in two-dimensional systems than the hyperfine coupling discussed
above.

Based on this estimate, we might actually expect spin-orbit coupling
to dominate over the energy scales $\musf^\pm$ and $kT$ we have been
discussing so far, in contrast to our observation of
spin-conservation in most cases. In zero dimensions, however,
several of the mechanisms which dominate spin-relaxation in two
dimensions are suppressed \cite{Khaetskii2000}, and so the effect on
antidot states may actually be significantly weaker.  The main
effect of spin-orbit coupling in zero-dimensions is a reorganisation
of the SP levels into admixtures of states with different spin and
orbital quantum numbers $m$ and $s$.  As for the familiar spin-orbit
coupling in atomic systems, mixing occurs predominantly between
zero-dimensional states with the same value of $j = m+s$
\cite{Pietilainen2006}, and this suggests an interesting difference
between antidots and quantum dots. In contrast to quantum dots, LLL
antidot states with the same $j$ do not cross as $B$ is increased.
As can be seen in \figvref{fig:SPladder}, when $\Ez$ is increased
relative to $\dEsp$, the first crossings between SP states occur
between the $\lvert m,s\rangle$ states $\lvert m,\frac{1}{2}\rangle$
and $\lvert m\!-\!1,-\frac{1}{2}\rangle$, which have $\Delta j=2$.
Subsequent crossings have even larger $\Delta j$.  These selection
rules are only strict for a system with perfect circular symmetry,
which of course our real device does not have, but this may still
help explain the relative weakness of spin-orbit coupling which we
observe.

\begin{figure}[tb]
    \centering
    \includegraphics[]{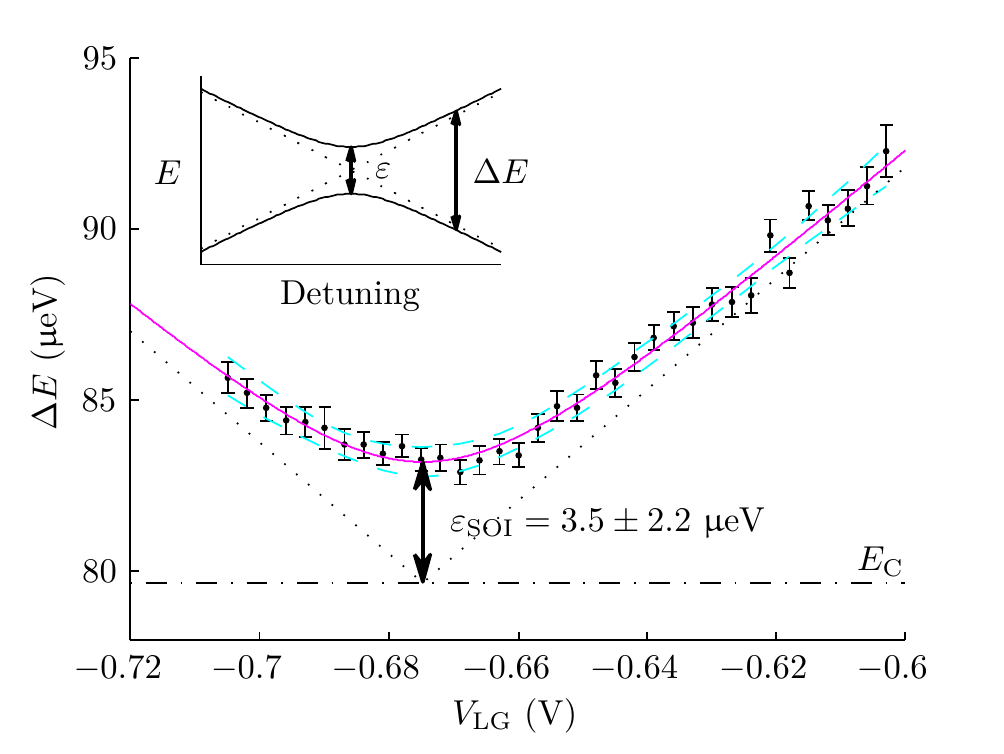}
    \caption[Anticrossing transmission resonances]{%
Energy separation of a pair of transmission resonances from
\figref{fig:SpinConservation}, with best-fit function (magenta
curve) and 95\% confidence intervals (dashed cyan curve) of the
anticrossing form given by \eqnref{eq:anticrossingfn}. As shown, the
spin-orbit coupling parameter $\eSOI$ is the difference between the
constant charging energy $\Ec$ (with best-fit value given by the
horizontal dot-dashed line) and the closest approach of the two
states.  The best-fit value of $\eSOI$ and errors at 95\% confidence
are given. Separations in $\Vad$ are converted to energy using the
`lever-arm' scaling
$\alpha_\mathrm{G}=\unit{0.022}{\electronvolt\per\volt}$ measured
from \figvref{fig:NLIVdata}.  The uncertainty in this scaling is
much less than those of the fit parameters.  Shown in the inset is
the general form of anticrossing states, ignoring $\Ec$, defined by
an anticrossing energy $\varepsilon$ at zero detuning, and
asymptotic unperturbed states at large detuning (dotted lines), to
show the relationship between the energy eigenvalues and their
separation $\Delta E$. \label{fig:anticrossing}}
\end{figure}

To obtain a quantitative measure of the spin-mixing in our device
for comparison with the energy scales $\varepsilon_\mathrm{HF}$ and
$\eSOI$, we can measure the evolution of the resonance peak
positions through a region in which spin-flips occur.  The data in
\figref{fig:anticrossing} is obtained from the positions of the
peaks in \figref{fig:Greflected}, for the resonances which mix
around $(\Vad,\Vlg)=(\unit{-1.13}{\volt},\unit{-0.67}{\volt})$.
The peak positions are converted to energy using the `lever-arm'
scaling discussed in \secref{sec:SequentialTransport}, which is
obtained from non-linear transport measurements.  Their
energy-separation shows a clear anticrossing, which is well-matched
by the standard hyperbolic function
\begin{equation}\label{eq:anticrossingfn}
  \Delta E(V) = \Ec+\varepsilon\sqrt{1+\left(\frac{V-V_0}{\beta}\right)},
\end{equation}
where $\varepsilon$ is the spin-mixing strength, $\Ec$ is the
charging energy in the constant-interaction model, and $V_0$,
$\beta$ are additional free parameters in the fit. We expect that
the effect of making $\Vlg$ more negative is to decrease $\dEsp$ by
flattening the slope of the antidot potential in the lower
constriction, through the `geometric' effect discussed in
\chapref{chap:Geometry}. Similar fits to the other crossings in
\figref{fig:Greflected} give consistent results, although the fits
are less constrained due to the limited range of data available. The
measured value of $\varepsilon =
3.5\pm\unit{2.2}{\micro\electronvolt}$ (with errors at 95\%
confidence) is clearly inconsistent with the much smaller hyperfine
energy $\eHF$, implying that the spin-mixing we observe is due to
the spin-orbit interaction. Furthermore, since the spin-orbit
coupling is large enough to keep states separated by at least
$\eSOI\gg\eHF$, the hyperfine effect is
unlikely to couple spins at all in our device.\footnote{%
Typical lead-antidot tunnel couplings in our device are
$\Gamma\approx\unit{500}{\mega\hertz}$, corresponding to
lifetime-broadening energies of
$\hbar\Gamma\approx\unit{0.5}{\micro\electronvolt}$, which therefore
also satisfy $\hbar\Gamma\ll\eSOI$.} %
The measured value of $\eSOI$ is also consistent with our
requirement that
\begin{equation}\label{eq:SpinConsCond}
  \eSOI<\musf^\pm\lesssim \Etherm
\end{equation}
in the regions where spin is conserved, since $\Etherm\approx
10$\nbd\unit{15}{\micro\electronvolt}.  As discussed above, we find
that \eqnref{eq:SpinConsCond} holds for the majority of the
parameter space we explored, while the condition
\begin{equation}
  \label{eq:SpinFlipCond}
  \musf^\pm<\eSOI<\Etherm
\end{equation}
holds at the positions of crossings between antidot states of
opposite spin.

This concludes our discussion of spin-mixing.  In
\chapref{chap:SpinTransportModel} we will concentrate on the regime
in which spin is conserved, in an effort to explain why
$\musf^\pm\lesssim \Etherm$ in all our measurements. Although it is
beyond the scope of this work, it would certainly be interesting to
further investigate the spin-orbit effect in antidots.  In
particular, it is highly likely that spin-orbit coupling
renormalises the effective $g$-factor for antidot states, which
could explain the somewhat enhanced value which we observe in our
measurements.

\subsection{Non-linear measurements: excitation spectra\label{sec:NLIV}}

While the spin-selective linear-response measurements discussed in
the previous section provide crucial new information about spin
selection rules for antidot transport resonances, they do not give
us a direct measure of the antidot energetics.  To convert
experimental adjustments of the external fields $\Vad$ or $B$ into
changes in the antidot chemical potential, we need the capacitive
couplings,\footnote{%
Since the magnetic field changes the size of antidot states, it
changes the antidot `effective charge' as described in
\secref{sec:ABeffect} and so can be treated in the same way as a
capacitively-coupled surface gate.} %
or `lever-arm' scalings, which can be obtained from the slope of the
ground-state resonance positions as a function of drain-source bias
$\Vds$, as described in \secref{sec:SequentialTransport}.  This
conversion factor then enables us to measure changes in the energy
separation between resonances, as in \figref{fig:anticrossing}, and
to estimate the electron temperature in our device, by fitting the
resonance line-shapes to a thermally broadened Fermi-function
derivative, \eqnvref{eq:FermiDerivative}.  In addition to the
lever-arm scalings, non-linear $I$-$V$ measurements of $\nuAD=2$
transmission resonances show much additional structure, which we can
use to reconstruct the \emph{excitation spectrum} of the antidot.
These serve as an important set of complementary measurements to our
spin-selective experiments at zero bias, since we have already seen
that the energy-spacing of antidot states plays a crucial role in
determining the spin-selectivity of transport.  Any model we propose
to explain the spin-selective measurements must also reproduce the
observed excitations, and vice versa.\footnote{%
While it would certainly be desirable to employ the two techniques
together, i.e.\ to use the selective injection/detection technique
to measure the spin-selectivity of excited states, in practice we
find that selective edge-mode population is only effective for
chemical potential differences of
$\lesssim\unit{30}{\micro\electronvolt}$, and so we generally
perform non-linear measurements with the injector and detector gates
fully open so that all incident modes are equilibrated.} %

\begin{figure}[tb]
    \centering
    \includegraphics[]{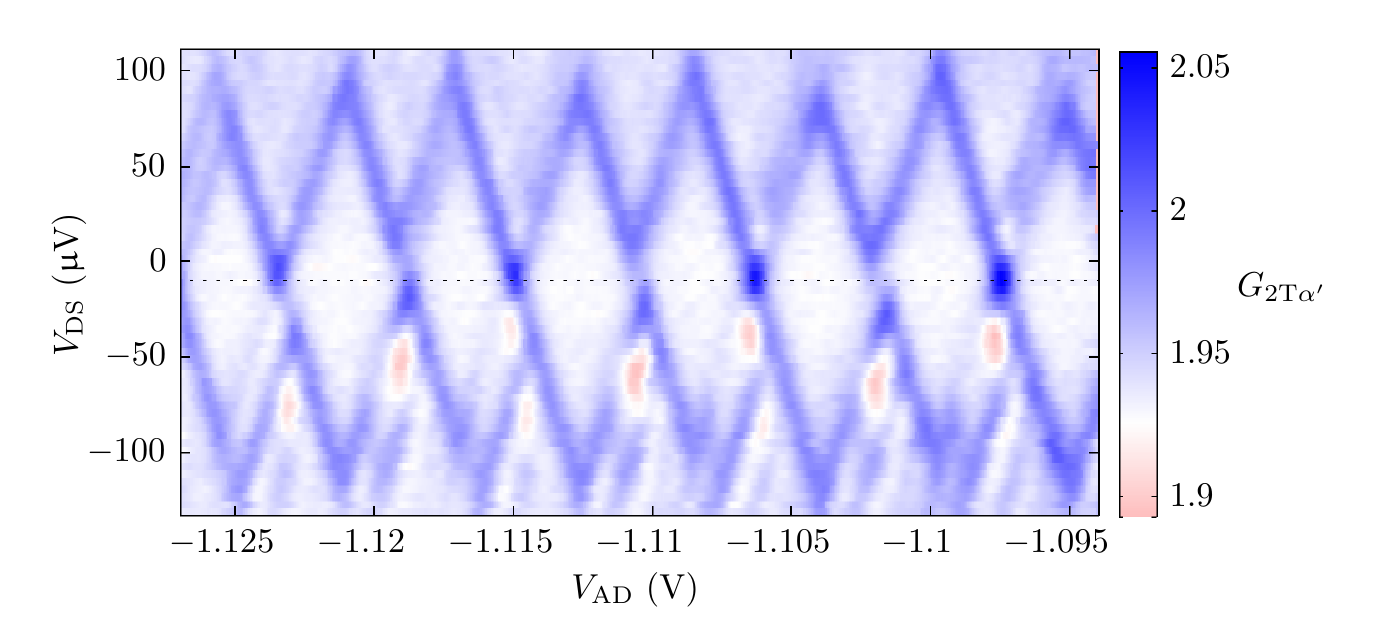}
    \caption[$\nuAD =2$ excitation spectrum]{%
Two-terminal differential conductance of antidot transmission peaks
on the $\nuAD=2$ plateau, in units of $e^2/h$. DC bias is applied to
the drain contact $\alpha^\prime$, and the injector is fully open.
Colours are scaled such that white corresponds to the $2e^2/h$
conductance plateau, which is shifted from the ideal value by series
resistances in the circuit. Individual sweeps are shifted
horizontally using interleaved calibration sweeps at zero bias to
account for device drift during the run.  External fields $B$,
$\Vug$, $\Vlg$ are set to similar values as for the measurements
shown in \figsref{fig:SpinConservation} and \ref{fig:Greflected}.
The horizontal dotted line corresponds to the zero of the current
preamplifier, as determined from simultaneous measurements of the DC
component of the transport current. Measurements are taken at
$B=\unit{1.21}{\tesla}$. \label{fig:NLIVdata}}
\end{figure}

An example of such non-linear conductance measurements is shown in
\figref{fig:NLIVdata}.
We clearly observe the Coulomb blockade diamond pattern of the
ground-state lines familiar to studies of single quantum dots, as
well as additional structure in the transport regions outside the
diamonds.  The additional lines parallel to the Coulomb diamonds
correspond to excitation energies, measured though their separation
in $\Vds$, in the range $\Eex\approx 15$\nbd\unit{50}{\mueV}. It is
important to note that we would not observe distinct lines for
states separated by less than $\approx\unit{15}{\mueV}$, since these
would not be resolved by the thermally broadened peaks. Still, the
clear presence of `gaps' of up to $\approx\unit{50}{\mueV}$ implies
either an antidot energy spectrum with these spacings, or that
strict selection rules prohibit transport through states at
intermediate energies.

\begin{figure}[tp]
    \centering
    \includegraphics[]{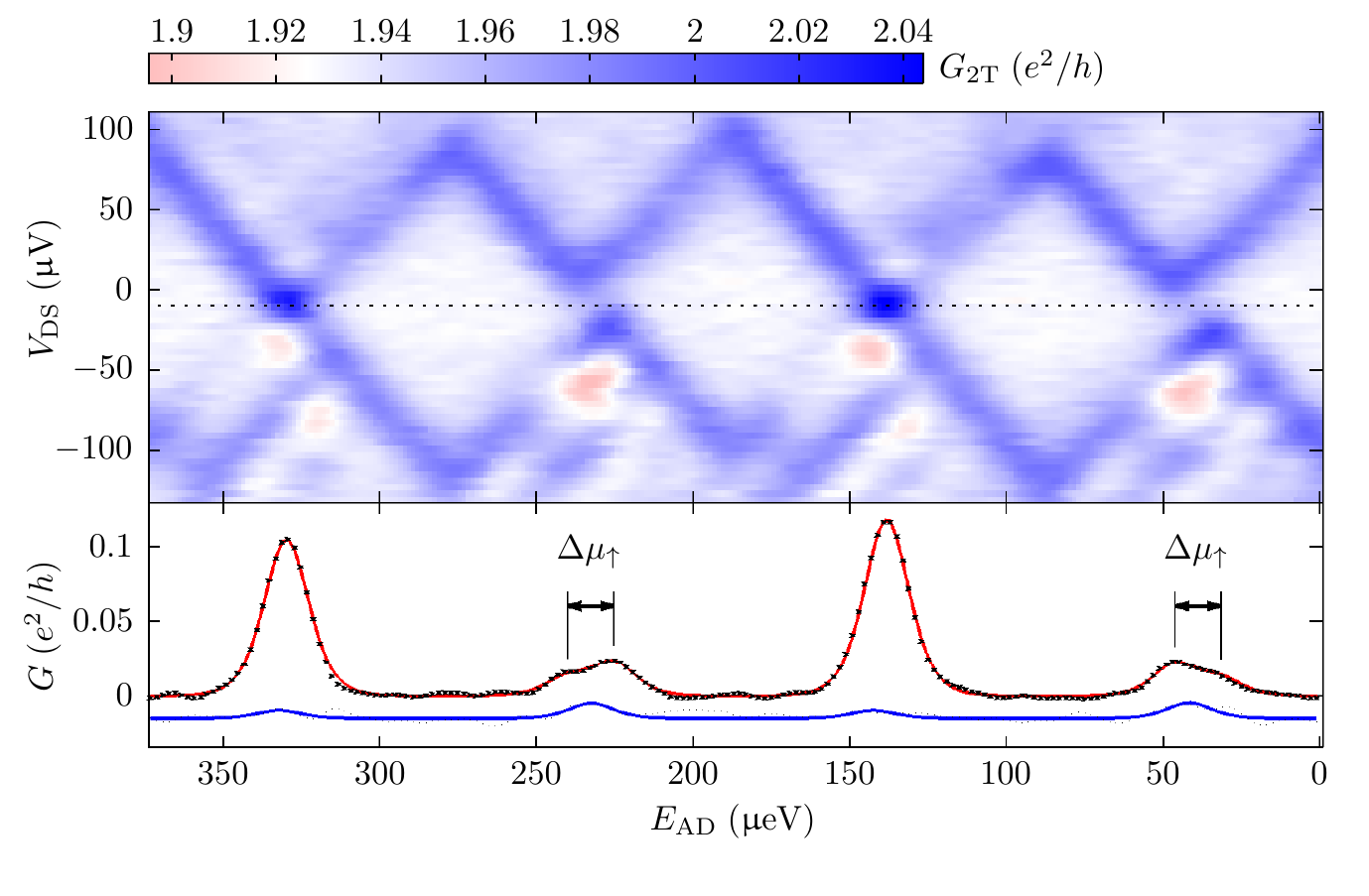}
    \caption[`Broken' ground-state lines]{%
Upper panel --- Close-up view of non-linear conductance measurements
from \figref{fig:NLIVdata} between $\Vad=\unit{-1.117}{\volt}$ and
\unit{-1.1}{\volt}, with $\Vad$ scaled to $\Ead$ using the lever-arm
factor $\alpha_\mathrm{G} =
\numprint{2.2e4}$~\micro\electronvolt\per\volt. Lower panel ---
Spin-resolved conductance at zero bias (along the dotted horizontal
line in the upper panel) with best-fit curves to the separate
contributions $\Gup$ (red curve) and $\Gdn$ (blue curve).  The
fitting function consists of thermally-broadened peaks
(Fermi-function derivatives) with a common width parameter
corresponding to the electron temperature, with best-fit value
$T_\mathrm{elec} = 54.9\pm\unit{0.5}{\milli\kelvin}$.  Counting from
the left, the second and fourth resonances in $\Gup$ clearly require
two peaks for a good fit; their best-fit positions are marked above
by vertical lines, and their separation is the energy scale
$\dmuup\approx\unit{15}{\mueV}$ as described in the text.  The data
and fit of $\Gdn$ is vertically offset for clarity.
\label{fig:NLIVandZBG}}
\end{figure}

In addition to the excited state lines, we observe two additional
important features in \figref{fig:NLIVdata}:
\begin{itemize}
  \item{%
The regions which appear red, predominantly at negative $\Vds$,
correspond to the occurrence of negative differential conductance
(NDC), in which the transmitted current actually drops as the bias
is increased.\footnote{%
In the case of a $\nuAD=2$ antidot, the conductance always remains
positive due to the background conductance of the LLL edge modes
which propagate freely through the constrictions.  By checking in
turn the conductance of the constrictions alone and by employing the
selective detection technique at low bias, we can show that the dips
in conductance below $2e^2/h$ are due to a reduction in the
transmitted current through the antidot rather than an effect in the
constrictions.} %
As is well-known in the quantum dot literature
\citep[e.g.,][]{Johnson1992,Weis1993} the observation of NDC implies
the presence of a `slow' state which becomes accessible with
increasing DC bias, and which competes with the remaining transport
channels in such a way as to reduce the average rate at which
electrons pass through the system.  Such slow states are usually
associated with some form of `spin blockade' \cite{Weinmann1995}, in
which selection rules and/or spin-selective barriers affect the
rates associated with different transitions.
  }
  \item{%
Furthermore, we notice that many of the ground state lines with
positive slope, which track the chemical potential of the drain,
appear to be `broken' around $\Vds=0$.  We are not aware of any
previous observations of this behaviour in the quantum dot
literature, suggesting that it reflects a unique property of the
system we are probing. We often observe features like this near
regions where spin-mixing occurs, as in \figref{fig:Greflected},
although they do not coincide precisely with the locations of state
crossings.  When we zoom in on a few of these `broken' lines and
compare them to corresponding spin-resolved measurements at zero
bias, we see that the line-shape of the \spinup\ zero-bias
conductance peak is strongly asymmetric, requiring a model with two
separate peaks for a good fit, as shown in \figref{fig:NLIVandZBG}.
In between the two \spinup\ peaks we find a smaller single peak in
the \spindn\ channel.
  }
\end{itemize}
We explain these features below in terms of a spin-dependent
asymmetry in the tunnel barriers which, combined with suitable DC
bias, results in a dynamic `pumping' of the antidot
spin-configuration.

\begin{figure}[tp]
\centering
\includegraphics[]{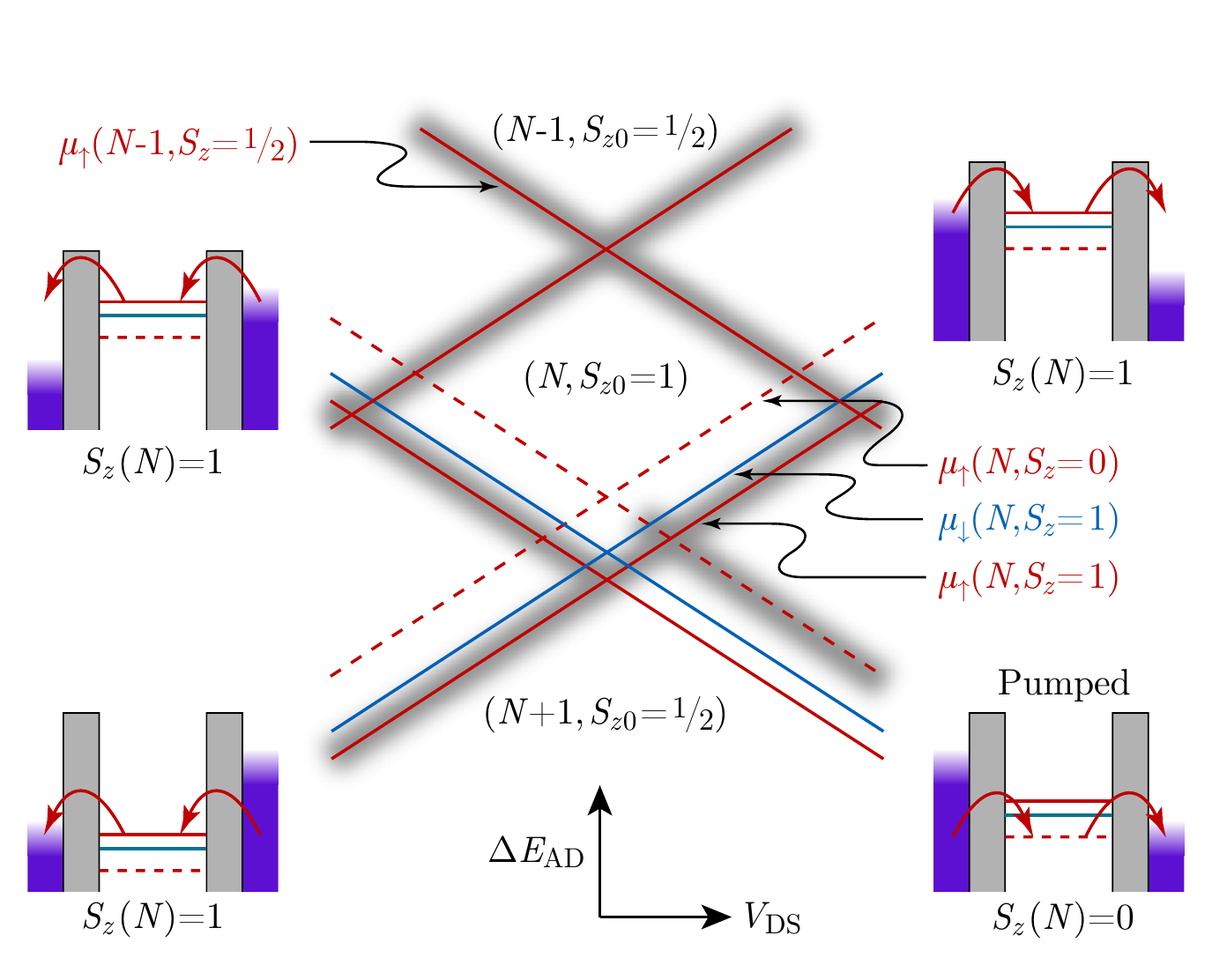}
\caption[Schematic of dynamic spin pumping]{%
Schematic of the dynamic spin-pumping mechanism which results in
`broken' ground-state lines near $\Vds=0$.  The locations where the
antidot chemical potentials align with either the source or drain
chemical potential are shown as a function of $\Vds$ and the
capacitive energy $\Delta\Ead$ added by the antidot gate voltage.
The ground-state \spindn\ transition $(N,1)\leftrightarrow
(\Npone,1/2)$, shown in blue, is suppressed by a very large tunnel
barrier for \spindn\ between the antidot and the drain, so transport
occurs instead through the two neighbouring \spinup\ transitions,
shown in red. For each `branch,' tunneling diagrams show, from left
to right, the configuration of the source, antidot, and drain
chemical potentials. The blurred lines show the resulting positions
of the observed Coulomb-blockade boundaries.  In this case, dynamic
pumping of the branch in the lower-right results in the appearance
of a broken line around zero bias.  The upper-right branch is not
strongly pumped because the source cannot maintain an occupation in
$(N,0)$ when it is aligned with $\muup(N,0)$, as it can when the
drain is aligned with $\muup(N,0)$. The adjacent ground-state
resonance ($N-1,1/2)\leftrightarrow (N,1)$) involves \spinup, and
therefore behaves normally as a single resonance split by
drain-source bias. \label{fig:SpinPumping}}
\end{figure}

We begin by showing how dynamic spin-pumping can result in
dislocations in the Coulomb-blockade boundaries near zero bias. From
the non-equilibrium measurements in \secref{sec:ZeroBiasExpts} we
have already seen that the \spindn\ tunnel couplings are highly
asymmetric, with $\gDdn\ll\gSdn$.  An analysis of the amplitudes of
the resonances in \figsref{fig:SpinConservation} and
\ref{fig:Greflected} suggests that
$\gSdn\simeq\gSup\simeq\gDup\approx\unit{500}{\mega\hertz}$, and
that $\gDdn$ is roughly an order of magnitude lower.  As shown in
\figref{fig:SpinPumping}, this asymmetry in combination with a DC
drain-source bias affects the dynamic equilibrium of the antidot
$(N,S_z)$ states, and can change the maximally-occupied state.
When $\Vds$ is positive (negative), the suppressed rate for \spindn\
electrons leaving (entering) the antidot enhances the occupation
probabilities of states with lower (higher) $S_z$. If the spin-flip
energies $\musf^\pm$ are small for both the $N$ and $\Npone$
ground-state configurations, then a small bias may be sufficient to
change the maximally-occupied $S_z$ configurations by $\pm1$,
causing a shift in the resonance position which reflects the new
chemical potential. These conditions may be met when the
ground-state transition involves the blocked \spindn\ transport
channel, and we instead observe nearby \spinup\ resonances as the
borders of the Coulomb-blockade region.  The specific configuration
shown in \figref{fig:SpinPumping} is only an example; the same
essential features can occur in several ways.  In most cases, the
spin-configuration is efficiently pumped only in one of the four
`branches,' since the bias must supply both the asymmetry to pump
$S_z$ and the energy to maintain the excited-state population at the
position of the new resonance.  The resonances shown as blurred
lines in \figref{fig:SpinPumping} are a simplified approximation,
since the position of the actual resonance peak depends on the
overlapping contributions from several transitions, and these
contributions change as a function of bias within each branch.  But
even with these complications, the appearance of a dislocation in
the line tracking the drain chemical potential is maintained, and
occasionally the instability is such that both \spinup\ transitions
can be thermally-activated at zero bias, resulting in an asymmetric
or double peak as observed in \figref{fig:NLIVandZBG}.  We will see
in \chapref{chap:SpinTransportModel} how this behaviour may be
reproduced in a sequential tunneling model which incorporates
asymmetric tunnel couplings.

With this interpretation of the features observed in our non-linear
transport data, we can identify the energy-separation of the
double-peaks in \figref{fig:NLIVandZBG} as the difference between
two \spinup\ chemical potentials.  This represents an important
additional energy scale which we must reproduce with a model of the
system:
\begin{equation}
  \label{eq:deltamuup}
  \begin{split}
  \dmuup(N,S_z) & = \muup(N,S_z+1)-\muup(N,S_z),\\
   & = \musf^+(\Npone,S_z+\tfrac{1}{2}) - \musf^+(N,S_z).
  \end{split}
\end{equation}
Within the SP model it is easy to show that $\dmuup = \dEsp$. The
observed value of $\dmuup\approx\unit{15}{\mueV}$, however, is too
small to produce the excitation spectrum we observe, which requires
$\dEsp\approx 30$\nbd\unit{50}{\mueV}.  We discuss the implications
of this inconsistency in \chapref{chap:SpinTransportModel}.

A similar consideration of the transport channels which become
available at different locations in the $(\Vds,\Vad)$ plane explains
the observation of NDC in the non-linear transport data. Taking the
configuration depicted in \figref{fig:SpinPumping} as an example, we
note that the chemical potential $\mudn(N,S_z=0)$ enters the
transport window at $\Vds>0$ just to the right of the
$\muup(N,S_z=1)$ line. This results in a \emph{decrease} in
transmitted current, since \spindn\ electrons will enter the dot
through $(N,0)\rightarrow(\Npone,-1/2)$ but then become trapped by
the large exit barrier.  Analogously, the chemical potential
$\mudn(N,S_z=2)$ becomes available to the left of $\muup(N,S_z=0)$
line at negative $\Vds$, leading to NDC through a similar effect.
Most likely, the NDC is mainly observed at $\Vds<0$ in the
experimental data because the source-drain asymmetry changes at
positive bias, probably due to a bias-induced charging event of the
nearby impurity.

\subsection{Confirmation from a second antidot}

As described in \secref{sec:EdgeScattering}, spin-selective
measurements of transmission resonances are limited to a relatively
narrow range of $B$ by the conflicting requirements that
\onetothree\ and \twotofour\ coupling should be suppressed along the
edge but present at the antidot.  Furthermore, the observation of
state-crossings and spin-flip transport as described in
\secref{sec:SpinFlipTransport} suggests that we are close to a
degeneracy point in the antidot configuration energy $U(S_z)$. Based
on the tilted-field measurements of \chapref{chap:TiltedB}, taken
later from the same device, we believe this to be the first set of
LLL state-crossings, i.e.\ $S_{z0}=0\rightarrow S_{z0}=1$ for
even-$N$ occupation numbers.  In the SP model, this transition
occurs when $\dEsp = \Ez$, and although we have already shown that
several details of the non-linear measurements seem inconsistent
with a non-interacting model, we might wonder whether the behaviour
we observe (in particular, that $\musf^\pm\lesssim \Etherm$) is
simply the result of a coincidental degeneracy between the SP energy
scales.

\begin{figure}[tp]
    \centering
    \includegraphics[]{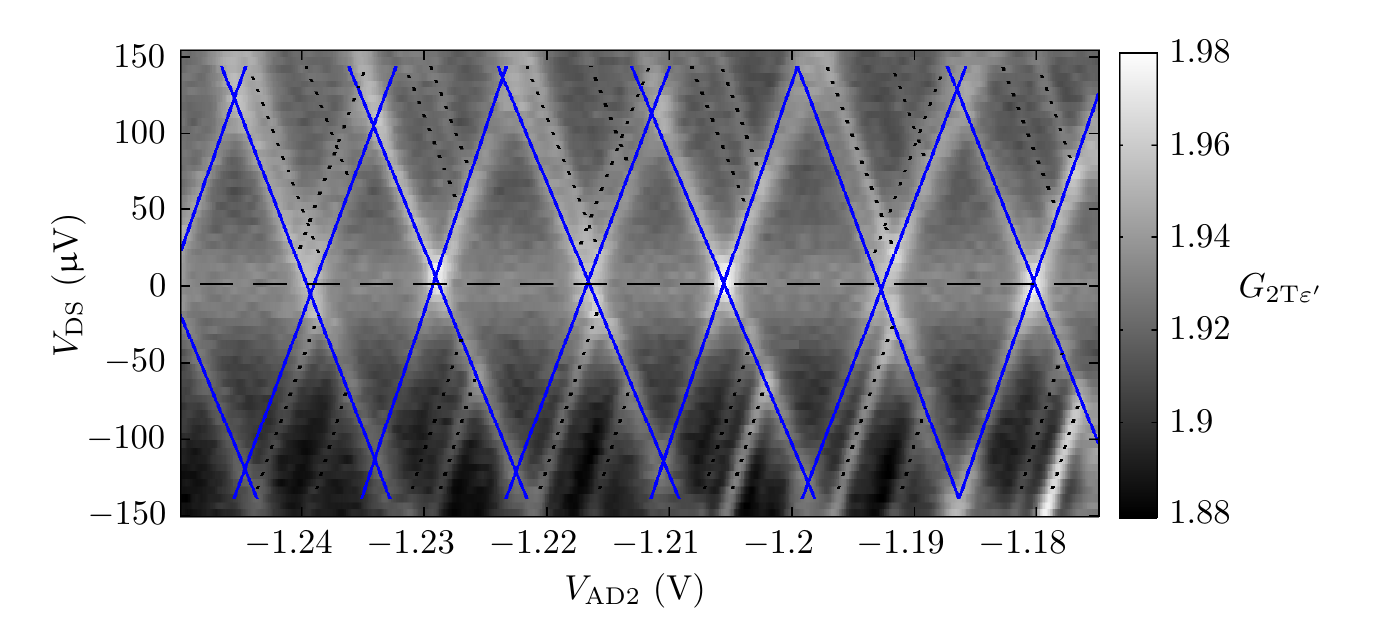}
    \caption[Excitation spectrum of AD2]{%
Non-linear transport measurements of the second antidot on our
device, with bias applied to the drain contact $\varepsilon^\prime$.
Ground-state and excited-state transitions are marked as solid blue
and dotted black lines, respectively.  Measurements are taken at
$B=\unit{0.75}{\tesla}$.\label{fig:AD2NLIVdata}}
\end{figure}

From \eqnref{eq:dEspapprox} we see that
$\dEsp\sim(BR_\mathrm{AD})^{-1}$, so we can change the SP energy
spacing through either $B$ or $\Vad$, which controls the antidot
size.  For measurements of the central antidot, we were able to vary
both $B$ and $R_\mathrm{AD}$ by $\approx20$\%, but this is probably
not enough to ensure that $\dEsp$ is changed by more than
$E_\mathrm{therm}\approx\unit{15}{\mueV}$.  The second antidot on
our device, however, is lithographically much smaller, with a
diameter of \unit{200}{\nano\metre} compared to
\unit{300}{\nano\metre} for AD1.  The excitation spectrum we
observe, shown in \figref{fig:AD2NLIVdata}, contains a very regular
pattern of excitation energies, with alternating values of
$\approx\unit{20}{\mueV}$ and $\approx\unit{40}{\mueV}$.
These data seem fully consistent with the SP model, within which we
identify the excitations as $\Ez$ and $\dEsp-\Ez$, respectively,
such that $\Ez=\unit{20}{\mueV}$, $\dEsp=\unit{60}{\mueV}$, and
$\Ec\approx\unit{100}{\mueV}$.  This agrees with the expected value
for the Zeeman energy at $B=\unit{0.75}{\tesla}$ of
$\abs{g}\mu_\mathrm{B}B=\unit{19}{\mueV}$.

\begin{figure}[tb]
    \centering
    \includegraphics[]{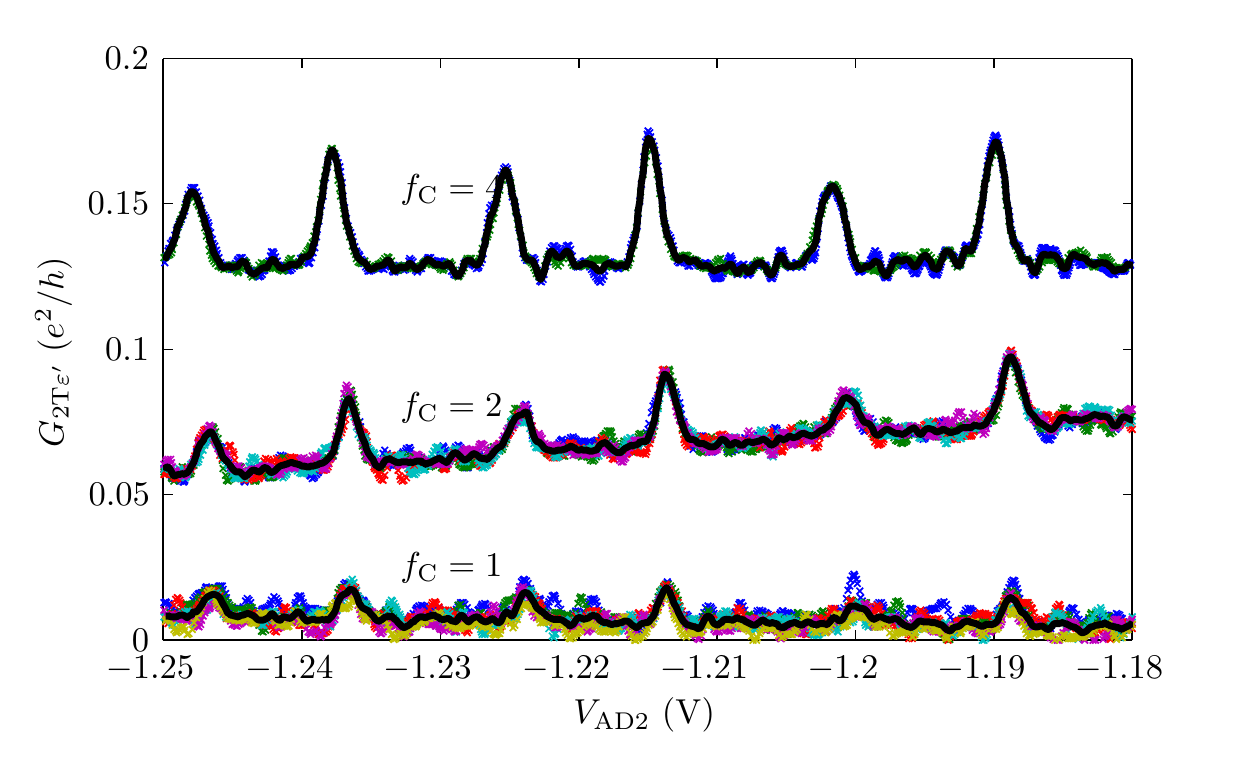}
    \caption[Selective injection to AD2]{%
Two-terminal conductance measured from contact $\varepsilon^\prime$
as a function of the filling factor of the central constriction
$f_\mathrm{C}$.  With $\Vug$ set to a large negative voltage such
that the upper antidot constriction is pinched off, the lower
constriction functions as a standard injector QPC to selectively
populate modes with the excitation signal applied to contact
$\alpha^\prime$.  At the low field used for these measurements
($B=\unit{0.75}{\tesla}$), equilibration between edge-modes
\onetothree\ and \twotofour\ is significant, but spin-scattering is
still suppressed, so setting $f_\mathrm{C}=1$ results in a
population of only \spinup\ modes $1,3,5,\dotsc$ at the antidot.
Solid black lines represent the average of several sweeps at each
$f_\mathrm{C}$ setting as shown, and each set of traces has an
arbitrary vertical offset for clarity. \label{fig:AD2Injection}}
\end{figure}

With both $\Ez,\dEsp > \Etherm$, we would expect to observe nearly
complete spin-polarised transport based on the SP model, but that is
not what we observe.  While we do not have enough QPCs to perform
full injection/detection measurements of AD2, we can use one channel
of AD1 as an injector to partially populate the edge-modes incident
on AD2.  Due to its smaller size, we were forced to use much lower
fields of $B\approx 0.6$\nbd\unit{0.8}{\tesla} in order to observe
transmission resonances.  At these fields significant equilibration
occurs between edge modes of the same spin, but calibration
measurements similar to those presented in
\secref{sec:EdgeScattering} show that spin-scattering is still
suppressed below $\approx5$\nbd10\%, so by setting central QPC to
$f_\mathrm{C}=1$, we populate only the \spinup\ modes.  Measurements
of a few $\nuAD=2$ transmission resonances corresponding to the data
in \figref{fig:AD2NLIVdata} are shown in \figref{fig:AD2Injection},
for injector filling factors $f_\mathrm{C}=1$ (\spinup\ partially
populated), 2 (both spins partially populated) and 4 (both spins
fully populated).
It is difficult to know exactly what are the populations incident on
AD2 in order to interpret the amplitude variations in
\figref{fig:AD2Injection}, but the important result is that, at
$f_\mathrm{C}=1$ when only one spin is injected, all resonances
remain with roughly similar amplitudes. Thus the transmitted current
is not spin-polarised, even though the observed SP energy scales
seem to imply $\musf^\pm>\Etherm$.  This is strong evidence that
additional physics is playing an important role `behind the scenes'
in our system, somehow softening the spin-excitation energy $\musf$
while preserving SP-like excitation energies.  In
\chapref{chap:SpinTransportModel} we describe a model of
electron-electron interactions with exactly these properties and
compare simulated data incorporating this model with our
measurements.

\section{Conclusions}

In this chapter we have demonstrated the power of the
non-equilibrium selective injection/detection technique to extract
the individual transmission coefficients of a mesoscopic device in
the quantum Hall regime. By using this technique to measure the
spins of electrons transmitted through LLL antidot states, we have
discovered that, contrary to expectation, the individual resonances
are not spin-polarised. Generally we find that spin is conserved
during transport, although we can detect certain regimes in which
this is not the case, which coincide with (anti)crossings between
antidot energy levels. We attribute the non-conservation of electron
spin to the spin-orbit effect, which mixes opposite spin states that
pass within a characteristic energy $\eSOI\approx\unit{3}{\mueV}$ of
each other.  In the measurements in which spin is conserved, the
observed lack of spin-polarisation still requires that the spin-flip
excitation energy of the antidot remain small, $\musf^\pm\lesssim
\Etherm\approx\unit{15}{\mueV}$, but we observe clear excitation
energies in non-linear transport experiments which do not satisfy
this condition.  Measurements of a second, much smaller antidot on
the same device prove that the lack of spin-selectivity is not a
coincidence of SP energy scales, but rather a more general feature
of antidots in this low-$B$ regime, which requires an explanation
including physics beyond the non-interacting model.  In
\chapref{chap:SpinTransportModel} we explain these results by
modeling the antidot as a maximum density droplet of holes, as
described in \secref{sec:HFtheory}, and explore the implications of
our measurements for the physics of interacting electrons.

\chapter[Spin-Resolved Transport: Modeling and Discussion]{Spin-Resolved Transport:\\ Modeling and Discussion\label{chap:SpinTransportModel}}
\chaptermark{Spin Resolved Transport: Modeling}

\ifpdf
    \graphicspath{{Chapter5/Figures/PNG/}{Chapter5/Figures/PDF/}{Chapter5/Figures/}}
\else
    \graphicspath{{Chapter5/Figures/EPS/}{Chapter5/Figures/}}
\fi


In this chapter we discuss the implications of the experimental
results presented in \chapref{chap:SpinTransport}.  We propose a
model to explain these results in terms of an interacting theory of
antidot ground states within the Hartree-Fock framework.  In order
to test this proposal, we develop a computational model of
spin-resolved sequential transport through an antidot, for an
arbitrary antidot energy-spectrum and set of selection rules,
including spin- and energy-dependent tunnel barriers.  We find
excellent agreement between our experimental observations and these
transport calculations when the antidot ground-state is modeled as
an interacting maximum density droplet (MDD) of holes in the lowest
Landau level (LLL).  We consider the limitations of the Hartree-Fock
model, particularly with regards to excitation spectra of the MDD,
and propose an effective theory for the antidot edge excitations
which agrees with our experiments and encompasses most of the
relevant physics.

\section{Maximum density droplets\label{sec:MDDmodel}}

A brief introduction to the theory of MDDs within a Hartree-Fock
approach is presented in \secref{sec:ADIntTheory} of this thesis.
The concept of an MDD forms an integral part of our understanding of
interactions in quantum Hall fluids, and is particularly relevant
for the theoretical description of quantum dots at high magnetic
fields. The review by Reimann and Manninen \cite{Reimann2002}
provides a useful discussion of both experimental and theoretical
efforts to understand the electronic structure of quantum dots, in
which the MDD phase is covered in some detail.  The description of
an antidot in terms of an MDD of `holes' within the LLL has been
developed in previous theoretical works \cite{Hwang2004,Sim2004},
but these have mainly focused on effects observed at higher magnetic
fields than those considered here.

\subsection{The exchange effect}

We begin with a few simple arguments based on dimensional analysis
to introduce the basic physics considered in this chapter. Recall
that the major experimental results which we seek to explain are the
following: the antidot excitation spectrum shows clear evidence of
excitation energies $\Eex \gg kT$, consistent with the SP picture of
orbital states, but transport at zero bias is not spin-selective.
Since spin is conserved during transport, the total antidot spin
$S_z$ must be a good quantum number, but the spin-excitation energy
$\musf^\pm$ defined in \eqnvref{eq:musfDefn} must be smaller than
that predicted by the SP model, to satisfy $\musf^\pm \lesssim
\Etherm$ at every resonance.

Within the SP model described in \secref{sec:SPnu=2} for a $\nuAD=2$
antidot, the LLL energy spectrum consists of two `ladders' of
orbital states with approximately uniform spacing $\dEsp$, and with
opposite spin.  The spacing between the ladders is the Zeeman
energy, $\Ez$.  When $\dEsp>\Ez$ as we typically expect, the
ground-state total spin alternates between $S_z=0$ for even values
of the occupation number $N$, and $S_z = \frac{1}{2}$ for odd $N$,
and the lowest-energy excitations for these configurations are
$\dEsp - \Ez$ and $\Ez$, respectively. Consider the form of the
spin-configuration energy $U(S_z)$ for fixed $N$ within this model.
Starting for example from $S_z=0$ with even $N$, we obtain the
lowest-energy configurations at higher $S_z$ by promoting successive
electrons from the upper ladder of \spindn\ electrons into the
lowest available state of the lower \spinup\ ladder. Each time we do
this, we gain the Zeeman energy by moving between ladders, but have
to pay orbital energy costs of $\dEsp,\: 3\dEsp,\: 5\dEsp,\dotsc$,
to reach the next available state.  Thus the configuration energy is
given by
\begin{equation}\label{eq:Usz1}
\begin{split}
  U(S_z) & = \sum_{i=1}^{S_z}\Bigl[(2i-1)\dEsp-\Ez\Bigr]\\
   & = \dEsp S_z^2 - \Ez S_z.
\end{split}
\end{equation}
It is easy to show that $U(S_z)$ takes this basic form for all
available values of $S_z$, and for both odd and even $N$.  Up to an
irrelevant constant, \eqnref{eq:Usz1} is equivalent to
\begin{equation}\label{eq:Uszquad1}
  U(S_z) = \dEsp(S_z - S_z^\ast)^2,
\end{equation}
where the minimum $S_z^\ast=\Ez/2\dEsp$ is between 0 and
$\frac{1}{2}$ when $\dEsp>\Ez$.  In general, the spin-flip energy is
therefore of order $\musf^\pm \simeq U(S_z^\ast \pm1) - U(S_z^\ast)
= \dEsp$, and it takes a special coincidence (e.g., $\dEsp \approx
\Ez$) to have $\musf^\pm \approx kT \ll \dEsp$.

When we include interactions within Hartree-Fock theory, as
described in \secref{sec:HFtheory}, the configuration energy of each
Slater orbital contains both a positive contribution from the
`direct' Coulomb term, which describes the usual repulsive effect of
the Coulomb interaction, and also a negative contribution from the
`exchange' term, which favours parallel alignment of spins. Assuming
that the antidot potential is strong enough to compensate for the
direct term and preserve the MDD as the ground state, the
predominant new contribution to $U(S_z)$ arises from the exchange
term, which we can approximate in terms of the `overlap' between SP
states at the spin-polarised edge.  The scale of the exchange
interaction is the Coulomb energy scale
\begin{equation}\label{eq:CoulScale}
    J \approx \frac{e^2}{4\pi\epsilon\epsilon_0\ellB},
\end{equation}
and if we approximate the SP states as annuli with width $\ellB$ and
radius $\ellB\sqrt{2m}$, for orbital quantum numbers $m\gg1$, the
width of the spin-polarised edge is $w = 2S_z\Delta r$, where the
separation of successive orbital states is $\Delta r \approx
\ellB^2/\Rad$. For small values of $S_z$, such that $w\ll\ellB$, the
exchange effects therefore reduce the configuration energy by an
amount $\approx J w/\ellB$ for each of the $2S_z$ electrons in the
spin-polarised edge.  Thus the total contribution to $U(S_z)$ is of
order (dropping factors of two which are not significant within this
approximation)
\begin{equation}\label{eq:Eex1storder}
  \Delta E_\mathrm{exchange} \approx -J \frac{\Delta r}{\ellB}S_z^2
  =  -K S_z^2,
\end{equation}
where
\begin{equation}\label{eq:Kscale}
  K \approx \frac{e^2}{4\pi\epsilon\epsilon_0\Rad}
\end{equation}
is the energy scale of the exchange interaction.  As $S_z$ increases
and the width of the spin-polarised region increases such that
$w\gtrsim\ellB$, the exchange contribution begins to saturate and
\eqnref{eq:Eex1storder} ceases to be a good approximation.  We can
account for this by adding higher-order terms to the spin
functional, and since the Coulomb energy is independent of the spin
\emph{direction}, the next available term is a quartic:
\begin{equation}
  \Delta E_\mathrm{exchange} \approx -K S_z^2 + \beta S_z^4,
\end{equation}
where $\beta\ll K$.  Non-parabolicity of the confinement potential
may also contribute to the magnitude of the higher-order terms, but
this must also be an even functional of $S_z$.

\begin{figure}[tb]
    \centering
    \includegraphics[]{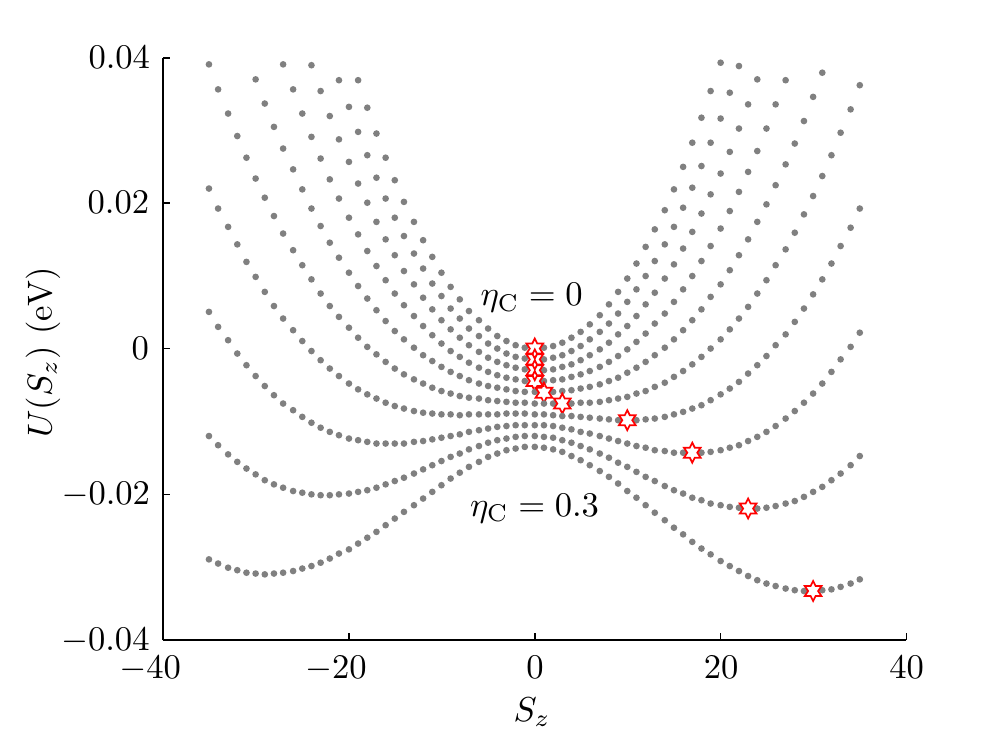}
    \caption[MDD configuration energy]{%
Configuration energy $U(S_z)$ for MDDs as the Coulomb interaction is
`turned on,'  via uniform steps of the interaction strength $\etaC$.
A red star denotes the ground state $S_{z0}$ configuration for each
calculation.  Calculations for different $\etaC$ are offset
vertically for clarity.  Parameters used in this calculation:
$B_\perp =\unit{0.65}{\tesla}$, $\dEsp = \unit{100}{\mueV}$, $\Ez =
\unit{40}{\mueV}$, $\Rad = \unit{400}{\nano\metre}$ ($N=130$ holes).
\label{fig:UofSz}}
\end{figure}

Putting all this together, we obtain a spin functional of the form
\begin{equation}\label{eq:Usz2}
  U(S_z) = -\Ez S_z + (\dEsp - K)S_z^2 + \beta S_z^4 + \dotsb,
\end{equation}
which we can approximate as a quadratic about the minimum $S_z^\ast$
as in \eqnref{eq:Uszquad1},
\begin{equation}
  U(S_z)\approx \alpha(S_z-S_z^\ast)^2.
\end{equation}
Several examples of the configuration energy $U(S_z)$ are shown in
\figref{fig:UofSz}, calculated for antidot MDD states using
\eqnvref{eq:ConfigEmatform}.  These demonstrate the evolution from a
non-interacting configuration ($\dEsp\gg K$) to a
strongly-interacting one ($K\gg\dEsp$) as the Coulomb interaction
strength $\etaC$ is increased.
For most of this parameter range, the spin-flip scale
$\musf\approx\alpha$ is dominated by the quadratic term of
\eqnref{eq:Usz2}, such that $\alpha\approx\dEsp$ when $K\ll\dEsp$
and $\alpha\approx2K$ when $K\gg\dEsp$.  But in the crossover regime
where $K\approx\dEsp$, the curvature near $S_z=0$ is dominated by
the quartic term, such that $\alpha\approx\beta$ is systematically
suppressed.  Since the small parameter $\beta\ll\dEsp$ is likely to
satisfy $\beta\lesssim kT$, this is a regime in which we would not
expect to observe spin-selectivity in transport.  For
$\Rad=\unit{400}{\nano\metre}$ as determined from the Aharonov-Bohm
period in our device, \eqnref{eq:Kscale} yields
$K\approx\unit{300}{\mueV}$.  This is larger than the orbital energy
spacing $\dEsp\approx50$\nbd\unit{100}{\mueV} we observe in
transport measurements, but is a similar order of magnitude.  We
also expect that \eqnref{eq:Kscale} significantly over-estimates the
strength of exchange, since in \eqnref{eq:CoulScale} we did not
account for the extent of the wave function either around the
antidot or perpendicular to the 2DES, or for screening by other
parts of the device.  These factors are roughly included through the
multiplicative parameter $\etaC$ which we vary to control the
strength of the Coulomb interactions.

\begin{figure}[p]
    \centering
    \includegraphics[]{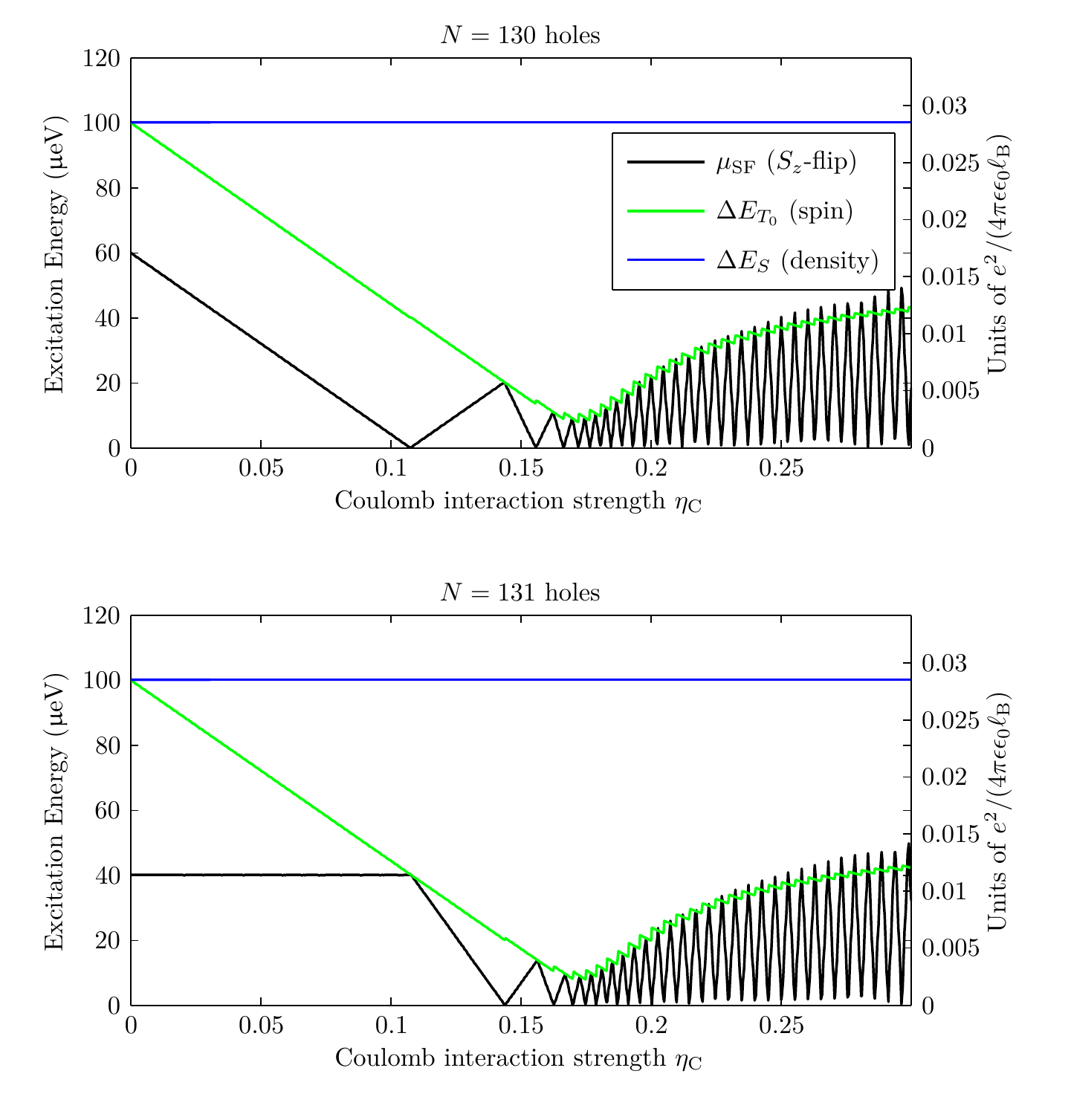}
    \caption[Edge excitation energies versus $\etaC$]{%
MDD edge excitation energies as a function of the Coulomb
interaction strength $\etaC$, for both even (top) and odd (bottom)
occupation numbers.  As labeled in the legend, we plot the spin-flip
energy $\musf = \min(\musf^+,\musf^-)$, and the lowest-energy
$S_z$-preserving excitations for the outermost holes of the MDD,
corresponding to spin (triplet, $\dEt$) and density (singlet,
$\dEs$) excitations. The Coulomb interaction suppresses the spin
excitation energies (both $\musf$ and $\dEt$), eventually leading to
a series of $S_z$-flips of the ground state, but leaves the density
excitation $\dEs$ unchanged.  Parameters used in this calculation:
$B_\perp =\unit{0.65}{\tesla}$, $\dEsp = \unit{100}{\mueV}$, $\Ez =
\unit{40}{\mueV}$, $\Rad = \unit{400}{\nano\metre}$.
\label{fig:EexVetaC}}
\end{figure}

While the spin-excitation energy may be significantly suppressed by
exchange, especially when $\dEsp\approx K$, a key feature of this
model is that the orbital excitation energy is \emph{preserved}.
Fundamentally, this is because the orbital excitations represent
modulations of the electron density at the edge of the MDD, in which
total spin is conserved.  The exchange interaction only affects
energy-differences between states with different spin, and so it
leaves these density-excitations unaltered.  This is demonstrated in
\figref{fig:EexVetaC}, in which we plot the lowest-energy
edge-excitations as well as the spin-flip chemical potential $\musf$
as functions of the Coulomb interaction strength.
We will discuss the MDD excitation spectrum in more detail in
\secref{sec:MDDexspectra}, but the two edge-excitations shown in
\figref{fig:EexVetaC} may be simply interpreted as the singlet
($\dEs$) and $T_0$ triplet ($\dEt$) components of the possible
excitations involving the outermost \spinup\ and \spindn\ particles
of the MDD.  The remaining $T_\pm$ triplet states belong to other
$S_z$ subspaces, so these correspond to the spin-flip potentials
$\musf^\pm$. Exchange interactions affect the energy of the
$S_z$-conserving triplet state along with the spin-flip states, but
not the singlet state.  The `crossover' region $K\approx\dEsp$
occurs at $\etaC\approx0.15$, coinciding with the first few
spin-flips of the even-$N$ ground state.

This exchange-driven \emph{spin-charge separation} is the essence of
our explanation for the observations described in
\chapref{chap:SpinTransport}.  In the following sections we will
discuss further details of the MDD configuration, and explain the
model we use to incorporate other aspects of the experiment, such as
the spin-pumping effect discussed in \secref{sec:NLIV} due to
asymmetric and spin-dependent tunnel barriers.  This transport model
enables us to make detailed comparisons between theoretical
predictions and our experimental results.

\subsection{Stability of the MDD\label{sec:MDDstability}}

\begin{figure}[tbp]
    \centering
    \includegraphics[]{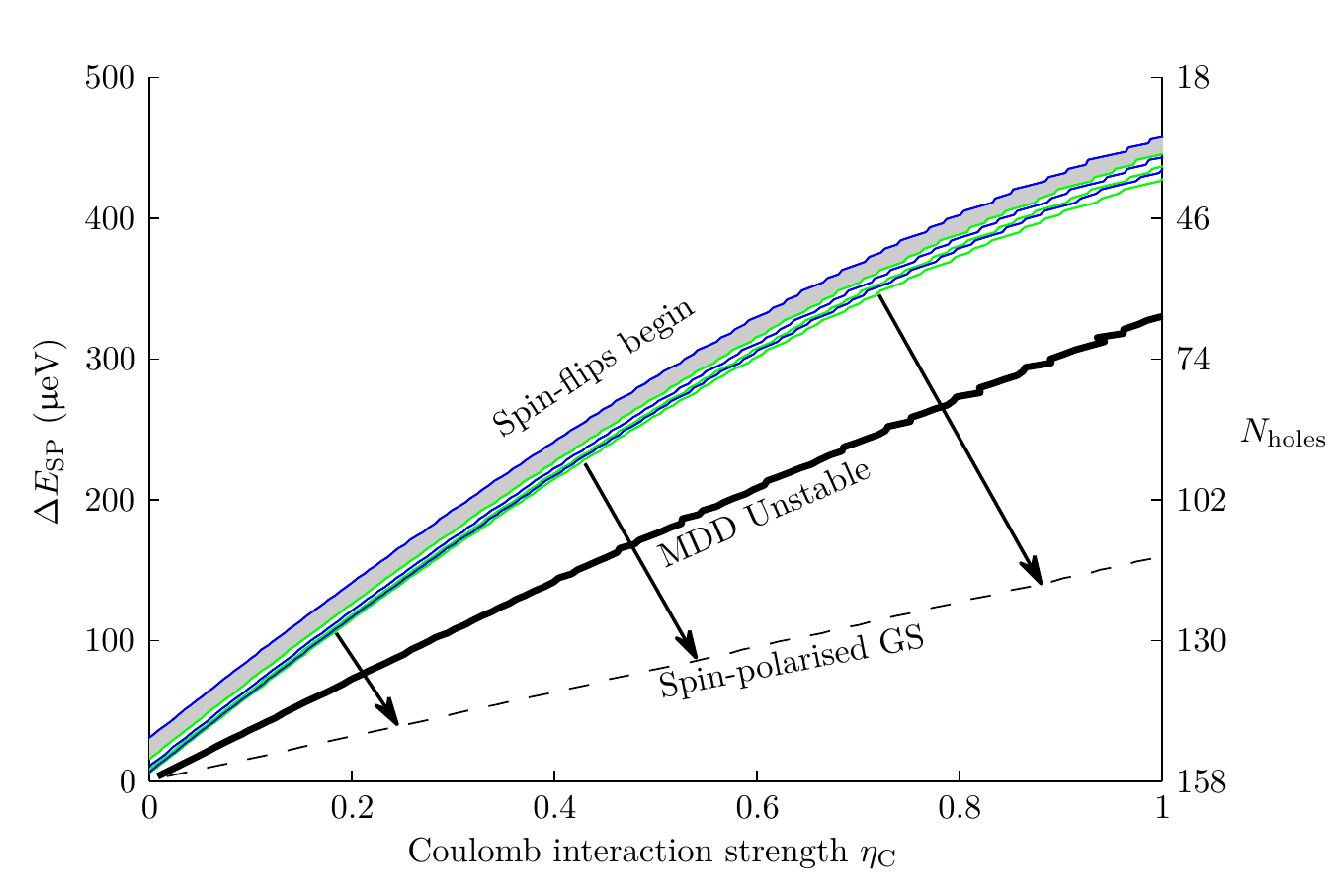}
    \caption[MDD phase diagram]{%
Phase diagram of the MDD ground state for a parabolic antidot as a
function of the Coulomb interaction strength $\etaC$ and $\dEsp$.
The first few transitions of $S_z$ are shown for both odd-$N$
(green) and even-$N$ (blue) configurations, and arrows note the
direction of increasing $S_z$.  Below the dashed black curve, the
ground-state MDD is fully spin polarised.  The grey shaded region
between the first two spin flips denotes the regime in which both
$S_{z0}=\frac{1}{2}$ for odd-$N$, and $S_{z0}=1$ for even-$N$
configurations, as required for the spin-pumping mechanism described
in \secref{sec:NLIV}.  The thick black curve marks the boundary of
the MDD phase, below which MDDs are unstable.  For a parabolic
potential, the number of holes within the fixed radius
$\Rad=\unit{400}{\nano\metre}$ varies as a function of $\dEsp$ as
marked on the right-hand axes, since the length scale $\ell =
\ellB/\sqrt{b}$ depends on the confinement, in terms of the
confinement parameter $b$ defined by \eqnvref{eq:effectiveb}. Other
parameters used in this calculation: $B_\perp =\unit{0.65}{\tesla}$,
$\Ez = \unit{30}{\mueV}$. \label{fig:MDDPhaseDiagram}}
\end{figure}

As discussed in \secref{sec:MDDs}, the MDD is the stable ground
state configuration only if the confinement potential\footnote{%
N.B. For an antidot this refers to the `confinement' of holes, since
the antidot potential is repulsive for electrons.} %
is strong enough to overcome the repulsive action of the Coulomb
force.  For a given set of parameters, it is therefore important to
check the stability of the MDD, using
\eqnvref{eq:MDDstablecondition}.  A phase diagram showing the
locations of ground-state spin-flips and the instability point for a
parabolic potential is given in \figref{fig:MDDPhaseDiagram}.
In order to compare the model with our experimental data, we are
interested in the region around the first few transitions in the
ground-state spin $S_{z0}$ away from the spin-unpolarised state.
These spin-flips correspond to the `kinks' in $\musf$ which appear
in \figref{fig:EexVetaC}.  In particular, in order to model the
spin-pumping behaviour observed in the non-linear transport
measurements presented in \secref{sec:NLIV}, we require the regime
in which $S_{z0}=\frac{1}{2}$ for odd-$N$ configurations, but
$S_{z0} = 1$ for even-$N$ configurations, which is shaded in grey in
\figref{fig:MDDPhaseDiagram}.  These calculations show that the MDD
is the stable ground state throughout the low-$S_z$ parameter range
in which we are interested, and remains so up to a significant value
of the ground-state polarisation.

The choice of potential has only a very small effect on the dynamics
of the edge, especially for energy scales at which we can
approximate the radial gradient as locally linear, but it can affect
the stability of the MDD phase.  For a parabolic potential, the MDD
first becomes unstable at the \emph{centre} of the dot, as shown in
\figvref{fig:MDDstability}, since the central holes have the largest
contribution from the Coulomb interaction.  If we consider instead a
more realistic `bell-shaped' potential, with the gradient at $\Rad$
set appropriately to match $\dEsp$ at the edge, the added depth of
the confining potential at the centre counteracts this effect and
causes the MDD phase to be stable for an even greater range of
parameter space.  If the bell shape becomes so narrow that the
gradient at the edge is rapidly flattening with increasing $R$, the
opposite is true: the MDD first becomes unstable at the \emph{edge},
since the confinement for the outer spin drops rapidly with
increasing $S_z$.  We expect that the most realistic situation for
typical device parameters is a bell-shaped potential with
characteristic `width' similar to $\Rad$.  Such a potential is
nearly the ideal with regards to exploring MDD physics, since this
is the shape for which the MDD phase is at its most stable.

\subsection{Excitation spectra\label{sec:MDDexspectra}}

For circularly-symmetric potentials, the $z$-projection of the total
orbital angular momentum (AM), given by $M = \sum_i L_{zi}$, where
$L_{zi}$ is the orbital quantum number of the $i^\mathrm{th}$
particle, is a good quantum number of the multi-particle antidot
state. Given a fixed number of holes $\Nh$ and the spin-projection
$S_z$, the minimum value which $M$ may take due to Fermi exclusion
is
\begin{equation}
\begin{split}
  \min(M) & = \tfrac{1}{4}\Nh(\Nh-2)+S_z^2 \\
  & = \tfrac{1}{2}\Nhup(\Nhup-1)+\tfrac{1}{2}\Nhdn(\Nhdn-1).
  \end{split}
\end{equation}
The MDD is the single Slater orbital with this minimum value of $M$,
and is therefore an eigenstate of the interacting Hamiltonian.  If
the MDD is stable, then it will be the sole ground state
configuration, $\ket{\mathrm{MDD}}=\ket{\Nh,S_z}_0$.  We can
therefore sort the \emph{excitations} of this MDD ground state by
the number of additional AM quanta, $\Delta M = M - \min(M) =
1,2,\dotsc$, required to form each state.  We write these states in
the form $\ket{N,S_z,\Delta M,p}$, where $p=1,2,\dotsc,d$ labels the
states within each subspace of definite AM, and the dimension $d$ of
each subspace is the number of ways of distributing $\Delta M$
quanta of AM amongst the particles in the MDD, accounting for Fermi
exclusion.

\begin{figure}[tp]
    \centering
    \includegraphics[]{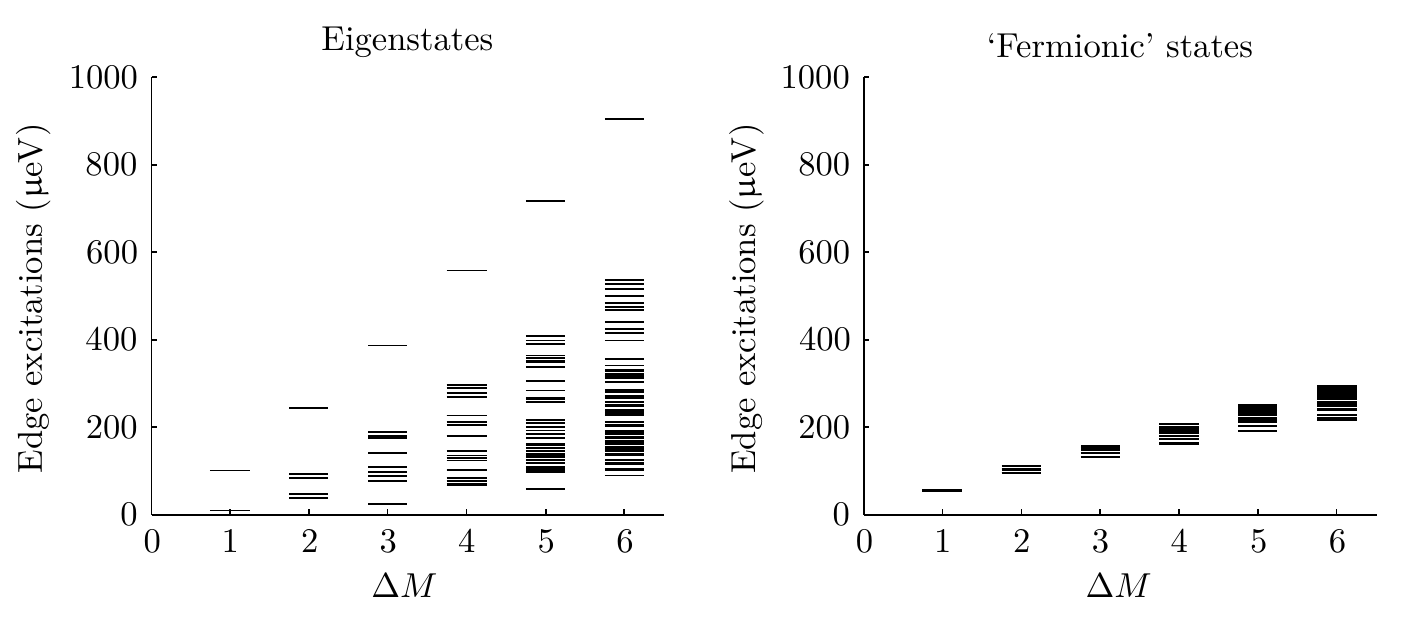}
    \caption[Edge excitation energies]{%
Comparison of edge excitation energies obtained by diagonalising the
interacting Hamiltonian within subspaces of definite AM (left), and
the corresponding excitations within the `fermionic' picture
obtained by promoting individual particles to higher orbitals
(right).  The subspaces are labeled by $\Delta M$, the number of AM
quanta added to the MDD state.  Within each AM subspace, the
`fermionic' excitation energies are simply the diagonal terms of the
Hamiltonian matrix. Parameters used in this calculation: $B_\perp
=\unit{0.65}{\tesla}$, $\dEsp = \unit{100}{\mueV}$, $\Ez =
\unit{40}{\mueV}$, $\Rad = \unit{400}{\nano\metre}$, $\etaC = 0.17$.
\label{fig:ExEnergies}}
\end{figure}

We can obtain excited-state Slater orbitals with definite $\Delta M$
by promoting individual particles of the MDD to obtain such states,
but these `fermionic' configurations, defined by occupation numbers
$(\nhupvec,\nhdnvec)$, are not eigenstates of the Hamiltonian in
general. This is because the individual AM quantum numbers $L_{zi}$
of the SP states are not conserved by the interaction term in the
Hamiltonian, which produces nonzero matrix elements between pairs of
Slater orbitals which differ by exactly two occupation numbers
\cite{Bethe1986}.  It is possible, however, to determine the
eigenspectrum by constructing the full set fermionic basis states
for a given value of $\Delta M$ and then diagonalising the resulting
matrix Hamiltonian. The excitations plotted in \figref{fig:EexVetaC}
are examples of this procedure for the $\Delta M=1$ subspace, whose
eigenspace consists of a `spin' and a `density' excitation. In the
fermionic basis, the excitations consist of an individually-excited
\spinup\ or \spindn\ particle, which have roughly similar energies.
After diagonalising the $2\times2$ matrix constructed from these
states, we find that the eigenstates are a spatially-symmetric
`triplet' excitation with a small energy gap, corresponding to a
change in $S^2$, and a spatially-antisymmetric `singlet' excitation
corresponding to a modulation of the electron density along the
edge, with a much larger energy gap.  Excitation energies for larger
values of $\Delta M$ are shown in \figref{fig:ExEnergies}, with the
corresponding sets of fermionic excitations provided for comparison.
For each $\Delta M$ subspace, the largest excitation corresponds to
a singlet transition which is essentially a pure density modulation
similar to the `bosonic' modes of a Tomonaga-Luttinger liquid for
spinless electrons \cite{Stone1992}.  The calculations in
\figref{fig:ExEnergies} correspond to the `crossover' region of
\figref{fig:EexVetaC} at $\etaC=0.17$, and we find in this regime
that the singlet mode is well-separated from the remaining
transition energies for all $\Delta M$.  The lower-energy modes
consist of combinations of spin and density excitations.

\section{Model of non-linear transport\label{sec:TransportModel}}

To directly compare the predictions of the MDD model with our
experimental results, we need a flexible method to compute the
antidot conductance, preferably in both the linear and non-linear
regimes.  The easiest method by far is to construct and solve a
rate-equation matrix for the steady-state occupation probabilities
of the antidot, within the sequential-transport framework presented
in \secref{sec:SequentialTransport}.  With this method, the physics
of the antidot enters the calculation in the form of
\begin{enumerate}
  \item{A set of eigenstates and associated energy spectrum, and}
  \item{A set of matrix elements for transitions between states.  In
  the simplest case these are simply a set of boolean `selection rules.'}
\end{enumerate}
The remaining parts of the calculation, incorporating the tunnel
barriers (possibly spin- and/or energy-dependent) and bias (possibly
mode-dependent through the non-equilibrium population of edge
modes), form a `shell' within which we can explore essentially
arbitrary models of the antidot physics.  Limitations of the model
stem mainly from the perturbative approximation to the tunneling
rates, which is strictly valid only in the `weak coupling' regime
where all the tunnel couplings $\Gamma_{r\sigma}$ satisfy
$h\Gamma_{r\sigma}\ll kT$.  The amplitude of the transmission
resonances in our experiments imply typical couplings of order
$\Gamma\approx\unit{500}{\mega\hertz}$\nbd\unit{1}{\giga\hertz},
with corresponding lifetime broadening
$h\Gamma\approx2$\nbd\unit{4}{\mueV}.  With an electron temperature
around $T_\mathrm{elec}\approx\unit{50}{\milli\kelvin}$, we have
$kT\approx\unit{5}{\mueV}$ and so we are close to the edge of the
model's range of validity.  We still find very close agreement
between the features observed in the experiment and the predictions
of our model, but it is important to bear in mind that the
sequential transport model does not include lifetime broadening due
to quantum fluctuations, or higher order `cotunneling' processes
which may be present in the experiment.  Further documentation on
the model discussed in this section, along with the programs
themselves, implemented in the \textsc{MATLAB}\textregistered\
programming language, are available in the open repository
\emph{DSpace@Cambridge}.\footnote{%
\emph{DSpace@Cambridge} is the institutional repository of the
University
of Cambridge, available at \url{http://www.dspace.cam.ac.uk/}.} %

For a given model of the antidot spectrum, only a subset of
configurations participate in transport due to energy
considerations, but in general it can be a difficult problem to
determine which states to keep, especially for systems of many
particles which have a very large number of possible configurations.
Given a set of values for the external fields (gate voltages,
magnetic field, and drain-source bias), we can determine the
ground-state configuration, but if the energy spacing between states
is small, or if $\Vds$ is large, the steady-state solution will
contain significant populations of many exited states as well. Our
solution to this problem is to start with a relatively small subset
of states, chosen to be the ground-state configuration plus all of
the excited states which are `accessible' through a single tunneling
event, i.e., the states with energy $\varepsilon_i$ such that the
chemical potential
\begin{equation}
  \mu_i = \varepsilon_i - \varepsilon_\mathrm{GS}
\end{equation}
is within the `energy window' defined by the chemical potentials of
the leads:
\begin{equation}
\bigl[\min(\muS,\muD)-\Etherm,\,\max(\muS,\muD)+\Etherm\bigr],
\end{equation}
where $\Etherm\approx 4kT$.  If we find that many of these excited
states have significant occupation probabilities, we can add to this
set all of the states which are `connected' to the significant
excited states through the same rule, in terms of the chemical
potentials for transitions from each excited state.  By continuing
to expand the set of states in this way, we will eventually reach a
situation in which all of the newly-added states have sufficiently
low occupation probability for convergence of the transport current
to a desired tolerance.  For both conceptual and computational
purposes, it is useful to think of the system in terms of an
abstract graph, in which the set of configurations correspond to
nodes and the selection rules define the connections, or `edges.'
The procedure described above always results in a graph which is
\emph{fully-connected}, meaning that it is possible to move between
any two nodes of the graph given enough transitions.  It is
important to maintain this property throughout any manipulations of
the model, to avoid a situation in which the population can become
artificially `trapped' in a disconnected portion of the graph.

To make comparisons with our spin-selective measurements, we need a
method of organising the antidot configurations which allows us to
keep track of the \emph{spin} of each electron which tunnels into or
out of the antidot.  Below we outline the procedure we use to
accomplish this, using the fermionic configurations defined by
occupation vectors $(\nupvec,\ndnvec)$ as an example.\footnote{%
For the purposes of this discussion we consider electron occupation
numbers since this is the standard picture for quantum dot
transport, but these are simply related to the hole occupation
numbers by $\mathbf{n}^\mathrm{h}_\sigma =
\mathbf{1}-\mathbf{n}^\mathrm{e}_\sigma$.} %
To begin, we consider only transitions between ground-state
configurations with different occupation numbers $N$ at zero bias,
with chemical potentials $\mu_0(N)$. Given a set of capacitances as
described in \secref{sec:SequentialTransport}, the condition
$\mu_0(N)=0$ defines the value of the gate voltage $\Vg$ at which
charge degeneracy occurs for the $N\leftrightarrow\Npone$ transition
at zero bias.  In between these resonance positions, the condition
$\mu(\Nmone) = -\mu(N)$ defines an `inflection point' within each
Coulomb blockade region. On one side of the inflection point we need
only consider configurations with occupation numbers $(\Nmone,N)$,
while on the other we consider only $(N,\Npone)$ states.  In the
plane of $(\Vg,\Vds)$ these become inflection lines which pass
vertically through the centre of each Coulomb diamond, and divide
the calculation region by the occupation numbers involved.  This
means that we cannot fully model the transport at high bias above
the crossings of adjacent ground state lines, since we would then
need to consider more than two sets of occupation states of the
antidot.

Next, we divide the configurations within each region (defined by
occupation numbers $N$, $\Npone$) by their total spin projection
$S_z$. Suppose the ground-state spin for the $N$-particle state is
$\Szz$ and for the $\Npone$ particle state is $\Szz-\frac{1}{2}$.
Given these values, we begin by constructing the vector of
configurations:
\begin{equation}
    \lbrace{\ket{\Psi_\mathrm{AD}}\rbrace} =
  \begin{pmatrix}
    \lbrace\ket{\Npone,\Szz\!-\!\frac{3}{2}}\rbrace \\
    \lbrace\ket{N,\Szz\!-\!1}\rbrace \\
    \lbrace\ket{\Npone,\Szz\!-\!\frac{1}{2}}\rbrace \\
    \lbrace\ket{N,\Szz}\rbrace \\
    \lbrace\ket{\Npone,\Szz\!+\!\frac{1}{2}}\rbrace \\
    \lbrace\ket{N,\Szz\!-\!1}\rbrace
  \end{pmatrix},
\end{equation}
where each $\lbrace\ket{N,S_z}\rbrace$ corresponds to a vector of
individual states $\ket{N,S_z,q_\uparrow,q_\downarrow}$, where
$q_\sigma$ labels the configuration of the spin-$\sigma$ particles.
In the presence of interactions these states are not true
eigenstates of the Hamiltonian, but they provide a qualitative
approximation to the excitation gaps in most cases.  In
\secref{sec:EffModel} later in this chapter we consider an effective
model to better capture the physics of the edge excitations.  For
any model, the number of excited states to include is determined
through a consideration of the chemical potentials for transitions
to or from the ground states with spin $S_z\pm\frac{1}{2}$ as
described above. With this arrangement for the configurations, the
selection rules take the block-matrix form
\begin{equation}\label{eq:Selrules}
  \begin{pmatrix}
    0 & W^{+\downarrow}_{\Szz-1} & & & \cdots & 0 \\
    W^{-\downarrow}_{\Szz-\frac{3}{2}} & 0 & W^{-\uparrow}_{\Szz-\frac{1}{2}} & & & \vdots \\
     & W^{+\uparrow}_{\Szz-1} & 0 & W^{+\downarrow}_{\Szz}  & & \\
     & & W^{-\downarrow}_{\Szz-\frac{1}{2}} & 0 & W^{-\uparrow}_{\Szz+\frac{1}{2}} & \\
     \vdots & & & W^{+\uparrow}_{\Szz} & 0 & W^{+\downarrow}_{\Szz+1} \\
     0 & \cdots & & & W^{-\downarrow}_{\Szz+\frac{1}{2}} & 0 \\
  \end{pmatrix},
\end{equation}
where, assuming the vectors of states $\lbrace\ket{N,S_z}\rbrace$
are listed as subsequent groups of \spinup\ states (labeled by
$q_\uparrow$) for each \spindn\ state (labeled by $q_\downarrow$),
the sub-matrices $W^{\pm\sigma}_{S_z}$ are given by
\begin{subequations}
\begin{align}
  W^{\pm\uparrow}_{S_z} & = \mathbf{1}_\downarrow \otimes M^{\pm\uparrow}_{S_z}, \\
  W^{\pm\downarrow}_{S_z} & = M^{\pm\downarrow}_{S_z} \otimes
  \mathbf{1}_\uparrow.
\end{align}
\end{subequations}
The matrices $M^{\pm\sigma}_{S_z}$ contain the selection rules for
transitions in the spin-$\sigma$ configuration individually, and are
easily worked out by comparing the occupation vectors
$\mathbf{n}_\sigma$ of the initial and final states.  For example,
$M^{+\uparrow}_{ij}=1$ whenever the $q_\uparrow=i$ state of the
$\Npone$ configurations results from adding a single \spinup\
particle to the $q_\uparrow=j$ state of the $N$ configurations,
which we can write as
\begin{equation}
  M^{+\uparrow}_{ij} = \begin{cases}
    1\quad\text{if}\quad
    \mathbf{n}_\uparrow(i)\cdot[\mathbf{1}-\mathbf{n}_\uparrow(j)]=1,\\
    0\quad\text{otherwise}.
  \end{cases}
\end{equation}
Similar relations determine the selection rules for other types of
processes.

The rate matrix has a similar form to \eqnref{eq:Selrules}, where
the nonzero selection rules are replaced by the transition rates
\begin{equation}
  R^{\pm\sigma}_{ij} =
  \sum_{r=\mathrm{S,D}}\Gamma^r_\sigma(\mu_{ij})W^{\pm\sigma}_{ij}f_r^\pm(\mu_{ij}),
\end{equation}
where $f^+_r = f_r$ is the Fermi function of lead $r$, and $f^-_r =
1-f_r$.  As described in \secref{sec:SequentialTransport}, we then
add diagonal elements to the rate matrix to impose a net balance of
rates in equilibrium, and an extra row of ones to enforce
normalisation, constructing the master equation in the form of
\eqnvref{eq:MasterEqn}. The solution to this equation gives the
steady state occupation probability of each state
$\ket{N,S_z,q_\uparrow,q_\downarrow}$, which we then use to compute
the current flowing through the system. The current is most easily
computed using \eqnref{eq:Ir1}, by isolating the transition rate
involving only a single lead, e.g.\ for the source,
\begin{equation}
  S^{\pm\sigma}_{ij}=\GammaS_{\sigma}(\mu_{ij})W^{\pm\sigma}_{ij}f_\mathrm{S}^\pm(\mu_{ij}).
\end{equation}
Including signs to account for the direction of current flow, we can
then write
\begin{equation}
  I = e\sum_{ij}T_{ij}P_j,
\end{equation}
where $P_j$ are the equilibrium occupation probabilities and
\begin{equation}
  T = \begin{pmatrix}
    0 & S^{+\downarrow}_{\Szz-1} & & & \cdots & 0 \\
    -S^{-\downarrow}_{\Szz-\frac{3}{2}} & 0 & -S^{-\uparrow}_{\Szz-\frac{1}{2}} & & & \vdots \\
     & S^{+\uparrow}_{\Szz-1} & 0 & S^{+\downarrow}_{\Szz}  & & \\
     & & -S^{-\downarrow}_{\Szz-\frac{1}{2}} & 0 & -S^{-\uparrow}_{\Szz+\frac{1}{2}} & \\
     \vdots & & & S^{+\uparrow}_{\Szz} & 0 & S^{+\downarrow}_{\Szz+1} \\
     0 & \cdots & & & -S^{-\downarrow}_{\Szz+\frac{1}{2}} & 0 \\
  \end{pmatrix}.
\end{equation}
This procedure is easily generalised to account for additional
effects.  For example, we can include spin-conserving relaxation of
excited states within each set $\lbrace\ket{N,S_z}\rbrace$ by adding
block matrices describing these processes to the main diagonal of
\eqnref{eq:Selrules}.  Spin non-conserving relaxation due to
spin-orbit coupling or the hyperfine interaction could also be
included by adding terms to the next off-diagonal blocks (connecting
states $\lbrace\ket{N,S_z}\rbrace$ with
$\lbrace\ket{N,S_z\!\pm\!1}\rbrace$). Note, however, that this model
only obtains the steady-state ($t\rightarrow\infty$) configuration,
so we are not able to investigate coherent effects due to
spin-precession with this procedure.

In calculations, we iterate this procedure until we reach
convergence, adding additional $S_z$-configurations and excited
states until the occupation probability of the `outermost' states
falls below a given threshold.  We can see at this point how it is
easily possible to produce very large matrices, since the total
number of configurations is given roughly by
\begin{multline}
  \text{dim} = (\text{\# \spinup\ states per $S_z$})
      \times(\text{\# \spindn\ states per $S_z$})\\
      \times(\text{\# $S_z$ configurations}).
\end{multline}
Luckily, the rate matrix is also very sparse, which makes solving
the rate equations numerically tractable in most cases.  In some
important cases, however, the problem becomes so large that solving
the master equation is computationally prohibitive.  Unfortunately,
this is often the case around the `crossover' region discussed in
\secref{sec:MDDmodel}, since we need to consider a large number of
$S_z$ configurations when the $S_z$-excitation energy is small, and
this is exactly the regime we wish to investigate.  The
computation-limiting step is almost always the Gaussian elimination
procedure used to solve the master equation, which scales badly with
the dimension of the rate matrix. Even though we take advantage of
parallel-computing resources provided by
\emph{CamGrid},\footnote{%
\emph{CamGrid} is a distributed computing resource coordinated by
the Cambridge eScience Centre.  More information is available at
\url{http://www.escience.cam.ac.uk/projects/camgrid/}.} %
we run into insurmountable memory limitations with matrices above a
given size.

To solve this problem, we have developed a number of routines which
attempt to shrink the system in an intelligent way, keeping only the
states which contribute to transport for a given set of external
parameters, since this is often a small subset of the total number
of possible states determined from energy considerations.  These
routines are inspired by the picture of the rate matrix as an
abstract graph, and rely on several computationally-efficient
algorithms from the field of graph theory (see the book by Bollabas
\cite{Bollabas1985} for a good introduction). After solving the
master equation for an initial subset of states as described above,
we determine the `important' states by thresholding the probability
vector.  We then use several graph algorithms, particularly
Dijkstra's shortest-path algorithm \cite{Dijkstra1959}, to determine
the minimal connected graph which incorporates the important nodes,
and then `expand' these nodes into additional excited states as
needed to achieve convergence.  The level of approximation involved
in this procedure may be carefully controlled, for instance by
changing the threshold used to determine which states are retained,
or the number of new states which are added in each iteration. Our
tests have shown, for several important cases where the full
matrices become impractically large, that we can use these methods
to speed up the calculation by more than an order of magnitude with
a loss of accuracy in the computed current much less than one
percent. These routines were used in several of the calculations
presented in this chapter, and they have been of critical importance
in enabling a practical investigation of many of the features we
discuss.

\section{Comparisons with experimental results}

At this point we are ready to use the transport calculations
described in the previous section together with the MDD model of
\secref{sec:MDDmodel} to compare the predictions of our theory with
the experimental results of \chapref{chap:SpinTransport}.  Below we
show how the MDD model successfully reproduces both the observed
spin-resolved conductance in linear response and the energy scales
observed in the non-linear transport experiments, while
non-interacting models fail to do so. We also investigate the
`spin-pumping' mechanism in more detail and consider an effective
theory to describe the edge-excitations of the MDD.

\subsection{Non-linear transport}

We begin with a consideration of the non-linear transport
experiments discussed in \secref{sec:NLIV}.  The important features
of these experiments, including the associated spin-resolved
measurements at zero bias, are summarised by
\figref{fig:NLIVandZBG}, which we use in this section as a reference
for comparison of theoretical results. Recall that the observations
are characterised by three important energy scales inherent to the
antidot physics:
\begin{itemize}
  \item{%
The spin-flip excitation energy $\musf^\pm = U(S_z\!\pm\!1)-U(S_z)$,
which must always satisfy $\musf^\pm\lesssim
\Etherm\approx\unit{10}{\mueV}$ such that transport is not
spin-selective.}
  \item{%
The orbital excitation energy $\Eex$ inferred from the extra lines
in non-linear transport measurements.  The separations of these
lines imply that $\Eex\approx50$\nbd\unit{60}{\mueV}.}
  \item{%
The spin-pumping energy scale $\dmuup =
\muup(N,S_z\!+\!1)-\muup(N,S_z)$, which determines the size of the
`break' in ground-state lines due to dynamic pumping of $S_z$ caused
by the combination of DC bias and asymmetric spin-dependent tunnel
barriers.  This is observed to be midway between $\musf$ and $\Eex$,
with characteristic spacings around
$\dmuup\approx20$\nbd\unit{30}{\mueV}.}
\end{itemize}
Within the SP model, these are given by (see
\eqnsref{eq:dEspSPmodel}, \eqref{eq:musfSPmodel}, and the discussion
on page \pageref{eq:deltamuup} for more details)
\begin{subequations}
  \begin{align}
      \musf^\pm & = (1\pm 2S_z)\dEsp - \Ez,\\
      \Eex & = \pm s\Ez + j\dEsp,\\
      \dmuup & = \dEsp,
  \end{align}
\end{subequations}
in terms of the SP energy scales $\dEsp$ and $\Ez$.  It is
relatively easy to see that these relations are inconsistent with
the values observed in the experiment, but we provide a few specific
examples of calculations for further illumination.

\begin{figure}[tb]
    \centering
    \includegraphics[]{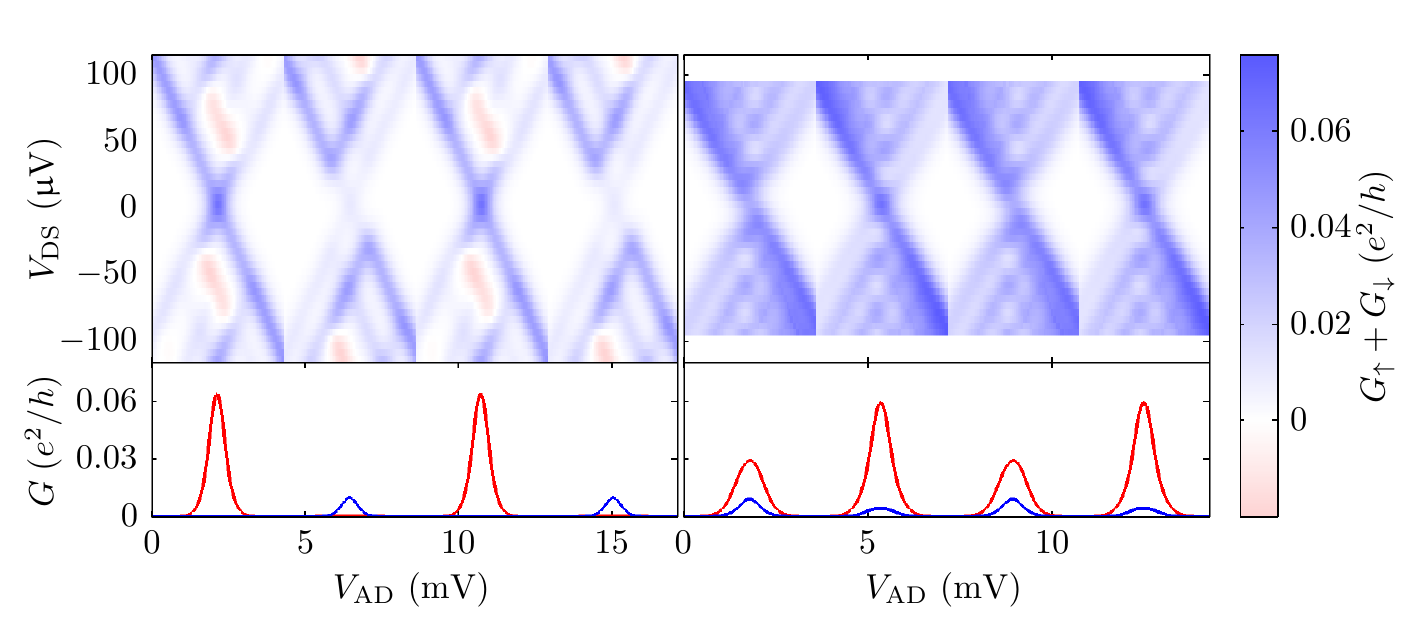}
    \caption[Non-linear transport calculations: SP model]{%
Transport calculations for the single-particle model, for comparison
with the data in \figvref{fig:NLIVandZBG}.  Total conductance
$\Gup+\Gdn$ is plotted in the upper panels, with the zero-bias
spin-resolved conductance shown in the lower panels, where red
(blue) curves represent $\Gup$ ($\Gdn$).  Left --- If we choose
$\dEsp=\unit{60}{\mueV}$ to match the excitations in
\figref{fig:NLIVandZBG}, the zero-bias conductance is spin-polarised
in contradiction to the experiment.  Right --- If we set
$\dEsp=\unit{20}{\mueV}$ instead to match $\musf$, we reproduce the
splitting of the `broken' ground-state lines, but the excitations
observed in the transport window are clearly much more
closely-spaced than in the experimental data, and we observe no NDC.
Parameters used in this calculation: $\Ez=\unit{30}{\mueV}$,
$\Rad=\unit{400}{\nano\metre}$, $T=\unit{55}{\milli\kelvin}$,
$\gSup=\gDup=\gSdn=\unit{600}{\mega\hertz}$,
$\gDdn=\unit{50}{\mega\hertz}$. \label{fig:NLIVcalcsSP}}
\end{figure}

In \figref{fig:NLIVcalcsSP} we show transport calculations within
the non-interacting model for two different choices of $\dEsp$, with
$\Ez=\unit{30}{\mueV}$ as expected at $B=\unit{1.2}{\tesla}$.
The other parameters in the simulation, such as the capacitances
between the antidot states and external voltages and the strength
and asymmetry of tunnel barriers, are chosen based on the slopes and
amplitudes of the lines observed in \figref{fig:NLIVandZBG} as
described in \secref{sec:SequentialTransport}.  In the first
calculation (left panels) we choose $\dEsp=\unit{60}{\mueV}$ in
order to match the observed excitation spectrum.  This is mostly
successful (the negative differential conductance (NDC) is more
pronounced then in the experiment but this is affected by our
particular choice of tunnel barriers), but the simulation clearly
fails to reproduce the spin-resolved conductance at zero bias.  As
expected when $\musf\approx\unit{30}{\mueV}>\Etherm$, the model
predicts spin-selective resonances with alternate polarisation for
successive peaks, contrary to our observations.  If we instead
choose $\dEsp=\unit{30}{\mueV}$ (right panels) to match the observed
value of $\dmuup$, then we observe `breaks' in the ground-state line
tracking the drain chemical potential, as predicted by the
spin-pumping model (see \figvref{fig:SpinPumping}). The spin-flip
energy $\musf$ is also much reduced, and so the spin-resolved
conductance is consistent with experiments. The excitation spectrum,
however, is clearly inconsistent with our measurements, with lines
spaced much too closely together and no clear NDC.

We therefore proceed to add interactions to the model, including the
Hartree-Fock contribution to the configuration energy of each state,
which we compute for the Slater-determinant wave function
corresponding to the \emph{empty} (hole) orbitals of the fermionic
state $\ket{N,S_z,q_\uparrow,q_\downarrow}$.  To fix the strength of
the Coulomb interactions, controlled by $\etaC$, we use excitation
energy diagrams like \figref{fig:EexVetaC} and phase diagrams like
\figref{fig:MDDPhaseDiagram}.  Given a value for the orbital
excitation scale, which is set by the constant $\dEsp$ for a
parabolic potential,\footnote{%
N.B. We are only probing the physics of the \emph{edge}, so assuming
the potential varies slowly on the scale of $\ellB$, the details of
the potential are not important.  The potential shape is mainly
important for the stability of the MDD phase, as discussed in
\secref{sec:MDDstability}.} %
we choose $\etaC$ appropriately to reach a point in the crossover
region where $S_z$-flips just start to occur.  Some fine tuning is
often required to reproduce specific features like the spin pumping
and NDC observed in the experiment, but this is to be expected since
these effects rely on a delicate balance between different
excitation scales of the antidot and strengths of the various
tunnel-couplings, and we have tuned the experimental system quite
substantially in order to observe them in the first place.

\begin{figure}[tb]
    \centering
    \includegraphics[]{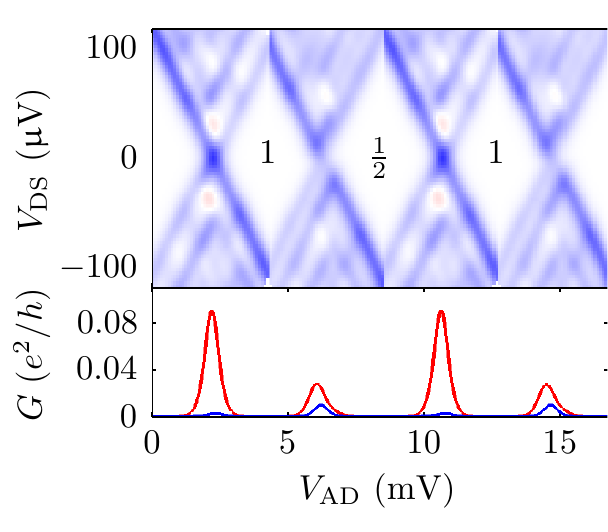}
    \includegraphics[]{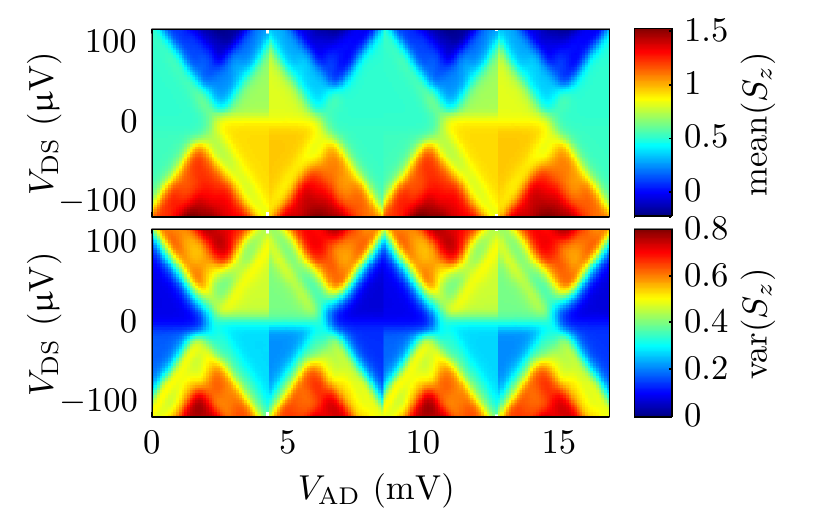}
    \caption[Non-linear transport calculations: MDD model]{%
Transport calculations for the antidot MDD model.  Top left ---
Total non-linear conductance $\Gup+\Gdn$, with colour scale as in
\figref{fig:NLIVandZBG}.  Numbers printed in each Coulomb diamond
show the ground-state spin configuration.  Bottom left
--- Zero-bias spin-resolved conductances $\Gup$ (red curve) and
$\Gdn$ (blue curve).  Right panels --- Mean spin configuration (top)
showing bias-induced pumping, and the standard deviation (bottom),
corresponding to the non-linear transport calculations in the top
left.  Parameters used in this calculation:
$\dEsp=\unit{70}{\mueV}$, $\Ez=\unit{40}{\mueV}$,
$\Rad=\unit{400}{\nano\metre}$, $T=\unit{55}{\milli\kelvin}$,
$\gSup=\gDup=\unit{900}{\mega\hertz}$,
$\gSdn=\unit{300}{\mega\hertz}$, $\gDdn=\unit{60}{\mega\hertz}$,
$\etaC=0.08$. \label{fig:NLIVcalcsMDD}}
\end{figure}

An example of the calculated transport including interactions is
shown in \figref{fig:NLIVcalcsMDD}.
Most of the parameters are the same as in \figref{fig:NLIVcalcsSP},
although we have chosen the tunnel barriers more carefully to
reproduce the peak amplitudes in \figref{fig:NLIVandZBG}.
Qualitatively, this model does an excellent job of reproducing our
observations.  The excitations have the correct scale, and the
exchange interactions reduce $\musf$ sufficiently to break the
spin-selectivity at zero bias.  Moreover, the interactions suppress
$\dmuup$ as well, such that the breaks in ground-state lines have
the right magnitude.  Upon close inspection, it is also apparent
that the \spinup\ conductance peaks corresponding to these
`frustrated' resonances are widened and asymmetric in qualitative
agreement with the experiment.

\subsection{Spin pumping}

Within our transport model, we can gain further insight into the
spin-pumping mechanism which causes the breaks in ground-state lines
near zero bias.  In particular, we can compute both the average spin
configuration and the standard deviation from the occupation
probabilities:
\begin{align}
  \mean(S_z) & = \sum_i P_i S_z(i), \\
  \stdev(S_z) & = \biggl[\sum_i P_i \Bigl(S_z(i)\Bigr)^2
      - \Bigl(\mean(S_z)\Bigr)^2\biggr]^{1/2},
\end{align}
as shown in the right-hand panels of \figref{fig:NLIVcalcsMDD}.
Plots of the average spin show the pumping directly; in this case a
positive (negative) bias drives the system to a lower (higher) spin
than in equilibrium due to the asymmetric tunnel barriers.  The
standard deviation gives a measure of the level of `frustration' at
a given configuration, and increases in regions where the pumping
mechanism competes with energy considerations to determine the
steady-state occupations.  Note that the vertical discontinuities
visible within the Coulomb blockade diamonds correspond to the
points at which we switch between a consideration of
$\Nmone\leftrightarrow N$ to $N\leftrightarrow\Npone$ transitions.
The available pumping mechanisms change across these boundaries,
affecting the standard deviation of $S_z$ in particular.  This does
not affect the reliability of the transport results, however, mainly
reflecting the fact that the steady-state occupations are not
particularly well-defined in the Coulomb-blockaded regions, since no
current is flowing to provide equilibration.

At this point we briefly consider another experiment which
demonstrates the spin-pumping mechanism.  In standard non-linear
transport the pumping is provided by the combination of the bias and
spin-dependent tunnel barriers which are more strongly asymmetric
for one spin than the other.  A similar effect occurs if we supply a
spin-dependent \emph{bias}, using the selective-injection technique.
The quantum Hall edge modes cannot withstand a large non-equilibrium
bias without suffering significant equilibration, but we have
determined through injection/detection measurements with the quantum
point contacts on our device that the $\nu=3$ and $\nu=4$ modes can
maintain differences in chemical potentials of up to
$\approx30$\nbd\unit{40}{\mueV}. Since this is greater than the
scale we expect for the spin-flip potentials $\musf^\pm$,
spin-pumping experiments using this technique are feasible.

To apply a non-equilibrium bias without driving large currents
through the constrictions on either side of the antidot, we apply a
DC potential to the \emph{bulk modes} via contact $\alpha$ on the
source side of the device, including the \spindn\ mode $\nu=4$. This
is accomplished by setting the injector constriction to $\finj=3$,
and connecting the injector ohmic contact $\delta$ to DC ground,
such that the modes $\nu=1$\nbd3 reach the antidot with
$\mu_{\mathrm{S}\uparrow}=0$, while modes $\nu\geq4$ have the
chemical potential $\mu_{\mathrm{S}\downarrow}-e\Vs$.  To measure
the effect of this non-equilibrium potential on the transport, we
apply a small AC excitation to modes $\nu=1$\nbd3 through the
injector.\footnote{%
It is also possible to apply the excitation to modes $\nu\geq4$ with
the DC bias, but this gives a much weaker signal since the tunnel
barriers are much higher for \spindn\ transport than for \spinup.} %
Since the $\nu=4$ mode does not feel the excitation, the antidot
transmission resonances we measure are due to \spinup\ transport
only.  With this experiment we are therefore probing the effect on
$\Gup$ of a change in the chemical potential
$\mu_{\mathrm{S}\downarrow}$, while the remaining
$\mu_{\mathrm{S}\uparrow}
=\mu_{\mathrm{D}\uparrow}=\mu_{\mathrm{D}\downarrow}=0$.

\begin{figure}[p]
    \centering
    \includegraphics[]{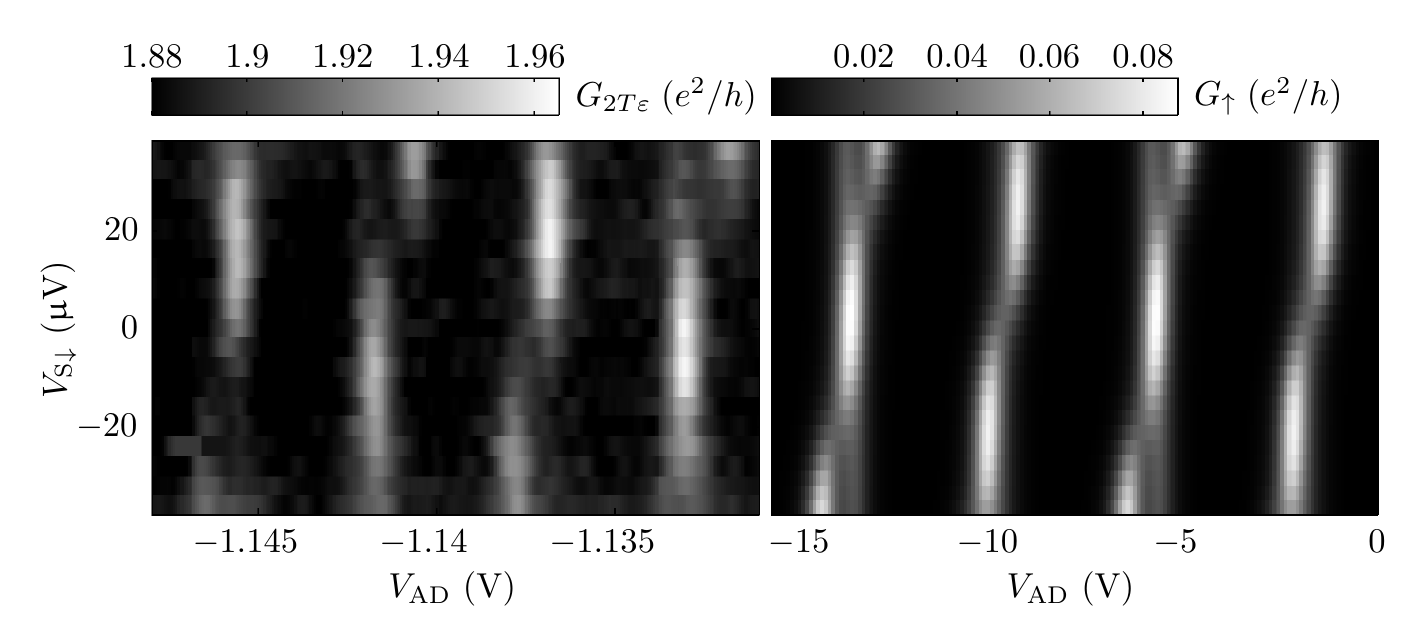}\\
    \includegraphics[]{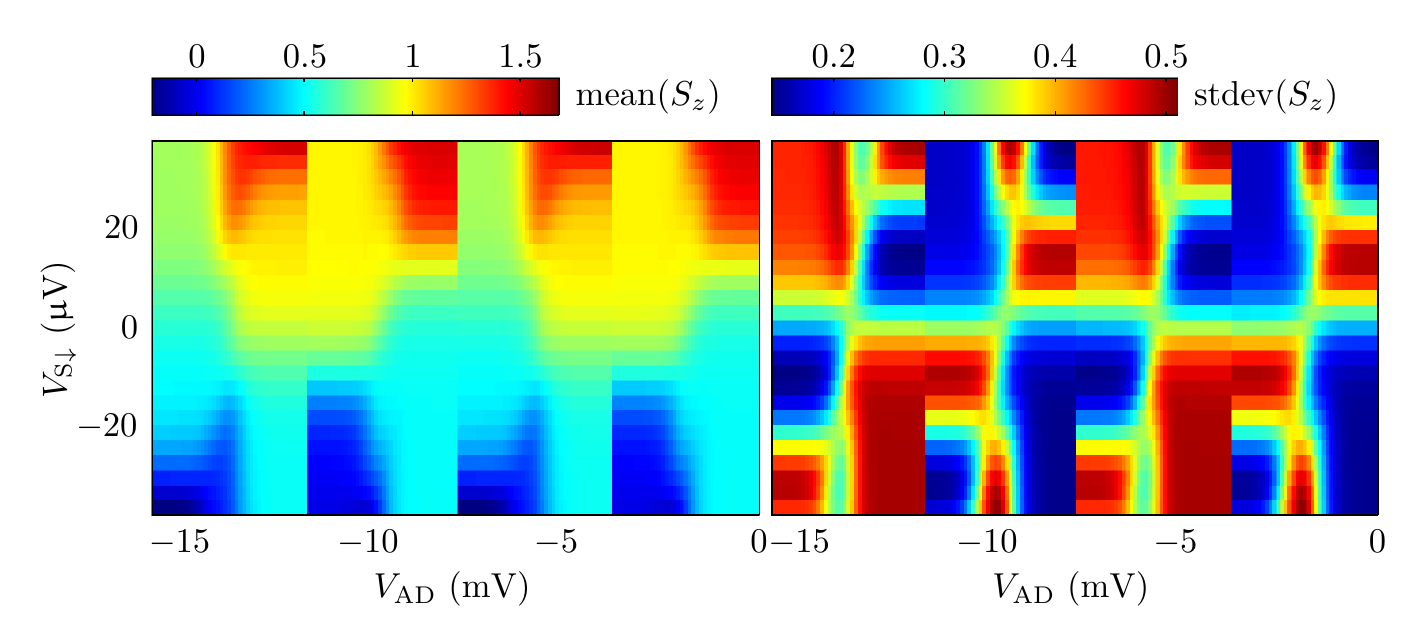}
    \caption[Spin pumping with non-equilibrium bias]{%
Top left --- Experimental data showing two-terminal differential
conductance at the detector (contact $\varepsilon$) with
$\fdet=\finj=3$, when DC bias is applied only to the modes $\nu\geq
4$ through contact $\alpha$, and AC bias is applied to modes
$\nu\leq 3$ through contact $\delta$ in the injector.  Top right ---
Spin-$\uparrow$ conductance computed in our model of this
measurement, with DC bias applied to \spindn\ only, and AC bias
applied to \spinup\ at the source.  Bottom panels --- Steady state
values of the average (left) and standard deviation (right) of $S_z$
within the calculation, showing the effects of spin pumping. The
discontinuities in the Coulomb blockade regions result from the
different pumping mechanisms available given the occupation numbers
$(N,\Npone)$ we consider in each region.  These will be smoothed out
in the real system. Parameters used in this calculation:
$\dEsp=\unit{50}{\mueV}$, $\Ez=\unit{30}{\mueV}$,
$\Rad=\unit{400}{\nano\metre}$, $T=\unit{55}{\milli\kelvin}$,
$\gSup=\gDup=\unit{900}{\mega\hertz}$,
$\gSdn=\unit{300}{\mega\hertz}$, $\gDdn=\unit{60}{\mega\hertz}$,
$\etaC=0.05$. \label{fig:NeqBias}}
\end{figure}

The results of this experiment are depicted in the top-left panel of
\figref{fig:NeqBias}.
The bias induces clear transitions at specific values of $\Vs$ for
each resonance, providing direct measurements of the antidot
spin-flip potentials $\musf^\pm$.  Note that, aside from the
stepwise dislocations, the resonance positions are independent of
$\Vs$.  This implies that the capacitive coupling between the
antidot states and the external voltage $\Vs$ is very small.  This
is to be expected since the $\nu\geq 4$ states are separated from
the antidot by the $\nu=3$ mode, and provides further evidence that
the difference in chemical potentials $\mu_{\mathrm{S}\uparrow}$ and
$\mu_{\mathrm{S}\downarrow}$ is maintained at the position of the
antidot.

The remaining panels in \figref{fig:NeqBias} describe a simulation
of the experiment using our transport model.  Clear discontinuities
are observed in the resonances with spacings corresponding to
$\dmuup(N,S_z)$, as the steady-state occupation of the dot changes
due to the non-equilibrium bias.  This behaviour is reproduced by
the transport model, calculated for the same set of system
parameters as in \figref{fig:NLIVcalcsMDD}, but with the bias
applied only to the \spindn\ electrons in the source lead.  In this
case, a positive (negative) bias corresponds to a reduced
(increased) chemical potential for \spindn\ in the source, and so a
net flow of \spindn\ electrons out of (into) the dot, thereby
increasing (reducing) the steady-state $S_z$.  The
maximally-occupied spin is likely to flip once the bias reaches the
appropriate spin-flip potential $\musf^\pm$, but the resonance will
only shift when \emph{both} the $N$ and $\Npone$ configurations
undergo a spin-flip.  It is thus a complicated procedure to extract
particular energy scales from experiments like this, but the
potential to do so clearly exists, and for this purpose the
simulations prove an invaluable tool for comparisons.

\subsection{An effective model for excitations\label{sec:EffModel}}

As discussed in \secref{sec:MDDexspectra}, the excited states which
appear in our transport calculation are not eigenstates of the
Hamiltonian when interactions are included.  They do seem to have
approximately the right energies, judging by the good qualitative
agreement between the calculations in \figref{fig:NLIVcalcsMDD} and
the experimental data in \figref{fig:NLIVandZBG}, but it would be
better to have a justification for using them, or preferably to
isolate the important characteristics which make the predictions of
the model consistent with our experiments.  We can compute the
actual eigenspectrum by diagonalising the Hamiltonian for each set
of fermionic orbitals at a given level of excitation, $\Delta M$,
but the selection rules for tunneling then become much more
complicated. Furthermore, effects beyond Hartree-Fock mean-field
theory are likely to be important for a full description of the
excited states.  Electron \emph{correlation}, in particular, which
is ignored in the Slater-orbital picture we use here, has been shown
to modify the excitation of quantum-dot MDDs significantly
\cite{Korkusi'nski2004}.  But rather than introducing further
complications to a model which already seems to work, we would
prefer instead to simplify it, in order to distill the essential
ingredients which are necessary to reproduce our results. In this
section we therefore consider an `effective' model for the
excitation spectrum of the edge and a set of simple selection rules
based on Luttinger Liquid theory, which are consistent with our
experiments and succinctly describe the important physics of the
system.

We have already seen that the $\Delta M=1$ subspace consists of a
pair of excitations which we identify respectively with a `spin' and
`density' mode of the MDD edge.  If we consider these to be the
fundamental modes of excitation for the edge, then we can write a
general antidot state in the form $\ket{N,S_z,\nL,\nS}$, where $\nL$
and $\nS$ are the number of excitations in the orbital (density) and
spin modes, respectively.  The energy of such a state is then simply
\begin{equation}
  \hat{H}\ket{N,S_z,\nL,\nS} = \bigl(U_\mathrm{MDD}(N,S_z)+\nL\El +
  \nS\Es\bigr)\ket{N,S_z,\nL,\nS},
\end{equation}
where $\El$ and $\Es$ are the orbital- and spin-excitation energies
associated with these edge modes, and $U_\mathrm{MDD}$ is the
configuration of the unexcited MDD.  Due to the spin-charge
separation introduced by exchange, we expect that $\Es$ is
significantly lower than $\El$.  With a glance back to the
eigenspectra in \figref{fig:ExEnergies}, we observe that this model
will qualitatively reproduce the Hartree-Fock excitations, capturing
the separation of scales between the high-energy singlet modes and
the low-energy pure spin modes, and with combinations of the two
filling the region in between.

The selection rules for transport are equally important to the
dynamics of the system.  For orbital excitations, we have no
particular reason to limit possible transitions, and so we allow
\begin{equation}
  \Delta\nS = 0,\pm1,\pm2,\dotsc,\qquad\text{such that $\nS^\mathrm{final}\geq0$.}
\end{equation}
The changes in spin, however, must be supplied by the
spin-$\frac{1}{2}$ electrons passing to and from the antidot.  With
the definition $\nS = S-S_z$, where $S$ is the total spin, we
therefore allow $\Delta S=\pm\frac{1}{2}$ for each event, while
$\Delta S_z$ is determined by the spin-projection ($\uparrow$ or
$\downarrow$) of the electron involved.  Since $\Delta\nS = \Delta
S-\Delta S_z$, this means that
\begin{equation}
  \Delta\nS \in \begin{cases}
    \lbrace 0,-1\rbrace,\quad\text{if $\Delta S_z=+\frac{1}{2}$},\\
    \lbrace 0,+1\rbrace,\quad\text{if $\Delta S_z=-\frac{1}{2}$},\\
  \end{cases}
\end{equation}
where again we require that $\nS^\mathrm{final}\geq0$.

\begin{figure}[tb]
    \centering
    \includegraphics[]{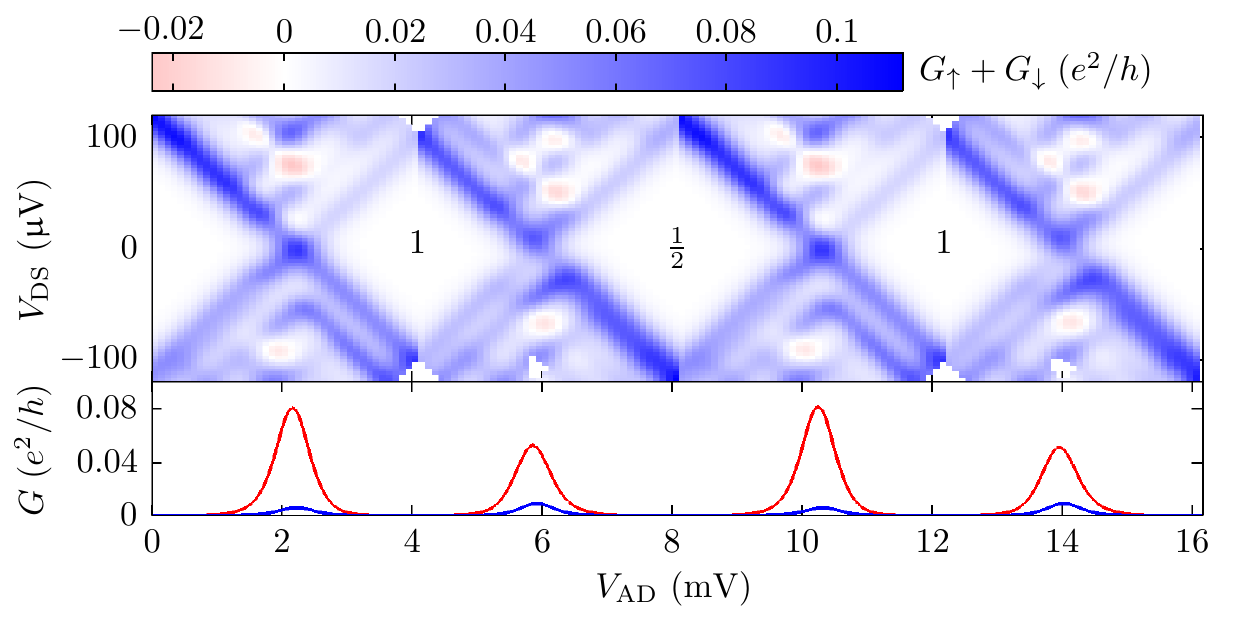}
    \caption[Non-linear transport calculations: effective model]{%
Transport calculations for the effective model of edge excitations,
with $\El = \unit{100}{\mueV}$ and $\Es=\unit{27}{\mueV}$. Total
differential conductance $\Gup+\Gdn$ is shown in the top panel, and
spin-resolved conductances $\Gup$ (red curve) and $\Gdn$ (blue
curve) at zero bias are shown in the lower panel. Numbers printed in
each Coulomb diamond show the ground-state spin configuration.  The
parameters used in this calculation are similar to those in
\figref{fig:EexVetaC}: $\dEsp=\unit{100}{\mueV}$,
$\Ez=\unit{30}{\mueV}$, $\Rad=\unit{400}{\nano\metre}$,
$T=\unit{55}{\milli\kelvin}$, $\gSup=\gDup=\unit{900}{\mega\hertz}$,
$\gSdn=\unit{300}{\mega\hertz}$, $\gDdn=\unit{60}{\mega\hertz}$,
$\etaC=0.14$. \label{fig:NLIVcalcsEffMod}}
\end{figure}

By replacing the fermionic chemical potentials and selection rules
described in \secref{sec:TransportModel} with those of this
effective theory, we can use our transport model to simulate
non-linear measurements and explore the observable consequences of
the theory. An example of the calculated non-linear transport for an
antidot described by the effective model is shown in
\figref{fig:NLIVcalcsEffMod}.  With a very small amount of tuning,
we again find very close agreement with the experiment.  We
therefore conclude that this approximate model of edge excitations
captures the all the qualitatively important aspects of the antidot
physics.   Notice that the excited state lines with energies
$\approx\unit{30}{\mueV}$ observed in the simulation correspond to
\emph{spin} excitations. This choice of energy scales is required to
reproduce the observed combination of excitation lines and negative
differential conductance observed in the measurements, which results
from a combination of spin-dependent tunnel barriers and the strict
selection rules which apply to spin transitions.

\section{Conclusions}

Combining the various results of the modeling described in the
previous sections, we arrive at the following conclusions regarding
the role of interactions in our measurements:
\begin{itemize}
  \item{%
We have shown that the energy scales observed in our experiments are
inconsistent with the SP model.  In particular, the excitation
energies visible in non-linear transport require a relatively large
value for $\dEsp$, which does not agree with either the consistently
small value of $\musf$ or the small observed splitting, $\dmuup$, of
transport lines due to spin pumping.}
  \item{%
The interacting MDD model correctly captures the \emph{spin-charge
separation} of the excitations at the antidot edge, in which both
$\musf$ and $\dmuup$ are softened by exchange while orbital
excitation energies are preserved.  By combining the MDD antidot
model with the dynamic effects of bias and asymmetric tunnel
barriers, we can qualitatively reproduce the experimental
measurements of spin-resolved transport at zero bias, standard
non-linear transport, and non-equilibrium bias measurements.
Calculations of the steady-state spin populations within our model
provide further insight into the interplay between the antidot
energetics and the spin pumping induced by asymmetric tunnel
barriers and/or bias.}
  \item{%
Using an effective model for the edge excitation-spectrum, we have
shown that the particular details of the eigenspectrum are not
crucial for qualitative comparisons, and that our observations are
consistent with the spectrum of spin and density modes predicted by
Hartree-Fock theory for the antidot eigenstates.  Strict spin
selection-rules appear to be a requirement, however, in order to
explain the combination of excitation lines and negative
differential conductance we measure. }
\end{itemize}
We have also presented the details of a method for modeling
spin-resolved transport through a zero-dimensional system such as a
quantum dot or antidot.  It provides the flexibility to directly
compare the predictions of arbitrary physical models of the
dot/antidot with experimental results, incorporating the real-world
importance of spin- and energy-dependent tunnel couplings.
Considering all the subtle and interesting effects we have
discovered in the low-$B$ regime, we believe this combination of
experimental and theoretical techniques presents great potential to
explore spin-related effects in more complicated regimes, such as at
the breakdown of the MDD phase at higher $B$ or for an antidot at
$\nuAD=1$, where the existence of Skyrmions or other non-trivial
ground states is an intriguing possibility.

\chapter{Tilted-Field Measurements\label{chap:TiltedB}}

\ifpdf
    \graphicspath{{Chapter6/Figures/PNG/}{Chapter6/Figures/PDF/}{Chapter6/Figures/}}
\else
    \graphicspath{{Chapter6/Figures/EPS/}{Chapter6/Figures/}}
\fi


In this chapter we present several additional measurements of the
device studied in \chapsref{chap:SpinTransport} and
\ref{chap:SpinTransportModel}.  They are taken in the same low-field
regime of the central antidot, at filling factor $\nuAD=2$, and with
essentially the same experimental setup, but with the addition of a
sample holder offering in situ control of the inclination angle of
the device.  This allows us to independently control the
perpendicular component of the magnetic field, $\Bperp$, which
determines the orbital properties of the electrons in the two
dimensional electron system (2DES), from the total magnitude of the
field, $\Btot$, which to a good approximation affects spin
properties only. For the $\nuAD=2$ lowest Landau level (LLL) antidot
states, this means that we can vary the Zeeman energy, $\Ez$,
separately from the single-particle orbital energy spacing, $\dEsp$.
This extra degree of freedom offers valuable additional information
about the antidot eigenspectrum, providing further confirmation of
the model of low-$B$ antidot physics we develop in
\chapref{chap:SpinTransportModel} in terms of maximum density
droplets (MDDs) in the LLL.  Here we describe the calibration and
operation of the rotating sample holder and the results of
preliminary measurements. Unfortunately a long series of technical
delays meant that the rotator was only functional for our final
measurement run, but we were still able to explore its capabilities
with `proof of principle' experiments, and to gain some additional
insight into the physics of our antidot in the process.

Also in this chapter, we present experimental evidence for
`molecular antidot states' formed in the presence of unintentional
impurities close to the antidot.  We investigated the effects of
these impurities in some detail in the course of our experiments,
mainly in an effort to isolate their contributions to our
measurements, in order to be confident that they did not influence
our conclusions about the physics of the main antidot.  Many of our
measurements of impurity effects are interesting in their own right,
however, and here we provide some examples of fully controllable,
coherent coupling between the main antidot in our device and a small
impurity in one of the side channels.

\section{A coherently-coupled `antidot molecule'\label{sec:ADmolecule}}

Molecular antidot states have been experimentally detected in a few
previous investigations. Gould~et~al.~\cite{Gould1996} designed and
fabricated a double antidot for the purposes of exploring such a
system, and their measurements provided a few surprises with regards
to the frequency of Aharonov-Bohm oscillations.  These were
interpreted through a model of charging-dominated resonance
conditions in which the molecular `spectator modes' only provide a
transport path, without significantly affecting the resonance
structure. While measuring an antidot in the fractional quantum Hall
regime, Maasilta and Goldman \cite{Maasilta2000} noticed `phase
slips' and strange resonance shapes, which they attributed to
coherent quasiparticle tunneling between their intentional antidot
and a nearby impurity. Such phase slips have occasionally been
observed in antidot Aharonov-Bohm oscillations in the integer
quantum Hall regime and it has long been suspected that they result
from the background disorder potential, but as far as we are aware
no one has previously investigated the anomalous resonances as a
function of a second parameter.  In our measurements we usually
prefer to sweep the antidot voltage, $\Vad$, rather than $B$, since
the magnetic field strongly affects equilibration between edge
modes, which can obscure the interpretation of the non-equilibrium
measurements described in \chapref{chap:SpinTransport}. Changing
$\Vad$ necessarily affects the size of the antidot depletion region,
making it more likely to encounter features of the 2DES background
potential, and it could be for this reason that we noticed more
impurity effects in our experiments than most previous studies.
Additionally, our device was fabricated on a wafer with lower
electron density ($n_e=\unit{1.1}{\centi\metre\rpsquared}$) than in
most antidot experiments.  Since disorder tends to be more
significant for lower densities due to the reduced screening ability
of the 2DES, it is possible that we are more likely to find strong
impurities near the antidots in our devices.

\begin{figure}[tp]
    \centering
    \includegraphics[]{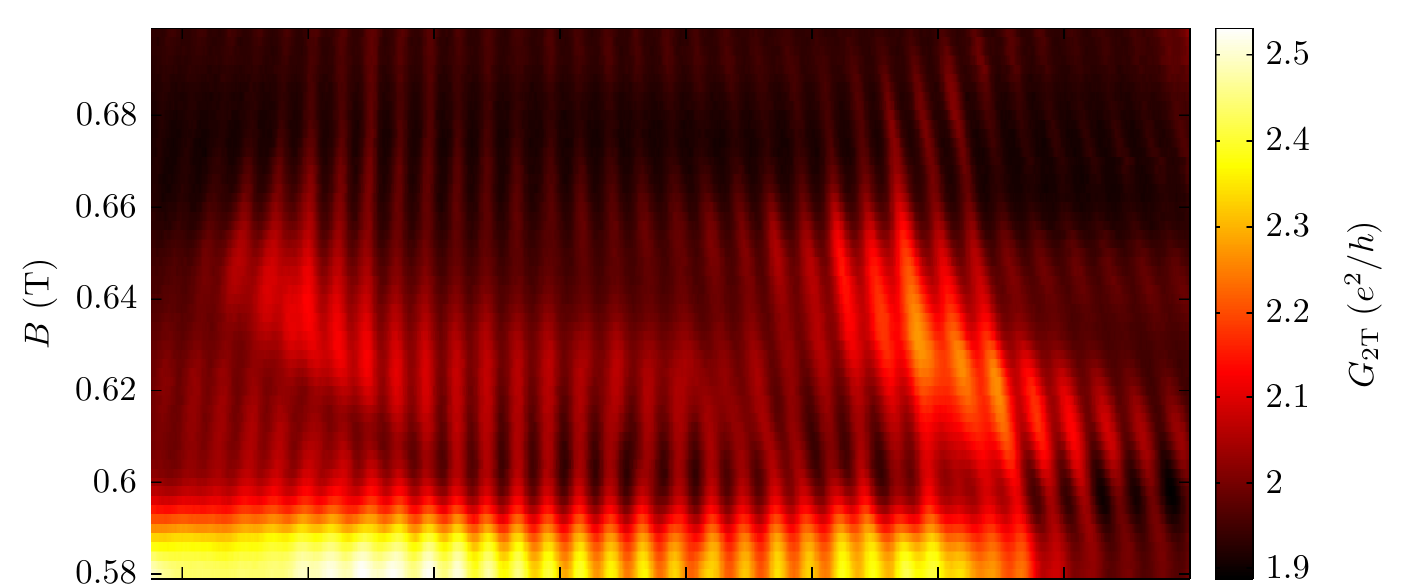}\\
    \includegraphics[]{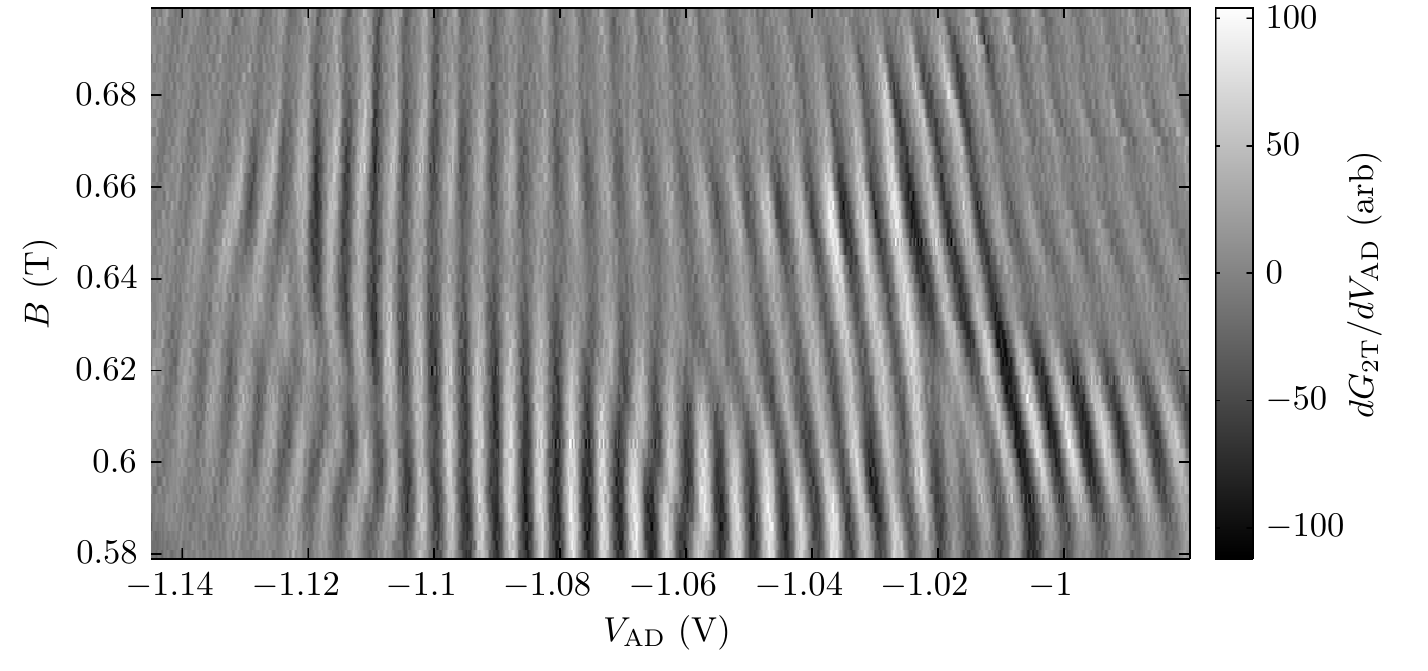}
    \caption[Molecular states in the antidot resonance spectrum]{%
Top --- Two-terminal antidot conductance as a function of $\Vad$
(swept) and $B$ (stepped), in the vicinity of anomalous resonances
due to antidot-impurity coupling.  Horizontal offsets have been
applied to individual sweeps to `straighten' the resonance pattern
and remove an overall slope in the $(\Vad,B)$ plane, to make it
easier to track resonance positions by eye.  Bottom --- Derivative
of the data in the top panel with respect to $\Vad$, in which the
evolution of resonance positions is clearer.  These data were taken
at zero tilt angle, so $B=\Bperp$.
 \label{fig:ImpBvVad}}
\end{figure}

In our $\nuAD=2$ transmission resonances, we initially noticed a few
strangely-shaped resonances in sweeps of $\Vad$ or $B$, which were
often accompanied by abnormal spacings with neighboring conductance
peaks.  Usually the strange line shapes were asymmetric peaks, but
occasionally we observed overlapping but clearly-resolved double
peaks.  Eventually we measured the antidot transmission in the plane
of $\Vad$ and $B$, and found that the strange resonances correspond
to locations where states seem to `disappear' from the antidot
spectrum as $B$ is increased, as shown in \figref{fig:ImpBvVad}.
Broadly, this behaviour seems to make sense, since the increasing
magnetic field causes the width of both the impurity and antidot
states, given approximately by the magnetic length
$\ellB=\sqrt{\hbar/eB}$, to shrink, thereby reducing the spatial
overlap of the wave functions.  Thus an impurity state which was
incorporated into the antidot as a molecular wave function at low
$B$ should eventually decouple as the field is increased.  We know
from basic quantum mechanics that a molecule resulting from the
coupling of two different spatial states should consist of two
molecular wave functions with different eigenenergies; in the
simplest case these are the symmetric and antisymmetric combinations
of the original states.  Our transport measurements only probe the
states which are connected to the antidot, so we should see both
resonances of the molecular state, but will only observe one for the
antidot state when the impurity has fully decoupled.

Despite this qualitative agreement with our expectations, there are
confusing features of the measurements in \figref{fig:ImpBvVad}.
First, it is surprising that the impurity resonances disappear from
the antidot spectrum within such a small range of $B$, especially
given the relatively weak $1/\sqrt{B}$ dependence of the wave
function width on magnetic field. Also, we are surprised that the
impurity only seems to affect a single antidot resonance, with the
others simply shifting their positions slightly to account for the
added or subtracted state.  The $B$-dependence of both charging and
antidot single-particle energies depends inversely on antidot area,
as described in \secref{sec:ABeffect}.  We expect the impurity to be
much smaller than the lithographically-defined antidot with
$\Rad\approx\unit{400}{\nano\metre}$, and therefore for its
eigenenergies to have a much stronger dependence on magnetic
field.\footnote{%
For example, if the impurity background results from the
electrostatic perturbations produced by nearby ionised donors, we
expect the potential to vary on the scale of the spacer layer, which
in this case is \unit{60}{\nano\metre}.  Therefore, if the impurity
has $R_\mathrm{Imp}\approx\unit{100}{\nano\metre}$, the $1/R^2$
dependence of $dU/dB$ means that energies of the impurity states
should vary 10\nbd20 times faster with $B$ than
those of the large antidot.} %
We would also expect a weaker capacitive coupling between the
impurity energies and $\Vad$ than for the main antidot states.
Keeping in mind that our linear response conductance measurements
probe resonances only when they are at the Fermi level, we would
therefore expect the impurity states to show a vastly different
trajectory in the $(\Vad,B)$ plane, and to pass through several
normal resonances while the two wave functions are weakly coupled.
The difference between our expectations as outlined above and the
observations remains unexplained at this time.

\begin{figure}[p]
    \centering
    \includegraphics[]{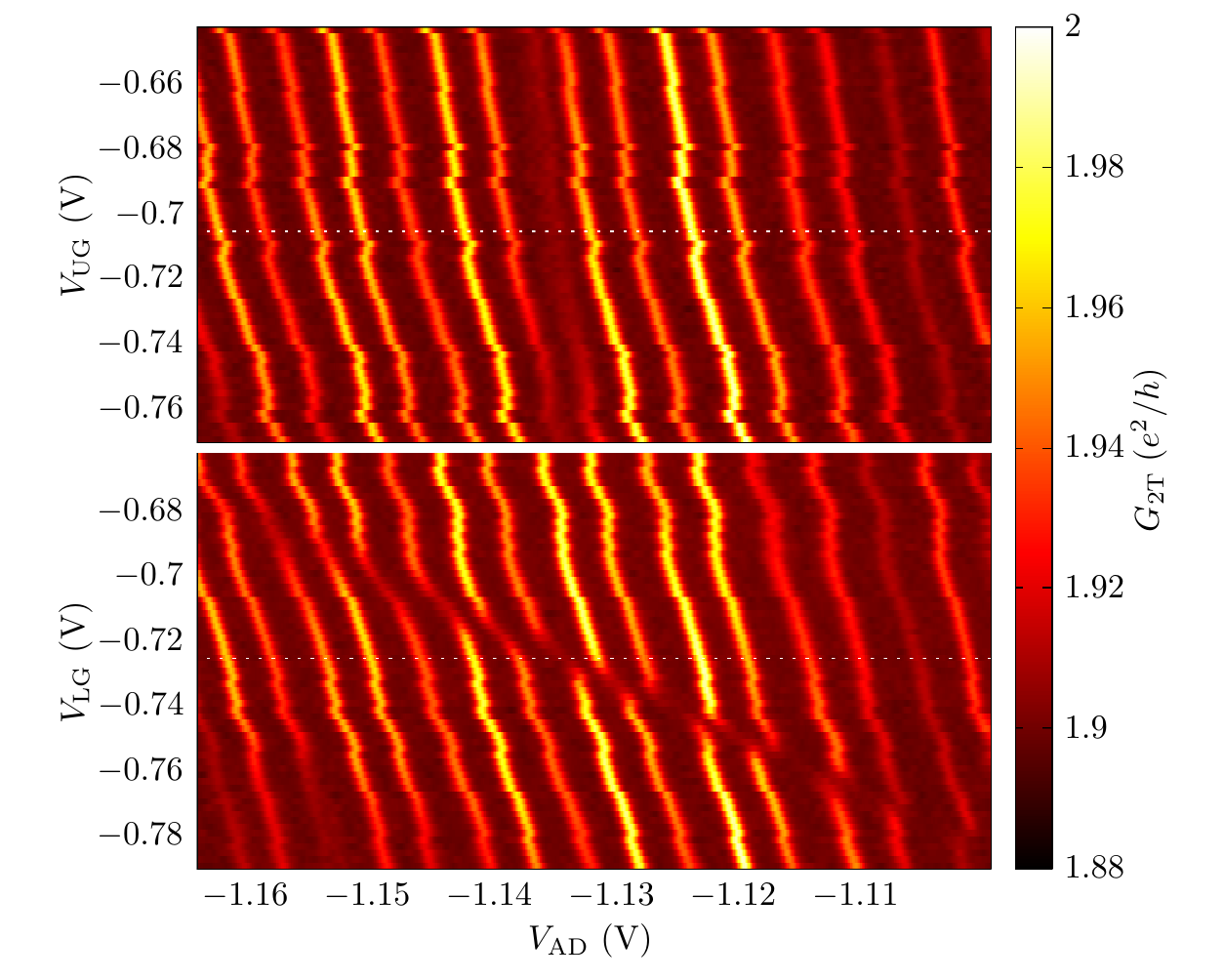}
    \caption[Detection of a nearby impurity]{%
Two terminal antidot conductance as a function of $\Vad$ (swept) and
stepped side-gate voltages $\Vug$ (top) and $\Vlg$ (bottom).  Dotted
white lines show the `intersection' of these two datasets, marking
the fixed value of each side gate voltage which is used for the
opposite set of measurements. \label{fig:SteppingSGs}}
\end{figure}

While the effects of changing magnetic fields on the
antidot-impurity coupling are somewhat confusing, measurements at
constant $B$ are more easily explained. Both the impurity and
antidot states have small capacitive couplings to other gates on the
device in addition to the antidot gate.  In some cases, this
provides a means of \emph{locating} the impurity on the device, at
least approximately. Shown in \figref{fig:SteppingSGs} are two more
`resonance maps,' with two terminal conductance measured as a
function of $\Vad$, with stepped voltages applied in turn to the
nearby side gates $\Vug$ and $\Vlg$ (see \figvref{fig:Device} for a
device photo and gate labels).
As a function of $\Vug$ (upper panel), the resonances shift with a
uniform negative slope, whose magnitude is determined by the ratio
of the antidot states' capacitive couplings to the antidot gate and
upper gate, respectively.  When $\Vlg$ is varied, on the other hand,
as shown in the lower panel, we observe a series of shifts in the
resonances, along a line with much flatter slope in the
$(\Vad,\Vlg)$ plane. Following lines from bottom to top in the lower
panel of \figref{fig:SteppingSGs}, these shifts evolve from
`dislocations' of the resonance position toward more \emph{positive}
$\Vad$ on the right-hand side of the plot to smooth `deformations'
towards more \emph{negative} $\Vad$ on the left-hand side, which
steadily get weaker. This means that at some point a state is added
to the antidot spectrum (i.e., if we pick a `connected' resonance on
the far left and another on the far right and count the number of
lines between them, we find one more along the top edge of the plot
than along the bottom).

We identify the anomalous line in \figref{fig:SteppingSGs} with an
impurity in or near the lower channel, which has a much larger
capacitive coupling to $\Vlg$ than the normal antidot states and a
correspondingly shallower slope.  As we follow it from bottom right
to top left in the figure, the antidot potential is increasing (as
$\Vad$ becomes more negative) to account for the decreasing
potential from $\Vlg$, in order to keep the impurity state at the
Fermi level.  The antidot is getting larger along this line, and so
the antidot-impurity coupling increases.  At the bottom right, the
two wave functions are mostly uncoupled, but the antidot states
shift in response to the discrete charging of the nearby impurity,
which is occupied by an electron above the impurity line and
unoccupied below.  As the antidot becomes larger and the coupling
increases, we start to see clear anticrossings between the states.
In this regime, the right-to-left shifts are produced by two
anticrossings in quick succession, as each antidot state evolves
continuously into an `impurity-like' state following the trajectory
of the impurity line, and then into the neighbouring antidot state.
All of these transitions are measured at the Fermi level, as the
changing gate voltages cause the antidot energy spectrum to `slide
past' the impurity state.

\begin{figure}[p]
    \centering
    \includegraphics[]{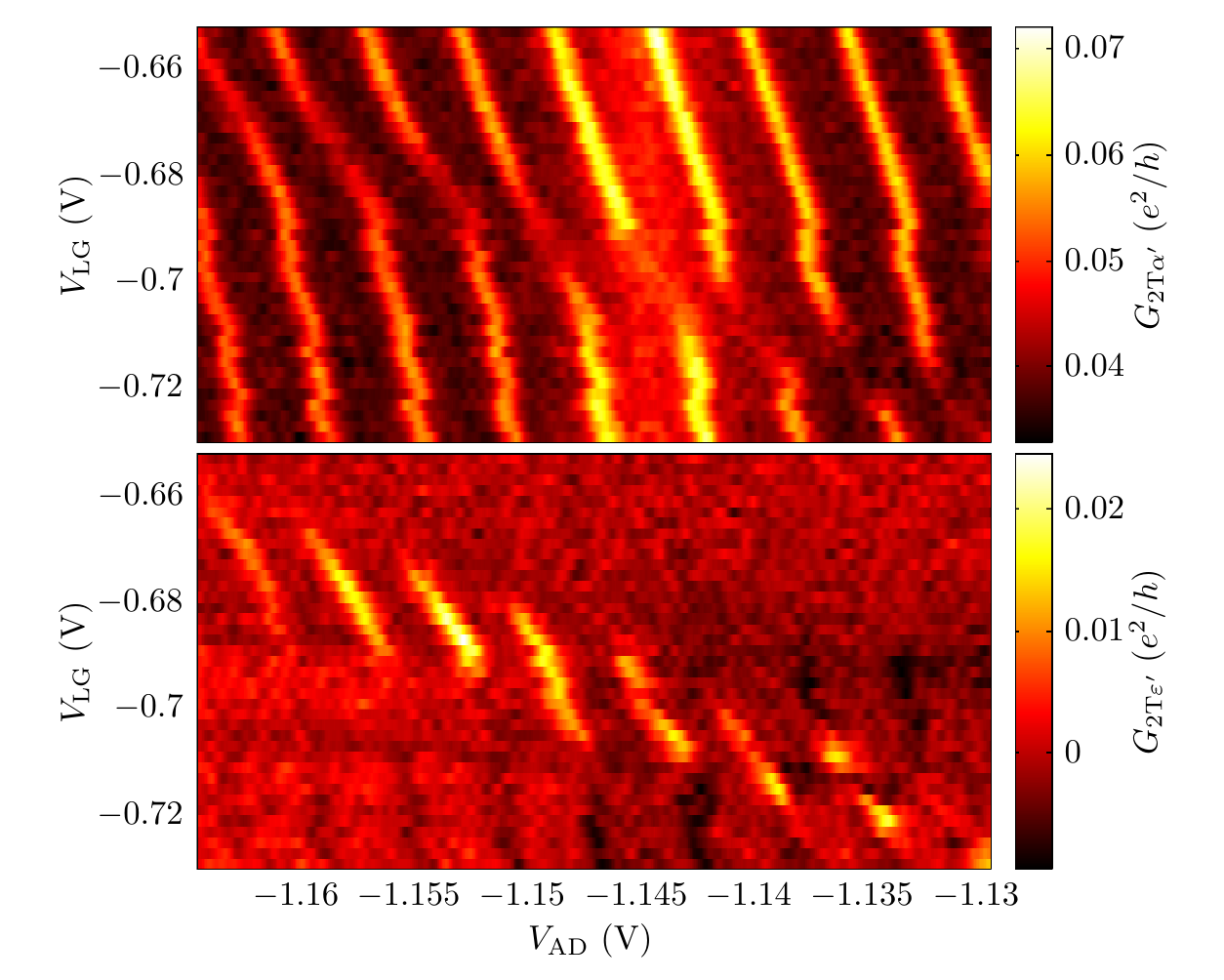}
    \caption[Channel impurity: selective detection measurements]{%
Two terminal conductances reflecting the currents \emph{transmitted}
to the drain contact $\alpha^\prime$ (top) and \emph{reflected} to
the lower-left detector contact $\varepsilon^\prime$ (bottom), as a
function of $\Vad$ and $\Vlg$, with $\fdettwo=2$.  Since
equilibration between edge modes between the antidot and the
detector varies strongly as a function of $\Vlg$, obscuring the
resonance structure, the mean value of each horizontal sweep has
been subtracted in the lower panel. \label{fig:ImpNeqDet}}
\end{figure}

The mode-selective injection/detection technique presented in
\chapref{chap:SpinTransport} provides a nice method to probe the
changing `character' of states in this region.  In the top panel of
\figref{fig:ImpNeqDet} 
we show transmission resonances in the anticrossing regime similar
to those in \figref{fig:SteppingSGs}, while in the bottom panel we
show simultaneous measurements of the \emph{reflected} signal in
edge modes 1 and 2. This is obtained by using one constriction of
the bottom-left antidot (the other is pinched off) as a detector
with filling factor $\fdettwo=2$ and measuring the current flowing
into contact $\varepsilon^\prime$.  Wherever the resonances become
more `impurity-like,' we observe a strong signal in this reflected
current, while the transmitted signal becomes weaker.  The faint
negative signal in the reflected current (dark lines corresponding
to antidot resonance positions) result from significant
$\Vlg$-dependent equilibration between the edge modes traveling from
the antidot to the detector.  Thus we conclude that the impurity
state provides a `link' between the antidot states and the normally
unperturbed LLL edge modes flowing through the lower constriction,
redirecting current from the drain to the bottom-left detector.

From the measurements in \figsref{fig:SteppingSGs} and
\ref{fig:ImpNeqDet}, it seems that the impurity states do not
demonstrate any particular `preference' for coupling to individual
antidot states.  This observation provides further evidence for the
lack of spin-selectivity in the antidot resonances, the impurity
resonances, or in both.  Since the single-particle spacing satisfies
$\dEsp\sim 1/R$, we expect a larger orbital energy spacing for the
impurity than for the antidot, and so these may very likely be
spin-selective.  But if the antidot resonances actually result from
transmission through several states including both spins, we would
expect to see the effects of antidot-impurity coupling on every
resonance line, as observed in the experiment.  Of course it is
possible that something more complicated is taking place to form the
many-body antidot-impurity molecular states, but without
spin-selective antidot states as a reference it is difficult to
extract further information from the measurements.

This concludes our discussion of molecular antidot states, for the
purposes of understanding the observable effects of antidot-impurity
coupling in our device, but there is still scope for much further
investigation.  In particular, the details of the `appearance' of
molecular states in the resonance spectrum as $B$ is reduced remains
to be understood.  We spent some effort in `tracking' individual
impurity resonances like those in \figsref{fig:SteppingSGs} and
\ref{fig:ImpNeqDet} through the parameter space of $(B,\Vad,\Vlg)$,
and our measurements seem to suggest that the splittings observed in
\figref{fig:ImpBvVad} are actually unrelated to this `channel
impurity.' With careful analysis, small dislocations running through
the resonance map of \figref{fig:ImpBvVad} may be observed to
coincide with the expected locations of impurity states from
measurements like \figref{fig:SteppingSGs} (for example, such a
feature is visible in \figref{fig:ImpBvVad} starting from the bottom
around $\Vad\approx\unit{-1.06}{\volt}$ and running towards the top
left). It is therefore possible that these features represent an
entirely different phenomenon, although at this point it seems most
likely that they reflect interactions between the antidot and an
impurity somewhere else in the device.

\section{Operation of an in situ rotation unit}

\begin{figure}[tb]
    \centering
    \includegraphics[bb=0 0 640 480,width=4.267truein]{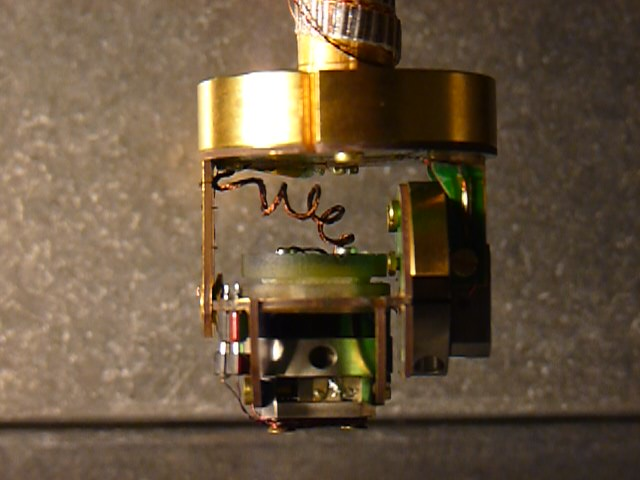}
    \caption[Rotating sample holder]{%
Photograph of the rotating sample holder used for the measurements
described in this chapter.  The sample, mounted in a standard LCC
package, sits on a stage sandwiched between two PCB layers
(horizontal in the photograph).  The twenty pads on the bottom of
the package are pressed against a set of pogo pins mounted in the
stage, which make electrical contact to the bundle of copper wire
seen in the picture, which connects to a loom of twisted pair wiring
in the fridge at the mixing chamber plate.  The sample holder is
shown mounted on the end of a gold-plated copper tailpiece, in the
orientation used for measurements.  Two attocube\textregistered\
piezoelectric rotators provide in situ positioning: one on the right
side of the photograph controls the inclination angle, and a second
mounted directly below the sample controls the azimuthal rotation.
In our measurements, only the inclination rotator was functioning.
\label{fig:RotatorPic}}
\end{figure}

A photograph of our rotating sample holder is shown in
\figref{fig:RotatorPic}.
It was designed and assembled by \emph{Cambridge Magnetic
Refrigeration}, but after several years of delays in fixing flaws
with the original product, we eventually completed the intricate
wiring and final assembly ourselves.  In its final form, it provides
dual-axis in situ angular positioning through a pair of
attocube\textregistered\ ANR50 piezoelectric rotation units. These
operate on a `slip-stick' mechanism, in which the piezoelectric
components (`piezos') are driven by a sawtooth-shaped voltage pulse.
Within the riser of each pulse, the piezos expand or shrink to move
the mechanism, but then snap back to their original position with
the sudden return to ground potential, applying a force which
overwhelms the static friction between the axle and the piezo,
leaving the stage at a new orientation.  Although they are very
fragile to mechanical stress and must be handled with care, the
piezos perform efficiently even at base temperature of a dilution
refrigerator and in high magnetic fields. In practice, the
necessarily large driving voltages lead to enormous pickup in the
experimental circuit, and the resulting currents produce significant
heating of the fridge, so we cannot take measurements while sweeping
the angle; instead we must step to a desired orientation and wait
for the system to settle before taking data. In our experiments, we
only used the inclination rotator, to adjust $\Bperp$ relative to
$\Btot$ (the other rotator was defective), and we achieved suitable
low-temperature operation with a sawtooth wave with amplitude
60\nbd\unit{70}{\volt} (the maximum supplied by the pulse generator
is \unit{70}{\volt}) and frequency \unit{50}{\hertz}.

The angular response to the motor is not uniform or symmetric,
however (i.e., $N$ steps in one direction followed by $N$ steps in
the opposite direction generally does not return the stage to its
original orientation), so we require a separate sensor to determine
the orientation inside the fridge.  Since the sample is already
mounted at the centre of a solenoid magnet, it is convenient to use
Hall effect sensors for this purpose.  These give the component of
$B$ perpendicular to the sensor through a measurement of the Hall
voltage $V_\mathrm{H}$ due to the the (classical) Hall effect,
\begin{equation}
  R_\mathrm{H} = \frac{V_\mathrm{H}}{I} = \frac{B_\perp}{ned},
\end{equation}
where $n$ is the three-dimensional electron density and $d$ is the
thickness of the sensor.  A Hall sensor is mounted in the lid of the
sample holder which holds the chip package in place, with an
orientation perpendicular to that of the sample.  It would have been
ideal to mount two such Hall sensors on the sample holder to gain
dual-axis measurement capability, but unfortunately we are limited
to one due to wiring constraints of the sample holder design.  In
our case, however, we can use the device itself as a second sensor,
since near zero magnetic field the quantum Hall effect is weak and
$R_\mathrm{H}$ is approximately linear in $B$.  By mounting the
sensor perpendicular to the sample, we retain the ability to measure
both rotation angles in most cases.

\begin{figure}[tb]
    \centering
    \includegraphics[]{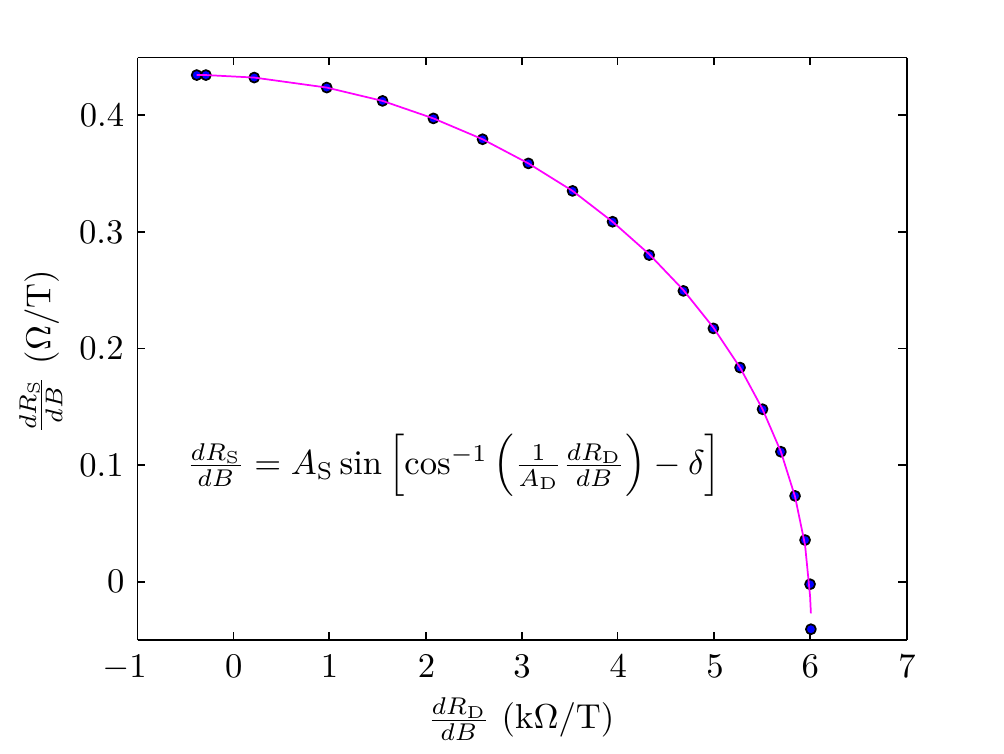}
    \caption[Calibration of the rotating sample holder]{%
Hall-resistance measurements (blue circles) of the the device and
probe-mounted Hall sensor, with the best-fit function (magenta
curve) of the form printed inside the axes.  The parameters
$(A_\mathrm{S},A_\mathrm{H},\delta)$ define the calibration function
for the Hall sensor, \eqnref{eq:HScalibration}, giving the
inclination angle of the device in terms of $\Rs/B$.  The best-fit
values are $A_\mathrm{S}=\unit{0.435}{\ohm\per\tesla}$,
$A_\mathrm{D}=\unit{6.005}{\kilo\ohm\per\tesla}$, and
$\delta=\unit{3.6}{\degree}$. }{{}\label{fig:RotatorCal}}
\end{figure}

Calibration of the Hall sensor to measure the inclination angle in
this orientation is relatively straightforward.  In terms of the
inclination angle, $\theta$, the Hall resistances $\Rs$ and $\Rd$ of
the sample and device, respectively, are given by
\begin{subequations}
\begin{align}
  \Rd & = A_\mathrm{D}B\cos(\theta)\qquad\text{near $\Bperp=0$} \\
  \Rs & = A_\mathrm{S}B\sin(\theta-\delta),
\end{align}
\end{subequations}
in terms of constants $A_\mathrm{S}$ and $A_\mathrm{D}$, and a
possible misalignment angle between the device and the sensor,
$\delta$.  Taking the derivative with respect to $B$ and eliminating
$\theta$ from these equations, we obtain the relation
\begin{equation}\label{eq:CalibrateRs}
  \frac{d\Rs}{dB} =
  A_\mathrm{S}\sin\left[\cos^{-1}\left(\frac{d\Rd}{dB}\frac{1}{A_\mathrm{D}}\right)-\delta\right],
\end{equation}
which we can use to determine the constant parameters
$A_\mathrm{S}$, $A_\mathrm{D}$, and $\delta$.  We step the rotator
through the full \unit{90}{\degree} range, stopping occasionally to
take simultaneous measurements of the slopes of $R_\mathrm{H}$ for
both the sensor and the device.\footnote{%
Note that in order to obtain an accurate value of $R_\mathrm{H}$ it
is preferable to measure the diagonal resistance at both positive
and negative field.  The true Hall resistance is then given by
$[R(B)-R(-B)]/2$, independent of longitudinal effects.} %
An example of these calibration measurements, with a best-fit
function in the form of \eqnref{eq:CalibrateRs}, is given in
\figref{fig:RotatorCal}.
With the calibration parameters determined from such a fit, we can
then determine $\theta$ during our experiment through a measure of
$\Rs$, with the relation
\begin{equation}\label{eq:HScalibration}
  \theta = \sin^{-1}\left(\frac{\Rs}{A_\mathrm{S}B}\right)+\delta.
\end{equation}
Near $\theta\approx\unit{90}{\degree}$, \eqnref{eq:HScalibration}
becomes sensitive to small changes in $\Rs$, and in this regime it
is preferable to use low-field measurements of the device, for which
\begin{equation}
  \theta = \cos^{-1}\left(\frac{\Rd}{A_\mathrm{D}B}\right)
\end{equation}
is better constrained.

\section{Independent control of the Zeeman energy\label{sec:TiltedBexpts}}

There are several potentially interesting applications of
tilted-field measurements, but the most obvious involve the control
of the Zeeman energy, $\Ez=g\mu_\mathrm{B}\Btot$, independently of
$\Bperp$, to separate the effects of spin from the orbital wave
functions.  In order to accomplish this, we step $\theta$ and
$\Btot$ together such that $\Bperp=B\cos(\theta)$ remains constant.
We can measure the effects of changing $\Ez$ through standard
linear-response conductance measurements, since the evolution of
resonance positions reflects changes in the ground-state chemical
potentials of the antidot.

\begin{figure}[p]
    \centering
    \includegraphics[]{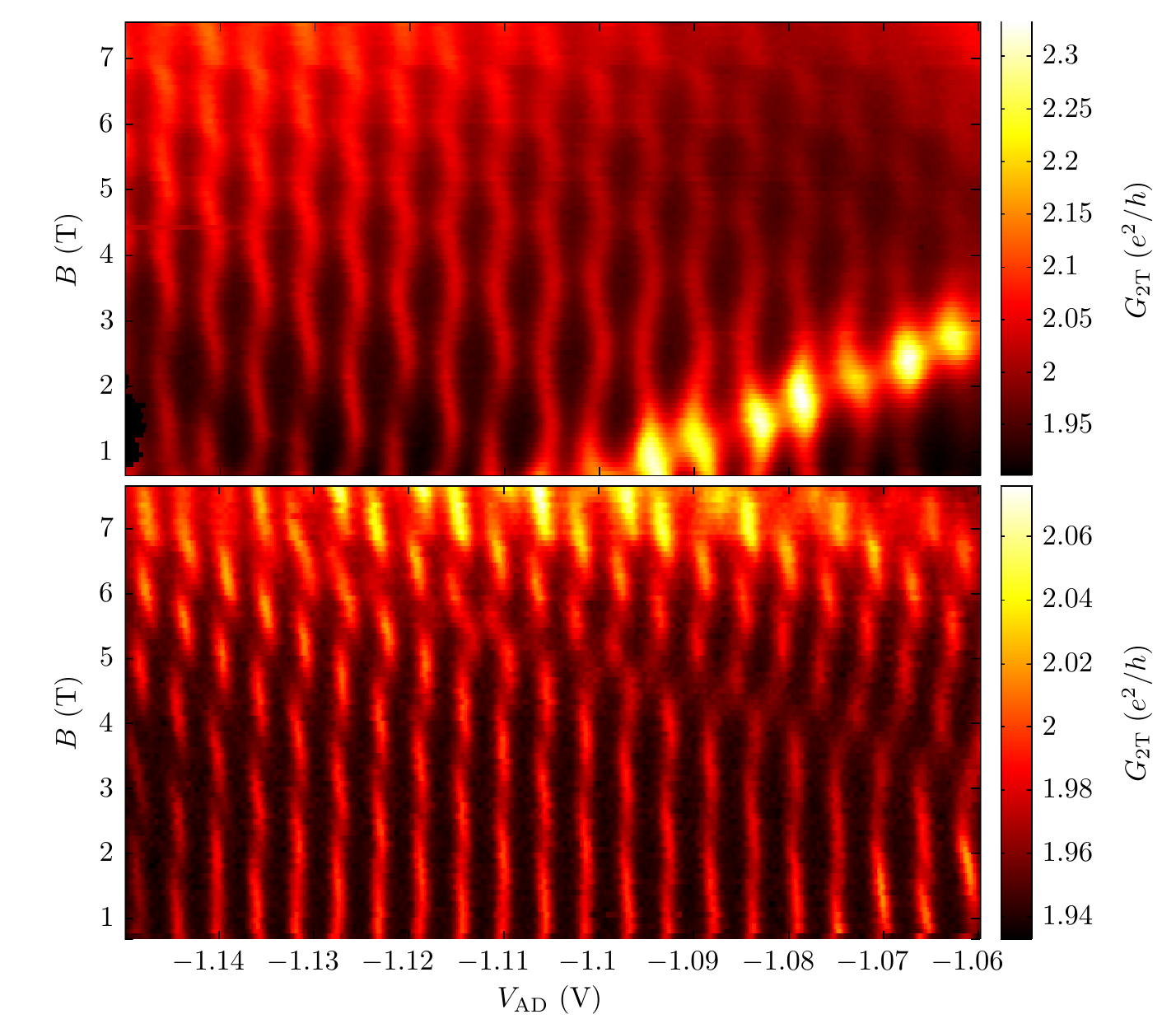}
    \caption[Tilted-field measurements]{%
Antidot transmission resonances as a function of $\Vad$ and $\Btot$,
with $\Bperp$ fixed at \unit{0.62}{\tesla} and \unit{0.72}{\tesla}
in the top and bottom panels, respectively, obtained by stepping the
tilt angle and sweeping $\Vad$. Small horizontal offsets are applied
to individual sweeps to correct for slight mis-calibrations of the
tilt angle. \label{fig:Tiltdp}}
\end{figure}

In \figref{fig:Tiltdp} we show two examples of this measurement, in
which sweeps of $\Vad$ across the $\nuAD=2$ transmission resonances
are taken at a series of angles chosen to produce uniform steps in
$\Btot$, with $\Bperp$ fixed.
For most inclination angles, a measurement of the Hall sensor
resistance is sufficient to determine $\theta$ accurately enough for
these experiments, but when $\theta\geq\unit{80}{\degree}$ the
uncertainty in \eqnref{eq:HScalibration} becomes unacceptably large,
and we must use a measurement of the device to determine the angle,
and hence the appropriate value of $\Btot$ to use at each step. This
can be the device Hall resistance near
$\Bperp\approx\unit{0}{\tesla}$ as explained in the previous section
or another $\Bperp$-dependent feature.  In the measurements shown in
\figref{fig:Tiltdp},  we use the position of a riser in between
quantum Hall plateaux to determine $\theta$ without sweeping the
field all the way to zero between each $\Vad$ trace.  Even so, small
errors in the measurement of $\theta$ for each sweep combined with
device drift tend to obscure the resonance pattern, and so we have
used a series of peak-fitting routines to `line up' the resonances
in \figref{fig:Tiltdp}.

Several features are immediately apparent in these measurements. The
pattern of (anti)crossing resonances is unmistakable; especially in
the upper panel of \figref{fig:Tiltdp}, taken at a lower value of
$\Bperp$, sets of right- and left-moving states can be followed
clearly through the pattern.  This is obviously consistent with the
behaviour we expect from the `ladder-states' of the single-particle
model (see \secref{sec:SPstates} for details), in which a steady
increase of $\Ez$ causes \spinup\ and \spindn\ states to cross, as
the ground-state spin-projection of the occupied states, $S_z$,
increases. In transport measurements the resonances do not actually
cross each other because the are separated by the charging energy
required to add additional electrons to the antidot.  We also
notice, particularly in the lower panel at higher $\Bperp$, that
states moving left have larger amplitude than those moving right.
The left-moving states are traveling \emph{down} the antidot energy
spectrum as $\Ez$ increases, so they correspond to \spinup\
eigenstates, and the difference in amplitude simply reflects the
spin-asymmetric tunnel barriers we discovered in the measurements of
\chapref{chap:SpinTransport}.  As $S_z$ increases with $\Ez$, the
physical `gap' between the inner \spinup\ and outer \spindn\ edges
should increase, which we might expect to enhance the \spindn\
tunneling instead, but we should bear in mind that a similar spatial
separation occurs for the $\nu=3$ and 4 edge modes, and that the
combined effect could be different than we might na\"{i}vely expect.
We can also not fail to notice the `envelope' of large-amplitude
resonances which passes through the upper panel with much shallower
slope than the resonances themselves.  This corresponds to one of
the impurity states discussed in \secref{sec:ADmolecule}, which we
can infer from the positive slope to be \spindn.  The shallow slope
results from the smaller capacitive coupling between the impurity
state and $\Vad$ compared to the main antidot states.  A slightly
more subtle feature occurs in the lower panel above
$B\approx\unit{3}{\tesla}$, in which a new left-moving state seems
to `appear' in the spectrum, causing the nearby resonances to shift
and rearrange around it.  This is presumably a \spinup\ impurity
state, which gets incorporated into the antidot as a molecular state
as the size of the antidot increases (towards more negative $\Vad$).
Finally, we observe that both the amplitude and width of resonances
increases with $\Btot$.  This probably reflects the changes in
tunnel barriers which occurs as both the source and drain edge modes
and the antidot edges rearrange spatially with changing $\Ez$.

While the structure we observe in measurements like
\figref{fig:Tiltdp} seems consistent with the single-particle model
excluding interactions, we note that it is also completely
consistent with the interacting MDD theory described in
\chapref{chap:SpinTransportModel}.  As described in
\secref{sec:MDDmodel}, in the MDD theory the Zeeman energy simply
serves to break the degeneracy in the configuration energy
functional $U(S_z)$, determining the ground-state spin $S_{z0}$,
while the combination of $\dEsp$ and exchange energies determine the
shape of the minimum, and in particular the energy scale for
spin-excitations.  Thus changing $\Ez$ in the MDD model simply
results in successive changes in $S_{z0}$, and with changes in the
ground-state chemical potentials which mimic those of the
single-particle model.  In future experiments, it would be
interesting to investigate the changes in the excitation spectrum
which result from adjustments of $\Ez$, through detailed non-linear
transport spectroscopy measurements.  In our device, such
measurements were not particularly enlightening, mainly because of a
large electron temperature which obscured most of the excitation
lines.\footnote{%
The rotating sample holder is slightly too large to fit in the
standard heat-shield of our dilution refrigerator.  We tried to use
a slightly larger one, but the tolerances are so small that we could
not avoid it touching the \unit{4}{\kelvin} inner vacuum can and
generating a large heat load on the dilution unit.  Running without
the heat shield, we measured electron temperatures of
$\approx\unit{150}{\milli\kelvin}$, increased by a factor of three
over our previous run, which made it impossible to resolve the
excitations we hoped to observe.} %

Besides an investigation of excitations as a function of $\Ez$, it
would clearly also be interesting to apply the spin-selective
measurement technique to these experiments.  In particular, one
could perform measurements of spin-flip events which likely occur
due to spin-orbit coupling near the (anti)crossings of opposite-spin
states, to see if the strength of the coupling depends on the pair
of orbital states involved.

For now, we present only this first glimpse of the range of
experiments possible with a rotating sample holder.  We hope that
further experiments will follow up on some of the unanswered
questions posed above regarding the spin structure of antidots, and
that future researchers will use the rotator with other devices to
add an extra degree of freedom to their experiments.

\chapter*{Conclusions and Suggestions for Further Work\markboth{CONCLUSIONS}{}}
\addcontentsline{toc}{chapter}{Conclusions}

In this work we have studied the physics of the lowest Landau level
($\nuAD=2$) eigenstates of a quantum antidot in the low magnetic
field regime, through a series of experiments and theoretical
models. We carefully consider the non-interacting single-particle
model which has been used to understand low-field antidot
measurements in the past, in an effort to understand the emergence
of effects at higher fields which appear to require an interacting
physical description. We conclude that, while most \emph{orbital}
effects appear to be consistent with non-interacting physics, the
\emph{spin} structure of the low-field antidot is not. Through
extensive measurements of \emph{spin-resolved transport}, which
exploit the properties of edge modes in the quantum Hall regime, we
discover a spectrum of antidot excitations which demonstrates
\emph{spin-charge separation} between the energy scales for spin and
density excitations of the antidot edge.  We interpret these results
in terms of a model of the $\nuAD=2$ antidot as a maximum density
droplet (MDD) of `holes' in the lowest Landau level, and analyse its
expected excitations within Hartree-Fock mean-field theory.  Using a
transport model we have developed to simulate experimental data for
a wide range of theoretical antidot models, we find excellent
agreement between our observations and the predictions of the MDD
picture.  Thus our experiments are of general importance to the
wider community interested in MDDs as they relate to quantum Hall
physics and quantum dots at high magnetic fields.  In several
important ways, our antidot experiments are more versatile and
powerful than previous experimental investigations of MDD
configurations in quantum dots, and we believe there is much
potential remaining for the techniques presented in this work to
address further important questions about the MDD phase and any
other many-particle states which may become important in different
regimes.

In the course of nearly any Ph.D.\ research, unanticipated side
avenues appear, many of which warrant significant further
investigation.  Occasionally these side avenues even develop into
the central thrust of the work, which is exactly what happened in
this case.  We designed the spin-selective experiment, not sure if
the edge modes would actually be spin-selective at such low fields
but fully expecting the antidot states to be.  Instead we quickly
found the opposite, and thus ensued a long endeavour to answer the
question: why?  In the process, we noticed several other intriguing
effects, some of which we investigated and some not.  A few of these
extra measurements were presented in \chapref{chap:TiltedB},
including a few preliminary experiments with the rotating sample
holder which we had expected to form an integral part of most of our
experiments, before it was broken and remained unusable and out of
our hands for three years.  Thus many potentially interesting
experiments remain, a few of which we briefly discuss below.  The
`field' of antidot research is very small, but we have shown with
this work that the potential impact of antidot experiments is
somewhat wider than commonly assumed, being of particular importance
to the much larger group interested in quantum dot physics.  We
therefore hope that some of these projects may soon come to
fruition.

First, the obvious extensions of our central results concerning
spin-resolved transport experiments are to carry out similar
experiments in other regimes of the antidot.  In particular it would
be interesting to investigate the breakdown of the MDD at higher
fields, since the transition between the MDD and a `lower density
droplet' has been the subject of much theoretical work.  Such
measurements would also shed further light on the structure of the
high-field antidot configurations responsible for `double-frequency'
Aharonov-Bohm resonances, which have provoked debate within the
community in the past and are still not fully understood. Also of
interest would be measurements at filling factor $\nuAD=1$, since
the potential exists to detect nontrivial ground states with
`canted' spin order (e.g., Skyrmions) which have been proposed
theoretically.

Furthermore, we believe that potential may still exist for the
`spin-filter' application employing an antidot with spin-selective
resonances, despite our experiments showing the contrary.  Our
analysis of the energy scales implied by our measurements and
comparisons with the interacting model suggest that our device was
very near the `crossover' regime in which the antidot ground state
first starts to develop a spin-polarisation.  We therefore believe
that it should be possible to engineer a new device which moves far
enough towards the non-interacting model to behave in the way
required for the spin-filter application.  Mainly this requires a
larger orbital energy scale $\dEsp$, which could be obtained through
a variety of means, for example by reducing the antidot size,
increasing the electron density (since this may cause the potential
slope at the Fermi energy to be steeper), or by directly increasing
the potential slope induced by the antidot gate by depositing it in
an etched region closer to the 2DES.  We have actually already
fabricated some devices with a few of these modifications, so these
experiments could begin immediately.

Although we did not focus on it due to our preoccupation with
finding descriptions for the `usual' observations of spin-conserved
transport, we have included in this work the first observations of
spin-nonconservation due to spin-orbit coupling in an antidot. Since
antidot states have such a uniform spectrum and are so easily
controlled, there is certainly potential to investigate these
spin-orbit effects further and possibly to develop useful
applications.  One more ambitious idea would be to use the coherent
spin-evolution provided `for free' by the spin-orbit coupling to
manipulate spins controllably in electronic devices based on quantum
Hall topology.  This would require the temporal control of electrons
tunneling to and from the antidot, to reproducibly achieve a desired
spin rotation.  Potential for this control possibly exists in the
form of a `pump,' such as the one proposed \cite{Simon2000} for the
purposes of measuring the charge of fractional quantum Hall
quasiparticles.  Electrons could be controllably added to the
antidot by pulsing one of the side gates with a suitable bias
applied to the edge modes to make it likely for an electron to
tunnel on but not off from the antidot.  After a desired waiting
time for rotation, the other side gate could be pulsed to remove the
electron, and its spin measured through selective detection at a
quantum point contact.  This experiment requires pulse times to be
much faster than the rotation time scale of the spins on the
antidot.  Based on our estimate for the spin-orbit coupling strength
of $\eSOI\approx\unit{3}{\mueV}$, we require pumping frequencies
faster than $\eSOI/h\approx\unit{1}{\giga\hertz}$ which is near the
limit of current experimental capabilities, but it may be possible
to tune the coupling strength to a lower value, either by varying
the Rashba coefficient through a top or bottom gate, or by finding
pairs of antidot states with intrinsically weaker coupling.

As discussed in \secref{sec:ADmolecule}, while several aspects of
the `molecular antidot' measurements seem to have straightforward
explanations, some of the observations remain to be understood.  The
details of the states which `appear' in the antidot resonance
spectrum as the magnetic field is reduced do not seem to agree with
our expectations.  This could simply be a result of a mis-estimation
of the relevant sizes and field-dependence of the impurity states,
and some additional modeling could help to clear up the discrepancy.
The Green's function calculations used in \chapref{chap:Geometry}
would be particularly amenable to this task, since it would be
straightforward to create a simulated potential for an antidot
molecule and investigate what effects may be expected due to
non-interacting physics only.

Finally, it would clearly be desirable to extend the tilted-field
experiments we presented in \chapref{chap:TiltedB} along the lines
discussed at the end of \secref{sec:TiltedBexpts}.  In particular,
we discovered tantalising evidence of strange new behaviour in a few
non-linear spectroscopy experiments at large tilt angle (where $\Ez$
is significantly enhanced over its usual value) taken at the very
end of our measurement run.  The features are faint due to weak
coupling and the elevated electron temperature, but the
Coulomb-blockade pattern seems to be severely distorted, with
splittings of lines appearing at many values of DC bias and no clear
`zero' at which lines merge onto single peaks.  It is almost as if
the Coulomb blockade is completely broken in this regime, although
we have no idea why this should be.  Further measurements of the
evolution of these spectra as $\Ez$ is increased would hopefully
help to answer some of these questions.  Of course there is always
the chance that more measurements will rather pose more questions
than they manage to answer, but such is the fun of exploring quantum
devices with the lens provided by a new experimental technique.

\appendix
\chapter{Wafer Properties\label{app:Wafers}}

The device studied in
\chapsref{chap:SpinTransport}\nbd\ref{chap:TiltedB} was fabricated
on the wafer T792, grown by Ian Farrer of the molecular beam epitaxy
team in the Semiconductor Physics group.  It is essentially a
standard Si-doped GaAs/AlGaAs high-electron-mobility transistor,
with a slightly larger spacer than usual between the two-dimensional
electron gas (2DEG) and the doping layer to further enhance
mobility. Its important properties are included in the table below.
The values marked (light) show the result of heavy illumination with
a red light-emitting diode.  We tried illuminating our device but
found that the gates became ineffective or unstable, so all the
experiments presented in this thesis were performed in the dark.

\vspace{12pt}%
\begin{center}
\begin{tabular}{l|l}
\hline%
\hline%
\multicolumn{2}{c}{T7972 properties} \\%
\hline%
Electron density (dark) & \numprint{1.1e11}~\centi\metre\rpsquared \\%
Electron density (light) & \numprint{1.8e11}~\centi\metre\rpsquared \\%
Mobility (dark) & \numprint{2.56e6}~\centi\metre\per(\volt\usk\second) \\%
Mobility (light) & \numprint{3.51e6}~\centi\metre\per(\volt\usk\second) \\%
Spacer thickness & \unit{60}{\nano\metre} \\%
Total depth of 2DEG & \unit{276.7}{\nano\metre} \\%
\hline%
\end{tabular}
\end{center}

\chapter{Derivation of Tunneling Rates\label{app:TunRates}}

The expression
\begin{equation}
    \label{eq:FGRrate2}
  W^p_{s^\prime\chi^\prime\rightarrow s\chi}\simeq
      \frac{2\pi}{\hbar}\Bigl\lvert\langle \chi s\rvert H_\mathrm{tun}
          \lvert \chi^\prime s^\prime\rangle\Bigr\rvert^2
      \delta(E_s - E_{s^\prime} + E_\chi - E_{\chi^\prime}+p\mu_r),
\end{equation}
is derived directly from Fermi's golden rule for tunneling between
an isolated state on the quantum dot $\ket{\ls}$ and one of a
continuum of states $\ket{k\sigma}$ in the non-interacting reservoir
$r$. In \secref{sec:MasterEqn} we require an expression for the
total rate for the dot transition, given by the trace over all
states in the reservoirs,
\begin{equation}
\label{eq:sumoutleads2}
  \gamma^p_\sptos = \mspace{-18mu} \sum_{\substack{\chi\chi^\prime \\ N(\chi^\prime) =
  N(\chi)+p}}
      \mspace{-18mu} W^p_{s^\prime\chi^\prime\rightarrow s\chi}
      \rho^\mathrm{eq}_\mathrm{res}(\chi^\prime),
\end{equation}
which we proceed to derive in this appendix.

The density of states $\rho^\mathrm{eq}_\mathrm{res}$ of a reservoir
with chemical potential $\mu$ is given by the grand canonical
distribution
\begin{equation}
  \rho^\mathrm{eq}_\mathrm{res}(\chi) =
  \frac{e^{-\beta(E_\chi - \mu \hat{n}(\chi))}}{\mathcal{Z}_\mathrm{F}(\mu,V_r,T)}
   = \frac{e^{-\beta\sum_\alpha(\varepsilon_\alpha - \mu)n_\alpha}}
   {\prod_{\alpha^\prime}(1 + e^{-\beta(\varepsilon_{\alpha^\prime}-\mu)})},
\end{equation}
where $\beta = 1/kT$ and $\mathcal{Z}_\mathrm{F}$ is the partition
function for fermions.  The reservoir configuration $\chi$ is
represented by a set of occupation numbers $n_\alpha$ for the
single-particle states $\ket{k\sigma}$ in the reservoirs, which we
sometimes label by a single index $\alpha$ for notational
simplicity. For tunneling to and from a given lead, the matrix
elements in \eqnref{eq:FGRrate2} are given by
\begin{multline}\label{eq:TunMatEls}
  \bigl\lvert\bra{\chi
  s}H_\mathrm{tun}\ket{\chi^\prime\sp}\bigr\rvert^2
   = \sum_\kls \sum_{k^\prime\ell^\prime\sigma^\prime}
   \Big[
   T_\kls T^\ast_{k^\prime\ell^\prime\sigma^\prime}
   \bra{\chi}a^\dag_{k\sigma}\ket{\chi^\prime}
   \bra{s}a_\ls\ket{\sp}
   \bra{\chi^\prime}a_{k^\prime\sigma^\prime}\ket{\chi}
   \bra{\sp}a^\dag_{\ell^\prime\sigma^\prime}\ket{s} \\
   +
   T^\ast_\kls T_{k^\prime\ell^\prime\sigma^\prime}
   \bra{\chi}a_{k\sigma}\ket{\chi^\prime}
   \bra{s}a^\dag_\ls\ket{\sp}
   \bra{\chi^\prime}a^\dag_{k^\prime\sigma^\prime}\ket{\chi}
   \bra{\sp}a_{\ell^\prime\sigma^\prime}\ket{s}
   \Bigr].
\end{multline}
The first term above represents tunneling of an electron from the
dot to the reservoir ($p=-1$), while the second term represents the
opposite process ($p=+1$), so only one term contributes to
\eqnref{eq:FGRrate2} given a choice of $p$.

To evaluate \eqnref{eq:sumoutleads2}, we first perform the sum over
$\chi$.  For the first term of \eqnref{eq:TunMatEls} this yields
\begin{equation}
    \sum_\chi
    \bra{\chi}a^\dag_{k\sigma}\ket{\chi^\prime}
    \bra{\chi^\prime}a_{k^\prime\sigma^\prime}\ket{\chi}
     = \delta_{0,n^\prime_{k\sigma}}
     \delta_{kk^\prime}
     \delta_{\sigma\sigma^\prime},
\end{equation}
and so, performing also the sums over $k^\prime$ and
$\sigma^\prime$, we obtain
\begin{equation}\label{eq:gminus1}
  \gamma^-_\sptos = \frac{2\pi}{\hbar}
  \sum_{\chi^\prime}\sum_{k\sigma}
  \rho^\mathrm{eq}_\mathrm{res}(\chi^\prime)
  \delta_{0,n^\prime_{k\sigma}}
  \sum_{\ell\ell^\prime}T_\kls T^\ast_{k\ell^\prime\sigma}
  \bra{s}a_\ls\ket{\sp}
  \bra{\sp}a^\dag_{\ell^\prime\sigma}\ket{s}
  \delta(E_s - E_\sp + \varepsilon_{k\sigma}),
\end{equation}
where we have used the relation
\begin{equation}\label{eq:EnConsRelation}
  E_{\chi^\prime} - E_\chi = p(\varepsilon_{k\sigma}+\mu),
\end{equation}
to simplify the energy-conserving $\delta$\nobreakdash-function. We
can now perform the sum over $\chi^\prime$, equivalent to a sum over
all possible configurations $\lbrace n^\prime_\alpha\rbrace$, with
the result
\begin{equation}\label{eq:SumChiPrime}
\begin{split}
  \sum_{\chi^\prime}
  \rho^\mathrm{eq}_\mathrm{res}(\chi^\prime)
  \delta_{0,n^\prime_{k\sigma}}
  & =
      \sum_{\lbrace{n^\prime_\alpha}\rbrace}
      \frac{e^{-\beta\sum_\alpha(\varepsilon_\alpha - \mu)n^\prime_\alpha}}
      {\prod_{\alpha^\prime}\left(1 +
      e^{-\beta(\varepsilon_{\alpha^\prime}-\mu)}\right)}
      \delta_{0,n^\prime_{k\sigma}} \\
  & =
      \frac{\prod_\alpha\left(1 + e^{-\beta(\varepsilon_\alpha - \mu)}(1-\delta_{\alpha,k\sigma})\right)}
      {\prod_{\alpha^\prime}\left(1 +
      e^{-\beta(\varepsilon_{\alpha^\prime}-\mu)}\right)} \\
  & =
      \frac{1}{1+e^{-\beta(\varepsilon_{k\sigma}-\mu)}}
  = 1 - f_\mu(\varepsilon_{k\sigma}).
\end{split}
\end{equation}
Energy conservation requires
\begin{equation}
  \varepsilon_{k\sigma} = p(E_s - E_\sp),
\end{equation}
and so by combining \eqnsref{eq:gminus1} and \eqref{eq:SumChiPrime}
we obtain the desired result for the transition rate as a result of
tunneling from the dot to reservoir $r$,
\begin{equation}
  \gamma^-_{r,\sptos} = \sum_\llps \Gamma^r_\llps(E_\sp - E_s)
        \bra{s}a_\ls\ket{\sp} \bra{\sp}a^\dag_\lps\ket{s}
        \Bigl[1-f_r(E_\sp - E_s)\Bigr],
\end{equation}
in terms of the spectral function $\Gamma^r_\llps(E)$ defined by
\eqnvref{eq:spectralfn}.

For the second term in \eqnref{eq:TunMatEls}, a similar analysis of
the sums over $\chi$, $k^\prime$, and $\sigma^\prime$ yields
\begin{equation}\label{eq:gplus1}
  \gamma^+_\sptos = \frac{2\pi}{\hbar}
  \sum_{\chi^\prime}\sum_{k\sigma}
  \rho^\mathrm{eq}_\mathrm{res}(\chi^\prime)
  \delta_{1,n^\prime_{k\sigma}}
  \sum_{\ell\ell^\prime}T^\ast_\kls T_{k\ell^\prime\sigma}
  \bra{s}a^\dag_\ls\ket{\sp}
  \bra{\sp}a_{\ell^\prime\sigma}\ket{s}
  \delta(E_s - E_\sp - \varepsilon_{k\sigma}),
\end{equation}
and we obtain
\begin{equation}\label{eq:SumChiPrime2}
\begin{split}
  \sum_{\chi^\prime}
  \rho^\mathrm{eq}_\mathrm{res}(\chi^\prime)
  \delta_{1,n^\prime_{k\sigma}}
  & =
      \frac{\prod_\alpha\left((1-\delta_{\alpha,k\sigma}) + e^{-\beta(\varepsilon_\alpha - \mu)}\right)}
      {\prod_{\alpha^\prime}\left(1 +
      e^{-\beta(\varepsilon_{\alpha^\prime}-\mu)}\right)} \\
  & =
      \frac{e^{-\beta(\varepsilon_\alpha - \mu)}}{1+e^{-\beta(\varepsilon_{k\sigma}-\mu)}}
  = f_\mu(\varepsilon_{k\sigma}).
\end{split}
\end{equation}
for the sum over $\chi^\prime$.  Therefore, the transition rate for
an electron tunneling into the dot from reservoir $r$ is given by
\begin{equation}
  \gamma^+_{r,\sptos} = \sum_\llps \Gamma^r_\llps(E_s-E_\sp)
        \bra{s}a^\dag_\ls\ket{\sp} \bra{\sp} a_\lps \ket{s}
        f_r(E_s - E_\sp).
\end{equation}

\renewcommand{\bibname}{References} 
\bibliography{C:/Research/References/SPdatabase} 
\addcontentsline{toc}{chapter}{References} 

\end{document}